\documentclass[12pt, a4paper]{book} % for a long document
\usepackage[utf8]{inputenc} % set input encoding to utf8
\usepackage{amsmath}
\usepackage{amssymb}
\usepackage{graphicx}
\usepackage{float}
\usepackage{color}
\usepackage{breqn}

\usepackage{booktabs} % for much better looking tables

\usepackage{longtable}
\usepackage{spverbatim}

\usepackage[margin=1.00 in]{geometry}

\usepackage{hyperref}
 \hypersetup{
     colorlinks=true,
     linkcolor=blue,
     filecolor=blue,
     citecolor = black,      
     urlcolor=red,
     }

\usepackage{graphicx}
\usepackage{lipsum}

\title{\textbf{Physics of Soil and Other Natural Porous Media}}
\author{Tairone Paiva Leão\\ \small
\href{mailto:tleao@unb.br}{tleao@unb.br} \\  Draft 0.3} 
\begin{document}

\maketitle
 
\tableofcontents % the asterisk means that the contents itself isn't put into the ToC
\listoffigures

%\setcounter{chapter}{10}

% !TEX TS-program = pdflatex
% !TEX encoding = UTF-8 Unicode

% Example of the Memoir class, an alternative to the default LaTeX classes such as article and book, with many added features built into the class itself.

\addcontentsline{toc}{chapter}{Preface}
\chapter*{Preface}

The purpose of this monograph is to review the early theoretical basis of what is known today as soil physics and to serve as a textbook for intermediate level porous media physics or transport in porous media graduate courses. In writing this material we tried to avoid as much as possible the presentation and discussion of empirical methods lacking a rigorous physical basis. Because much of agricultural soil physics is still based on empirical methods, it was not possible to avoid discussing water retention and unsaturated hydraulic conductivity equations, which still lack a rigorous theoretical basis. The fully empirical soil resistance equation is also discussed in a few sections. We also avoided relying on soil physics books opting to consult original sources in fluid mechanics, mathematics, engineering and porous media physics, from which much of soil physics is largely derived. Because of that we understand that ``Physics of Soil and Other Natural Porous Media'' is a more appropriate title than ``Soil Physics'', which imply restriction in methods and applications.   

%We are deeply indebted in writing this manuscript to the early predecessors that laid the foundations for porous media physics and fluid dynamics, namely the mathematicians L. Euler, G.G. Stokes, H. Lamb, G.I. Taylor and G.K. Batchelor, the physicists I. Newton, L.F. Richardson, E. Buckingham, M. Muskat, L.A. Richards, L.D. Landau, E.M. Lifshitz, A.E. Scheidegger, the engineers H. Darcy, J. Bear and his collaborators, G. de Josselin de Jong, J. Philip. We also became aware of a profound contributions to transport phenomena in porous media of a class unknown to me and largely unsung in the soil science community, the chemical engineers, listing here R. Aris, H. Brenner, F.A.L. Dullien, J.C. Slattery, R.B. Bird, W.E. Stewart and E.N. Lightfoot. Not forgotten is the dedication of many  mathematicians, computer scientists and physicist who worked on numerical and computational methods. Of course there are many people with significant contributions to theoretical soil physics and transport phenomena in porous media, scattered in many publications for at least two centuries, too many to mention.       

We plan to gradually introduce newer literature and expand the mathematical foundations in later editions, always aiming at maintaining the physical basis of the processes and phenomena and avoiding purely empirical methods.  

Because of the nature of this material and the amount of mathematics required, many mistakes and misspellings are expected in earlier drafts. Potential users are encouraged to reports mistakes and to provide suggestions and corrections to improve the material. Drafts will be indicated as fractional numbers on the cover, while main releases are indicated as integers.

\chapter{Mathematical preliminaries}
\label{ch1}

\section{Variables, parameters and functions}
A quantity is said to be a scalar if it is not associated with both magnitude and direction. One example of a scalar is the temperature at a given point in space. If you are interested in the temperature at a given point at a given time, the measurement is associated with a single value. In this case, if we use the symbol T to represent temperature and the measured value is in  Kelvin units, it can be represented as
\begin{equation}
T = 376.7 K
\end{equation}  
If you record several instances the temperature in space and/or time you have a set of temperatures and it can now be treated as a variable.  
\begin{equation}
T = \left\{376.7 , 375.6 , 370.7 , 369.9  , ..., N \right\} 
\end{equation}  
where $N$ is the $n^{th}$ temperature measurement. In general you would want to associate the variable of interest with a system of spacial coordinates and/or time measurements. Imagine that your thermometer is placed in a fixed position on a weather station and records temperature at regular intervals throughout the year. Each temperature measurement is now associated with a time measurement, usually the date and the time in hours, minutes and seconds. A table of such measurements would look like Table \ref{ch1_table1}.

\begin{table}[h]
\centering
\caption{Temperature measurements recorded at different time intervals.}
\begin{center}
\begin{tabular}{lllllll}
\hline
Year & Month & Day & Hours & Minutes & Seconds & Temperature  (ºC) \\ \hline
2020 & 10    & 24  & 10    & 05      & 50      & 25.8              \\ 
2020 & 10    & 24  & 10    & 06      & 00      & 25.9              \\ 
2020 & 10    & 24  & 11    & 06      & 10      & 25.9              \\ 
2020 & 10    & 24  & 11    & 06      & 20      & 26.0                \\ 
...  &       &     &       &         &         &                   \\ \hline
\end{tabular}
\end{center}
\label{ch1_table1}
\end{table}

Now we can assign the temperature as a function of time at each point in space. In reality, temperature is not a function of time, nothing really is a function of time, temperature varies because the position of the sun in the sky and its influence on incident radiation, season of the year, cloud cover, convective air flow and many other factors, but sometimes it is easier for us to model it simply as a function of time.  Temperature as a function of time can be represented as $T = f(t)$ or simply as $T(t)$. One class of functions which can be used to model temperature is

\begin{equation}
T(t) = T_{0} + \alpha \cos{(\omega t + \phi)} 
\end{equation}  

\begin{figure}[ht]
\centering
 \includegraphics[width=0.75\textwidth]{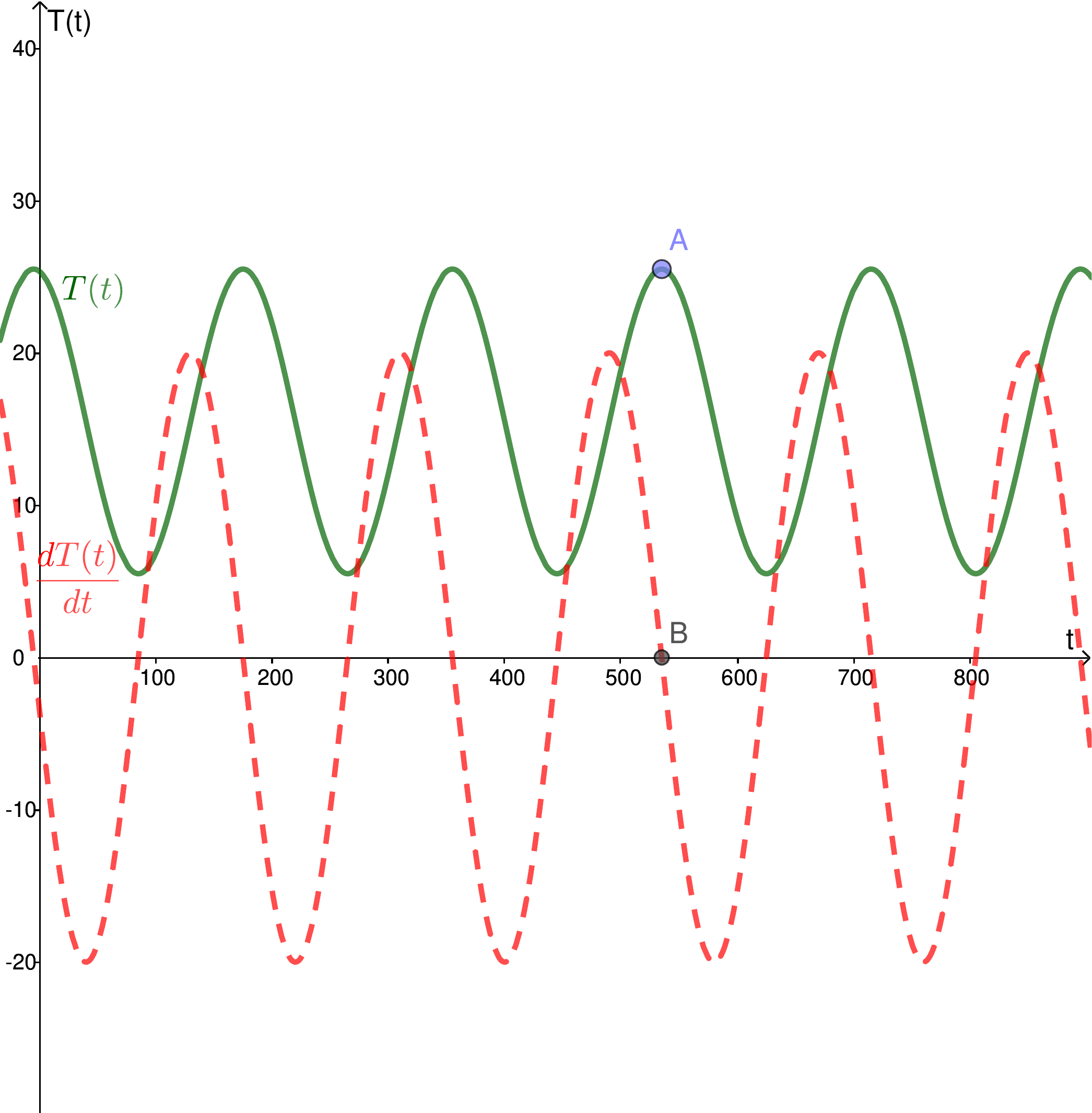}
\caption{Example of a sinusoidal function of time with parameters $T_{0}$ = 15.5, $\alpha$ = 10, $\omega$ = 2 and $\phi$ = 10.}
\label{ch1_fig1}
\end{figure}

\noindent
In this equation temperature and time are variables since each represents a set of values, while $T_{0}$, $\alpha$, $\omega$ and $\phi$ are parameters which can be obtained by fitting the equation to data. Fitting can be achieved using statistical techniques such as least-squares regression or could also be done visually for such a simplistic equation, as long as one is aware of the potential limitations of doing so. The values of the parameters obtained by fitting the equation to observed data are usually refer to fitting parameters or empirical parameters. In this case each has a specific mathematical meaning. Parameter $\alpha$ is the amplitude of oscillation of the sinusoidal wave produced, $T_{0}$ is the value in which the predictions are centered, $\omega$ is the angular frequency and $ \phi $ is the phase lag. Equation 1.3 is said to be univariate as temperature $T$ depends on time alone. Notice that the parameters determine the shape of the function but $T$ is not a function of the parameters since they do not vary. $T$ is a functions of time $t$ alone in this case (Figure \ref{ch1_fig1}).   

Given any parameterization of Equation 1.3, one might be interested in the rate of variation of temperature as a function of time. The rate of variation of a given variable as a function of the same independent variable can be calculated using the first derivative of the function. There are many symbolic representations for the derivative of a function. For example, given a generic function $y = f(x) $ its first derivative can be represented using Lagrange's notation as  $y'(x)$ and by Leibniz's notation as  $dy/dx$.  Newton's \emph{dot} notation, common in classical mechanics is usually reserved for position as a function of time,  $x(t)$ for example, such that its velocity is  $v = \dot{x}(t) $  and its acceleration is   $ a = \dot{v}(t)= \ddot{x}(t) $.  In classical mechanics $x$, $v$ and $a$ are usually vector quantities which were not denoted as such here because we haven't talked about vectors yet. We shall use Leibniz's notation in this book and as such the first derivative of Equation 1.3 is  

\begin{equation}
\frac{ dT(t)}{dt} =-A \omega \sin{(\omega t + \phi)} 
\end{equation}  
 
 \noindent
If one is interested on the rate of variarion of the rate of variation of the temperature as a function of time, what is needed is the second derivative of the function, represented using the same notation as 

\begin{equation}
\frac{ dT^{2}(t)}{dt^{2}} =-A \omega^{2} \cos{(\omega t + \phi)} 
\end{equation}  
and so forth. For now you don't have to worry about the rules of differentiation, we will get to that later, what you should understand first is the geometrical meaning of the derivative. Observe the graph of the first derivative of Figure \ref{ch1_fig1}.   

The dashed line in Figure \ref{ch1_fig1} is the rate of change of temperature as a function of time. Notice that at the maximum and minimum values of the original function (Point A), the value of the derivative is zero (Point B) while at the inflection points, where the curvature of the first graphs changes from upwards to downwards and vice-versa, the rate of change, or the derivative function, is at its absolute maximum values.  

Let's now analyze a simple example from classical mechanics. Suppose an object is relased from rest from a height of 10 meters. Following Newton's laws the governing equation is  
 
\begin{equation}
Y(t) = Y_{0} + v_{0} t - \frac{1}{2}g t^{2} 
\end{equation}  
 
The only variables in this equation are $Y$ and $t$, everything else is either a constant or can assumed to be constant if the height of fall is small compared to the radius of Earth, as is the case with gravitational acceleration. As before, we are treating gravitation acceleration as a scalar, since we are assuming it as a constant and are only interested in its effect in the $Y$ direction. $Y_{0}$ is the initial height from which the object is released, $v_{0}$ is the initial velocity, Y is the predicted height and t is time. If the initial height is $Y_{0} = 10 \ m $ and if the object is released from rest, such that $v_{0} = 0 $, we have    

\begin{equation}
Y(t) = 10 - \frac{1}{2}g t^{2} 
\end{equation}  

\noindent
Notice that the height $Y$ decreases as a function of time, as it is expected, since the object is falling on Earth's gravitational field. At one point $Y$ will reach zero, which is when the object reaches the surface. It is also important to notice that the equation \textit{predicts} negative heights. This means only that the object has passed a surface that we assumed as a reference point, such that all gravitational potential energy considerations are made with reference to that point, if we were interested in that. It turns out we aren't right now but that will be fundamental when we study the energy of water in porous media. If we plot $Y(t)$ as a function of time you will see that position $Y$ decreases quadratically with time, since time is raised to the power of two in this equation. $Y$ is a polynomial function of time in this case. Extracting the roots of such polynomials can provide important information in physics problems, as we shall discuss later.   For now suppose you want the velocity as a function of time. We know that the velocity is the rate of variation of distance over time and this can be expressed in derivative form as

\begin{equation}
\frac{dY(t)}{dt} = v(t) = - g t 
\end{equation}  
The velocity is a linear function of time, meaning that it increases at a constant rate as the object falls. As you might expect the rate with which the velocity increases is the gravitational acceleration of Earth, $g$. To further make that clear let us calculate the rate of change of velocity with time, or the acceleration

\begin{equation}
\frac{dY^{2}(t)}{dt^{2}} = \frac{dv(t)}{dt} = a = - g 
\end{equation}  
With that, one can be convinced that an object falling under Earth's gravitational field falls with an acceleration equal to Earth's gravitational acceleration. 

Suppose now that all you known is that an object is falling under a gravitational field equal to $g$ and you have no equation of motion. From Equation 1.9, we know that the rate of variation of velocity of the object is   
\begin{equation}
\frac{dv(t)}{dt} =  - g 
\end{equation}  
As before we also know that the object was released from a height of 10 m with initial velocity equal to zero. We can rearrange Equation 1.10 by multiplying both sides of the equality by $dt$  
\begin{equation}
dv(t) =  - g dt
\end{equation}  
This is one of the simplest possible cases of what is called a \emph{differential equation} and to remove the differential from both sides all we need to do is integrate it
\begin{equation}
\int dv(t) =  \int - g dt = - g  \int dt
\end{equation}  
Since g is a constant. As this is an indefinite integral, the solution includes a constant of integration $C_{1}$.  
\begin{equation}
v(t) =  - g t + C_{1}
\end{equation}  
Constants will routinely appear in the solutions of differential equations in physical problems. To find the value of the constant one needs specific conditions of the problem under evaluation. These can be called \emph{initial conditions} or, in many applications, \emph{boundary conditions}. We know here that when the velocity is zero, the time is also zero, in other words when the experiment was set, the very moment when the object was released from rest it was at a height of 10 m. Therefore, our initial conditions are 
\begin{equation}
v = 0,\   t = 0
\end{equation} 
Substitution in Equation 1.13 results 
\begin{equation}
0 =  - g 0 + C_{1} \Rightarrow C_{1} = 0
\end{equation}  
Now from Equations 1.8, 1.13 and 1.15
\begin{equation}
dY(t) =  v(t)dt = - g t dt
\end{equation}  
By repeating the integration procedure we arrive at 
\begin{equation}
Y(t) =  - \frac{1}{2}g t^{2} + C_{2}
\end{equation}  
Since we know that the height is 10 m when time is zero we arrive at
\begin{equation}
Y(t) =  - \frac{1}{2}g t^{2} + 10 = 10 - \frac{1}{2}g t^{2}
\end{equation}  
Which is the exact same equation we started with (Equation 1.7).

The objective of the steps from Equations 1.6 to 1.18 were twofold, first to show the relationship between derivation and integration and how such operations can appear on the solutions of physical problems, and second, mainly on the steps covered from Equations 1.10 to 1.18, to show the logic behind what most students outside the fields of physics and mathematics understand by \emph{mathematical modeling}. In a broad sense, mathematical modeling is viewed as the solution of a differential equation applied to a set of initial or boundary conditions. As physicists and mathematicians know well, simple solutions to these problems are rare if they exist at all. Nevertheless, the strategy used here can be applied to simplifications of mechanics and dynamics problems, such as uniform flux through an isotropic medium with constant boundary conditions.

Derivation and integration are at the core of physics and applied mathematics. These techniques were developed along with the disciplines of physics and mathematics during the ``Scientific Revolution"  on around the years 1600-1700s to solve real problems in mechanics by the likes of Newton and Lebniz, followed by improvements and innovations by mathematicians such as Laplace, Lagrange and Gauss. The discipline of fluid dynamics was developed shortly after, based on the theoretical insights provided by Isaac Newton. Leonard Euler, contemporary of Laplace and Lagrange laid the foundation of much of what is now known as fluid dynamics, followed by fundamental developments by  George G. Stokes and Claude-Louis Navier which fathered the equation at the heart of fluid mechanics, the \emph{Navier-Stokes equations}.

% What we have studied in the last examples are univariate functions. An univariate function is one where the  

\section{Differentiation and integration}
Now that we have an idea how differentiation and integration can be used to solve physical problems, lets look at some of the basic rules and symbols associated with these techniques. For a function $f(x)$ the following basic rules of differentiation apply.
 
The derivative of a constant is zero
\begin{equation}
\frac{d}{dx}c = 0
\end{equation} 
In other words, there can be no rate of variation of a constant function $f(x) = c$,  the value remains \emph{constant} with time and is equal to c. The most basic rule of differentiation can be written as  

\begin{equation}
\frac{d}{dx} x^{n} = nx^{n-1}
\end{equation} 
For any $ n \in \mathbb{R} $.   A constant multiplied by any function can be written out of the differentiation
\begin{equation}
\frac{d}{dx} cf(x) = c \frac{d}{dx} f(x)  
\end{equation} 
For  $ c $ constant. If we have two functions of $x$, say $u(x)$ and $v(x)$ the following sum, product and quotient rules apply. 
\begin{equation}
\frac{d}{dx} (u + v) = \frac{du}{dx} + \frac{dv}{dx}     
\end{equation} 
\begin{equation}
\frac{d}{dx} (uv) =   u\frac{dv}{dx} + v\frac{du}{dx} 
\end{equation} 
\begin{equation}
\frac{d}{dx} (\frac{u}{v}) =     \frac{v\frac{du}{dx} - u\frac{dv}{dx}}{v^{2}}
\end{equation} 
A few examples are

\begin{equation}
\frac{d}{dx}2 = 0
\end{equation} 
\begin{equation}
\frac{d}{dx} x^{3} = 3x^{3-1} = 3x^{2} 
\end{equation} 
\begin{equation}
\frac{d}{dx} 2f = 2 \frac{d}{dx}x^{3} = 6x^{2}  
\end{equation} 
\begin{equation}
\frac{d}{dx} (2x^{3} + 3x^{2}) = 6x^{2} + 6x     
\end{equation} 
\begin{equation}
\frac{d}{dx} (x^{4}x^{3}) = x^{4}3x^{2} + 4x^{3}x^{3} = 3x^{6} + 4x^{6} = x^{4}x^{3}     
\end{equation} 
It  should be obvious that $x^{4}x^{3}$ could have been written as $x^{7}$ and derived as such, but we wish to limit our examples to the derivatives of univariate functions. We will use a similar approach below for simplicity.  
\begin{equation}
\frac{d}{dx} (\frac{x^{4}}{x^{3}}) =  \frac{x^{3}4x^{3}-x^{4}3x^{2}}{(x^{3})^{2}} = \frac{x^{6}}{x^{6}} = 1
\end{equation} 
Other simple derivatives that might not directly follow these rules are 
\begin{equation}
\frac{d}{dx} \sin{x}  = \cos{x} 
\end{equation}    
\begin{equation}
 \frac{d}{dx} \cos{x}  = -\sin{x}
\end{equation}    
\begin{equation}
 \frac{d}{dx} \tan{x}  =  sec^{2} x
\end{equation}
\begin{equation}
 \frac{d}{dx} \exp{x}  = \exp{x} 
\end{equation}
\begin{equation}
 \frac{d}{dx} \log{x}   = \frac{1}{x}
\end{equation}    
\begin{equation}
\frac{d}{dx} \frac{1}{x}  = -\frac{1}{x^{2}}
\end{equation}

Our objective here is not to teach elementary calculus, as you should have been exposed to that in intro math courses, but to review the symbols that should be necessary in later chapters. The relationship between integration and differentiation is summarized by the \emph{fundamental theorem of calculus}
\begin{equation}
F(x) = \int_{a}^{x} f(t)dt 
\end{equation} 
\begin{equation}
\frac{d}{dx}F(x) = \frac{d}{dx} \int_{a}^{x} f(t)dt  = f(x) 
\end{equation} 
and
\begin{equation}
\int_{a}^{b} f(x)dx = F(a) - F(b)
\end{equation} 
For $f(x)$ continuous on the interval $[a,b]$ and $F(x)$ continuous on the interval $[a,b]$ and differentiable in $(a,b)$. $F(x)$ in this case is called the primitive of $f(x)$. In other words $F(x)$ is any function that when differentiated results in $f(x)$. For our purposes, what you have to take from the theorem is that, for a function $f(x)$ which follows the conditions stated above, its integral is $F(x)$, which can be reverted back to $f(x)$ by differentiation. We have demonstrated an example of this process in Equations 1.6 to 1.18, but bear in mind that this discussion is as far as possible from a mathematical proof, which we will not provide here. The interested reader can find more about proofs in mathematical books on the field of \emph{analysis}\footnote{See for example the classics \emph{Undergraduate Analysis} by Serge Lang and \emph{Principles of Mathematical Analysis} by W. Rudin}.        
   
Now, intuitively, integration can be understood as the ``inverse path" of differentiation, so if the derivative of any function $x^{n}$ is given by 
\begin{equation}
\frac{d}{dx} x^{n} = nx^{n-1}
\end{equation} 
for $n \ne -1$, we can retrieve the original function by adding 1 on the exponent and dividing the derivative by $n$ such that    

\begin{equation}
 \frac{nx^{n-1+1}}{n-1+1} = x^{n}
\end{equation}     
and thus, at least operationally, the integral of a function $x^{n}$ can be computed as    
    
\begin{equation}
 \int x^{n} dx = \frac{x^{n+1}}{n+1} + C
\end{equation}     
Where $C$ is an integration constant and $dx$ indicates that the integration is performed on the variable $x$. The integral in equation (1.42) is said to be unbounded because there are no limits of integration. By rule (1.19) any constant when differentiated will be equal to zero and in this case if we move from right to left in equation (1.42) by differentiation is not difficult to see that the original function will result if differentiation is carried out. Rarely in physics and applied mathematics we will encounter such simple functions. In many cases the functions to be integrated are much more complicated and often, especially when dealing with differential equations, might not have an analytical or closed form solution. In water and other transport phenomena such equations are the rule, and they are solved using a different approach that uses approximations and computational methods. These numerical solutions are at the core of not only fluid dynamics in the form of computational fluid dynamics (CFD), but  are also an essential component of modern soil physics, hydrology and hydrogeology. 

For simple univariate integrals the rules and techniques below apply

\begin{equation}
 \int x^{n}dx = \frac{x^{n+1}}{n+1} + C
\end{equation}     
\begin{equation}
 \int x^{2}dx = \frac{x^{3}}{3} + C
\end{equation}    
\begin{equation}
 \int dx = x + C
\end{equation}
\begin{equation}
 \int x^{3}dx = \frac{x^{4}}{4} + C
\end{equation}    
\begin{equation}
 \int \sin{x} \ dx = -\cos{x} + C 
\end{equation}    
\begin{equation}
 \int \cos{x} \  dx = sin{x} + C
\end{equation}    
\begin{equation}
 \int \tan{x} \ dx = -\log{\cos{x}} + C 
\end{equation}
\begin{equation}
 \int e^{x} \ dx = e^{x} + C 
\end{equation}
\begin{equation}
 \int \log{x}  \ dx = x \log{x} - x + C
\end{equation}    
\begin{equation}
 \int \frac{1}{x}  \ dx = \log{x} + C
\end{equation}    

In introductory calculus books you learn about integration as a Riemann sum and as the area under a curve
\begin{equation}
\lim_{n\to\infty} \sum_{k=1}^{n} f(c_{k}) \Delta{x} = J = \int_{a}^{b} f(x) dx
\end{equation}    
 
In this equation, when the length of the intervals $\Delta{x}$ goes to zero and the number of $k$ intervals goes to infinite, the sum can be interpreted as an integral over the entire interval from $a$ to $b$ and the result is the area under a curve, $J$.  We will not discuss this right now but it is important that you get familiarized with sums and their role in calculus, because they make frequent appearance in properties of granular materials and on the solutions of differential equations in advanced porous media physics.  
    
\begin{center}
\begin{figure}[ht]
 \includegraphics[width=0.75\textwidth]{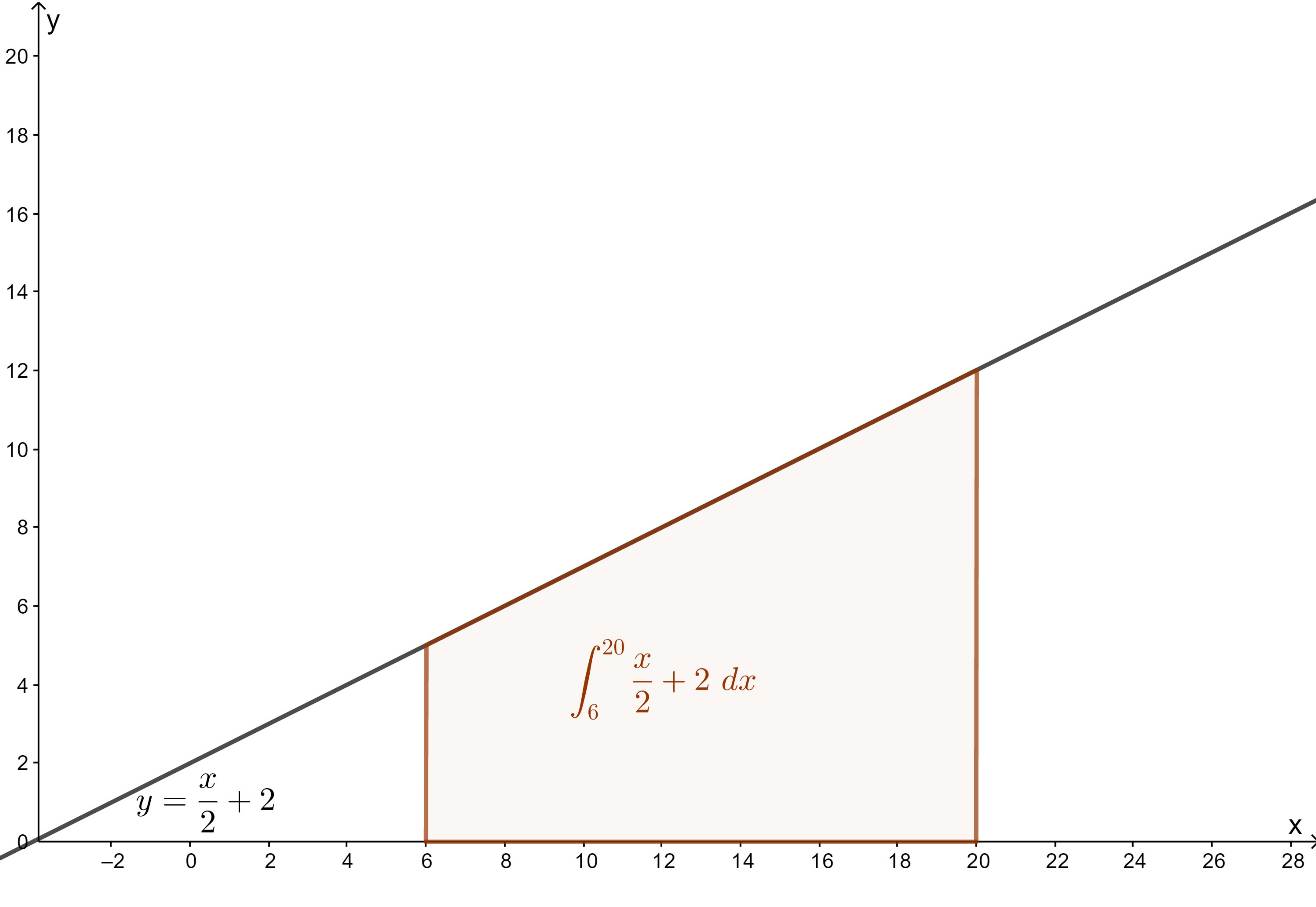}
  \caption{Graph of the function $f(x) = y = \frac{x}{2} + 2 $ .}
  \label{ch1_fig2}
\end{figure}
\end{center}

Consider a very simple example of a line described by Figure \ref{ch1_fig2}
\begin{equation}
f(x) = y = \frac{x}{2} + 2
\end{equation}    

If you wish to calculate the area of the curve on the interval from $x = 6$ to $x = 20$, you could simply calculate the area of the trapeze bounded on its upper end by the line and in its lower end by the x axis and at its left and right limits by the lines $x = 6$ and $x = 20$ resulting in 504. Notice that you would get the same result if you divide the interval from 6 to 20 in 14 equal intervals of length 1 in $x$, calculate the area of each and sum the areas. This is exactly what it is done by the left hand side of equation (1.53). Now by integration, the area under the line is given by Figure \ref{ch1_fig2}

\begin{equation}
J = \int_{6}^{20} 2x + 10 dx = [x^{2} + 10x]\Biggr|_{6}^{20} =  [20^{2} + 10\times 20] - [6^{2} + 10 \times 6] = 504 
\end{equation}    
    
It is obvious from the previous discussion that you don't need integration to calculate the area under a straight line, but it is also clear that you cannot use the formulas created for regular geometric figures for calculating the area under curves and this is the appeal of integration. You should be familiarized with integration techniques from your intro math courses such as substitution, integration by parts and partial fractions.\footnote{If you haven't had any college level math you are urged to take at least a semester to study it before any serious endeavor in soil (and porous media) physics. True soil physics requires vector calculus and mathematical methods of physics, and whoever tells you  - ``you don't need these subjects to understand soil physics" - does not know what soil physics is} When integration is performed in a bounded interval such as in equation (1.55), the integral is said to be a \emph{definite integral} and there is no need for a constant of integration as is the case with \emph{indefinite integrals}. Some properties of the indefinite integral are %(Thomas et al., ) 

\begin{equation}
\int_{b}^{a} f(x) dx = -\int_{a}^{b} f(x) dx  
\end{equation}    

\begin{equation}
\int_{a}^{a} f(x) dx = 0  
\end{equation}    

\begin{equation}
\int_{a}^{b} kf(x) dx = k\int_{a}^{b} f(x) dx  
\end{equation}    

\begin{equation}
\int_{a}^{b} (f(x) \pm g(x))dx =  \int_{a}^{b} f(x) dx \pm \int_{a}^{b} g(x) dx    
\end{equation}    

\begin{equation}
\int_{a}^{b} f(x) dx + \int_{b}^{c} f(x) dx = \int_{a}^{c} f(x) dx 
\end{equation}    
Where $f(x)$ and $g(x)$ are both univariate functions of $x$, and $k$ is a constant.
\section{Vectors}

You must have learned in high school physics that a vector differs from a scalar as it has both direction and magnitude. Now if we think of soil physical properties we might think of a range of properties that are scalars such as soil bulk density, soil temperature and soil water content. Now when we think of flow of water and gases inside porous media, most often than not it is as important to know the magnitude of the water or gas being transported, or in other words the bulk mass being transported by unit of time, as it is important to know the direction in which they are being transported. Suppose you have an irrigation project and you calculate exactly the volume of water that need to be provided for each plant for adequate growth and productivity, now suppose that you provide this water below the root zone so that it is mostly lost by downwards flow, you would be then providing the right magnitude of water but the direction of the flow would be inadequate. Suppose you have a rainfall and you are concerned with aquifer recharge and sediment loss by erosion. If most of the water flows vertically it is likely that groundwater recharge would be maximized and erosion minimized, and thus you wanna take measures for this to happen such as conserving natural vegetation or using sustainable land management practices.    

Let's say $x$ is a scalar, if we say that $x = 10$ that is all we need to describe it, meaning that its magnitude is 10.\footnote{Most often you'll need a unit of measurement as well} Now before talking about direction is important to define a system of coordinates. The familiar system of coordinates that you might be familiar with is the two dimensional Cartesian space or the plane. Suppose you and your friend are pushing a large box on an empty room. Imagine that you are pushing the box by applying a force parallel to two of the walls and your friend is applying a force perpendicular to the force you are applying. Let's call the direction which you are applying the force x so that the wall parallel to that directions has the coordinates of x, while the direction in which your friend is applying the force is y and the wall parallel to that direction defines the y axis. If none of the forces applied is zero, the object will move at an angle to each of the applied forces because the vector sum of the forces has a resulting force $\mathbf{F}$ defined as
\begin{equation}
\mathbf{F} = F_{x} \mathbf{i} + F_{y} \mathbf{j} 
\end{equation}    
In this equation $F_{x}$ and $F_{y}$ are the components of the force F in the directions x and y or the forces applied by you and your friend. The symbols $\mathbf{i}$ and $\mathbf{j}$ are used to indicate unit vectors, and their role is to indicate in which directions the components of the vector are acting. The unit vectors have magnitude 1 and can be represented by different notations. If we have three forces being applied in a three dimensional space, with coordinates x, y and z equation (1.61) can be written as  
\begin{equation}
\mathbf{F} = F_{x} \mathbf{i} + F_{y} \mathbf{j} + F_{z} \mathbf{k} 
\end{equation}    
In fluid dynamics is not unusual that the unit vectors are represented using a different notation. A velocity vector in a three dimensional space is usually written as

\begin{equation}
\mathbf{v} = v_{x} \mathbf{x} + v_{y} \mathbf{y} + v_{z} \mathbf{z} 
\end{equation}          
Where $\mathbf{i} = \mathbf{x}$, $\mathbf{j} = \mathbf{y}$ and $\mathbf{k} = \mathbf{z}$. 
Now in your handwritten notebook or in a board it is not convenient to write letters as boldface as notation for representing vectors. In handwritten form but also in a few books, vectors are represented with arrows draw above the symbol while unit vectors  are represented by the circumflex accent over the symbol, known in mathematics and colloquial language as hat ``\^{}''. Using this notation, equations (1.62) and (1.63) can be written as
\begin{equation}
\vec{F} = F_{x} \hat{i} + F_{y} \hat{j} + F_{z} \hat{k} 
\end{equation}    

\begin{figure}[ht]
 \includegraphics[width=0.8\textwidth]{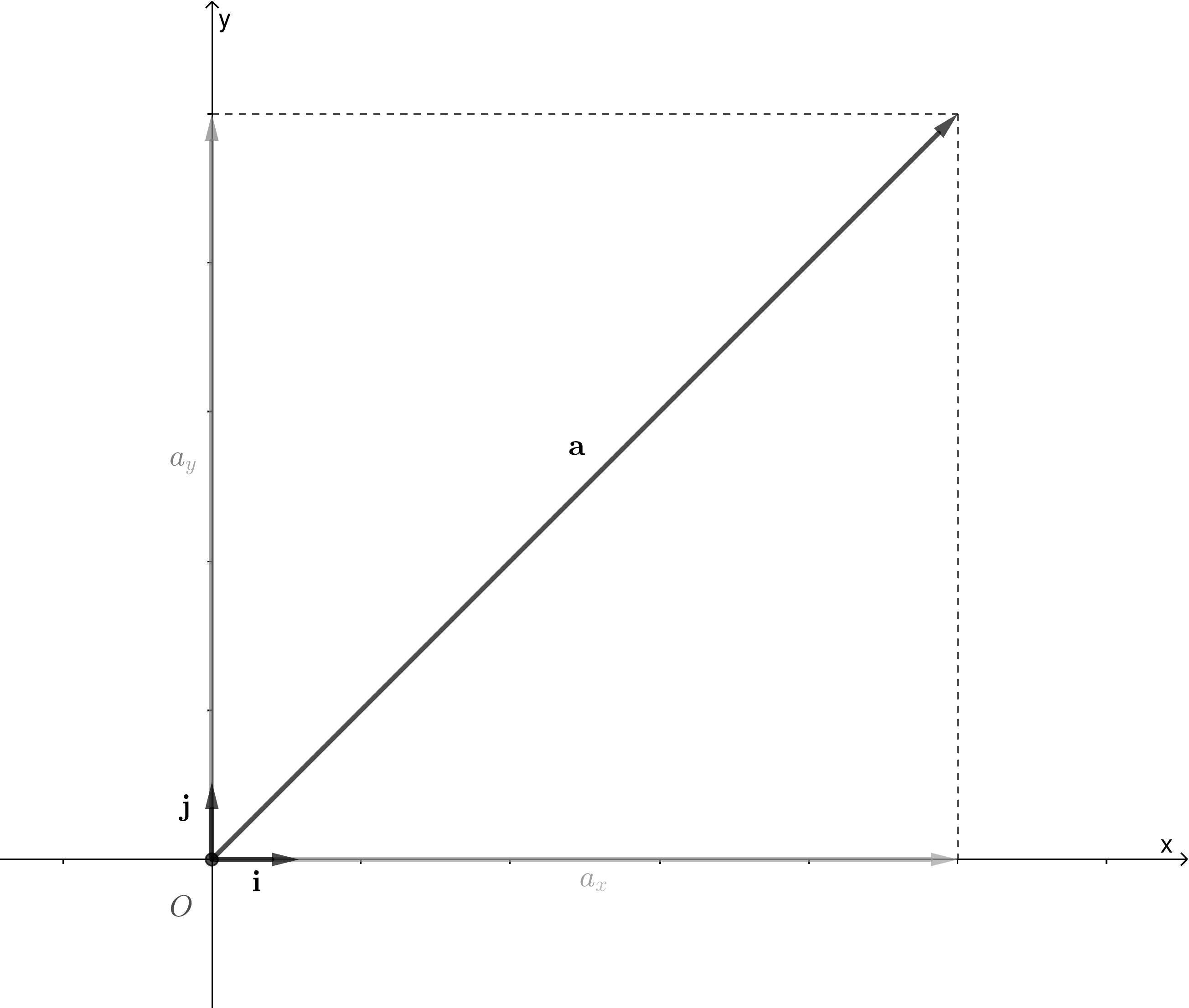}
  \caption{ Vector $\mathbf{a}$ with components $\mathbf{a_{x}}$ and $\mathbf{a _{y}}$.}
  \label{ch1_fig3}
\end{figure}

Another notation commonly seen in fluid dynamics is 
\begin{equation}
\vec{v} = v_{x} \hat{x} + v_{y} \hat{y} + v_{z} \hat{z} 
\end{equation}          
In this material I will use the boldface representation because I feel it looks more elegant in print and electronic format. 
In more advanced physics and mathematics books these vectors could also be represented as 
\begin{equation}
\mathbf{F} = \langle F_{x}, F_{y}, F_{z} \rangle 
\end{equation}    
\begin{equation}
\mathbf{v} = \langle v_{x}, v_{y}, v_{z} \rangle 
\end{equation}          
and we might make use of this notation for higher dimension vectors (i.e. higher than three dimensions).
Suppose now that in Cartesian space there is a vector a defined as 
\begin{equation}
\mathbf{a} = a_{x} \mathbf{i} + a_{y} \mathbf{j} 
\end{equation}          
as represented in Figure \ref{ch1_fig3}. 
 
You will notice that the components of the vector are the projections of $\mathbf{a}$ over the axis in the directions $\mathbf{i}$ and $\mathbf{j}$. The length of the vector or its magnitude is given by  

\begin{equation}
|\mathbf{a}| = \sqrt{a_{x}^2 + a_{y}^2} 
\end{equation}          
The proof is immediate from elementary mathematics and the user is encouraged to consult calculus and linear algebra books for further information. A few properties of vectors are listed below

\begin{equation}
\mathbf{a} + \mathbf{b} = \mathbf{b} + \mathbf{a}   
\end{equation}  
\begin{equation}
\mathbf{a} + \mathbf{0}  = \mathbf{a}   
\end{equation}  
\begin{equation}
k(\mathbf{a} + \mathbf{b}) = k\mathbf{a} + k\mathbf{b}   
\end{equation}  
\begin{equation}
cd\mathbf{a}  = c(d\mathbf{a})   
\end{equation}  
\begin{equation}
\mathbf{a} + (\mathbf{b} + \mathbf{c}) = (\mathbf{a} + \mathbf{b}) + \mathbf{c}     
\end{equation}  
\begin{equation}
\mathbf{a} + (-\mathbf{a}) = \mathbf{0}     
\end{equation}  
\begin{equation}
(c + d)\mathbf{a}  = c\mathbf{a} + d\mathbf{a}      
\end{equation}
\begin{equation}
1\mathbf{a}  = \mathbf{a}      
\end{equation}
Where $\mathbf{a}$, $\mathbf{b}$ and $\mathbf{c}$ are vectors and and $c$ and $d$ are scalars. 

In all vectors shown until now the components are treated as scalars as the direction is specified by the unit vectors. In many cases the components themselves are functions in x, y and z such that     
\begin{equation}
\mathbf{V}(x,y,z)  = u(x,y,z) \ \mathbf{i} + v(x,y,z) \ \mathbf{j} +  w(x,y,z) \ \mathbf{k}      
\end{equation}
or 
\begin{equation}
\mathbf{V}(x,y,z)  = \langle u(x,y,z), v(x,y,z), w(x,y,z) \rangle      
\end{equation}
In rectangular coordinates each of these components can represent an equation which is itself a function of $x$, $y$ and $z$. Equations 1.78 and 1.79 represent \emph{vector fields} where in each point in three dimensional space a vector can be drawn specifying the field. A simple example would be a vector field where 
\begin{equation}
u(x,y,z)  = 2xyz      
\end{equation}
\begin{equation}
v(x,y,z)  = x^{2} + yz      
\end{equation}
\begin{equation}
w(x,y,z)  = x - y + 2z      
\end{equation}
which can be written as 
\begin{equation}
\mathbf{V}(x,y,z)  = 2xyz \ \mathbf{i} + (x^{2} + yz) \ \mathbf{j} +  (x - y + 2z) \ \mathbf{k}      
\end{equation}

If we define two vectors $\mathbf{a}$ and $\mathbf{b}$
\begin{equation}
\mathbf{a}  = a_{x} \mathbf{i} + a_{y} \ \mathbf{j} +  a_{z} \ \mathbf{k}         
\end{equation}

\begin{equation}
\mathbf{b}  = b_{x} \mathbf{i} + b_{y} \ \mathbf{j} +  b_{z} \ \mathbf{k}         
\end{equation}
regular mathematical operations of sum and subtractions of these two vectors and multiplication by scalars are defined by properties (1.70) to (1.77), see for example property (1.76) \\
\begin{align*}
(c + d)\mathbf{a} = (c + d)(a_{x} \mathbf{i} + a_{y} \ \mathbf{j} +  a_{z} \ \mathbf{k}) 
\\
= (c + d)a_{x} \mathbf{i} + (c + d) a_{y} \ \mathbf{j} +  (c + d)a_{z} \ \mathbf{k} 
\\
= ca_{x} \mathbf{i} + c a_{y} \ \mathbf{j} +  ca_{z} \ \mathbf{k} + da_{x} \mathbf{i} + d a_{y} \ \mathbf{j} +  da_{z} \ \mathbf{k} 
\\
= c \mathbf{a} +  d \mathbf{a} 
\end{align*}

You might have noticed that nothing is said about multiplication of vectors by vectors in those properties. This is because regular multiplication and division operations used for scalars are not valid for vectors, they require their own rules. The first of such rules is the \emph{scalar} or \emph{dot} product, represented by the symbol ``$\cdot{}$''
\begin{equation}
\mathbf{a} \cdot \mathbf{b}  = a_{x}b_{x} + a_{y}b_{y} +  a_{z}b_{z}         
\end{equation}
The demonstration requires some further explanation as you might not be familiar or might not recall the properties of unit vectors
   
\begin{align*}
\mathbf{a} \cdot \mathbf{b}  = 
(a_{x}\mathbf{i} + a_{y}\mathbf{j} +  a_{z}\mathbf{k}) \cdot (b_{x}\mathbf{i} + b_{y}\mathbf{j} +  b_{z}\mathbf{k})  \\
= a_{x}\mathbf{i} \cdot b_{x}\mathbf{i} +
a_{x}\mathbf{i} \cdot b_{y}\mathbf{j} +
a_{x}\mathbf{i} \cdot b_{z}\mathbf{k} +
a_{y}\mathbf{i} \cdot b_{x}\mathbf{i} +
a_{y}\mathbf{i} \cdot b_{y}\mathbf{j} + \\ 
+ a_{y}\mathbf{i} \cdot b_{z}\mathbf{k} + 
a_{z}\mathbf{i} \cdot b_{x}\mathbf{i} + 
a_{z}\mathbf{i} \cdot b_{y}\mathbf{j} +
a_{z}\mathbf{i} \cdot b_{z}\mathbf{k} \\
= a_{x}b_{x}\mathbf{i} \cdot \mathbf{i} +
a_{x}b_{y}\mathbf{i} \cdot \mathbf{j} +
a_{x}b_{z}\mathbf{i} \cdot \mathbf{k} +
a_{y}b_{x}\mathbf{i} \cdot \mathbf{i} +
a_{y}b_{y}\mathbf{i} \cdot \mathbf{j} + \\ 
+ a_{y}b_{z}\mathbf{i} \cdot \mathbf{k} + 
a_{z}b_{x}\mathbf{i} \cdot \mathbf{i} + 
a_{z}b_{y}\mathbf{i} \cdot \mathbf{j} +
a_{z}b_{z}\mathbf{i} \cdot \mathbf{k} \\
 = a_{x}b_{x} + a_{y}b_{y} +  a_{z}b_{z} 
\end{align*}
This is because the dot product operates following a distributive property, and for the unit vectors defined as $i = \langle 1,0,0 \rangle$, $j = \langle 0,1,0 \rangle$ and $k = \langle 0,0,1 \rangle$ 
\begin{equation}
\mathbf{i} \cdot \mathbf{i}  = \mathbf{i} \cdot \mathbf{i} = \mathbf{i} \cdot \mathbf{i} = 1          
\end{equation}
\begin{equation}
\mathbf{i} \cdot \mathbf{j}  = \mathbf{i} \cdot \mathbf{k} = \mathbf{j} \cdot \mathbf{k} = 0          
\end{equation}
A special case of the dot product in two dimensions is the Pythagoras theorem, which as we already seen it provides the length of a vector
\begin{equation}
\mathbf{a} \cdot \mathbf{a}  = |\mathbf{a}|^{2} = a_{x}^2 + a_{y}^2          
\end{equation}
Back to three dimensions, the dot product can be written as 
\begin{equation}
\mathbf{a} \cdot \mathbf{b}  = |\mathbf{a}|  |\mathbf{b}| \cos{\theta}        
\end{equation}
Where $\cos{\theta}$ is the angle between the two vectors. There is much more information about the geometric interpretation of the dot product in calculus and linear algebra books. This information is important for the interpretation of vector quantities in soil physics and groundwater hydrology but it would be beyond the scope of this book to present it. What is important now is to understand the definitions of these vector operators because they are required for understanding vector calculus operators and the fundamental transport equations. 
The second operator is the \emph{cross} or \emph{vector} product, represented by the symbol ``$\times$''
\begin{equation}
\mathbf{a} \times \mathbf{b}  = 
(a_{y} b_{z} - a_{z} b_{y}) \mathbf{i} + 
(a_{z} b_{x} - a_{x} b_{z}) \mathbf{j} + 
(a_{x} b_{y} - a_{y} b_{x}) \mathbf{k} 
\end{equation}
or
\begin{equation}
\mathbf{a} \times \mathbf{b}  = \
\langle a_{y}b_{z} - a_{z}b_{y}, a_{z}b_{x} - a_{x}b_{z}, a_{x}b_{y} - a_{y}b_{x} \rangle
\end{equation}
The demonstration can be done in a similar way as we did for the dot product

\begin{align*}
\mathbf{a} \times \mathbf{b}  = 
(a_{x}\mathbf{i} + a_{y}\mathbf{j} +  a_{z}\mathbf{k}) \times (b_{x}\mathbf{i} + b_{y}\mathbf{j} +  b_{z}\mathbf{k})  \\
= a_{x}\mathbf{i} \times b_{x}\mathbf{i} +
a_{x}\mathbf{i} \times b_{y}\mathbf{j} +
a_{x}\mathbf{i} \times b_{z}\mathbf{k} +
a_{y}\mathbf{i} \times b_{x}\mathbf{i} +
a_{y}\mathbf{i} \times b_{y}\mathbf{j} + \\ 
+ a_{y}\mathbf{i} \times b_{z}\mathbf{k} + 
a_{z}\mathbf{i} \times b_{x}\mathbf{i} + 
a_{z}\mathbf{i} \times b_{y}\mathbf{j} +
a_{z}\mathbf{i} \times b_{z}\mathbf{k} \\
= a_{x}b_{x}\mathbf{i} \times \mathbf{i} +
a_{x}b_{y}\mathbf{i} \times \mathbf{j} +
a_{x}b_{z}\mathbf{i} \times \mathbf{k} +
a_{y}b_{x}\mathbf{j} \times \mathbf{i} +
a_{y}b_{y}\mathbf{j} \times \mathbf{j} + \\ 
+ a_{y}b_{z}\mathbf{j} \times \mathbf{k} + 
a_{z}b_{x}\mathbf{k} \times \mathbf{i} + 
a_{z}b_{y}\mathbf{k} \times \mathbf{j} +
a_{z}b_{z}\mathbf{k} \times \mathbf{k} \\
\end{align*}
and because the geometrical properties of the cross product are different than those of the dot product such that 
\begin{equation}
\mathbf{i} \times \mathbf{i}  = \mathbf{j} \times \mathbf{j} = \mathbf{k} \times \mathbf{k} = 0          
\end{equation}
\begin{equation}
\mathbf{i} \times \mathbf{j}  = \mathbf{k}          
\end{equation}
\begin{equation}
\mathbf{j} \times \mathbf{k}  = \mathbf{i}          
\end{equation}
\begin{equation}
\mathbf{k} \times \mathbf{i}  = \mathbf{j}          
\end{equation}
\begin{equation}
\mathbf{j} \times \mathbf{i}  = -\mathbf{k}          
\end{equation}
\begin{equation}
\mathbf{k} \times \mathbf{j}  = -\mathbf{i}          
\end{equation}
\begin{equation}
\mathbf{i} \times \mathbf{k}  = -\mathbf{j}          
\end{equation}
we have
\begin{align*}
\mathbf{a} \times \mathbf{b}  = 
a_{x}b_{x} 0 +
a_{x}b_{y}\mathbf{k} +
a_{x}b_{z}(-\mathbf{j}) +
a_{y}b_{x}(-\mathbf{k}) +
a_{y}b_{y}0 + \\ 
+ a_{y}b_{z}\mathbf{i} + 
a_{z}b_{x}\mathbf{j} + 
a_{z}b_{y}(-\mathbf{i}) +
a_{z}b_{z}0 \\
= (a_{y} b_{z} - a_{z} b_{y}) \mathbf{i} + 
(a_{z} b_{x} - a_{x} b_{z}) \mathbf{j} + 
(a_{x} b_{y} - a_{y} b_{x}) \mathbf{k} 
\end{align*}
Notice that the result of scalar product is, as the name implies, a scalar while the vector product is, as you might have guessed, a vector. In practice the vector product might be calculated from 
\begin{equation}
  \mathbf{a} \times \mathbf{b} =  \begin{vmatrix} 
   \mathbf{i} & \mathbf{j} & \mathbf{k}  \\
   a_{x} & a_{y} & a_{z}  \\
   b_{x} & b_{y} & b_{z}  \\
   \end{vmatrix} 
\end{equation}
such that
\begin{align*}
\mathbf{a} \times \mathbf{b}  = 
a_{y}b_{z} \mathbf{i} +  
a_{z}b_{x} \mathbf{j} +  
a_{x}b_{y} \mathbf{k} -  
a_{z}b_{y} \mathbf{i} -  
a_{x}b_{z} \mathbf{j} -  
a_{y}b_{x} \mathbf{k} \\
=  (a_{y} b_{z} - a_{z} b_{y}) \mathbf{i} + 
(a_{z} b_{x} - a_{x} b_{z}) \mathbf{j} + 
(a_{x} b_{y} - a_{y} b_{x}) \mathbf{k} 
\end{align*}
Note that there are different approaches to calculating the determinant of a matrix so we are omitting a few steps. You can consult any calculus or linear algebra book you if you are unfamiliar with the procedure. Notice that in many countries such procedures are taught in secondary school so you might want to consult such books as well for a more didactic introduction.  The vector product can also be written as 
\begin{equation}
\mathbf{a} \times \mathbf{b}  = |\mathbf{a}|  |\mathbf{b}| \sin{\theta}        
\end{equation}
Where $\sin{\theta}$ is the sin of the angle between the two vectors. As with the dot product the reader is encouraged to consult its geometrical properties and interpretation. The properties of the scalar and vector products are summarized below  
\begin{equation}
\mathbf{a} \cdot \mathbf{a} = |\mathbf{a}|^{2} 
\end{equation}
\begin{equation}
\mathbf{a} \cdot (\mathbf{b} + \mathbf{c}) = \mathbf{a} \cdot \mathbf{b} + \mathbf{a} \cdot\mathbf{c}
\end{equation}
\begin{equation}
\mathbf{a} \cdot \mathbf{b}  = \mathbf{b} \cdot \mathbf{a}
\end{equation}
\begin{equation}
(c\mathbf{a}) \cdot \mathbf{b} = c(\mathbf{a} \cdot \mathbf{b}) = \mathbf{a} \cdot (c\mathbf{b}) 
\end{equation}
\begin{equation}
\mathbf{0} \cdot \mathbf{a} = 0 
\end{equation}
\begin{equation}
\mathbf{a} \times \mathbf{b} = -\mathbf{b} \times \mathbf{a} 
\end{equation}
\begin{equation}
(c\mathbf{a}) \times \mathbf{b} = c(\mathbf{a} \times \mathbf{b}) = \mathbf{a} \times (c\mathbf{b}) 
\end{equation}
\begin{equation}
\mathbf{a} \times (\mathbf{b} + \mathbf{c}) =  
\mathbf{a} \times \mathbf{b} + \mathbf{a} \times \mathbf{c} 
\end{equation}
\begin{equation}
(\mathbf{a} \times \mathbf{b}) + \mathbf{c} =  
\mathbf{a} \times \mathbf{c} + \mathbf{b} \times \mathbf{c} 
\end{equation}
\begin{equation}
\mathbf{a} \cdot (\mathbf{b} + \mathbf{c}) =  
(\mathbf{a} \times \mathbf{b})  \cdot \mathbf{c} 
\end{equation}
\begin{equation}
\mathbf{a} \times (\mathbf{b} + \mathbf{c}) =  
(\mathbf{a} \cdot \mathbf{c})\mathbf{b} - (\mathbf{a} \cdot \mathbf{b})\mathbf{c}  
\end{equation}

\section{Vector calculus operators}
Before delving into vector calculus operators we need to take a step back and look what how derivation and integration work when you have a multivariate function. As we've seen, a univariate function has a dependent variable that it is a function of a single independent variable as is the case with the equation $y = x$. A multivariate function is one which is a function of two or more independent variables, for example, each component of the vector field in (1.83) is a multivariate function of the variables $x$, $y$ and $z$ as shown in (1.80), (1.81) and (1.82). You already know a few basic rules to differentiate univariate functions, but how to proceed with multivariate functions? The tools for solving such problems are in the realm of multivariate calculus and are the partial derivatives and multiple integrals. Most engineering, mathematics and physics undergraduate programs devote an entire course for these subjects, usually calculus III or multivariate calculus. 
The idea behind partial differentiation is apparently simple if we ignore rigorous mathematical proofs and theorems behind it as we will do in this material. If you have a multivariate function you can take the derivative with respect to a single variable and treat the other variables as constants as in (1.21). When taking partial derivatives the symbol ``$\partial$'' is used instead of ``$d$''. For example, if we wish to differentiate (1.80) with respect to $x$ we have
\begin{equation}
\frac{\partial}{\partial x} u(x,y,z) = \frac{\partial}{\partial x}2xyz = yz \frac{\partial x}{\partial x} 2x = 2yz
\end{equation}
Now let's say you want to derivate (1.81) with respect to $y$
\begin{equation}
\frac{\partial}{\partial y} v(x,y,z) = \frac{\partial}{\partial y} (x^{2} + yz) = \frac{\partial}{\partial y} x^{2} + \frac{\partial}{\partial y} yz = 0 + z = z 
\end{equation}
Because here $x$ is a constant and any constant squared is still a constant, and $z$ is another constant. Other representations of a partial derivative are, for example
\begin{equation}
(\frac{\partial}{\partial x} u)_{y,z}  
\end{equation}
and
\begin{equation}
\frac{\partial}{\partial x} u \Bigm| _{y,z}  
\end{equation}
Where in both cases the differentiation is performed with respect to $x$ with $y$ and $z$ held constant. Such notation is most commonly used Thermodynamics, a field which makes extensive use of multivariate calculus and  interlaces with both fluid dynamics and soil physics. Earlier we saw that the first derivative is the rate of variation of a function, intuitively, for a straight line
\begin{equation}
y = ax + b  
\end{equation}
The first derivative is the slope of the line
\begin{equation}
\frac{dy}{dx} = a  
\end{equation}
For a second degree polynomial, the rate of variation is not a constant, but it is a straight line so that the slope of the quadratic curve varies with $x$ 
\begin{equation}
y = ax^{2} + bx + c  
\end{equation}
The first derivative is the slope of the line
\begin{equation}
\frac{dy}{dx} = 2ax + b  
\end{equation}  
Notice that the derivatives can be used to identify critical points of functions as you might learn in any intro calculus books. A function in two or more dimensions will have slopes in each particular direction as represented by the partial derivatives. Lets consider a function $w(x,y,z)$ 
\begin{equation}
w = x^{2} + 2xy + z^{3}  
\end{equation}
with partial derivatives in $x$, $y$ and $z$ 
\begin{equation}
\frac{\partial w}{\partial x} = 2x + 2y  
\end{equation}  
\begin{equation}
\frac{\partial w}{\partial y} = 2x  
\end{equation} 
\begin{equation}
\frac{\partial w}{\partial z} = 3z^{2}  
\end{equation} 
Which are the slopes of the function in the directions $x$ and $y$.  From our previous discussions it should be apparent that these partial derivatives represent the slopes of the function on the directions $ \mathbf{i}$, $ \mathbf{j}$ and $ \mathbf{k}$ and as such we can represent these partial derivatives as a vector function at any point in space (if it exists) as
\begin{equation}
\nabla{w(x,y,z)} = 
\frac{\partial w}{\partial x} = (2x + 2y)  \mathbf{i} +
\frac{\partial w}{\partial y} = (2x)  \mathbf{j} +
\frac{\partial w}{\partial z} = (3z^{2})  \mathbf{k}
\end{equation} 
Where the symbol ``$\nabla$'' is called ``nabla'' or ``del'' and is a representation for a mathematical operator called \emph{gradient}. The gradient of a scalar or vector function is a vector function and can be generalized as 
\begin{equation}
\boxed{\nabla{} = 
\frac{\partial w}{\partial x}   \mathbf{i} +
\frac{\partial w}{\partial y}   \mathbf{j} +
\frac{\partial w}{\partial z}   \mathbf{k}}
\end{equation} 
applied to any scalar function $f$ results 
\begin{equation}
\boxed{\nabla{f} = 
\frac{\partial w}{\partial x}f  \ \mathbf{i} +
\frac{\partial w}{\partial y}f  \ \mathbf{j} +
\frac{\partial w}{\partial z}f  \ \mathbf{k}}
\end{equation} 
Now for a vector 
\begin{equation}
\mathbf{u} = u  \ \mathbf{i} + v \  \mathbf{j} + w \  \mathbf{k}
\end{equation} 
The gradient in rectangular coordinates is a \emph{second rank tensor}
\begin{equation}
\nabla{\mathbf{u}} = 
\begin{pmatrix}
\frac{\partial u}{\partial x} & \frac{\partial v}{\partial x} & \frac{\partial w}{\partial x}\\
\frac{\partial u}{\partial y} & \frac{\partial v}{\partial y} & \frac{\partial w}{\partial y} \\
\frac{\partial u}{\partial z} & \frac{\partial v}{\partial z} & \frac{\partial w}{\partial z}
\end{pmatrix}
\end{equation} 
Which in proper terminology is the Jacobian of the vector $\mathbf{u}$. Although this vector gradient sometimes appears in hydrodynamic equations, it is often not necessary to expand it. The reader of this material should focus on the gradient of scalar functions whose importance will be apparent when we deal of water flux in porous media.
Now the dot product of the gradient operator over a vector, say $u$, defines the \emph{divergence} of the vector  
\begin{equation}
\boxed{\nabla{\cdot \mathbf{u}} = 
\frac{\partial u}{\partial x} +
\frac{\partial v}{\partial y} +
\frac{\partial w}{\partial z} }
\end{equation} 
Which is a scalar. The demonstration is straightforward from the definition of $\nabla$ and $\mathbf{u}$ and from the dot product properties
\begin{align*}
\nabla{\cdot \mathbf{u}} = 
(\frac{\partial }{\partial x} \mathbf{i} +
\frac{\partial }{\partial y} \mathbf{j} +
\frac{\partial }{\partial z} \mathbf{k})  \cdot 
 (u   \mathbf{i} + v   \mathbf{j} + w   \mathbf{k}) \\
  = \frac{\partial u}{\partial x}  \mathbf{i} \cdot\mathbf{i} + 
\frac{\partial v}{\partial y} \mathbf{j}  \cdot \mathbf{j}+
\frac{\partial w}{\partial z} \mathbf{k} \cdot \mathbf{k} =
\frac{\partial u}{\partial x} +
\frac{\partial v}{\partial y} +
\frac{\partial w}{\partial z} 
\end{align*}
Following the same logic we can apply the vector product of the gradient to any vector function yielding the curl of a vector, $\mathbf{u}$ for example 
\begin{equation}
\boxed{\nabla{\times \mathbf{u}} = 
(\frac{\partial w}{\partial y}-\frac{\partial v}{\partial z}) \mathbf{i} + 
(\frac{\partial u}{\partial z}-\frac{\partial w}{\partial x}) \mathbf{j} +
(\frac{\partial v}{\partial x}-\frac{\partial u}{\partial y}) \mathbf{k}}
\end{equation} 
Which is a also vector field. As we learned with the dot product the determinant form is the easiest way to remember how to calculate the curl of a vector function 
\begin{equation}
\nabla{\times \mathbf{u}} = 
\begin{vmatrix}
\mathbf{i} & \mathbf{j} & \mathbf{k}\\
\frac{\partial }{\partial x} & \frac{\partial }{\partial y} & \frac{\partial }{\partial z} \\
u & v & w
\end{vmatrix}
\end{equation} 
As with the scalar and vector products there are several properties that define operations for the gradient, divergence and curl, including operations among them. One such operation which appears very frequently in applied mathematics and physics is the divergence of a gradient.
\begin{equation}
\boxed{\nabla{\cdot (\nabla{f)} = 
\frac{\partial^{2} f}{\partial x^{2}} +
\frac{\partial^{2} f}{\partial y^{2}} +
\frac{\partial^{2} f}{\partial z^{2}} }}
\end{equation} 
The demonstration is also somewhat trivial
\begin{align*}
\nabla{\cdot (\nabla{f})} = 
(\frac{\partial}{\partial x} \mathbf{i}+
\frac{\partial}{\partial y} \mathbf{j}+
\frac{\partial}{\partial z} \mathbf{k}) \cdot 
(\frac{\partial f}{\partial x} \mathbf{i}+
\frac{\partial f}{\partial y} \mathbf{j}+
\frac{\partial f}{\partial z} \mathbf{k}) \\
  = \frac{\partial^{2} f}{\partial x^{2}}  \mathbf{i} \cdot\mathbf{i} + 
\frac{\partial f ^{2}}{\partial y^{2}} \mathbf{j}  \cdot \mathbf{j}+
\frac{\partial w^{2}}{\partial z^{2}} \mathbf{k} \cdot \mathbf{k} \\
= \frac{\partial^{2} f}{\partial x^{2}} +
\frac{\partial^{2} f}{\partial y^{2}} +
\frac{\partial^{2} f}{\partial z^{2}} 
\end{align*}
The operator
\begin{equation}
\nabla{}^{2} = \nabla{} \cdot \nabla{} 
\end{equation}
is called the Laplace operator because of the equation of the form 
\begin{equation}
\nabla{}^{2}f = \frac{\partial^{2} f}{\partial x^{2}} +
\frac{\partial^{2} f}{\partial y^{2}} +
\frac{\partial^{2} f}{\partial z^{2}}  = 0
\end{equation}
which is of the utmost importance not only for soil physics and fluid mechanics but to most areas of physics. The Laplace operator also operates in vectors such that 
\begin{equation}
\nabla{}^{2} \mathbf{u}= \nabla{}^{2} u \ \mathbf{i} + \nabla{}^{2} v  \ \mathbf{i} + \nabla{}^{2} w \ \mathbf{i} 
\end{equation}
Understanding the gradient, divergence, curl and Laplacian is fundamental for the understanding of fluid mechanics and therefore soil physics and underground hydrology. Understanding these operators along with their properties and interrelations and the fundamental theorems of vectors calculus is almost all that is needed before presenting the fundamental equations of fluid mechanics. If you decide to consult the literature be aware that in many older books the operators are presented as del($f$), grad($f$) or grad $f$ for the gradient of a scalar function $f$, div($\mathbf{u}$) or div $\mathbf{u}$ for the divergence of a vector function $\mathbf{u}$, curl($\mathbf{u}$), curl $\mathbf{u}$, rot($\mathbf{u}$) or rot $\mathbf{u}$ for the curl of a vector $\mathbf{u}$ and div(grad $f$) for the Laplacian of a scalar function $f$. The ``rot'' in curl is because in many countries the curl is called \emph{rotational} or one its translated similar words. The word rotational as you might imagine comes from rotation and the divergence is obviously related to something that diverges. The meaning is exactly as implied, these operators were developed in the realm of the physics of fluids. Most of the mathematics used in fluid dynamics was used to describe the pervasive ether (or \ae ther) before Einstein and his contemporaries proved that it was not a necessary entity for the description of the the cosmos. 
A few vector operators identities are offered without proof
\begin{equation}
\nabla{(fg)} = f (\nabla{g}) + g (\nabla{f}) 
\end{equation}
\begin{equation}
\nabla{(\mathbf{A} \cdot \mathbf{B} )} = \mathbf{A} \times (\nabla \times \mathbf{B}) + \mathbf{B} \times (\nabla \times \mathbf{A}) + (\mathbf{A} \cdot \nabla) \mathbf{B} + (\mathbf{B} \cdot \nabla)\mathbf{A} 
\end{equation}
\begin{equation}
\nabla \cdot (f \mathbf{A}) = f(\nabla \cdot \mathbf{A}) + \mathbf{A} \cdot (\nabla{f})
\end{equation}
\begin{equation}
\nabla \cdot (\mathbf{A} \times \mathbf{B}) = 
\mathbf{B} \cdot (\nabla \times \mathbf{A}) - \mathbf{A} \cdot (\nabla \times \mathbf{B}) 
\end{equation}
\begin{equation}
\nabla \times (f\mathbf{A}) = 
f(\nabla \times \mathbf{A}) - \mathbf{A} \times (\nabla f) 
\end{equation}
\begin{equation}
\nabla \times (\mathbf{A} \times \mathbf{B}) = (\mathbf{B} \cdot \nabla)\mathbf{A} - (\mathbf{A} \cdot \nabla) \mathbf{B} + \mathbf{A}(\nabla \cdot  \mathbf{B}) -  \mathbf{B}(\nabla \cdot \mathbf{A}) 
\end{equation}
\begin{equation}
\nabla \cdot (\nabla \times \mathbf{A}) = 0 
\end{equation}
\begin{equation}
\nabla \times (\nabla f) = 0 
\end{equation}
\begin{equation}
\nabla \times (\nabla \times \mathbf{A}) = \nabla(\nabla \times \mathbf{A}) - \nabla^{2}\mathbf{A} 
\end{equation}
For now it is also important to say that integration can also be performed in multidimensional space, we have seen that integration in one dimension can be represented as 
\begin{equation}
\int x dx = \frac{x^{2}}{2} +  C
\end{equation}
now integration in two dimensions can be understood in terms of a surface integral, for example
\begin{equation}
\int\int xy dx dy  
\end{equation} 
Similar for what was discussed for partial derivatives, the integration is performed on each variable per step, considering the others as a constant
\begin{align*}
\int\int xy dx dy  = \frac{x^{2}}{2}\int y dy + C_{1} \\
= \frac{x^{2}}{2}\frac{y^{2}}{2} + C_{1}y + C_{2} \\
= \frac{x^{2}y^{2}}{4} + C_{1}y + C_{2} 
\end{align*}
The same principles being applied to simple integrals in three or more dimensions. An integral in three dimensions is an integral in volume, for example 
\begin{equation}
\int\int\int xyz dx dy dz =  C_{1} x y + C_{2} y + C_{3} + 1/8 x^{2} y^{2} z^{2}  
\end{equation} 
Of course integrals of such simple functions are rarely seen in physics and applied mathematics. More often than not more advanced integration techniques and system of coordinates transformation are needed for solving the equations. The most common systems of coordinates used being polar, cylindrical and spherical coordinates.  The user is encouraged to consult the literature for transformations in system of coordinates and for the gradient, divergence, curl and Laplacian in those coordinate systems. Another integral that is common in physics and applied math is the integral over a curve or line integral, in which the integration is performed over a curve  
\begin{equation}
\int_{a}^{b} \mathbf{v} \cdot d \mathbf{l}  
\end{equation} 
which for a closed loop is usually represented as 
\begin{equation}
\oint \mathbf{v} \cdot d \mathbf{l}  
\end{equation} 
Integrals over lines, regions (surface) and volumes (domains) can also be represented by the symbols
\begin{equation}
\oint_C, \ \ \int_R, \ \ \int_S, \ \ \int_V, \ \ \int_D,    
\end{equation} 
Such representations appear in three fundamental theorems of vector calculus which are used very often in fluid mechanics, electromagnetism and other areas of applied mathematics and physics \cite{thomas}, the divergence (or Gauss) theorem
\begin{equation}
\boxed{
\iint_S \mathbf{F} \cdot \mathbf{n} \ d\sigma =  \iiint_D \nabla \cdot  \mathbf{F} \ dV}
\end{equation} 
Green's theorem in normal form 
\begin{equation}
\boxed{
\oint_C \mathbf{F} \cdot \mathbf{n} \ ds =  \iint_R \nabla \cdot  \mathbf{F} \ dA}
\end{equation} 
and in tangential form
\begin{equation}
\boxed{
\oint_C \mathbf{F} \cdot d \mathbf{r} =  \iint_R \nabla \times  \mathbf{F} \ \cdot \mathbf{k} \ dA}
\end{equation} 
and Stokes' theorem 
\begin{equation}
\boxed{
\oint_C \mathbf{F} \cdot d \mathbf{r} =  \iint_S \nabla \times  \mathbf{F} \ \cdot \mathbf{n} \ d \sigma}
\end{equation}

%\section{Tensors***}
%In construction

\section{Einstein's summation convention}
When working with the algebra of general relativity and other theoretical physics topics, Albert Einstein introduced a notation aimed at simplifying the representation of summation terms. As you might know, a sum is represented by the Greek letter $\Sigma$, so a sum of three $x$ is terms can be represented as  
\begin{equation}
	y = \sum_{i=1}^n x_i = x_1 + x_2 + x_3 
\end{equation} 
Now imagine working through hundreds or thousands of equation and having to repeat this symbology over and over. To simplify the notation Einstein omitted the $\Sigma$ symbol from the sums. Thus, in Einstein's notation, the sum above can be simplified to   
\begin{equation}
	y = x_i = x_1 + x_2 + x_3 + ... 
\end{equation} 
as the number of terms in the sum can be omitted for most practical purposes. If there is a single index in the equation, say $i$, the sum is over $i$  
\begin{equation}
	y =a_i x_i = a_1 x_1 + a_2x_2 +a_2 x_3 + ... 
\end{equation} 
If there are two or more indexes, say $i$ and $j$i, the summation is on the repeated indexes, for example \cite{boas}
\begin{equation}
	y_i = a_{ij} x_j =  a_{i1} x_1  + a_{i2} x_2   + a_{i3} x_3   + ... 
\end{equation} 
\begin{equation}
	\frac{\partial u}{\partial x_i}\frac{\partial x_i}{\partial y} = \frac{\partial u}{\partial x_1}\frac{\partial x_1}{\partial y}+\frac{\partial u}{\partial x_2}\frac{\partial x_2}{\partial y}+   \frac{\partial u}{\partial x_3}\frac{\partial x_3}{\partial y}+ ...
\end{equation} 
Although Einstein's notation might take some time to get used to, it is very convenient to deal with a large number of equations, especially when dealing with tensors, and it will be used extensively in this material.

\section{Conclusion and advice}
As a final warning to the reader, if most or all of the content in this chapter is new to you, this material is far from enough for you to learn the subject as it condenses material that is usually seen in three one semester undergraduate calculus courses. You should study and review the subject on your own if you wish to be proficient in porous media physics. If you are learning soil physics for the first time as a soil scientist, agronomist or other life sciences field you should use this chapter to become familiar with the notation and language used in this book and you can consult it whenever necessity arises.

\section{List of symbols for this chapter}

\begin{longtable}{ll}
    	$ T $ & Temperature \\
    	$ f $ & Generalized function \\

	$ T(t) $ & Time dependent temperature \\
    	$ T_0  $ & Temperature where the function is centered \\
    	$ \alpha   $ & Amplitude \\
    	$ \omega  $ & Angular frequency \\
    	$ \phi   $ & Phase lag \\
    	$ t  $ & Time \\
    	$ Y  $ & Height \\
    	$ Y_0  $ &  Initial height \\
    	$ v_0  $ & Initial velocity   \\
    	$ g  $ &  Earth's gravitational acceleration   \\
    	$ k, c, C, C_1, C_2  $ & Constants \\
    	$ n, N $ & Integer number \\
    	$ x, y, v  $ & Generalized position and velocity scalar variables  \\
	$ \dot{x}  $ & First derivative of position, or velocity  \\
	$ \ddot{x}  $ & Second derivative of position, or acceleration $a$  \\
	$ F   $ & Generalized function \\
	$ F(x)   $ & Generalized function of x \\
	$ f(t)  $ &  Generalized function of t \\
	$ f(x), g(x)  $ & Generalized functions of x \\
    	$ J  $ & Area under a curve  \\
	$ \vec{F}  $ & Generalized vector function  \\
	$ F_x, F_y, F_z  $ & Components of $\vec{F}$ in Cartesian coordinates  \\
	$ \mathbf{i}, \mathbf{j}, \mathbf{k}, \hat{i}, \hat{j}, \hat{k}, \hat{x}, \hat{y}, \hat{z},   $ & Different representations of unit vectors in Cartesian coordinates  \\
	$ \vec{v}  $ & Velocity vector  \\
	$ v_x, v_y, v_z  $ & Components of $\vec{v}$ in Cartesian coordinates  \\
	$ u, v, w  $ & Alternative representation of components of $\vec{v}$ in Cartesian coordinates  \\
	$ \mathbf{a}, \mathbf{b}, \mathbf{A}, \mathbf{B}, \mathbf{V}   $ & Generalized vectors  \\
	$ c, d   $ & Generalized scalars  \\
	$ a_x, a_y   $ & Cartesian components of $\mathbf{a}$  \\

	$ |\mathbf{a}| $ & Norm of $\mathbf{a}$  \\
	$ |\mathbf{b}| $ & Norm of $\mathbf{b}$  \\

	$ d  $ & Univariate derivative operator  \\
    	$ \partial  $ & Partial derivative operator  \\
    	$ \int  $ &  Integral operator \\
    	$  \sum $ & Summation operator   \\
    	$ \nabla  $ & Gradient   \\
    	$  \nabla \cdot $ & Divergence   \\
    	$  \nabla \times  $ &  Curl \\
    	$  \nabla^2  $ &  Laplacian \\
	$  \mathbf{n}  $ & Unit normal vector   \\
	$  \mathbf{r}  $ & Generalized position vector   \\
	$  A  $ & Area element   \\
	$  V  $ & Volume  element   \\
	$  \sigma  $ & Surface area element   \\

\end{longtable}

%\include{ch1.1}
% !TEX TS-program = pdflatex
% !TEX encoding = UTF-8 Unicode

% Example of the Memoir class, an alternative to the default LaTeX classes such as article and book, with many added features built into the class itself.

\chapter{Natural porous media}
\label{ch2}

\section{Definitions}
The first requirement for a material to be considered a porous media is obviously that it is composed of a solid but not necessarily rigid matrix that is not continuous in space such that the spaces between the solid fractions are pores. Right now we are considering those pores as empty spaces, disregarding the possibility that they can be filled by a material of a different nature than that of the solid matrix. There is a large amount of natural and artificial porous materials on Earth and by a large extent on the universe. A common bath or kitchen sponge is a perfect example of a porous material whose matrix is not rigid and not granular. These set of distinctions regarding the mechanical nature of the matrix, i.e. rigid, elastic, plastic, porous, etc., and the nature of the pores, i.e. connected or non-connected, is fundamental for the understanding of the physical properties of the media and we will talk some of them throughout this book.  

Natural porous media includes tissues from animals, vegetables, fungi and a range of other organisms, while artificial porous media of significance for  large scale transport phenomena in nature include concrete, asphalt and other building materials. A lot of the theory discussed in this book might apply to these artificial materials, especially regarding water and contaminant transport and storage. Asphalt and concrete used for pavement can under certain circumstances be considered low porosity and low permeability porous media and thus control the infiltration of water in urban environments and roads. The understanding of water storage and dynamics in foods, animal feed, grains, paper, and a variety of material is fundamental for their conservation and theories regarding water storage and determination in soils and rocks can be readily applied to food products. These materials are however beyond the scope of this book. We are interested here in three types of natural porous materials that cover immense extensions in both area and volume on Earth and are the basis of water and chemical elements cycles, water storage, agriculture, and environmental impacting activities, the materials we are concerned are \emph{soils}, \emph{sediments} and \emph{rocks}. As with the last chapter a true introduction to these topics would require several one semester courses covering each of the subjects, including intro soil science and geology courses, soil genesis, mineralogy (of both soils and rocks), igneous and metamorphic petrology, and sedimentology and stratigraphy courses. Thus, we can only give a brief introduction and provide the reader with a terminology to be used in the rest of the book. While the last chapter might be superficial to the student of mathematics and physics and might be too advanced to the life sciences student, in this chapter the opposite is true, the challenge here is to present a palatable introduction on the subjects of soil science and geology for those who might not have had any introduction to the subject, as with the last chapter this is not without the risk of displeasing both groups. 

There are different types of porous media, as was mentioned before a common kitchen sponge is an example of a nonrigid and nongranular porous media. This means that the geometry and volume of the pores might not be constant as the material is wetted and dried, even excluding the possibility that the sponge is compressed by hand. Because water is a polar, almost non-compressible medium, it can apply forces to the sponge surfaces that might cause it to deform. In the saturated case, water can transmit forces and stresses on the material that holds it, and as it dries the water molecules exert capillary forces which might cause surfaces to come close together. Soils and sediments are a class of porous materials which are both granular and nonrigid, meaning that the grains can be reorganized in response to applied stresses, altering pore volume and pore sizes distribution. 
 
\section{Porous media as representation}
\begin{figure}[ht]
\centering
 \includegraphics[width=1.0\textwidth]{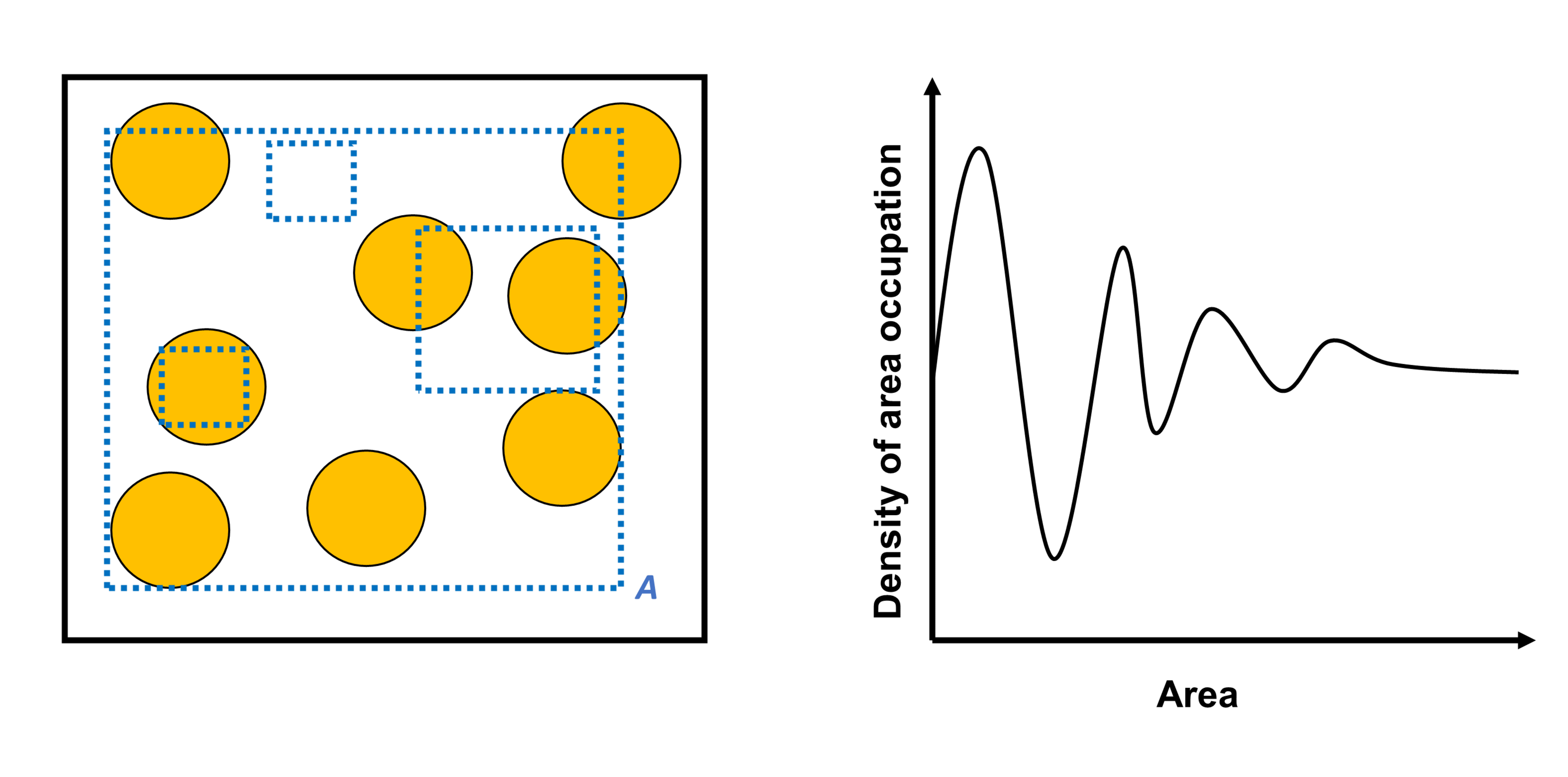}
\caption{Illustration of the Representative Elementary Volume (REV) concept. $A$ is the measured area and the circles represent solid particles. As the measured area increases, the areal density will tend to the average of the system.}
\label{ch2_fig1}
\end{figure}

Whether you subscribe to a philosophical view of an objective reality or not, this entire book deals with an idealized representation of what a porous medium is. Throughout it we will make a series of simplifications and assumptions in order to create a conceptual and physical representation of porous materials that is suitable for mathematical treatment. A representation is a simplification of a material which is often infinitely complex and impossible to be represented mathematically if every shape, every composition, every topological aspect is uniquely considered. One of the fundamental assumptions that lie at the core of solid and fluid mechanics and that is inherited into porous media physics is the \emph{continuum hypothesis}. Water, air, soils and rocks are composed of discrete particles. In the case of water, these particles are water molecules that could be further discretized into hydrogen and oxygen atoms which could be further discretized and so on\footnote{Elementary particles physics and quantum effects and uncertainty will not be dealt in this book unless in brief explanatory notes such as this one. For the most part, and maybe for the time being, soil physics can be entirely formulated within a classical mechanics framework, but not necessarily so porous media physics and condensed matter physics in general}. Soils, rocks and sedimentary deposits are for the most part composed of minerals, minerals on the other hand are composed of atoms occupying positions in a regular lattice. On a nanoscopic scales, if we measured a cross sectional area of a still frame of water or gas flowing or of a mineral and calculate the areal occupation of space we could fall into a molecule or atom, empty space or a boundary between the two. For most materials the areal density at this scale would be highly irregular depending on where the area is evaluated. As we increase the measured area, moving into macroscopic measured areas, the areal density will tend to a representation of the average areal occupation of the material (Figure \ref{ch2_fig1}). The principle holds if we consider a given property being measured over a volume of the material. The minimum volume in which the measured property is a representation of the statistical properties of the macroscopic area is the \emph{representative elementary volume} (REV) and corresponds to, by analogy, the approximate point on the plot in Figure \ref{ch2_fig1}  where the line oscillations are negligible. 

\section{Porous media as a system}

Compared to a pure liquid or gas, a porous medium is a much more complicated system as we often have multiple phases interacting. Each phase can be treated as a continuous medium on its own which interacts with the surrounding continuous phases. For some properties the entire system can be treated as a single continuous medium, including all phases. Often, when dealing with flow problems, fluids and gases are treated as dynamical phases while the solid matrix is treated as static, and the overall flow properties are defined by the interactions between the phases, while for measuring properties such as bulk mass, density, or temperature at a given time, for example, the individual composition of each phase might be neglected and the system might be treated as a whole. A system can be closed, isolated or open, closed systems do not interact with their surroundings, and there is no exchange of matter or energy, in closed systems there might be exchange of heat or other forms of energy through the walls of the system and its surroundings and in open systems there might be exchange of matter and energy of the system and its surroundings. Depending on the scenario, any representation can apply to natural porous media. When modeling transport phenomena, the system is usually considered open, such open system are sometimes understood to be analogous to the concept of control volume in engineering.   

We will use the traditional definitions of homogeneity and isotropy to classify porous media as a system. First a homogeneous system can be defined as one with only one component and one phase, and where its physical properties do not vary with space. Pure water in liquid state at 25 $^{o}$C is a homogeneous substance. Of course, achieving a complete state of purity in water might not be feasible, and defining a completely homogeneous density and temperature system for water might also not be realistic, for one because of the constraints of the continuum hypothesis discussed before. But this would lead us back to the philosophical argument which, if explored to the full extent, might making writing this book impossible. So we will established that a substance that can be thought as composed of only one constituent and a single phase is homogeneous. Substances composed of two or more phases and/or two or more separate constituents are heterogeneous. A mixture of water and ice is heterogeneous while a mixture of water and salt is homogeneous if temperature, density etc. do not vary within the system. Soils, rocks and sediments are invariable heterogeneous systems, although it is often useful to assume that soils where the physical in which the properties are at near equilibrium everywhere are homogeneous systems. Even dry sand composed only of quartz grains might be thought as heterogeneous because it is a mixture of quartz and air. In nature, most soils are heterogeneous mixtures composed of various minerals, amorphous inorganic compounds, organic matter, air and a solution composed of water and various dissolved substances. In later chapters we will discuss homogeneity in respect to dynamical physical properties, but that will be in a different context.  A porous medium can therefore be thought as a system which can be subdivided into subsystems. For many purposes these systems can be modeled using thermodynamics.

\section{Some thermodynamics I}

Classical thermodynamics is concerned with the macroscopic properties of matter. In this sense it usually deals with properties that represent an average of microscopic systems. The bridge between classical thermodynamics and the microscopic world is given by the discipline of \emph{statistical mechanics}. Many derivations in fluid mechanics and soil physics are grounded in classical thermodynamics. The entire theory of water retention in soils is derived from classical thermodynamics, although we are led to believe that it could be derived from first principles from statistical mechanics. Classical thermodynamics does not take into account quantum effects which can be studied in statistical mechanics. On a side note, general relativity can also be accounted in fluid mechanics, especially when dealing with gas ejects in astrophysics, in the discipline of \emph{relativistic fluid mechanics}. However, water and other fluids in porous media in Earth's gravitational fields, move slowly enough and are studied in scales large enough that classical mechanics and thermodynamics usually suffice. You could however use the theories presented in this book to study fluid flow in Mars or other planets, as long as the conditions of scale and velocity hold. In thermodynamics\footnote{Classical thermodynamics if not specified otherwise.} simple systems can be macroscopically characterized by extensive properties such as internal energy, volume and number of moles. What is meant by extensive variable is that if the system is subdivided into subsystems, the magnitude of the extensive variable into each subsystem is less than the total. Extensive variables depend on the extension of the system. Intensive variables can be defined as ratios of extensive variables and do not vary with extension or location where the measurements are taken in an homogeneous system in equilibrium. Thus, temperature can be defined as the partial derivative of internal energy, $U$,  with respect to entropy, $S$, while volume, $V$, and number of moles, N$_r$, of dissolved substances are held constant, represented as
\begin{equation}
T = \left( \frac{\partial U}{\partial S} \right)_{V, N_1, ..., N_r}
\end{equation}
The others being pressure, $P$, defined as the partial derivative of internal energy with respect to volume, V, with S and N$_r$ constant, and the electrochemical potential of a given component, defined as the internal energy divided by the number of moles of the component, N$_j$, while keeping the number of moles of other components, N$_r$, constant, along with $S$ and $V$.  
\begin{equation}
P = -\left( \frac{\partial U}{\partial V} \right)_{S, N_1, ..., N_r}
\end{equation}
\begin{equation}
\mu_j = -\left( \frac{\partial U}{\partial N_j} \right)_{S, N_1, ..., N_r}
\end{equation}
We can generalize the concept of intensive and extensive variables to commonly measured properties of materials, such as mass and volume, generating a number of well known intensive properties, the most common of which probably being density, represented in this book by the Greek letter $\rho$ (Appendix A), defined by the mass of a substance, $M$, divided by its volume, $V$  
\begin{equation}
\rho = \frac{M}{V}
\end{equation}
It is often convenient to define an area density as some extensive property divided by the area (as we did when explaining the REV concept), and a linear density as some extensive property divided by length. For example, a thin wire can have its linear density calculated by dividing its weight by its length. 

\section{Dimensions} %and the Buckigham $\pi$ theorem %} \section{Dimensions} 

Measurements and calculations of physical systems are often linked to some system of units. The most common systems are the International System of Units (SI), the CGS (centimeter-gram-seconds) and to some extent the British Imperial Units. In this book, whenever possible we will adhere to the international system of units. The SI uses seven base units, the second (s) for time, the meter (m) for length, the kilogram (kg) for mass, the ampere (A) for electric current, the kelvin (K) for temperature, the mole (mol) for amount of substance and the candela (cd) for luminous intensity. A number of  derived units can be defined from the SI, for example the newton (N = kg m s$^{-2}$) for force, the pascal (Pa = kg m$^{-1}$ s$^{-2}$) for pressure and stress and the joule (J = kg m$^{2}$ s$^{-2}$) for energy, work and heat.  
It is convenient, however, for many derivations and for checking consistency or units, to express the dimensions of physical quantities independent of a system of units. For elegance, in this book, whenever dimensions are considered when expressing a physical quantity, square brackets [] will be used. The elementary dimensions and symbols most often used in soil physics and groundwater hydrology are length [L], mass [M], time [T] and absolute temperature [$\Theta$]. Therefore, using the base length dimension L, length is represented as [L], area [L$^2$] and volume [L$^3$]. 

The consistence of dimensions and units can be calculated for a physical quantity. For pressure for example we have
\begin{equation}
P = \frac{\text{force}}{\text{area}} = \frac{N}{m^2} = \frac{kg~m~s^{-2}}{m^2} =  \frac{kg~m}{m^2~s^{2}} = \frac{kg}{m~s^{2}} = \text{Pa}   
\end{equation} 
Alternatively, in terms of dimensions
\begin{equation}
P =  \frac{[M~L~T^{-2}]}{[L^2]} =  \frac{[M~L]}{[L^2 T^{2}]} = \frac{[M]}{[L~T^{2}]} = [M L^{-1} T^{-2}]  
\end{equation} 
The last two terms on the right being representations of the same thing, any of which can be chosen for convenience. 

\section{Elementary intensive properties}

A range of intensive properties are defined in soil physics, soil mechanics and geology which are useful to characterize porous materials. We will, for the most part, use the formulas and definitions encountered in soil physics but several are analogous or are simple adaptations from soil mechanics or chemistry. 

Dry bulk density\footnote{because \emph{wet} bulk density is rarely considered in soil physics, the term ``bulk density" often refers to ``dry bulk density".} is the mass of dried material over volume. Dry bulk density can be measured by collecting an undisturbed sample of the material, by using a coring device and measuring the overall volume of the core. The sample is usually oven dried at 105 $^{o}$C for 24h to 48h to determine its dried mass be weighing. The measured volume includes pores and therefore, the more porous the material is the lower its bulk density is. 
\begin{equation}
\rho_b = \frac{m_s}{V_t}
\end{equation} 
In which $\rho_b$ is the dry bulk density in [M L$^{-3}$], $m_s$ is the dry soil mass [M] and V$_t$ is the total volume [L$^3$]. Bulk density can be modified by artificial compaction or consolidation due to applied loads or by natural processes such as pore filling with finer particles or diagenetic precipitation or consolidation in sedimentary deposits. Due to low natural bulk densities soils can be easily compacted by agricultural machinery, grazing, and other human activities, compaction being a process directly linked to soil degradation. Bulk density is determined by porosity, grain size and mineralogy of the solid particles. In most soils, bulk densities ranges from less than 1000 $kg~m^3$ in oxidic and organic soils under natural conditions to more than 1300 $kg~m^3$ in sandy or severely compacted clay soils.  In sedimentary deposits the density will depend on factors such as mineralogical composition, age of deposition and depth below overburden layers.

Particle density is analogous to the ``true" density of the material, considering only the mass and volume occupied by the solid particles. For accurate measurement of particle density, accurate measurement of solid particles volume is needed. This is usually achieved by liquid or gas displacement methods, where the volume occupied by solid particles is equivalent to the displaced volume of liquid or gas at a given temperature. 
\begin{equation}
\rho_s = \frac{m_s}{V_s}
\end{equation} 
In which $\rho_s$ is the (solid) particle density in [M L$^{-3}$], $m_s$ is the dry soil mass [M] and V$_p$ is the volume occupied by solid particles [L$^3$]. Particle density represents a physical averaging of the individual densities of the components of the solid phase. Soil and sediment particle density is often assumed as 2650 kg m$^{-3}$, the density of the mineral quartz, which is abundant in many soils and rocks. However the density can be lower in materials rich in organic particles and higher in materials rich in certain heavy iron and titanium oxides. 

The volumetric porosity is the volume of pores per unit total volume of soil including solids and pores. It is the fraction of the material that is occupied by voids. In this calculation we are not concerned if the voids are occupied by gas or water. 
\begin{equation}
\phi = \frac{V_v}{V_t}
\end{equation} 
In which $\phi$ is the total porosity [L$^{3}$ L$^{-3}$] (meaning volume of pores per total volume and thus the units are often not canceled), $V_v$ volume of voids, or pores [L$^3$] and V$_t$ is the total volume of the porous media [L$^3$]. Because it might be difficult to accurately measure the volume of voids in soil and as particle and bulk densities are routinely measured, it is convenient to calculate the total porosity as 
\begin{equation}
\phi = 1 - \frac{\rho}{\rho_s}
\end{equation} 
Equation 2.10 can be obtained by isolating the volumes in Equations 2.7 and 2.8, replacing into Equation 2.9 and assuming that the dry soil masses are the same. Bulk density and total porosity are sensitive to compaction, pore clogging and alterations in pore structure, either anthropogenic or caused by natural processes. Traffic by machinery, humans and animals and loads by buildings and roads can directly affect bulk density and porosity, which will in turn affect fluxes of matter and energy in porous materials. Particle distribution, shape and sorting can also affect bulk density and porosity. Density and porosity are important parts of the characterization of the solid matrix and pore space. 

Another important property used for the characterization of porous materials is the specific surface area (SSA). 
\begin{equation}
SSA = \frac{A_s}{m_s}
\end{equation} 
In which $A_s$ is the surface area of the particles and $m_s$ is the mass of particles.  The specific surface area is in [L$^{2}$ M$^{-1}$] and represents the surface area of solid particles per unit mass of particles. It is usually expressed in  $m^{2}~g^{-1}$ or $m^{2}~kg^{-1}$, although the former is usually preferred because it results in smaller numbers.  Surface area varies with grain size, roundness and surface rugosity, and with granulometric distribution of a material, and materials composed of finer particles will usually have larger specific surfaces areas. Surface area influences the reactivity of the material to water and contaminant retention via adsorption and other mechanisms and in the natural accumulation of organic matter.  Like many other properties discussed in this book, SSA is an operational concept that depends on the method used for its determination.  It can be determined by adsorption of nitrogen, water, ethylene glycol monoethyl ether (EGME) and other substances in vapor phase, and values in the range of 1 to 200 $m^{2}~g^{-1}$ can be found in soils depending on grain size distribution \cite{leaotuller14}. The most common method found in the literature is nitrogen ($N_2$) adsorption using BET\footnote{From the authors of the study that presented the adsorption isotherms used Brunauer, Emmett and Teller} method. However, BET could underestimate SSA in certain materials because the $N_2$ molecule might be too large to enter interlayer spaces and some very small pores  \cite{leaotuller14}. 

Equally important in soil physics in the quantification of the the amount and volume of the fraction occupied by fluid. In theory the fraction occupied by different fluids can be calculated in a similar fashion. Fluids of interest in porous media physics are, in addition to water, air, natural gas, petroleum, and a wide range of contaminants such as DNAPL\footnote{Dense non-aqueous phase liquids}, LNAPL\footnote{Light non-aqueous phase liquids}, industrial oils and a range of compounds dissolved in water at various concentrations. We will use the symbol $\theta$ for representing water mass and volume content in porous media because it has been traditionally used in soil physics, but the calculation can be extended to other dense fluids, and to some extent to gases in the case of volumetric saturation.   

The gravimetric water content is the mass of water per mass of dry soil or sediment material. It is usually measured by weighing (gravimetrically) a portion of material before and after drying in an oven at 105 $^{o}$C for 24h to 48h.  
\begin{equation}
\theta_g = \frac{m_w}{m_s}
\end{equation} 
In which $\theta_g$ is the gravimetric water content [M M$^{-1}$], meaning mass of water per mass of dry soil or sediment and thus the units are often not canceled, m$_a$ is the mass of water [M] and m$_s$ is mass of dry soil or sediment [M]. Gravimetric water content in excess of 1 kg kg$^{-1}$ is possible for highly porous materials such as volcanic ash and materials rich in organic matter. Because gravimetric water content does not take into consideration the volume available for water storage and total pore space, it is much more useful to define a volumetric water content, or the volume of water per unit volume of porous media 
\begin{equation}
\theta = \frac{V_w}{V_t}
\end{equation} 
In which $\theta$ is the volumetric water content [L$^{3}$ L$^{-3}$], meaning volume of water divided by total volume of soil or sediment and thus the units are often not canceled, V$_a$ volume of water [L$^{3}$] and V$_t$ is the total volume of dry soil or sediment [L$^{3}$], including pores. Defining the density of water as 
\begin{equation}
\rho_w = \frac{m_w}{V_w}
\end{equation} 
and isolating the volumes in Equations 2.8 and 2.14 and replacing into Equation 2.13 we have
\begin{equation*}
\theta = \frac{m_w/\rho_w}{m_s/\rho_b} = \frac{\rho_b}{\rho_w} \frac{m_w}{m_s}   
\end{equation*} 
which with Equation 2.9 can be written as 
\begin{equation*}
\theta =  \frac{\rho_b}{\rho_w} \theta_g   
\end{equation*} 
For simplicity, water content is usually calculated from this equation because bulk density and gravimetric water content are routinely measured in laboratory. The density of water can be calculated from temperature, as we will see in later chapters\footnote{Although it is often assumed as 1 g cm$^{-3}$ or 1000 kg m$^{-3}$, depending on the unit system adopted.}.
Another quantity seen in soil physics, soil mechanics and petroleum engineering is the degree of saturation. The degree of saturation is the volume of water divided by the volume of voids
\begin{equation}
S_w = \frac{V_w}{V_v} = \frac{\theta}{\phi}
\end{equation} 
The degree of saturation can theoretically vary from 0 in a dry media to 1 in a completely water saturated media and it is usually represented as a dimensionless number.  Another quantity used in soil mechanics is the void ratio
\begin{equation}
e = \frac{V_v}{V_s} 
\end{equation} 
In which V$_s$ is the volume of solids [L$^{3}$]. It is left to the reader to prove the following relationships
\begin{equation*}
e = \frac{\phi}{1 - \phi} 
\end{equation*} 
\begin{equation*}
\phi = \frac{e}{1 + e} 
\end{equation*} 
The void ratio is used in geotechnical engineering to evaluate volume changing in foundations under loading.

\section{A stochastic representation of porous media}

\section{List of symbols for this chapter}

\begin{longtable}{ll}
    	$ T $ & Temperature \\
    	$ U $ & Internal energy\\
 	$ S $ & Entropy\\
 	$ N_r $ & Number of moles of the r$^{th}$ substance  \\
 	$ P $ & Pressure \\
 	$ V$ & Volume \\
 	$ \mu_j$ & Chemical potential of the j$^{th}$ substance  \\
 	$ \rho$ & Generalized density  \\
	$ M$ & Generalized mass \\
	$ [L]$ & Generalized length unit \\
 	$ [M]$ & Generalized mass unit \\
 	$ [T]$ & Generalized time unit \\
 	$ [\Theta]$ & Generalized temperature unit \\
 	$ \rho_b$ & Dry bulk density  \\
 	$ m_s$ & Dry soil (or sediment) mass   \\
	$ V_t$ & Total volume \\
 	$ \rho_s$ & Solid particles density  \\
 	$ V_s$ & Solid particles volume  \\
  	$ \phi$ & Volumetric porosity  \\
	$ V_v$ & Volume of voids (pores)  \\
 	$ SSA$ & Specific surface area  \\
 	$ A_s$ & Surface area  \\
  	$ \theta_g$ & Gravimetric water content  \\
	$ m_w$ & Mass of water   \\
	$ \theta$ & Volumetric water content  \\
 	$ \rho_w$ & Water density  \\
	$ V_w$ & Volume of water   \\
	$ S_w$ & Degree of saturation   \\
	$ e$ & Void ratio
\end{longtable}

% !TEX TS-program = pdflatex
% !TEX encoding = UTF-8 Unicode

% Example of the Memoir class, an alternative to the default LaTeX classes such as article and book, with many added features built into the class itself.

\chapter{The solid phase}
\label{ch3}

\section{Soils as condensed matter}

Conceptually, a soil is a material that has been subjected to the process, or set of processes, known as pedogenesis. Large fractions of Earth's crust is composed of rocks, these rocks are usually formed under intense pressure and/or temperature conditions. Rocks are divided into three main groups igneous, metamorphic and sedimentary. Igneous rocks are formed when magma is cooled, either within Earth's crust or when it flows to the surface in the form of lava. Metamorphic rocks are formed when any type of rock is subjected to intense pressure and/or temperature conditions within the crust such that minerals are subject to mineralogical and petrological transformations without complete fusion of all minerals of the rock at high temperatures. Sedimentary rocks are those formed by sediments that are usually transported to an accumulation area and then subjected to pressure and temperature within ranges of values usually much lower than those required for the formation of metamorphic rocks, finally transforming into a sedimentary rock by a set of processes called diagenesis. Because rocks are formed at high temperatures and pressures when compared to those at Earth's surface they tend to be thermodynamically unstable when exposed to lower pressures, temperatures and chemical environment close to the surface. In addition, rocks near surface are often exposed to rainwater, yearly and daily temperature variation and a variety of compounds and forces produced by organisms in the interface with the atmosphere. 

The transformations related to the conditions at Earth's surface caused by all these processes is called \emph{weathering}. At the right conditions, weathering can evolve into pedogenesis and the minerals present in the rocks can be altered or reprecipitated into forms stable at atmospheric conditions. While rocks can be composed of a range of minerals, usually dependent on rock genesis, soils usually concentrate phylossilicate clays, iron and aluminum oxides and tectossilicates such as quartz and feldspars, depending on weathering conditions. These forms tend to be much more stable in soil environment than other minerals found in rocks. Of course soils are a complex material with a large variety of minerals of different species, but most soils will have a predominance of oxides, phylossilicate clays and quartz in different proportions in addition to the organic phase. Because soils and sedimentary deposits and to some extent partially weathered rocks of all types tend to be more porous, compressible and with an interconnected pore network than other rock types, the material presented in this book tends to apply more directly to these materials. However, fractures and pores of different geometries can be found in other types of rocks and with the necessary theoretical constraints, the models can be applied to these materials as well \cite{barenblattetal}. Soils and sedimentary deposits, in particular, are unconsolidated media and therefore, the individuals particles and aggregates can be easily disturbed by a series of processes such as compaction, penetration of roots and organisms, traffic, and many others, while rocks are consolidated materials being able to withstand mechanical stresses of larger magnitude. The degree of consolidation of rocks varies with genesis and mineralogical composition, but it is safe to say that most rocks under normal conditions will act as a rigid matrix  when considering flow and transport phenomena.      

We mentioned that soils are formed from weathering and pedogenesis of rocks due to the minerals present in rocks being thermodynamically unstable in Earth's superficial environment. Technically the process is a lot more complex, soils can not only form from rocks but also from sediments and from previously existing soils, and under influence of anthropogenic activities such as deposition of matter and superficial disturbance. A given soil is in equilibrium with its environment, including climate, natural vegetation and geomorphological setting. If any of these factors is significantly altered a soil can evolve to more stable forms, regress, be complete buried or completely destroyed by erosional processes over time. Likewise, sediment and rock formation is very complex and dependent on several geological and geomorphological processes and on factors such as Earth's internal dynamic and climate. Because soil, rock and sediment forming processes are very complex and covered in different fields of soil science, geology and physical geography, discussing each one of them is beyond the scope of this book.   

An important issue in natural porous media physics is the composition of solid phase where mechanical and transport phenomena are being considered. Rocks, sedimentary deposits and soils of interest are largely composed of minerals. A mineral is a compound that has a more or less constant chemical composition between samples of the same mineral, has an internal structure that is organized and is formed naturally in Earth or other astronomical objects. A rock is composed of one or more minerals and is consolidated, i.e. cannot be easily disassembled by hand or using manual tools such as spades or a geological hammer. From the soil physics point of view, soils are much more complicated materials than rocks because they are very heterogeneous in composition and are not consolidated, being easily deformed and disaggregated with much less effort than that needed for most rocks. Soils possess varying grain size distributions, proportions of inorganic mineral and amorphous materials, organic phases, and gas and solution. Large variability in the composition of the mineral phase can be found depending on climate, parent material and other soil forming processes. The other factors that makes soils so complex is that the colloidal solid phase is most often electrically charged, mainly due to inorganic and organic superficial groups and isomorphic substitution in crystalline lattices. Because of that, the small particles have the ability to interact with charged ions in solution. Modeling contaminant retention and transport can be much more challenging in soils than in sedimentary rocks composed of a single coarse grained mineral phase, such as in sandstones and sand deposits, where the density of superficial charges is negligible and the specific surface area is very small compared to that of most soils.    

Because soils are so complex, they cannot be fully characterized and studied in detail in laboratory or in lattice, physicochemical or thermodynamic simulations as geologists do with individual minerals and simple rocks and physicists do with synthetic crystals. Although soil physics is traditionally constrained to the domains of fluid phenomena and soil mechanics, much can be done by using the same approach and techniques used by condensed matter physicists. The fact that the study and characterization of soils has been almost entirely relegated to soil science and engineering during most of its history reflects the fact that the complexity and heterogeneity of the material prevents, for the most part, the physicist's approach which is often concerned with materials that are simple enough that they can be investigated and modeled using available laboratory and mathematical methods. With appropriate simplifications and with modern physical methods, soils can be studied in the framework of condensed matter physics. 

Although we mentioned that soils are more complex from the physical point of view than most rocks and sedimentary deposits, there are geological settings where the line between soils, rocks and sediments is blurred, either because of a physical mixing of different materials due to pedogenetic and geological conditions and/or because a system where mechanical stresses and transport phenomena is considered has interfaces between different materials. One example of a material intermediary between soils and rocks is the \emph{saprolite} which is a partially weathered rock theoretically in its way to become a soil. Another example is sedimentary rocks and deposits composed of phylossilicate clays. These materials can result in chemical behavior when interacting with contaminants more closely related to soils than to inert materials. It is common in natural environments that soils occupy the most superficial layers in Earth's crust followed by saprolite, slightly weathered or fragmented or fractured rock followed by consolidated rock. A dissolved contaminant, for example, will have to interact with soil and saprolite before it reaches deeper saturated layers in porous rock, which can compose aquifers where water is being extracted. Ultimately the mineralogical composition of soils, rocks and saprolite will influence, if not control, the interaction with contaminants, water transport and storage properties as well as mechanical behavior. Other than inorganic materials, organic deposits, which constitute some soils and geological deposits, present an unique challenge from the physics and engineering point of view, as they tend to be very loose, compressible, can be easily lost due to decomposition, and can be very reactive when it comes to interacting with chemicals, depending on the nature of the organic fractions involved. Water retention and transport is particularly challenging to be modeled in these materials as they tend to have very low densities, can be very porous and can absorb a water mass greater than the mass of solids, resulting on a mass base water content greater than one. We will focus on materials that are mostly inorganic throughout this book, as these are materials comprise the Earth's crust upper layer in most of the planet, while organic deposits are restricted to certain geological and pedogenetic settings.       

\section{Mineralogy and crystallography}
One of the disciplines in which condensed matter physics and mineralogy intersect is crystallography. Most of us have an intuitive knowledge of what a crystal is, it is usually associated with ``rock crystals'' such as quartz. Early physicists or \emph{natural philosophers} observed that many minerals had an external shape that was more or less constant between different samples of the same mineral \cite{hammond}. They speculated that the external shape is the result of the internal organization or the packing of minuscule particles which compose matter. Later, these minuscule particles would be recognized as atoms and the internal organization of these atoms would be experimentally confirmed by x-ray diffraction \cite{bragg}.  You might have noticed that most varieties of quartz that people collect are formed by elongated crystals. If you count the number of faces in these crystals and measure the internal angle between faces you will find six and approximately 120$^{o}$. This is reflects the internal atomic structure of quartz, which in ideal conditions is composed entirely of silicon and oxygen atoms. Crystals are then materials that have a highly organized internal structure which repeats in space, and crystallography is the branch of science dedicated to the study of crystals. Materials which have a random or non-organized arrangement of atoms are called \emph{amorphous}. Crystals can be synthesized in laboratory  or can be formed naturally such as in soil and rock forming processes where atoms reorganized in response to a change in thermodynamic conditions. These inorganic crystals which are formed under natural conditions and which can be identified as having a more or less constant chemical composition between different species, and whose internal arrangement and physical properties make it possible to recognized as a distinct species from other crystals in the environment are called \emph{minerals}.  Minerals in rocks originate from rock forming processes or they can be inherited from a different rock when they pass through the rock cycle. Rocks are primarily classified based on origin as igneous, sedimentary and metamorphic. Igneous rocks are formed from cooling of magma within Earth or lava in Earth's surface. As the materials cools, bonds are formed between atoms and the structure organizes itself based on the affinity between atoms, remaining concentration and atomic radius relationships. The issue of \emph{radius ratio} is usually modeled using the sphere models presented Chapter \ref{ch6}, but for atoms instead of mineral particles or aggregates as in that case. Because igneous rocks form from magma or lava which are molten materials, the minerals are newly formed when the rock finishes cooling.  This is not always the case in other types of rocks. Sedimentary rocks are formed from erosion, transport, deposition and consolidation of sediments which came from another source, usually either a rock that was weathered or a soil or sedimentary deposit which was eroded. In reality, sedimentary rocks can be also formed from precipitation of ions in solution with or without the influence of organisms.  Sedimentary rocks can be composed of minerals that were formed in other rocks or other conditions and were weathered and transported forming a new rock via consolidation without altering the chemical composition and crystalline structure or by minerals that were formed during diagenesis. Metamorphic rocks are formed by solid state transformations in which part or all of the minerals in a rock of any type are transformed into new minerals by the action of pressure and/or temperature at greater intensities than those that occur during the genesis of sedimentary rocks. During formation of metamorphic rocks there can be re-composition and/or reorganization of the internal structure of minerals, thus forming new species. Partial fusion of some minerals is allowed, but if the rock is completely molten and later recrystallized it would be classified as an igneous rock. 

\begin{figure}[h]
\centering
 \includegraphics[width=1.0\textwidth]{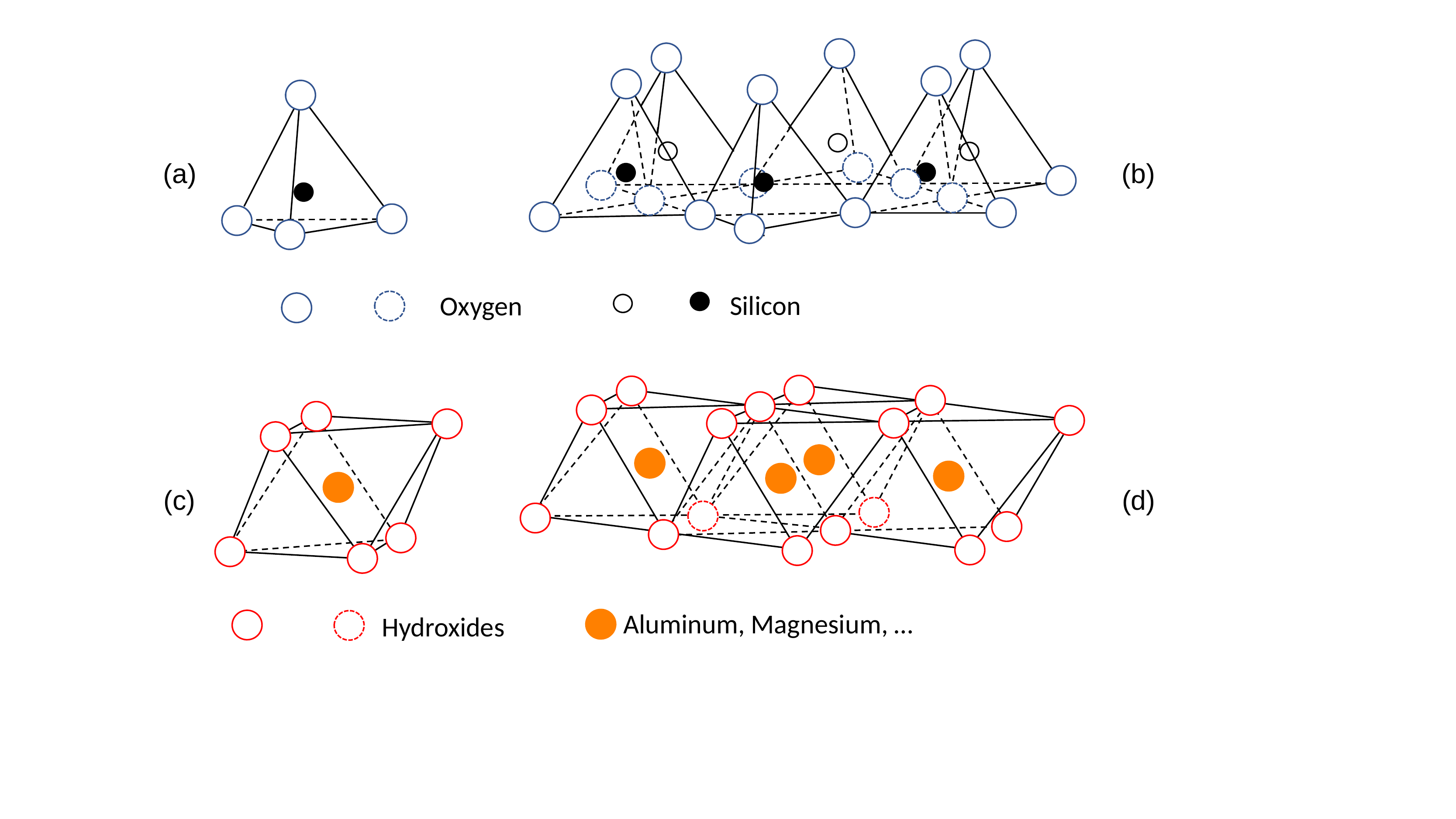}
\caption{Single silicon tetrahedron (a) and tetrahedral sheet (b) and sigle octahedra (c) and octahedral sheet (d) (Adapted from \cite{grim}).}
\label{ch3_fig1}
\end{figure}

Rocks are composed of minerals, there are rocks composed of only one mineral species, such as some marbles and sandstones and rocks composed of more than one type of mineral such as basalts and granites. There are minerals that are more commonly found in each type of rock and this is used in the classification and identification of rocks. The chemical composition of the most common minerals reflects the composition of Earth's continental crust in which oxygen, silicon, aluminum, iron, calcium, sodium, potassium and magnesium account for approximately 98\% of Earth's crust on a mass basis. Oxygen and silicon alone account for more than 70\% of the mass. Minerals are classified based on the chemical composition usually represented as a chemical formula. Although some minerals are composed by a single element bonded via ionic or covalent bond, diamond (C), graphite (C), sulfur (S), gold (Au) being a few examples, most minerals of interest in rocks, sediments and soils are composed of one or more cations bonded to an anion or anionic group. These cations and anions will bond to form stable configurations based on valence, affinity, abundance and radius ratio and according to thermodynamic laws. The two most common elements on Earth's crust are an anion and a cation, oxygen (O$^{2-}$) and silicon (Si$^{4+}$). Based on this configuration one would expect that a mineral with formula SiO$_2$ with one silicon bonding with two oxygens would be a stable configuration. Although SiO$_2$ is the formula of a common silicate mineral, the radius ratio precludes the coordination of only two oxygen atoms around the silicon. In reality, each silicon atom coordinates four oxygen atoms and shares one electron with each oxygen via a predominately covalent bond. This structure with one silicon coordinating four oxygen atoms is represented as a polyhedron called the silicon tetrahedron (Figure ~\ref{ch3_fig1}(a)) and forms the basis of the \emph{silicate minerals} which are extremely common in soils, sediments and rocks. Because each oxygen has only one electron shared with the silicon atom, one more electron needs to be shared in each corner of the tetrahedron for physicochemical stability to occur. In silicates the stability can be achieved by other cations forming ``bridges'' between tetrahedrons defining  the class of \emph{nesossilicates}, by sharing the apical oxygen between two tetrahedra forming a double tetrahedra in the \emph{sorossilicates}, by sharing two oxygens between adjacent tetrahedra forming ring of three or six tetrahedra in the \emph{cyclossilicates}, by sharing two or three oxygens forming chains in the \emph{inossilicates}, by sharing the three basal oxygens forming sheets of tetrahedrons in the \emph{phyillossilicates} (Figure ~\ref{ch3_fig1}(b)) or by sharing all four oxygens forming complex tridimensional networks in the \emph{tectossilicates}. 

The chemical formula in each of these classes is defined by the occurrence of Si$_n$O$_m$ in the formula in which $n$ and $m$ are whole numbers or fractions. The class of phylossilicates is particularly important in soils and a few sedimentary rocks as these clay minerals can define or heavily influence the physical and chemical behavior or the material in which they occur, they can also have surface charges and can interact with ions in solution. A few examples of silicates common in rocks are quartz (SiO$_2$), orthoclase feldspar (KAlSi$_3$O$_8$), plagioclase feldspars ranging from calcium the plagioclase anorthite (NaAlSi$_3$O$_8$) to the sodium plagioclase albite (CaAl$_2$Si$_2$O$_8$), olivine ((Mg, Fe)$_2$SiO$_4$) and various pyroxenes. These minerals are commonly found in igneous rocks and can appear in sedimentary and metamorphic rocks as an inheritance from the original material which either suffered diagenesis or metamorphism. Other mineral classes are found in many types of rocks usually in smaller quantities, a few examples being oxides such as zircon (ZrSiO$_4$), ilmenite (FeTiO$_3$) and rutile (TiO$_2$) and sulfides such as spharelite (ZnS) and galene (PbS). In sedimentary rocks, besides quartz which is common in sandstones and also in sedimentary deposits, it is also relatively common to find minerals that formed in solution via precipitation such as calcite and aragonite (both CaCO$_3$, but with different internal structures), dolomite (CaMg(CO$_3$)$_2$), gypsum (CaSO$_4\cdot$ 2H$_2O$), anhydrite (CaSO$_4$) and salts such as halite (NaCl) and sylvite (KCl). These minerals are usually composed primarily by ionic bonds and tend to be less resistant to weathering.

Soils are formed by weathering and pedogenesis over rocks, sedimentary deposits or either previously formed soils. They can inherit minerals from the soil forming material that have gone through no discernible mineralogical transformation except size reduction and fragmentation. These minerals are called \emph{primary minerals} and the amount in soil will depend on parent material and climate. Common primary minerals found in soils are quartz, feldspars, primary micas (a group of phylossilicates). Heavy minerals such as rutile, anatase (also TiO$_2$) and magnetite (Fe$_3$O$_4$), tend to occur in very minor quantities in some soils. More often than not new minerals are created during pedogenesis and these minerals control how the soil functions, its porosity, pore geometry, mechanical properties and how it interacts with water, gas and with solutes. The minerals that are formed during pedogenesis are often referred to as \emph{secondary minerals} while the minerals inherited from parent material are the primary minerals. On the tropics where average annual temperatures and precipitation are elevated, at least in part of the year, soils tend to accumulate iron and aluminum oxides, oxi-hydroxides and hydroxides as secondary minerals, commonly hematite (Fe$_2$O$_3$), goethite (FeOOH) and gibbsite (Al(OH)$_3$), and usually referred to as a group as iron and aluminum oxides. Iron oxides usually give strong red, orange and yellow hues to soils and are usually not restrictive to agricultural use of soils from a physical standpoint but can be severely restrictive from a soil fertility standpoint under natural conditions. Soils rich in oxides are challenging for modeling water and contaminant transport and storage because they usually are associated with unique pore geometry and distribution and because surface hydroxi groups (OH) these minerals are subjected to protonation and deprotonation according to solution pH following the reactions \cite{sparks}

\begin{equation}
\text{M} - \text{OH} + \text{H}^+  \rightleftharpoons  \text{M} - \text{OH}_2^+  
\end{equation}
\begin{equation}
\text{M} - \text{OH} \rightleftharpoons  \text{M} - \text{O}^- + \text{H}^+
\end{equation}

Thus, under acidic conditions the OH groups protonate, generating a positive charges surfaces with ability to attract anions while at higher pHs the OH groups deprotonate generating negatively charged surfaces with the ability to attract cations.  At an equilibrium point where there is an equilibrium between positive and negative sites the \emph{points of zero charge} (PCZ) of the compound can be found. In soils this behavior occurs in iron and aluminum oxides such as hematite, goethite and gibbsite where M represents iron (Fe) or aluminum (Al) metals usually on octahedral sites on the mineral structure. However these variable charged sites can be found in organic compounds and in silicates to some degree. Roughly speaking, the PZC of a soil represents a weighed average of PZCs of individual components. Cation exchange capacity (CEC) and anion exchange capacity (AEC) in soils are defined by the predominance of negative or positive surface sites. AEC is mainly observed in highly weathered oxidic soils in the tropics and can revert to CEC as soil pH increases. 

%In phylossilicate minerals the density of these sites is low and the number of variable charged sites is low when compared to permanent charges generated by other mechanisms.    

\begin{figure}[ht]
\centering
 \includegraphics[width=0.85\textwidth]{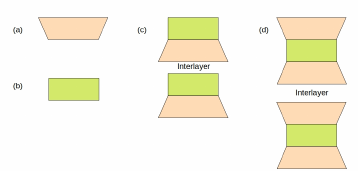}
\caption{Representation of the tetrahedral (a) and octahedral (b) sheets and 1:1 (c) and 2:1 (d) layer minerals.}
\label{ch3_fig2}
\end{figure}

The other major group of secondary minerals in soils are the phylossilicates or \emph{clay minerals}. Although these minerals can occur as primary minerals in sedimentary deposits and rocks, they occur in almost all soils in various quantities depending on parent material and pedogenetic conditions. Clay minerals along with iron and aluminum oxides can be found on the clay fraction of soils while primary minerals usually occur in the silt and sand fractions. Phylossilicate minerals in soils are layered or sheet minerals where there is an alternation of sheets formed by silica tetrahedrons and octahedral layers where metallic cations such as aluminum or magnesium coordinate six hydroxides forming an octahedral polyhedron of coordination (Figure \ref{ch3_fig1}(c)) in which some of its hydroxides are shared among adjacent octahedrons (Figure \ref{ch3_fig1}(d)).  Phylossilicate clays are classified according to the proportion of octahedral (Figure \ref{ch3_fig2}(a)) to tetrahedral (Figure \ref{ch3_fig2}(b)) layers. In 2:2 layer silicates, often simplified to 1:1, two octahedral layers bound to two tetrahedral layers (Figure \ref{ch3_fig2}(b)), while in 4:2 layer silicates, often simplified to 2:1, two tetrahedral layers are each bound to one octahedral layer forming two pairs of 2:1 structures (Figure ~\ref{ch3_fig2}(c)). The space between the pairs of sheets in the 1:1 configuration and pair of groups of sheets in the 2:1 configuration is called in the interlayer space and the distance between the groups of 1:1 or 2:1 sheets can vary according to environmental conditions in some types of minerals. 

\begin{figure}[ht]
\centering
 \includegraphics[width=0.6\textwidth]{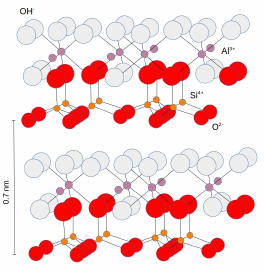}
\caption{Kaolinite structure (Adapted from \cite{dixon}).}
\label{ch3_fig3}
\end{figure}

The most common 1:1 layer mineral in tropical soils is kaolinite (Al$_2$Si$_2$O$_5$(OH)$_4$) and because the space between layers does not vary and because it has little \emph{isomorphic substitution} it has relatively low specific surface area, ion retention and exchange capacity.   The structure of  kaolinite  is presented in Figure \ref{ch3_fig3} illustrating the tetrahedral and octahedral sheets.  The distance between the basal planes of the two 1:1 layers does not vary with environmental conditions and is around 0.7 nm. On the octahedral sheets some of the hydroxides are replaced with oxygen and these are shared with silicon tetrahedrons forming the mineral structure (Figure \ref{ch3_fig3}).  Isomorphic substitution is a process that occurs during the mineral formation in which the central cation is replaced by a cation of similar atomic radios but with different valence. Thus if aluminum (Al$^{3+}$) is replaced by iron in its reduced state (Fe$^{2+}$) in specific point within the octahedral sheet, a charge deficit will appear due to the resulting unbalanced negative charges in hydroxides. These charges are the basis of the chemical reactivity and will control the dynamics of nutrients and contaminants in soils and other materials rich in phylossilicate clays. The most common 2:1 phylossilicate minerals in soils are from the smectite group, particularly montmorillonite ((Na,Ca)$_{0.33}$(Al,Mg)$_2$(Si$_4$O$_{10}$)(OH)$_2 \cdot$ nH$_2$O), and vermiculite ((Mg,Fe$^{2+}$,Fe$^{3+}$)$_3$[(Al,Si)$_4$O$_{10}$](OH)$_2 \cdot$ 4H$_2$O). Montmorillonite is highly expansible, meaning that the interlayer space can increase with the entrance of water molecules, has high specific surface area, is very reactive towards adsorption of water and solutes and can have shrinking swelling behavior according to soil water content, being extremely hard when dry and extremely plastic and sticky when wet.  Smectite's structure is illustrated in Figure \ref{ch3_fig3} showing the range of variation of the distance between the two 2:1 structures. Isomorphic substitution of the cations on the octahedral layer is common in smectites (Figure \ref{ch3_fig3}) and other 2:1 minerals and because of that, the surface charge and CEC of these minerals is much larger than oxides and 1:1 minerals. The interlayer can be filled with water and cations under wet conditions, expanding the mineral on a microscopic scale and consequently the material on a macroscopic scale. Another less common 2:1 mineral is vermiculite which is less expansive than montmorillonite. Illite  ((K,H$_3$O)(Al,Mg,Fe)$_2$(Si,Al)$_4$O$_{10}$[(OH)$_2$,(H$_2$O)]) can also occur in some soils  in minor amounts depending on genesis and parent material, but the interlayer space in this mineral usually does not vary and it tends to be less reactive than either montmorillonite or vermiculite. There are also hydroxi-interlayered minerals in which the interlayer space is occupied by octahedral sheets or natural polymers, but these are of restricted occurrence in terms of amount and representative and are of interest only in specific cases.  Anyone interested in soil physics or in unsaturated zone hydrogeology must be proficient in soil chemistry and mineralogy. It is impossible to study these soils and sediments without understanding how the solid phase interacts with fluids and gases, these disciplines cannot be dissociated. The reader is strongly recommended to consult books on the subject of soil chemistry and mineralogy \cite{sposito84, essington03}.    

\begin{figure}[ht]
\centering
 \includegraphics[width=0.6\textwidth]{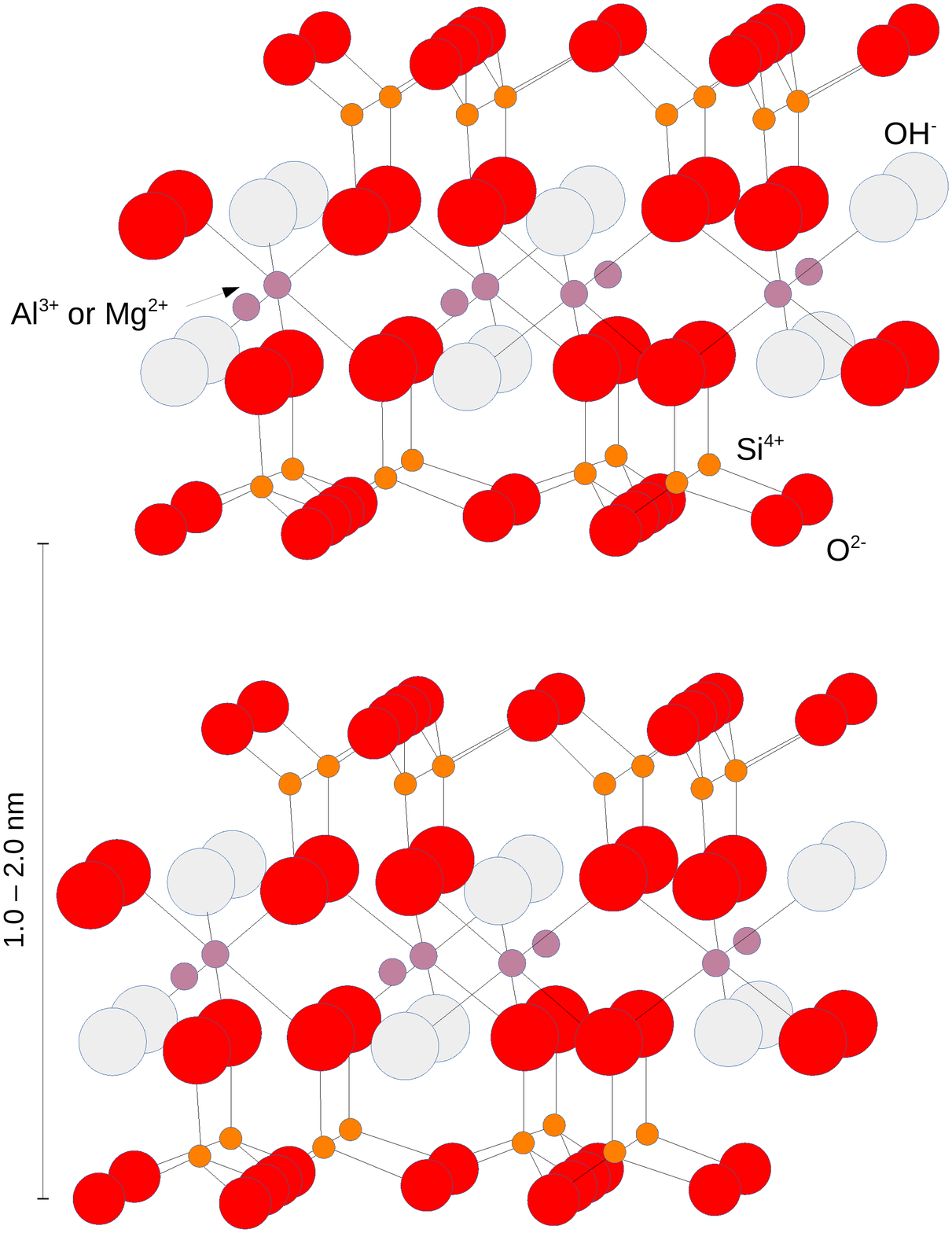}
\caption{Smectite structure (Adapted from \cite{borchardt}).}
\label{ch3_fig4}
\end{figure}

%\section{Bragg's law and X-ray diffraction analysis}

\section{Soil surface chemistry}
Because soil secondary minerals, colloidal organic matter and some amorphous compounds are charged, an electric field around those compounds results which can attract ions in solution. The liquid phase solution in this case is in the pores around solid particles and the volume occupied by it can range from complete saturation of the pores to thin films over solid surfaces. Ions in solution will form a distribution based on the equilibrium of the electrostatic attractive and repulsive forces in particle surfaces and the concentration gradient in solution. The interaction of the charged surfaces and ions results in a structure called Electric Double Layer (EDL). The presence of an EDL and its relationship with the different types of surface sites in soil minerals and organic compounds will influence the behavior of clay particles in aqueous solution as in the case of grain size analysis and in the stability of the soils in the field. The mechanisms of dispersion and coagulation play a fundamental role in soil dispersion both in laboratory and under field conditions. The surface chemistry will also determine parameters for solute transport and retention in soils. The solutes of interest include inorganic and organic compounds in fertilizers, polar and non-polar pesticides, viruses, bacteria and other pathogenic microorganisms, heavy metals, radionuclides and many other.  Solutes can also be transported attached to dispersed clays and other colloids in what is called \emph{colloid-facilitated transport}. 

The EDL can be modeled using the diffuse double layer theory, which assumes the mineral surfaces as a plane with an even distribution of charges. Under these conditions, the distribution of charges at a distance $x$ from the surface plane under equilibrium conditions follows the Boltzmann distribution \cite{babcock, gast}

\begin{equation}
n_i = n_{0i} \exp{(\frac{-z e \psi(x)}{ kT})}
\end{equation}
In which $n_i$ is the ion concentration of the i$^{th}$ species at a distance $x$ from the surface, $\psi(x)$ is the electrical potential at the same distance, $n_{0i}$ is the ion concentration of the i$^{th}$ species in bulk solution, $z$ is the ion valence, $e$ is the elementary charge, $k$ is the Boltzmann constant and $T$ is the absolute temperature.  The charge density at any point in the diffuse layer is given by 
\begin{equation}
\rho = \sum z_i e n_i
\end{equation}
In which $i$ indicates a summation term over every ion species. 
The variation of the electrical field strength with distance is given by a differential equation know as the Poisson equation
\begin{equation}
\frac{d^2 \psi}{dx^2} = - \frac{4 \pi}{\epsilon} \rho
\end{equation}
in which $\epsilon$ is the absolute permittivity of the medium.  For symmetrical electrolytes, the Poisson equation can be expressed in terms of ion concentration and charge density by combining it with the concentration distribution and charge density equations, resulting in the Poisson-Boltzmann equation 
\begin{equation}
\frac{d^2 \psi}{dx^2} = - \frac{4 \pi  }{\epsilon}  \sum z_i e n_{0i} \exp{(-\frac{z_i e \psi}{kT})} 
\end{equation}
For suspensions consisting of a single symmetrical electrolyte the summation expands to \footnote{This can be achieved by considering the valence of the positively charged ions as $z_{+} = +|z|$ and the valence of the negatively charged ions as $z_{-} = -|z|$. Note that a symmetrical electrolyte has the same proportion of cations and anions, the most simple example being NaCl which dissociates to Na$^{+}$ and Cl$^{-}$ , therefore, their concentrations in solution are also equal, i.e. $n_{0+} = n_{0-}$ }

\begin{equation}
\frac{d^2 \psi}{dx^2} = - \frac{4 \pi  |z_i| n_{0i}}{\epsilon}  [ \exp{(-\frac{|z_i| e \psi}{kT})}   - \exp{(\frac{|z_i| e \psi}{kT})}     ] 
\end{equation}
To solve this differential equation several simplifications and manipulations are necessary. The first is to define
\begin{equation}
y = \frac{|z_i| e \psi}{kT} 
\end{equation}
resulting in 
\begin{equation}
\frac{d^2 y}{dy^2} = - \frac{4 \pi  z_i^2 e^2 n_{0i}}{\epsilon k T}  [ \exp{(-y)}   - \exp{(y)} ] 
\label{eqn:ch3eq1}
\end{equation}
It is also useful to expand the exponential term using Taylor series \footnote{Note that $\exp{(-y)}   - \exp{(y)}$ can also be written as $-2 \sinh(y)  $ and $\sinh(y)$ can be written as a Taylor series expansion, as it will be used again later} 
\begin{equation}
 \exp{(-y)}   - \exp{(y)} = -2y - \frac{y^3}{3} - \frac{y^5}{60}  + ...   
\end{equation}
If the solution is dilute the Debye-Hückel approximation $-\frac{|z_i| e \psi}{kT} \ll 1$ is used and higher order terms on the series can be neglected. Under such conditions the expansion can be simplified to 
\begin{equation}
 \exp{(-y)}   - \exp{(y)} \approx -2y    
\end{equation}
Replacing the expanded exponential into the original equation
\begin{equation}
\frac{d^2 y}{dy^2} = - \frac{4 \pi  z_i^2 e^2 n_{0i}}{\epsilon k T}  (-2y) = \frac{8 \pi  z_i^2 e^2 n_{0i}}{\epsilon k T} y  
\end{equation}
We can now define 
\begin{equation}
\kappa^2 =  \frac{8 \pi  z_i^2 e^2 n_{0i}^2}{\epsilon k T}  
\end{equation}
The parameter $\kappa$ is an important parameter in surface chemistry and its inverse $\kappa^{-1}$ is the \emph{Debye length} or distance over which the electric potential decays to 1/e \footnote{e here is Euler’s number e = 2.71828...}. The differential equation to be solved then simplifies to 
\begin{equation}
\frac{d^2 y}{dy^2} = \kappa^2 y 
\end{equation}
This is a second order linear ordinary differential equation with general solution \footnote{The solution to this differential equation would require a lengthy detour into differential equations theory and will be provided in future editions}
\begin{equation}
y = c_1 \exp (\kappa x) + c_2 \exp (-\kappa x) 
\end{equation}
The constants can be determined by applying the boundary conditions to the problem. The first boundary condition states that the electrical potential $\psi = 0 $  at $x = \infty$. This implies that $y = 0$ at  $x = \infty$
\begin{equation*}
0 = c_1 \exp (\kappa \infty) + c_2 \exp (-\kappa \infty) = c_1 \exp (\kappa \infty) + 0   
\end{equation*}
\begin{equation*}
c_1 = 0   
\end{equation*}
The second is that the $\psi  = \psi_0$ and thus $y = y_0$  at x = 0 
\begin{equation*}
y_0 = 0 \exp (\kappa 0) + c_2 \exp (-\kappa 0) = c_2  
\end{equation*}
Replacing the constant on the general solution
\begin{equation}
y = y_0 \exp (-\kappa x) 
\end{equation}
Therefore
\begin{equation}
\psi = \psi_0 \exp (-\kappa x) 
\end{equation}
This is the electrical potential at the electric double layer following the Gouy-Chapman model with the Debye-Hückel approximation. Notice that this solution is valid for the potential in relation to an infinite plane. For different geometries and where the solute concentration is not low enough for the Debye-Hückel approximation to be valid, the solution for the Poisson-Boltzmann equation is not trivial (see for example \cite{sposito84} Chapter 5). The potential is $\psi_0$ at the surface and deceases exponentially as the distance $x$ from the surface increases. It is a simplistic model but it usually works well to explain double layer compression under high ionic strengths and the theory of surface forces balance that controls dispersion and coagulation of clays. This theory plays a fundamental role in the physicochemistry of salt affected soils, erosion and sedimentation, and grain size analysis. 

Let us now consider an alternative treatments to Equation \ref{eqn:ch3eq1}. From hyperbolic geometry identities, the hyperbolic sin function can be written in terms of exponential functions as
\begin{equation}\
\sinh y  =  -  \frac{\exp{(-y)}   - \exp{(y)}}{2}  
\end{equation}
Thus, replacing it into Equation  \ref{eqn:ch3eq1} results in
\begin{equation}
\frac{d^2 y}{dx^2} =  \frac{8 \pi  z_i^2 e^2 n_{0i}}{\epsilon k T} \sinh y 
\end{equation}
And from the definition of $\kappa$
\begin{equation}
\frac{d^2 y}{dx^2} =  \kappa^2 \sinh y 
\end{equation}
This is a nonlinear second-order differential equation and fortunately we only need to find how it reduces to a first-order equation on integration. The reason will become apparent in further steps. To reduce the equation to first order we multiply both sides by $dy/dx$ and integrate \cite{boas}
\begin{equation*}
\int \frac{d^2 y}{dx^2}\frac{d y}{dx} = \int \kappa^2 \sinh y \frac{d y}{dx} 
\end{equation*}
Substituting $u = dy/dx$ we have $du = d^2y/dx^2$ on the left side and making $v = y$ we have $dv = d^2y/dx^2$ on the left side 
\begin{equation*}
\int u du = \int \kappa^2 \sinh v dv 
\end{equation*}
which results in 
\begin{equation*}
\frac{u^2}{2} =  \kappa^2 \cosh v + C_1 
\end{equation*}
Where $C_1$ is an integration constant. Replacing u and v we have
\begin{equation*}
(\frac{d y}{dx})^2 =  2 \kappa^2 \cosh y + C_2 
\end{equation*}
in which $C_2 = 2C_1$ and it does not alter the solution to the problem. Applying the square root to both sides
\begin{equation*}
\frac{d y}{dx} = \sqrt{ 2 \kappa^2 \cosh y + C_2} 
\end{equation*}
Applying the boundary condition that the electrical potential and thus $y$ is zero when $x = \infty$ i.e. bulk solution, and that $dy/dx = 0$ at that point, we find that $C_2 = -2\kappa^2$ and
\begin{equation*}
\frac{d y}{dx} = \sqrt{ 2 \kappa^2 \cosh y - 2 \kappa^2} 
\end{equation*}
Finally we consider another geometric identity of the hyperbolic functions, the half-angle formula 
\begin{equation*}
\sinh \frac{y}{2} = \text{sgn}(y) \sqrt{ \frac{\cosh y - 1}{2}} 
\end{equation*}
In which sgn is the sign function with the following properties
\[
\text{sgn} (x)= 
\begin{cases}
-1 & \text{if}  ~ x < 0\\
0  & \text{if}  ~ x = 0\\
1  & \text{if}  ~ x > 0 
\end{cases}
\]
Considering that we are concerned with electrical potential and thus $y$ negative and rearranging and replacing the half-angle formula we have
\begin{equation}
\frac{d y}{dx} = - 2 \kappa \sinh \frac{y}{2} 
\end{equation}
This formula to will be useful to derive an equation for net charge on the diffuse layer, $\sigma$.  The net charge is the integral of the charge density on the $x$ interval of zero to infinity.  
\begin{equation}
\sigma = \int_0^\infty \rho dx
\end{equation}
Replacing the Poisson-Boltzmann equation into the surface charge equation we have
\begin{equation*}
\sigma = \int_0^\infty (- \frac{\epsilon}{4 \pi} \frac{d^2 \psi}{dx^2})  dx = - \frac{\epsilon}{4 \pi} \int_0^\infty  \frac{d^2 \psi}{dx^2}  dx 
\end{equation*}
The integration is over the second derivative, resulting
\begin{equation}
\sigma =   -\frac{\epsilon}{4 \pi}  \bigg(\frac{d \psi}{dx}\bigg)\bigg|_{x=0}^{x=\infty}   
\end{equation}
But the electrical potential  is constant at $x=\infty$ such that the term $d\psi/dx$ vanishes when evaluated at infinity, resulting 
\begin{equation}
\sigma =   \frac{\epsilon}{4 \pi}  \bigg(\frac{d \psi}{dx}\bigg)_{x=0}   
\end{equation}
Expressing sigma in terms of $y$
\begin{equation}
\sigma =   \frac{\epsilon k T }{4 \pi |z_i| e }  \bigg(\frac{d y}{dx}\bigg)_{x=0}   
\end{equation}
Now we can replace the equation for $dy/dx$ derived previously and the condition that $y = y_0$ at $x = 0$ resulting in
\begin{equation}
\sigma =   -\frac{\epsilon k T \kappa}{2 \pi |z_i| e }  \sinh \frac{y_0}{2}     
\end{equation}
Which in terms of electrical potential and replacing $\kappa$ is
\begin{equation}
\sigma =   -\frac{\epsilon k T }{2 \pi |z_i| e } \sqrt{\frac{8 \pi  z_i^2 e^2 n_{0i}^2}{\epsilon k T}}  \sinh \frac{\frac{|z_i| e \psi_0}{kT}}{2}  =    - \sqrt{\frac{2 \epsilon k T  n_{0i}^2}{\pi}}  \sinh \frac{|z_i| e \psi_0}{2kT}
\end{equation}
The surface charge, $\sigma_0$ has the same magnitude of the sum over the diffuse layer with opposite sign  
\begin{equation}
\sigma_0 =    \sqrt{\frac{2 \epsilon k T  n_{0i}^2}{\pi}}  \sinh \frac{|z_i| e \psi_0}{2kT}
\end{equation}
This is the Gouy-Chapman relationship for the surface charge density \cite{gast}. The charge density is a function of the surface potential and ion concentration among other factors. 

In variable charged oxides the surface potential is not constant and varies with pH. Considering small surface potential, the last equation can be simplified\footnote{Note that the hyperbolic sin expands to $ \sinh y = y + y^3/3! + y^5/5! + ...$ thus, for small values of $y$ higher order terms may be neglected } to  \cite{gast} 
\begin{equation}
\sigma_0 \approx   \frac{\epsilon \kappa}{4 \pi } \psi_0
\end{equation}
and the surface potential can be determined by the activity of the potential determining ion $a_{pdi}$ in relation to its activity at the \emph{point of zero charge} (pzc)
\begin{equation}
\psi_0 \pm    \frac{RT}{zF} \ln{\frac{a_{pdi}}{(a_{pdi})_{pzc}}}
\end{equation}
If $H^+$ is the potential determining ion as is the case in oxide surfaces, the relationship between pH and hydrogen activity is $ pH = - \log[H^+]$ and the equation can be written as   
\begin{equation}
\psi_0 =    \frac{-2.303 RT}{F} (pH - pH_{pzc})
\end{equation}
Combining the electrical potential and surface charge equations, the following equation is valid
\begin{equation}
\sigma_0 \approx    \frac{-2.303 RT \epsilon \kappa}{4\pi F} (pH - pH_{pzc})
\end{equation}
In which $F$ is the Faraday constant. 

On a side note, the rigorous derivation above was provided by Kenneth L. Babcock one of the fathers of soil chemistry \cite{babcock}. Much of soil chemistry interfaces with soil physics on the subjects of thermodynamics and transport and retention of solutes.  One of the most important soil chemists in recent history, Garrison Sposito, was a student of Kenneth Babcock. Sposito gave fundamental contributions to soil physics no only on thermodynamics and solute chemistry but also regarding water transport and scaling of soil properties and processes.

\section{Rocks versus soils}
We have illustrated that soils and rock differ in terms of their mineralogical composition and by the fact that rocks are consolidated materials. Rocks differ from soils in terms of genesis and mineralogy and by the fact that soils have organic matter in their composition. The amount of organic matter varies with soil type, but according to soil science criteria, mineral soils have around 5\% organic matter on mass basis for average soils. The concept of organic soils is a somewhat loose definition based on a more or less arbitrary limit of organic matter or carbon content \cite{soilsurvey}. The classification of organic soils varies according to engineering classification systems \cite{myslinska} and can vary depending on several specific criteria within a soil classification system used in soil science \cite{soilsurvey}. Because soils occupy only an thin layer of Earth's surface, usually from a meter or two to a few meters in depth at the most in most cases, they are under direct influence of the weather and organisms. Soils are usually rich in organisms, both macro and micro, and plants and are continually affected by their actions. Soils are dynamic and porosity and structure are continually being altered by the actions of roots, ants, termites and other burrowing animals of larger size. Rocks and sedimentary deposits can reach kilometers in depth on the Earth's crust and are very little affected by organisms and surface conditions in most cases. The transition zone between soils and subsurface materials is called the saprolite and can reach several meters in depth under extreme weathering conditions on the tropics over certain rocks. Soils constitute, for the most part, the unsaturated zone, where the pores are saturated by both gas and solution, although it is possible that the water table can be located within the soil profile. Below the water table is the saturated zone where the pores are saturated with water. Most aquifers are located in sedimentary deposits and to some extent porous or fractured rocks. Transport and retention of water, contaminants, oil and gas in rocks and sedimentary deposits is usually the object of study of hydrogeology and petroleum engineering while soil physics is often concerned with soils and other unsaturated media. However, the mathematical analytical and numerical framework between these disciplines is similar since the equations which govern transport processes in these materials have the same physical basis when they are not exactly the same. Much of the governing equations of transport processes originated from the realm of mechanics and later chemical engineering. In the next chapter we will try to provide a mathematical framework for the equations that govern transport processes in soils and other natural porous media.

\section{List of symbols for this chapter}

\begin{longtable}{ll}
    	$ \psi $ & Electric potential \\
    	$ \psi_0 $ & Electric potential at the surface of the charged plane\\
    	$ x $ & Distance from the charged plane\\
 	$ z $ & Ion valence\\
 	$ e $ & Elementary charge  \\
 	$ k $ & Boltzmann constant \\
 	$ T $ & Absolute temperature \\
 	$ \rho $ & Charge density  \\
 	$ \epsilon $ & Absolute permittivity of the medium  \\
	$ \kappa^{-1} $ & Debye length \\
	$ \text{e} $ & Euler's number \\
 	$ ln $ & Logarithm base e \\
 	$ log $ & Logarithm base 10 \\
 	$ c_1, c_2, C_1, C_2 $ & General constants for differential equations solutions \\
	$ y $ & General variable for mathematical manipulations \\
     	$ y_0 $ & Value of $y$ at $y = 0$ \\
   	$ u, v $ & Variables for integration by substitution \\
   	$ sgn $ & Sign function \\
   	$ \sigma $ & Net charge at the diffuse layer \\
   	$ \sigma_0 $ & Surface charge \\
   	$ R $ & Molar gas constant \\
   	$ F $ & Faraday constant \\
   	$ a_{pdi} $ & Activity of a potential determining ion (pdi) \\
   	$ pH_{pzc} $ & pH at the point of zero charge (pzc)   
\end{longtable}

% !TEX TS-program = pdflatex
% !TEX encoding = UTF-8 Unicode

% Example of the Memoir class, an alternative to the default LaTeX classes such as article and book, with many added features built into the class itself.

\chapter{The fundamental equations of fluid dynamics}
\label{ch4}

Up until now we have been discussing properties of the solid matrix of porous media. Some of these properties are directly affected by water content and can be altered by engineering, hydrological, agricultural practices as well as a by wide range of natural processes, both on short and long term. We have not discussed so far any type of flux of liquid or gaseous phases through porous materials. Detailed discussions of mass flow through porous media are reserved to the chapters dealing with saturated and unsaturated hydrodynamics of porous media, and other transport phenomena. Technically speaking, we will deal with another property of the solid matrix on the next chapter, the grain size distribution, so there should not be an immediate need for equations governing fluid flow. However, the majority of methods for determination of grain size distribution of porous media rely on settling in liquid and, occasionally, gaseous materials. To understand the dynamics of settling we need to understand the fundamental equations of fluid dynamics which control how settling occurs in a viscous fluid. The knowledge discussed in this chapter will be also the basis for modeling flow problems in open conducts and to a large extent, flow in porous media. This chapter will introduce you to the beauty and brutality that is the mathematics involved in fluid mechanics and in theoretical soil physics. If you are already familiar with tensor analysis in areas such as continuum mechanics or general relativity you should be able to follow this chapter without major issues. Consulting the bibliography might help with issues specific to fluid mechanics that might arise. If you are seeing the application of tensor analysis for the first time it might be worth consulting an introductory text on tensors.   

\section{Formulations of fluid flow}

When dealing with fluid flow, two formulations are usually discussed in the literature, the \textit{Lagrangian} and the \textit{Eulerian} specifications, after the great mathematicians Joseph-Louis Lagrange and Leonard Euler.\footnote{Although Euler was responsible for both formulations, see for example Landau and Lifshitz (1987, p. 5) \cite{landau}.}  The \textit{Lagrangian} formulation follows from the idea of tracking individual particles or the center of mass of a group of particles from classical mechanics. In the context of fluid flow the movement of a single water molecule could be tracked in relation to an initial time $t_0$ and position $\mathbf{s}$, where, as seen in Chapter \ref{ch1}.
\begin{equation}
\mathbf{s} = s_x \mathbf{i} + s_y \mathbf{j} + s_z \mathbf{k} 
\end{equation}   
Thus, in the \textit{Lagrangian} formulation, the velocity of a tracked particle or center of mass of a group of particles is  $\mathbf{v}(\mathbf{s}, t)$ or, considering the vector components $\mathbf{v}(s_x, s_y, s_z, t)$  . It is a consensus that the \textit{Lagrangian} formulation is not mathematically convenient for the majority of fluid flow problems and thus, it will not be discussed further. %\footnote{It would be useful to understand/discuss/explain this better}.
The formulation normally adopted is more akin to that used in electromagnetism in the sense that the flow quantities are defined as a vector field with independent variables, position in space and time, $\mathbf{x}$ and t. The vector velocity field in the \textit{Eulerian} formulation is traditionally represented as $\mathbf{u}(\mathbf{x}, t)  $, where       
\begin{equation}
\mathbf{u} = u_x \mathbf{i} + u_y \mathbf{j} + u_z \mathbf{k} 
\end{equation}   
often also represented as
\begin{equation}
\mathbf{u} = u \mathbf{x} + v \mathbf{y} + w \mathbf{z} 
\end{equation} 
such that $\mathbf{u}(\mathbf{x}, t)  $ can be written as $\mathbf{u}(u_x, u_y, u_z,t)  $ in the first case and $\mathbf{u}(u, v, w,t)   $ in the second. The difference between $\mathbf{v}$ and $\mathbf{u}$ is that in the latter we are now no longer interested in an individual particle or the center of mass of a group of particles, what we have is a velocity field in space and time. Now at any position in space and at a given time there is a velocity value assigned. Imagine a 2D river or stream seen from above, in the \textit{Lagrangian} river we would track the position of individual particles as a function of time, while on the \textit{Eulerian} river we are not worried about individual particles but we have a velocity field that covers the entire space of the river, such that if we chose an arbitrary point $\mathbf{x_1}$ and $t_1$ on the surface of the river, the velocity at that point is given by $\mathbf{u}(\mathbf{x_1}, t_1)$. One advantage of the \textit{Eulerian} formulation is now obvious as there is no need to track an exorbitant number of particles to model the flow of a river or stream. Another major advantage of the \textit{Eulerian} formulation is that it can be used to model many other important fluid properties such as density and pressure over space and time, however it is important to point out that these are scalar fields given by $\rho(\mathbf{x}, t)$ and $P(\mathbf{x}, t)$, for density and pressure, respectively, noting that $\mathbf{u}$ at any point in space and time is itself a vector, while $\rho$ and $P$ are scalars. From the \textit{Eulerian} formulation we now find the acceleration of a fluid element. The acceleration of the fluid element is not given directly by $ \partial \mathbf{u}/ \partial t $ as each fluid element travels through the vector field as times varies, being subjected to velocity variation in space along the vector field in addition to time. The correct formulation of the acceleration of a material element is given by    
\begin{equation}
\frac{\partial \mathbf{u}}{\partial t} + \mathbf{u} \cdot \nabla \mathbf{u}  
\end{equation} 
Using the notation introduced in Chapter \ref{ch1} and because $\mathbf{u}$ is a vector the equation above can be written as  

\begin{align*}
\frac{\partial }{\partial t} (u_x \mathbf{i} + u_y \mathbf{j} + u_z \mathbf{k}) + (u_x \mathbf{i} + u_y \mathbf{j} + u_z \mathbf{k}) \cdot  [\frac{\partial }{\partial x}(u_x \mathbf{i} + u_y \mathbf{j} + u_z \mathbf{k}) \mathbf{i} + \\ + \frac{\partial }{\partial y}(u_x \mathbf{i} + u_y \mathbf{j} + u_z \mathbf{k}) \mathbf{j} +  \frac{\partial}{\partial z} (u_x \mathbf{i} + u_y \mathbf{j} + u_z \mathbf{k})\mathbf{k}] 
\end{align*} 
which, considering the dot product $\mathbf{u} \cdot \nabla \mathbf{u}$, can be written as
\begin{align*}
\frac{\partial }{\partial t} (u_x \mathbf{i} + u_y \mathbf{j} + u_z \mathbf{k}) +  [u_x \frac{\partial }{\partial x}(u_x \mathbf{i} + u_y \mathbf{j} + u_z \mathbf{k}) +  u_y \frac{\partial }{\partial y}(u_x \mathbf{i} + u_y \mathbf{j}  + u_z \mathbf{k}) + \\ + u_z \frac{\partial  }{\partial z} (u_x \mathbf{i} + u_y \mathbf{j} + u_z \mathbf{k})] 
\end{align*} 
With some effort\footnote{As opposed to ``it can easily be shown''} it can be shown that the vector equation above can be written as a set of equations. You will see that most of the fluid dynamics equations in two and three-dimensional spaces are actually sets of equations and it is important that you familiarize yourself with the steps in converting from vector form to components and vice-versa when doable\footnote{In several fluid equations, multiple tensor components are involved, and it is not practical to do these operations without using vector and tensor identities}. Thus in Cartesian coordinates, the components in $x$, $y$ and $z$ are
\begin{align*}
\frac{\partial u_x}{\partial t}   +  [u_x \frac{\partial u_x}{\partial x}  +  u_y \frac{\partial u_x}{\partial y}  + u_z \frac{\partial  u_x}{\partial z}  ] \\ 
\frac{\partial u_y}{\partial t}  +  [u_x \frac{\partial u_y}{\partial x}  +  u_y \frac{\partial u_y}{\partial y}  + u_z \frac{\partial  u_y}{\partial z}  ] 
\\ 
\frac{\partial u_z}{\partial t}  +  [u_x \frac{\partial u_z}{\partial x}  +  u_y \frac{\partial u_z}{\partial y}  + u_z \frac{\partial  u_z}{\partial z}  ] 
\end{align*} 
%The quantity $\mathbf{u} \cdot \nabla \mathbf{u} $ is the dot product of a vector (first order tensor) and a second order tensor and is a vector, such that $  \frac{\partial \mathbf{u}}{\partial t} + \mathbf{u} \cdot \nabla \mathbf{u} $ is the sum of two vectors and is itself a vector. 

The derivative form found in Equation 4.4 has such an importance in fluid dynamics that it constitutes a mathematical operator often called the \emph{material} or \emph{substantial} derivative

\begin{equation}
\boxed{
\frac{D}{D t} = \frac{\partial }{\partial t} + \mathbf{u} \cdot \nabla 
}
\end{equation} 
such that for any scalar or vector quantity $f$ we have
\begin{equation}
\frac{D f}{D t} = \frac{\partial f}{\partial t} + \mathbf{u} \cdot \nabla f 
\end{equation} 
In which $ \partial f / \partial t $ is the temporal rate of change at a specific location and  $\mathbf{u} \cdot \nabla f $ is the convective rate of change due to transport of the fluid element from one point to another in space. If $f$ is constant throughout the space under consideration   
\begin{equation}
\frac{D f}{D t} = \frac{\partial f}{\partial t} + \mathbf{u} \cdot \nabla f = 0 
\end{equation} 
and there is no acceleration of the material elements of fluid.

\section{Conservation of mass}

The conservation equation, or mass conservation equation, is often referred to the \emph{continuity} equation on the literature. In agreement with G. K. Batchelor, we do not think that \emph{continuity} is an appropriate term and will refrain from using it in this text. The idea of conservation of mass, be it a gas, a fluid or a particulate solid, is a very intuitive one. Imagine a garden hose, if you consider a given length of the hose, say one meter, once you turn on the faucet and the flux out of the faucet becomes constant, and if there are no leaks along the hose, the amount of water that enters  on one end is equal to the amount that exits on the other end, per unit time. This condition in which the amounts in and out of the system are constant and equal is called \emph{steady state} and is very important in groundwater hydrology. This simple example is a basic application of the conservation equation, if there are no losses or gains within the system, matter is conserved. 

Suppose now you have a \emph{steady state} condition and there are leaks in your garden hose within the length considered. The mass conservation equation still applies, the amount in is equal to the sum of the amounts on the other end of the hose plus any leaks, per unit time. In real scenarios we often have to deal with sources or sinks within the volume considered, this is especially true in the case of reactive and/or volatile transport in which there can be loss of gain within the consider volume due to adsorption, condensation, degradation and many other physico-chemical processes. A simple example of a sink (loss) would be evaporation of water in a reservoir.

%%%%%%%%%%%%%%%%%% Not sure what you are trying to say here, the system is still steady-state
%Suppose now you have a water tank in your house that is half full. A pump is activated to refill the tank while at the same time you have the tap open in your kitchen sink. This condition is no longer \emph{steady state} since the flow of the pump into the water tank is much greater than the amount of water exiting the sink. The conservation laws still apply but now you have a net increase in the amount of water stored into the tank. This same logic applies to natural systems, be it a soil profile, an aquifer or a reservoir dam. The net increase (or decrease) in the amount of water in a given volume is the difference between the input and output per unit time. In real scenarios we often have to deal with sources or sinks within the volume considered, this is especially true in the case of reactive and/or volatile in which there can be loss of gain within the considered volume due to adsorption, condensation, degradation and many other physico-chemical processes. A simple example of a sink (loss) would be evaporation of water in a reservoir.                 

Let us now treat this problem from a mathematical point of view. There are at least two approaches to treat this problem, one is deceitfully simple and direct and is based on vector calculus. The other is based on algebra\footnote{As in algebraic manipulations in a broader sense} and appears to be very complex although being very simple.  The first one can be found in advanced fluid dynamics and mechanics, physics and applied mathematics books, while the other is preferred in introductory engineering and in soil physics books. The conceptual model is the same in both cases, an element of volume in space, in which flow is considered through an element of surface area. It is easy for us to consider water, but you will see later that the conservation equation applies to any substance in transport processes in porous media. In the vector treatment we are not initially concerned with the geometry of the volume, other than it should be a closed space. Initially, however, it should be much easier to imagine the volume as a cube or as a sphere or other similar closed surfaces. %\footnote{In mathematics, the topology of the space is of interest, and can have implications in terms of the mathematical description. This is, however, beyond the scope of this text}. 
Starting with a volume $V$ enclosed by a surface $A$ completely filled with a liquid of density $\rho$, the mass of water within the volume is given by   
\begin{equation}
\int \rho dV 
\end{equation} 
This should come of no surprise, as by integrating the volume elements $dV$ over the entire volume we are left with $V$, and obviously the water density in [M L$^{-3}$] multiplied by the total volume in [L$^{3}$] is the mass M of water within the enclosed volume. The amount of water flowing through an element of surface area $dA$   
\begin{equation}
\int \rho \mathbf{u} \cdot \mathbf{n} dA 
\end{equation} 
Where $\mathbf{n}$ is the unit normal in dA. Note that the velocity, $\mathbf{u}$, in [L T$^{-1}$] times water density, $\rho$, in [M L$^{-3}$], over the entire surface, $A$, in [L$^{2}$] is the rate of mass transfer across the surface in [M T$^{-1}$]. Now the rate of variation of the mass within the volume with time can be written in differential form as
\begin{equation}
 \frac{\partial}{\partial t} \int \rho dV 
\end{equation} 
in units of [M T$^{-1}$]. If this doesn't make much sense to you don't worry, you can find more information about this construction in vector calculus or calculus 3 books. Note that as we defined Equation 4.8 as the total mass within the volume, Equation 4.10 is nothing more than the variation of mass with time within the volume. Now, from our discussion of conservation laws we can agree that the variation of mass within a volume is the net rate of flow across its surface, and thus the conservation of mass implies that   
\begin{equation}
 \frac{\partial}{\partial t} \int \rho dV  = - \int \rho \mathbf{u} \cdot \mathbf{n} dA 
\end{equation} 
So that 
\begin{equation}
 \frac{\partial}{\partial t} \int \rho dV  + \int \rho \mathbf{u} \cdot \mathbf{n} dA = 0  
\end{equation} 
Now all the vector calculus tools discussed in Chapter \ref{ch1} will come in handy. We can use the divergence theorem to transform a surface integral into a volume integral. Applying it to the second term in Equation 4.12 results in  
\begin{equation}
 \frac{\partial}{\partial t} \int \rho dV  + \int \nabla \cdot (\rho \mathbf{u})  dV = 0  
\end{equation} 
Which can be written as
\begin{equation}
\int [ \frac{\partial \rho}{\partial t}   + \nabla \cdot (\rho \mathbf{u})] dV = 0  
\end{equation} 
Because the integration must hold for any volume, the integrand must vanish and we are left with one of the fundamental equations of fluid dynamics, the mass conservation (or continuity) equation
\begin{equation}
\boxed{ \frac{\partial  \rho}{\partial t}  + \nabla \cdot (\rho \mathbf{u}) = 0 } 
\end{equation} 

Now for the derivation that does not need vector calculus. Imagine a cube with sides $\Delta x$, $\Delta y$ and $\Delta z$. Consider a velocity component $u_x$ in the direction $x$. The area which is perpendicular to the velocity component $u_x$ is $\Delta y\Delta z$. In other words, the velocity component $u_x$ enters into the cube in the side $\Delta y\Delta z$. Now, the rate of mass accumulation in the direction $x$ is the amount of mass that enters on one end on the direction $x$ and leaves on the other, in that same direction. The amount of mass that enters in one end is equal to the velocity component $u_x$ times the density of the fluid $\rho $ times the area $ \Delta y \Delta z$, note that      
\begin{equation}
\frac{m_{in-x}}{t} = \rho u_x  \Delta y \Delta z = \frac{\mathrm{M}}{L^3} \frac{\mathrm{L}}{T}  L^2 = \frac{\mathrm{M}}{T}  
\end{equation} 
The amount that exits on the other end will depend on the variation of the velocity of the fluid through the direction $x$ within the volume, if the fluid slows down, the velocity decreases and vice-versa. Therefore, in the outflow we have to consider the rate of variation of velocity within the volume such that, the outflow mass in the $x$ direction is
\begin{equation}
\frac{m_{out-x}}{t} = ( \rho u_x  +  \frac{\partial \rho u_x}{\partial x}\Delta x) \Delta y \Delta z   
\end{equation}
Therefore, the amount of mass that accumulates or is lost per unit time due to flux in the $x$ direction is given by 
\begin{equation}
\frac{\partial}{\partial t} M_{x} = \rho u_x  \Delta y \Delta z - (\rho u_x  + \frac{\partial \rho u_x}{\partial x}\Delta x) \Delta y \Delta z   
\end{equation}
Which results in
\begin{equation}
\frac{\partial}{\partial t} M_{x} = - \frac{\partial \rho u_x}{\partial x}\Delta x \Delta y \Delta z   
\end{equation}
By symmetry it is not difficult to show that the mass lost or accumulated due to flux on the directions $y$ and $z$ is 
\begin{equation}
\frac{\partial}{\partial t} M_{y} = - \rho \frac{\partial \rho u_y}{\partial y}\Delta x \Delta y \Delta z   
\end{equation}
 \begin{equation}
\frac{\partial}{\partial t} M_{z} = - \rho \frac{\partial \rho u_z}{\partial z}\Delta x \Delta y \Delta z   
\end{equation}
and that the total mass accumulated or loss within the cube per unit time due to flow in the three directions is
\begin{equation}
\frac{\partial}{\partial t}  M_{xyz} = - (\frac{\partial \rho u_x}{\partial x}+ \frac{\partial \rho u_y}{\partial y}+ \frac{\partial \rho u_z}{\partial z})\Delta x \Delta y \Delta z   
\end{equation}
Since the density of the cube is $\rho$ and its volume is $\Delta x \Delta y \Delta z$ and rearranging $ \rho = M/V$ to $ M = \rho V$ we have 
\begin{equation*}
\frac{\partial}{\partial t}  \rho \Delta x \Delta y \Delta z = - (\frac{\partial \rho u_x}{\partial x}+\frac{\partial \rho u_y}{\partial y}+\frac{\partial \rho u_z}{\partial z})\Delta x \Delta y \Delta z   
\end{equation*}
or
\begin{equation*}
\frac{\partial \rho}{\partial t}  = - (\frac{\partial \rho u_x}{\partial x}+\frac{\partial \rho u_y}{\partial y}+\frac{\partial \rho u_z}{\partial z})
\end{equation*}
Using the vector operators, this equation can be written as
\begin{equation*}
\frac{\partial \rho}{\partial t}  + \nabla \cdot (\rho  \mathbf{u}) = 0
\end{equation*}
 which is the same as Equation 4.15. The derivation using geometric considerations is favored in soil physics and hydrology books while the vector calculus derivation is found in fluid mechanics and dynamics books for physicists and mathematicians. If the density of the fluid is constant, $\rho $ does not vary with time and $\partial \rho/\partial t = 0  $ and we arrive at an important case of the conservation equation 
\begin{equation}
\boxed{
\nabla \cdot \mathbf{u} = 0
}
\end{equation}
Stating in words, the divergence of the velocity field is zero for an incompressible fluid. For most practical purposes, water can be assumed as incompressible and Equation 4.23 will be valid for many flow problems. 

\section{Conservation of momentum}
Much like for mass, the conservation laws are valid for other extensive properties in fluid mechanics including energy and momentum and this will allow us to derive the equations at the core of fluid mechanics and transport phenomena. Recall from introductory physics that Newton's second law can be written either as 
\begin{equation}
 \mathbf{F}  = m\mathbf{a} 
\end{equation}
or as 
\begin{equation}
\mathbf{F}  = \frac{d \mathbf{p}}{dt} 
\end{equation} 
where the vector quantity $\mathbf{p}$ is the linear momentum and is equal to $ m \mathbf{v} $, for an object of mass $m$ moving at velocity $\mathbf{v}$. Considering an element of fluid of volume $V$, the momentum corresponding to the entire volume is given by 
\begin{equation}
\int \mathbf{u} \rho dV
\end{equation}
Notice that velocity is in [L T$^{-1}$], density is in [M L$^{-3}$] and volume is in [L$^{3}$]  such that the units are consistent with those of momentum [M L T$^{-1}$]. The rate of change of momentum for the elementary volume can be represented by the derivative of the relationship above with respect to time, the \emph{material derivative} in this case. If the density can be assumed as constant
\begin{equation}
\int \frac{D\mathbf{u}}{Dt} \rho dV
\end{equation}
Now, as specified by Newton's second law of motion, the rate of change of momentum in an object is equal to the sum of forces acting on the object. There are two important points here to be considered because we are dealing with fluid mechanics and not with a point particle. The first is that the total momentum is the sum of the individual momentum of each infinitesimal element of fluid. If the system was modeled as a system of individual particles the integral on the previous equations could be replaced by a summation of the momenta of the $n$ particles that compose the system. The second point is that the forces acting in a volume of fluid can be divided into body (or volume) forces such as those generated by gravity and electromagnetic fields and surface (of contact) forces, which act directly over the surface enclosing the volume element. Surface forces are ubiquitous in continuum mechanics and are an important concept in soil physics both in treating the solid and liquid phases and their interactions. To understand surface forces it is convenient to explore the concept of the \emph{stress tensor}. 

Imagine an element of volume of fluid  immersed in a fluid or other continuous matter, the matter around the element of fluid exerts a contact force across the entire surface of the fluid element. For the sake of simplicity, consider a volume element of cubic shape in Cartesian coordinates system. Each face can be subjected to forces that are perpendicular (i.e. normal) to the surface, in one direction, or parallel (tangential) to the surface in two directions. For example, the face that lies on the $x-y$ plane can be subjected to a perpendicular force on the $z$ direction and to two parallel forces on the $x$ and $y$ directions. Thus, the cube can be subjected to three forces in each face for each of its six faces.  If each surface of the cube is characterized by a surface area, we can define each component of the \emph{stress} acting on the surface as force over area. Thus, on the $z$ direction lies the $x-y$ plane with a normal stress component $\sigma_{zz}$ and two perpendicular stress components, one in the $x$ direction, represented by $\sigma_{zx}$ and another in the $y$ direction represented by $\sigma_{zy}$. This leaves us with three unique stress components in each direction, $x$, $y$ and $z$, with a total of nine stress components completely defining the stresses acting on the element. The nine elements define the \emph{stress tensor} 

\begin{equation} \sigma_{ij} = 
\begin{bmatrix}
\sigma_{xx} & \sigma_{xy}  & \sigma_{xz} \\
\sigma_{yx} & \sigma_{yy}  & \sigma_{yz} \\
\sigma_{zx} & \sigma_{zy}  & \sigma_{zz} \\
\end{bmatrix}
\end{equation}

The components where $i = j$ (i.e. $\sigma_{xx}$, $\sigma_{yy}$,  $\sigma_{zz}$) are the normal stress components, while the components where $i \ne j$ are the tangential stress components. The tangential stress components correspond to shearing stresses seen in solids which can cause isovolumetric deformations. It is convenient now to introduce the Kronecker delta operator

\begin{equation}
  \delta_{ij} =
    \begin{cases}
      0 & \text{if}  ~ i \ne j, \\
      1 & \text{if}  ~  i = j   \\
    \end{cases}       
\end{equation}

Such that the normal stresses on the fluid element can be represented as 

\begin{equation} 
\sigma_{ij} = \delta_{ij} \sigma_{ij} 
\end{equation}
It is important to point out that if the fluid element is at rest, the static fluid pressure is defined as  
\begin{equation} 
\sigma_{ij} = -p \delta_{ij}
\end{equation}
Applying this equation to the stress tensor results in 

\begin{equation} \sigma_{ij} = 
\begin{bmatrix}
-p & 0  & 0 \\
0 & -p  & 0 \\
0 & 0  & -p \\
\end{bmatrix}
\end{equation}
Which is nothing more than the well known principle from hydrostatics which states that for a small volume immersed in a fluid (such as in water for example), the pressure is constant across its surface. Keep in mind that for a moving fluid element the story is quite different. 
Now the total surface force exerted on the volume element is the sum of the stresses acting in each surface element with area $dS$ and normal $n_j$ and can be written as
\begin{equation}
\int \sigma_{ij} n_{j} dS
\end{equation}
which can be transformed to a volume integral using the divergence theorem
\begin{equation}
\int \frac{\partial \sigma_{ij}}{\partial x_j} dV
\end{equation}
From Newton's law expressed as conservation of momentum, for the element of volume of the fluid, we can write the total momentum as the sum of forces acting on the fluid, including the summation of all body forces and the surface forces
\begin{equation}
\int \frac{D u_i}{Dt} \rho dV  = \int F_i \rho dV + \int \frac{\partial \sigma_{ij}}{\partial x_j} dV
\end{equation}
As usual, the integral vanishes identically and we are left with 
\begin{equation}
 \frac{D u_i}{Dt} \rho  =  F_i \rho  +  \frac{\partial \sigma_{ij}}{\partial x_j} 
 \end{equation}
In which $F_i$ is the sum of all body forces acting on the fluid element, i.e. gravitational, electromagnetic, etc. Expanding this equation in three dimensional Cartesian coordinates using Einstein's summation convention we have
\begin{align*}
\frac{D u_x}{Dt} \rho  =  F_x \rho  +  \frac{\partial \sigma_{xx}}{\partial x} + \frac{\partial \sigma_{xy}}{\partial y} + \frac{\partial \sigma_{xz}}{\partial z}  
 \\ 
 \frac{D u_y}{Dt} \rho  =  F_y \rho  +  \frac{\partial \sigma_{yx}}{\partial x} + \frac{\partial \sigma_{yy}}{\partial y} + \frac{\partial \sigma_{yz}}{\partial z}  
 \\ 
 \frac{D u_z}{Dt} \rho  =  F_z \rho  +  \frac{\partial \sigma_{zx}}{\partial x} + \frac{\partial \sigma_{zy}}{\partial y} + \frac{\partial \sigma_{zz}}{\partial z}  
\end{align*} 
These equations show that the acceleration in each direction is caused by the forces acting on that direction and the components of the stress tensor acting on the face perpendicular to the direction of the acceleration. With some inspection it can become clear that the components of the stress tensor can cause deformation during motion depending of the nature of the shearing components (i.e. $i \ne j$). With that in mind, the stress tensor can be divided into two components, one equivalent to the static pressure experienced by a stationary volume of fluid and a non-isotropic component which originates due to motion of the fluid element
\begin{equation}
 \sigma_{ij}  =  -p \delta_{ij} + d_{ij} 
 \end{equation}
 Which can be represented as 
 \begin{equation}  
\begin{bmatrix}
\sigma_{xx} & \sigma_{xy}  & \sigma_{xz} \\
\sigma_{yx} & \sigma_{yy}  & \sigma_{yz} \\
\sigma_{zx} & \sigma_{zy}  & \sigma_{zz} \\
\end{bmatrix}
= 
\begin{bmatrix}
-p & 0  & 0 \\
0 & -p  & 0 \\
0 & 0  & -p \\
\end{bmatrix}
+
\begin{bmatrix}
d_{xx} & d_{xy}  & d_{xz} \\
d_{yx} & d_{yy}  & d_{yz} \\
d_{zx} & d_{zy}  & d_{zz} \\
\end{bmatrix}
\end{equation}
Now $d_{ij}$ is called the deviatoric stress tensor and is the non-isotropic, tangential, component of stresses. Considering the components of the stress tensor and expanding the material derivative operator, the unabridged equation of motion for an element of fluid in three-dimensional Cartesian coordinates is  
\begin{align*}
\rho (\frac{\partial u_x}{\partial t}+u_x \frac{\partial u_x}{\partial x}+u_y\frac{\partial u_x}{\partial y}+u_z\frac{\partial u_x}{\partial z})   =  F_x \rho -\frac{\partial p}{\partial x}  +  \frac{\partial d_{xx}}{\partial x} + \frac{\partial d_{xy}}{\partial y} + \frac{\partial d_{xz}}{\partial z}  
 \\ 
\rho (\frac{\partial u_y}{\partial t}+u_x \frac{\partial u_y}{\partial x}+u_y\frac{\partial u_y}{\partial y}+u_z\frac{\partial u_y}{\partial z})  =  F_y \rho  -\frac{\partial p}{\partial y}+  \frac{\partial d_{yx}}{\partial x} + \frac{\partial d_{yy}}{\partial y} + \frac{\partial d_{yz}}{\partial z}  
 \\ 
\rho (\frac{\partial u_z}{\partial t}+u_x \frac{\partial u_z}{\partial x}+u_y\frac{\partial u_z}{\partial y}+u_z\frac{\partial u_z}{\partial z})  =  F_z \rho   -\frac{\partial p}{\partial z} + \frac{\partial d_{zx}}{\partial x} + \frac{\partial d_{zy}}{\partial y} + \frac{\partial d_{zz}}{\partial z}  
\end{align*} 
The vector form, being simpler and more elegant, is referred to as the Cauchy momentum equation after Augustin-Louis Cauchy
\begin{equation}
\boxed{
 \rho \frac{D \mathbf{u}}{D t}   =  \mathbf{F} \rho     - \nabla p + \nabla \cdot \mathbf{d} 
 }
 \end{equation}
The full momentum equation is a highly nonlinear partial differential equation and knowledge of the stress tensor and body forces to which the fluid volume are subjected would be necessary to find a solution for this equation, if at all possible. Two of the most fundamental equations of fluid mechanics work with simplifications and assumptions over the momentum equation. The first developed from Newton's laws was derived by Euler and a more fundamental one which is thought to account for most fluid flow phenomena was developed by Claude-Louis Navier and Georges Gabriel Stokes less than a hundred years later.

\section{Euler equations}
The Euler equations were derived by Leonhard Euler himself in 1755 for the motion of an inviscid fluid, or a fluid without viscosity.  The complete set of equations of motion, including the viscosity term, were derived in the XIX century, culminating from the works of Claude-Louis Navier and Sir George Gabriel Stokes, after which they are named. The Euler equations can be understood as a special case of the Navier-Stokes equations and both sets of equations are applications of the Newton laws of motion to fluid flow, and both can be understood as special cases of the Cauchy equation of motion within the Newtonian framework. Up until now we have not discussed viscosity on any broader sense, most of us know the colloquial definition of viscosity used in daily language. A so called viscous fluid  has more resistance to \emph{flow} as we might have experienced with grease and honey as compared to water, for example. A higher viscosity reflects, in essence, higher internal friction on the fluid, or a larger magnitude of the forces between its elemental particles. 

Mathematically, viscosity can be represented as the coefficient of proportionality between a tangential stress and the rate of variation of velocity of a fluid in a direction perpendicular to that of the applied stress. The simplest explanation is that found on most of the literature and we can make use of the stress tensor already introduced for discussing it. Imagine you have a two-dimensional element of fluid within a system of horizontal and vertical axes $x$ and $y$. Now imagine that in the upper surface of the fluid element a stress $\sigma_{xy}$ parallel to the surface and perpendicular to the $y$ axis is applied. If the fluid has any internal friction, any imaginary fluid layer parallel to the surface layer where the stress is being applied will want to move with it. For a rigid material with infinite viscosity over a frictionless surface, every imaginary layer will move at the same velocity and the material will move as a block. Now if the material is not rigid the gradient of velocity in the $y$ direction will be proportional to the stress applied and as mentioned above, the viscosity is the coefficient of proportionality
\begin{equation}
 \sigma_{xy}  =  \mu \frac{d }{dy} u_x 
 \end{equation}
The more viscous the material, the more a tangential stress applied will propagate to the lower layers. One of the most important assumptions is that this relationship is linear, or alternatively, that the viscosity is constant in this equation. This is the case for the so called Newtonian fluids, if the relationship is nonlinear the fluid is called non-Newtonian and a different framework needs to be adopted for finding a general equation of motion. 
Now suppose an ideal fluid has no viscosity, in this case there is no sense in speaking of tangential components because there is no internal friction,  the deviatoric stress term in the Cauchy equation is zero and we are left with  
\begin{equation}
\boxed{
 \rho \frac{D \mathbf{u}}{D t}   =  \mathbf{F} \rho   - \nabla p  }
\end{equation}
This is Euler's equation of motion and is valid for ideal fluids. We can further simplify Euler's equation disregarding body forces, when possible, so that acceleration is caused only by surface forces 
\begin{equation}
 \rho \frac{D \mathbf{u}}{D t}   =  - \nabla p  
\end{equation}
Expanding the material derivative term
\begin{equation}
 \rho (\frac{\partial \mathbf{u}}{\partial t} + \mathbf{u} \cdot \nabla \mathbf{u})  =  - \nabla p  
\end{equation}
Euler's equation can  be also used to derive the equation for hydrostatic pressure in a fluid in equilibrium. In this cases there is no motion and the acceleration term represented by the material derivative is zero. For the hydrostatic pressure to exist it is necessary the action of a body force, represented here by gravitation force $ \mathbf{F} = \mathbf{g} $
\begin{equation}
 \nabla p  =   \rho \mathbf{g}    
\end{equation}
Expanding
\begin{equation}
 \frac{\partial p}{\partial x} \mathbf{i} + \frac{\partial p}{\partial y} \mathbf{j} + \frac{\partial p}{\partial z} \mathbf{k} =   \rho (g_x \mathbf{i} + g_y \mathbf{j} + g_z \mathbf{k})    
\end{equation}
And considering that Earth's gravitational force acts in $z$ only, and considering $z$ positive upwards
\begin{equation}
 \frac{\partial p}{\partial z}  =   -\rho  g    
\end{equation}
Which can be readily integrated as for $\rho $ a $g$ constant as
\begin{equation}
 \int \partial p  =   - \rho  g \int \partial z   
\end{equation}
Resulting in 
\begin{equation}
 p  =   - \rho  g  z + \text{constant}   
\end{equation}
Applying the boundary condition that $p = p_0$ when $z = 0$ (fluid surface) results
\begin{equation}
 p  =   p_0 - \rho  g  z    
\end{equation}
Which will increase considering that the absolute value of $z$ increases with depth and that $z$ is negative in the coordinate system adopted. The problem with Euler's equation, except for special applications, is that most liquids will have some degree of viscosity and the non-isotropic components of the stress tensor need to be considered in the analysis.

\section{Navier-Stokes equations}

The Navier-Stokes equations are the most general form of the governing equation of fluid motion and can be interpreted as a special case of Cauchy momentum equation. The transition from the Cauchy momentum equation to the Navier-Stokes equations requires that the non-isotropic stress tensor is expressed in a form amenable to the solution of the differential equation using analytical or numerical methods.  Following the language of Batchelor \cite{batchelor}, it is necessary to find a relationship between the deviatoric stress tensor and the local properties of the fluid. As with a lot of transport relations in physics, one possible simplification is to assume a linear relationship between the deviatoric stress and the velocity gradient. Keep in mind that this assumption carries profound implications in all areas of physics that deal with transport phenomena, either of matter, energy, momentum, electrical current, heat, ... and should not be taken lightly, I urge you to consult the references for details. The hypothesis of a linear relationship between the deviatoric stress tensor and the velocity gradients can be expressed analytically as
\begin{equation}
 d_{ij}  =  A_{ijkl}  \frac{\partial u_k}{\partial x_l} 
 \end{equation}
 Now $A_{ijkl}$ is a fourth order tensor with 81 individual components, from $A_{xxxx}$ to $A_{zzzz}$ in Cartesian $x$, $y$ and $z$ coordinates. The full expansion of this equation is not practical in this text due to space and formatting considerations, but a reduced form is provided, considering Einstein's summation convention
\begin{flalign*}
& d_{xx}  =  A_{xxxx}  \frac{\partial u_x}{\partial x}  + A_{xxxy}  \frac{\partial u_x}{\partial y} + A_{xxxz}  \frac{\partial u_x}{\partial z} + \cdots +  A_{xxzz}  \frac{\partial u_z}{\partial z}
\\
& d_{xy}  =  A_{xyxx}  \frac{\partial u_x}{\partial x}  + A_{xyxy}  \frac{\partial u_x}{\partial y} + A_{xyxz}  \frac{\partial u_x}{\partial z} + \cdots +  A_{xyzz}  \frac{\partial u_z}{\partial z}
\\ 
&  \vdots \\
& d_{zz}  =  A_{zzxx}  \frac{\partial u_x}{\partial x}  + A_{zzxy}  \frac{\partial u_x}{\partial y} + A_{zzxz}  \frac{\partial u_x}{\partial z} + \cdots +  A_{zzzz}  \frac{\partial u_z}{\partial z}
\end{flalign*} 
There are nine coefficients in each line and nine lines, one for each component of the deviatoric stress tensor, totaling the 81 coefficients of the $A_{ijkl}$ fourth-order tensor.  As we have done before, it is convenient to write the deviatoric stress tensor as the sum of its symmetric and antisymmetric counterparts  
\begin{equation}
 d_{ij}  =  A_{ijkl} d_{ij}^{s} + A_{ijkl} d_{ij}^{a}   
 \end{equation}
 or
\begin{equation}
 d_{ij}  =  A_{ijkl} e_{kl} + A_{ijkl} \xi_{ij}  
 \end{equation}
where 
\begin{equation}
 e_{kl}  =  \frac{1}{2}(\frac{\partial u_k}{\partial u_l} + \frac{\partial u_l}{\partial u_k})    
 \end{equation}
 is a symmetric second-order tensor, which in $x$, $y$ and $z$ Cartesian coordinates can be represented as
 \begin{equation}  
\frac{1}{2}
\begin{bmatrix}
(\frac{\partial u_x}{\partial x} + \frac{\partial u_x}{\partial x}) & (\frac{\partial u_x}{\partial y} + \frac{\partial u_y}{\partial x})  & (\frac{\partial u_x}{\partial z} + \frac{\partial u_z}{\partial x}) \\
(\frac{\partial u_y}{\partial x} + \frac{\partial u_x}{\partial y}) & (\frac{\partial u_y}{\partial y} + \frac{\partial u_y}{\partial y})  & (\frac{\partial u_y}{\partial z} + \frac{\partial u_z}{\partial y}) \\
(\frac{\partial u_z}{\partial x} + \frac{\partial u_x}{\partial z}) & (\frac{\partial u_z}{\partial y} + \frac{\partial u_y}{\partial z})  & (\frac{\partial u_z}{\partial z} + \frac{\partial u_z}{\partial z}) \\
\end{bmatrix}
\end{equation}
Verify that the quantities that are opposed to each other across the diagonal are the same, and this is what is mean be symmetric in this case.  The antisymmetric component is 
\begin{equation}
\xi_{kl}  =  \frac{1}{2}(\frac{\partial u_k}{\partial u_l} - \frac{\partial u_l}{\partial u_k})    
 \end{equation}
 and can be represented as 
 \begin{equation}  
\frac{1}{2}
\begin{bmatrix}
(\frac{\partial u_x}{\partial x} - \frac{\partial u_x}{\partial x}) & (\frac{\partial u_x}{\partial y} - \frac{\partial u_y}{\partial x})  & (\frac{\partial u_x}{\partial z} - \frac{\partial u_z}{\partial x}) \\
(\frac{\partial u_y}{\partial x} - \frac{\partial u_x}{\partial y}) & (\frac{\partial u_y}{\partial y} - \frac{\partial u_y}{\partial y})  & (\frac{\partial u_y}{\partial z} - \frac{\partial u_z}{\partial y}) \\
(\frac{\partial u_z}{\partial x} - \frac{\partial u_x}{\partial z}) & (\frac{\partial u_z}{\partial y} - \frac{\partial u_y}{\partial z})  & (\frac{\partial u_z}{\partial z} - \frac{\partial u_z}{\partial z}) \\
\end{bmatrix}
\end{equation}
Verify that the quantities across the diagonals are different (reverse sign) and this is what is meant by antisymmetric in this context. Verify also that the sum of the symmetric and antisymmetric counterparts reduces to the second-order tensor $ \partial u_k / \partial x_l$. The antisymmetric component can further be written as 
\begin{equation}
\xi_{kl}  =  -\frac{1}{2}A_{ijkl} \epsilon_{klm} \omega_{m}    
 \end{equation}
 The operator  $\epsilon_{klm}$ is a third-order tensor, with 27 components, where only six are non-zero, called the Levi-Civita symbol with the following property:
\begin{equation}
  \epsilon_{klm} =
    \begin{cases}
      +1 & \text{if} ~(k, l, m)~\text{is an even permutation}  \\
      -1 & \text{if} ~(k, l, m)~\text{is an odd permutation}  \\
      0 & \text{if any index is repeated}
    \end{cases}       
\end{equation}
Notice that the choice of $k, l, m$ is incidental to the problem, most often in the literature $i, j, k$ or numbers are used. Even permutations in this case are $(k,l,m)$, $(l,m,k)$  and $(m, k, l)$ and odd permutations are  $(m, l, k)$, $(l, k, m)$  and $(k, m, l)$. The vector $\omega_{m}$ has a fundamental role in fluid mechanics, and it is called the vorticity of the fluid. The components of the vorticity vector in $x$, $y$ and $z$ Cartesian coordinates are 
\begin{align}  
\omega_x = \frac{\partial u_z}{\partial y} - \frac{\partial u_y}{\partial z} \\
\omega_y = \frac{\partial u_x}{\partial z} - \frac{\partial u_z}{\partial x} \\
\omega_z = \frac{\partial u_y}{\partial x} - \frac{\partial u_x}{\partial y} 
\end{align}
Which can be obtained from the curl of the $\textbf{u}$ vector field (recall Chapter \ref{ch1})
\begin{equation}  
 \boldsymbol{\omega} = \nabla \times \mathbf{u}
\end{equation}
A fluid with zero curl is called irrotational. The mathematical concept of rotation is directly linked to the physical idea of rotation in fluid flow. The idea of rotational and irrotational flows is rarely discussed in soil physics, largely because of the assumptions adopted related to the nature of the flow as we will discuss later.  Now the deviatoric stress tensor can be rewritten as 
\begin{equation}
 d_{ij}  =  A_{ijkl} e_{kl} - \frac{1}{2} A_{ijkl} \epsilon_{klm} \omega_m  
 \end{equation}
In simple liquids, $A_{ijkl}$, is an isotropic tensor, that is, the molecular structure of the fluid is isotropic. In this case,  $A_{ijkl}$, can be written as a summation of products of delta tensors\footnote{Obviously, there is much more to this, the reader is encouraged to consult a book on tensor analysis and Aris \cite{aris}} 
\begin{equation}
A_{ijkl}  =  \mu \delta_{ik}\delta_{jl} + \mu' \delta_{il}\delta_{jk} + \mu'' \delta_{ij}\delta_{kl}
\end{equation}
Because $A_{ijkl}$ is symmtrical in $i$ and $j$, it is required that $\mu = \mu' $ such that
\begin{equation}
A_{ijkl}  =  2\mu \delta_{ik}\delta_{jl} + \mu'' \delta_{ij}\delta_{kl}
\end{equation}
The requirement that $A_{ijkl}$ is also symmetrical in $k$ and $l$ requires that the term $\boldsymbol{\omega}$ is dropped in the deviatoric stress tensor, so that by replacing into it we have 
\begin{equation}
 d_{ij}  =   2\mu \delta_{ik}\delta_{jl} e_{kl}  + \mu'' \delta_{ij}\delta_{kl} e_{kl} 
\end{equation}
Because the rules of manipulation of Kronecker delta tensors, the relationship can be simplified to 
\begin{equation}
 d_{ij}  =   2\mu  e_{ij}  + \mu''  \nabla \cdot e_{kk}~\delta_{ij}  
\end{equation}
 Where $e_{kk} = \nabla \cdot \mathbf{u} $ is the rate of expansion. Since the deviatoric stress tensor does not contribute to the mean normal stresses   
\begin{equation}
 d_{ii}  =   (2\mu    + 3\mu'')  e_{ii} = 0   
\end{equation}
Such that    
\begin{equation}
(2\mu    + 3\mu'')   = 0   
\end{equation}
Solving for $\mu''$ and replacing into the deviatoric stress tensor \footnote{C.L.} results
\begin{equation}
 d_{ij}  =   2\mu (e_{ij} - \frac{1}{3}  e_{ii}~\delta_{ij})    
\end{equation}
We now have an expression for the deviatoric stress tensor which can be used to define the stress tensor. Replacing into the stress tensor\footnote{C.L.}  results
\begin{equation}
 \sigma_{ij}  =  -p \delta_{ij}  + 2\mu (e_{ij} - \frac{1}{3}  e_{ii}~\delta_{ij})    
\end{equation}
Which can then be replaced into the equation of motion resulting in the set of equations 
%\small
\begin{multline}
\rho \frac{D u_x}{D t}   =  F_x \rho -\frac{\partial p}{\partial x}  + \frac{\partial}{\partial x}[ 2 \mu (e_{xx} - \frac{1}{3} e_{xx} \delta_{xx})]+  \\  \frac{\partial}{\partial y}[ 2 \mu (e_{xy} - \frac{1}{3} e_{xx} \delta_{xy})]+  \frac{\partial}{\partial z}[ 2 \mu (e_{xz} - \frac{1}{3} e_{xx} \delta_{xz})]  
\end{multline} 
\begin{multline}
\rho \frac{D u_y}{D t}  =  F_y \rho  -\frac{\partial p}{\partial y}+ \frac{\partial}{\partial x}[ 2 \mu (e_{yx} - \frac{1}{3} e_{yy} \delta_{yx})]+  \\ \frac{\partial}{\partial y}[ 2 \mu (e_{yy} - \frac{1}{3} e_{yy} \delta_{yy})]+  \frac{\partial}{\partial z}[ 2 \mu (e_{yz} - \frac{1}{3} e_{yy} \delta_{yz})] 
\end{multline} 
\begin{multline}
\rho \frac{D u_z}{D t}  =  F_z \rho  -\frac{\partial p}{\partial z}+ \frac{\partial}{\partial x}[ 2 \mu (e_{zx} - \frac{1}{3} e_{zz} \delta_{zx})]+ \\ \frac{\partial}{\partial y}[ 2 \mu (e_{zy} - \frac{1}{3} e_{zz} \delta_{zy})]+  \frac{\partial}{\partial z}[ 2 \mu (e_{zz} - \frac{1}{3} e_{zz} \delta_{zz})] 
\end{multline} 
%\normalsize
These are the Navier-Stokes equations and they account for the description of viscous fluid flow. The viscosity term in the full form presented above is not independent of position and the equation is valid for compressible fluids as well as incompressible fluids. If we consider that the viscosity is constant throughout the flow field, the term $\mu$ can be placed outside the derivative terms. Another simplification that can be used for modeling flow in porous media is incompressibility. For an incompressible fluid $\nabla \cdot \mathbf{u} = e_{ii} = 0 $ and the term $ e_{ii} \delta_{ij} $ is dropped, resulting in  
\begin{align}
\rho \frac{D u_x}{D t}  =  F_x \rho -\frac{\partial p}{\partial x}  + \mu\frac{\partial}{\partial x} (e_{xx})+\mu \frac{\partial}{\partial y}(e_{xy})+ \mu\frac{\partial}{\partial z} (e_{xz})
\\ 
\rho \frac{D u_y}{D t}  =  F_y \rho  -\frac{\partial p}{\partial y}+ \mu\frac{\partial}{\partial x}(e_{yx})+\mu\frac{\partial}{\partial y}(e_{yy})+ \mu\frac{\partial}{\partial z}(e_{yz}) 
 \\ 
\rho \frac{D u_z}{D t}  =  F_z \rho  -\frac{\partial p}{\partial z}+ \mu\frac{\partial}{\partial x}(e_{zx})+\mu\frac{\partial}{\partial y}(e_{zy})+ \mu\frac{\partial}{\partial z}(e_{zz}) 
\end{align} 
Note that, in this case, each partial derivative of  $e_{ij}$ can be written as
\begin{equation}
\frac{\partial }{\partial x_j} e_{ij}  =  \frac{\partial }{\partial x_j} \frac{1}{2}(\frac{\partial u_i}{\partial x_j}  + \frac{\partial u_j}{\partial x_i}) = \frac{1}{2}\frac{\partial^2 u_i}{\partial x_j\partial x_j}  + \frac{1}{2}\frac{\partial }{\partial x_i}\frac{\partial u_j}{\partial x_j}   = \frac{1}{2} \nabla^2 u_i + \frac{1}{2} \frac{\partial }{\partial x_i}(\nabla \cdot \mathbf{u}) 
\end{equation}
the incompressibility condition reduces the Navier-Stokes equations to
\begin{align*}
\rho \frac{D u_x}{D t}  =  F_x \rho -\frac{\partial p}{\partial x}  + \mu ( \frac{\partial^2 u_x}{\partial x^2} + \frac{\partial^2 u_x}{\partial y^2} + \frac{\partial^2 u_x}{\partial z^2})
\\ 
\rho \frac{D u_y}{D t}  =  F_y \rho  -\frac{\partial p}{\partial y} + \mu ( \frac{\partial^2 u_y}{\partial x^2} + \frac{\partial^2 u_y}{\partial y^2} + \frac{\partial^2 u_y}{\partial z^2})
\\ 
\rho \frac{D u_z}{D t}  =  F_z \rho  -\frac{\partial p}{\partial z} +  \mu ( \frac{\partial^2 u_z}{\partial x^2} + \frac{\partial^2 u_z}{\partial y^2} + \frac{\partial^2 u_z}{\partial z^2})
\end{align*} 
which can be written in vector form and by expanding the material derivative as
\begin{equation}
\boxed{
 \rho (\frac{\partial \mathbf{u}}{\partial  t} + \mathbf{u} \cdot \nabla \mathbf{u} )   =  \mathbf{F} \rho   - \nabla p +  \mu \nabla^2 \mathbf{u} 
 }
\end{equation}
This is the fundamental equation that governs the flow of incompressible fluids and many important results can be derived by solving it for a specified set of boundary conditions. We will used it over subsequent chapters to investigate settling and water flow phenomena.

\section{Stokes equation}
The Stokes equation is a special case of the Navier-Stokes equation for a set of specified conditions. Because we will use Stokes equation as a starting point for particle settling and flow in porous media derivations, it will be specified here. Under steady-state flow or flows in which the velocity variation is small enough that it can be considered constant and in conditions where the inertia is small as compared to the pressure and viscous terms, the acceleration term on the left side of the Navier-Stokes equation can be neglected. If the body forces can also be neglected due to being small or because of the direction of the flow in relation to a gravitational field for example, the governing equations become      
\begin{equation}
\boxed{
 \nabla p =  \mu \nabla^2 \mathbf{u} 
}
\end{equation}
\begin{equation}
\nabla \cdot \mathbf{u} = 0 
\end{equation}
The second of those being the conservation law described previously.

\section{Bernoulli equation}
Bernoulli equation can be derived from Euler equations by using the simplifying assumptions of steady and isentropic flow. Steady flow means that the velocity is constant in any particular point in the flow field, i.e. does not vary with time, such that $ \partial \mathbf{u} / \partial t = 0 $. Isentropic means that the entropy $s$ is constant throughout the flow field. To derive Bernoulli equation we need to first express Euler equation in terms of enthalpy. In thermodynamics, specific enthalpy can be expressed as
\begin{equation}
 dh = Tds + vdP
 \end{equation}
 But, because entropy is constant, $ds = 0 $ and
\begin{equation}
 dh =  vdp
 \end{equation}
 In this equation $h$ is the specific enthalpy or enthalpy per unit mass $h = H/M$, $s$ is the specific entropy, $s = S/M$ and $v$ is the specific volume $v = V/M$. Since we know that density of a fluid is mass per unit volume, the specific volume can be written as  $v = 1/\rho$ and
\begin{equation}
dh =  \frac{1}{\rho}dp
\end{equation}
 which can be generalized to
\begin{equation}
\nabla h =  \frac{1}{\rho} \nabla p
\end{equation}
and replaced into Euler equation resulting 
\begin{equation}
\frac{\partial \mathbf {u}}{\partial t} + \mathbf{u} \cdot \nabla \mathbf {u}  = - \nabla h + \mathbf{F} \rho 
\end{equation}
We now use a vector calculus identity referred to as the vector dot del operator
\begin{equation}
\mathbf {u} \cdot \nabla \mathbf{u}  = \frac{1}{2} \nabla u^2 - \mathbf{u}   \times (\nabla \times \mathbf {u} )
\end{equation}
By replacing it into the ``enthalpy'' Euler equation, and with the condition that the velocity does not vary in time specified previously, we have 
\begin{equation}
 \mathbf{u}   \times (\nabla \times \mathbf {u} ) = - \nabla (h + \frac{1}{2}  u^2) + \mathbf{F} 
\end{equation}
Flow conditions require that the curl terms on the left side is zero. We can also rewrite the force term as the negative of the gradient of a potential $\Psi$
\begin{equation}
0 = - \nabla (h + \frac{1}{2}  u^2)  - \nabla \Psi
\end{equation}
and lump the terms into a single gradient operator 
\begin{equation}
 \nabla (h + \frac{1}{2}  u^2 + \Psi) = 0
\end{equation}
Replacing the expression for the enthalpy previously defined and considering $\Psi$ as the force potential to Earth's gravitational field $\Psi = \rho g$, we are left with 
\begin{equation}
\boxed{
\frac{p}{\rho} + \frac{1}{2}  u^2 + \rho g = \text{constant}
}
\end{equation}
This equation states that the left hand sum is constant along any  \emph{streamline}. A streamline is a line whose tangent at any point gives the direction of the velocity vector at that point. Streamlines are defined by the system of equations
\begin{equation}
\frac{dx}{u_x} = \frac{dy}{u_y} = \frac{dz}{u_z}
\end{equation}

\section{Reynolds number}
The Reynolds number is known in soil physics  and applied hydraulics and hydrology as a means of establishing if a given flow regime is laminar or turbulent. It is much more than that, it derives from writing the Navier-Stokes equation in a dimensionless form and conveys important information about the flow regime and scaling of flow processes and can be used in laboratory to design experiments to measure flow parameters that could not be measured otherwise, on account of the law of similarity. We start with the Navier-Stokes equations in an incompressible and homogeneous fluid with regard to density. The body force is ignored, but the same principles applied to the other terms can be used if one wishes to proceed including the body force 
\begin{equation}
\rho(\frac{\partial u_i}{\partial t} + u_j \frac{\partial u_i}{\partial x_j} )  = -\frac{\partial p}{\partial x_i} + \mu \frac{\partial^2 u_i}{\partial x_j \partial x_i}
\end{equation}
Each quantity on the Navier-Stokes equation is now written in a dimensionless form by defining some representative length L and velocity U, with the new dimensionless variables $\mathbf{u}'$, $t'$ and $\mathbf{x}'$   
\begin{equation}
\mathbf{u}' = \frac{\mathbf{u}}{U}
\end{equation}
\begin{equation}
\mathbf{x}' = \frac{\mathbf{x}}{L}
\end{equation}
\begin{equation}
t' = \frac{tU}{L}
\end{equation}
Note that U/L has units of 1/time since [L T$^{-1}$] / [L] = [T$^{-1}$]. Pressure can also be normalized using
\begin{equation}
p' = \frac{p}{\rho U^2}
\end{equation}
Since $\rho U^2$ has units of [M L$^{-3}$] [L T$^{-1}$]$^2$ = [M L$^{-1}$ T$^{-2}$] which is equivalent to pressure.  Substituting the dimensionless quantities on the Navier-Stokes 

\begin{equation}
\rho(\frac{\partial u'_i U}{\partial t'L/U} + u'_j U \frac{\partial u'_i U}{\partial x'_j L} )  = -\frac{\partial p'\rho U^2}{\partial x'_i L} + \mu \frac{\partial^2 u'_i U}{\partial x'_j L \partial x'_i L}
\end{equation}
Recalling that constant terms can be brought out of the derivatives and reorganizing
\begin{align*}
\frac{U^2}{L}\rho(\frac{\partial u'_i}{\partial t'} + u'_j \frac{\partial u'_i}{\partial x'_j} )  = -\frac{U^2}{L}\frac{\partial p'\rho}{\partial x'_i} + \mu \frac{U}{L^2}\frac{\partial^2 u'_i}{\partial x'_j \partial x'_i} \\
\frac{\partial u'_i}{\partial t'} + u'_j \frac{\partial u'_i}{\partial x'_j}  = -\frac{\partial p'}{\partial x'_i} +  \frac{\mu}{\rho U L}\frac{\partial^2 u'_i}{\partial x'_j \partial x'_i} 
\end{align*}
The dimensionless quantity 
\begin{equation}
\boxed{
Re = \frac{\rho U L}{\mu}
}
\end{equation}
is called the Reynolds number, being introduced by Stokes and later named after Osborne Reynolds who popularized its use. The dimensionless form of the Navier-Stokes equation can be written in vector notation as   

\begin{equation}
\frac{\partial \mathbf{u}'}{\partial t'} + \mathbf{u}'\cdot \nabla \mathbf{u}'  = -\nabla p' + \frac{1}{Re} \nabla^2 \mathbf{u}'
\end{equation}

The Reynolds number is a ratio of inertial forces, that tend to keep the fluid in motion, and viscous forces, which resist movement. In fluids in which the Reynolds number is low, there is large contribution of inertial forces. If inertial forces are much larger than viscous forces the flow can become unstable and flow instabilities can develop. Under such conditions the assumption that the flow is smooth, constant and ``continuous'' is no longer valid and the flow can no longer be considered laminar. If inertial forces are very high in respect to viscous forces, fully turbulent flow might develop.

\section{List of symbols for this chapter}

\begin{longtable}{ll}
    	$ \mathbf{s}  $ & Vector position function  \\
    	$ \mathbf{u}, \mathbf{v}   $ & Vector velocity functions  \\
        $ \mathbf{i}, \mathbf{j}, \mathbf{k}, $ & Unit vectors in Cartesian coordinates  \\
        $ \mathbf{x}, \mathbf{y}, \mathbf{z}, $ & Unit vectors in Cartesian coordinates  \\
    	$ s_x, s_y, s_z $ & Components of the position vector in Cartesian coordinates  \\
        $ u_x, u_y, u_z $ & Components of the velocity vector in Cartesian coordinates  \\
    	$ u, v, w $ & Components of the velocity vector in Cartesian coordinates  \\
    	$ t $ & Time  \\
  	$ \frac{D}{D t}  $ & Substantial (or material) derivative  \\
    	$ f  $ & Generalized function  \\
    	$ \rho $ & Density  \\
    	$ V $ & Volume  \\
    	$ A $ & Area \\
    	$ \mathbf{n} $ & Normal unit vector  \\
    	$ m, M $ & Mass \\
    	$  x,  y,  z, $ & Generalized Cartesian coordinates  \\
    	$ \mathbf{F} $ & Force vector \\
    	$ \mathbf{a} $ & Acceleration vector\\
    	$ \mathbf{p} $ & Momentum vector \\
    	$ \sigma $ & Stress  \\
    	$ \delta_{ij} $ & Kroenecker delta  \\
    	$ p $ & Pressure  \\
    	$ S $ & surface  \\
    	$ d_{ij} $ & Deviatoric stress tensor  \\
    	$ \mathbf{d} $ & Deviatoric stress tensor, vector form  \\
    	$ \mathbf{g} $ & Gravitational body force vector \\
    	$ g_x, g_y, g_z $ & Components of the gravitational body force vector in Cartesian coordinates  \\
    	$ A_{ijkl} $ & Fourth-order tensor  \\
    	$ e_{kl} $ & Symmetric component of the deviatoric stress tensor \\
    	$ \xi_{kl} $ & Antisymmetric component of the deviatoric stress tensor \\
    	$ \epsilon_{klm} $ & Levi-Civita symbol  \\
    	$ \boldsymbol{ \omega } $ & Vorticity vector  \\
    	$ \omega_x, \omega_y, \omega_w  $ & Components of the vorticity vector in Cartesian coordinates  \\
        $ e_{kk}  $ & Rate of expansion  \\
        $ \mu $ & Viscosity  \\
    	$ h $ & Specific enthalpy  \\
    	$ s $ & Specific entropy  \\
    	$ v $ & Specific volume  \\
    	$ T $ & Temperature  \\
    	$ P $ & Pressure  \\
    	$ S $ & Entropy  \\
    	$ H $ & Enthalpy \\
    	$ \Psi $ & Potential  \\
    	$ \mathbf{u'} $ &  Dimensionless velocity vector\\
    	$ \mathbf{x'} $ &  Dimensionless position vector\\
    	$ t' $ &  Dimensionless time \\
    	$ p' $ &  Dimensionless pressure \\
    	$ Re $ & Reynolds number  \\
    	$ L $ & Length \\
    	$ U $ & Velocity  \\
 
\end{longtable}

% !TEX TS-program = pdflatex
% !TEX encoding = UTF-8 Unicode

% Example of the Memoir class, an alternative to the default LaTeX classes such as article and book, with many added features built into the class itself.

\chapter{Settling and grain size distributions}
\label{ch5}

\section{Stokes law of settling}

There are two basic methods for the determination of particle size distribution of soils and sediments. For coarser particles such as gravel and cobbles, and sands to some extent, grain size distributions or grain size fractions can be determined using sieves. In essence what is done is to disaggregate the particles, if they are aggregated at all, using physical and/or chemical methods, pass the material through a set of sieves of decreasing apertures, weigh the dried fraction retained in each sieve and express the amount of each fraction as a percent of the total dried mass of the material used.  The sieve aperture used for each class interval is a convention, and the textural classes vary between disciplines in soil science, Earth science and engineering. 

\begin{figure}[ht]
\centering
 \includegraphics[width=0.4\textwidth]{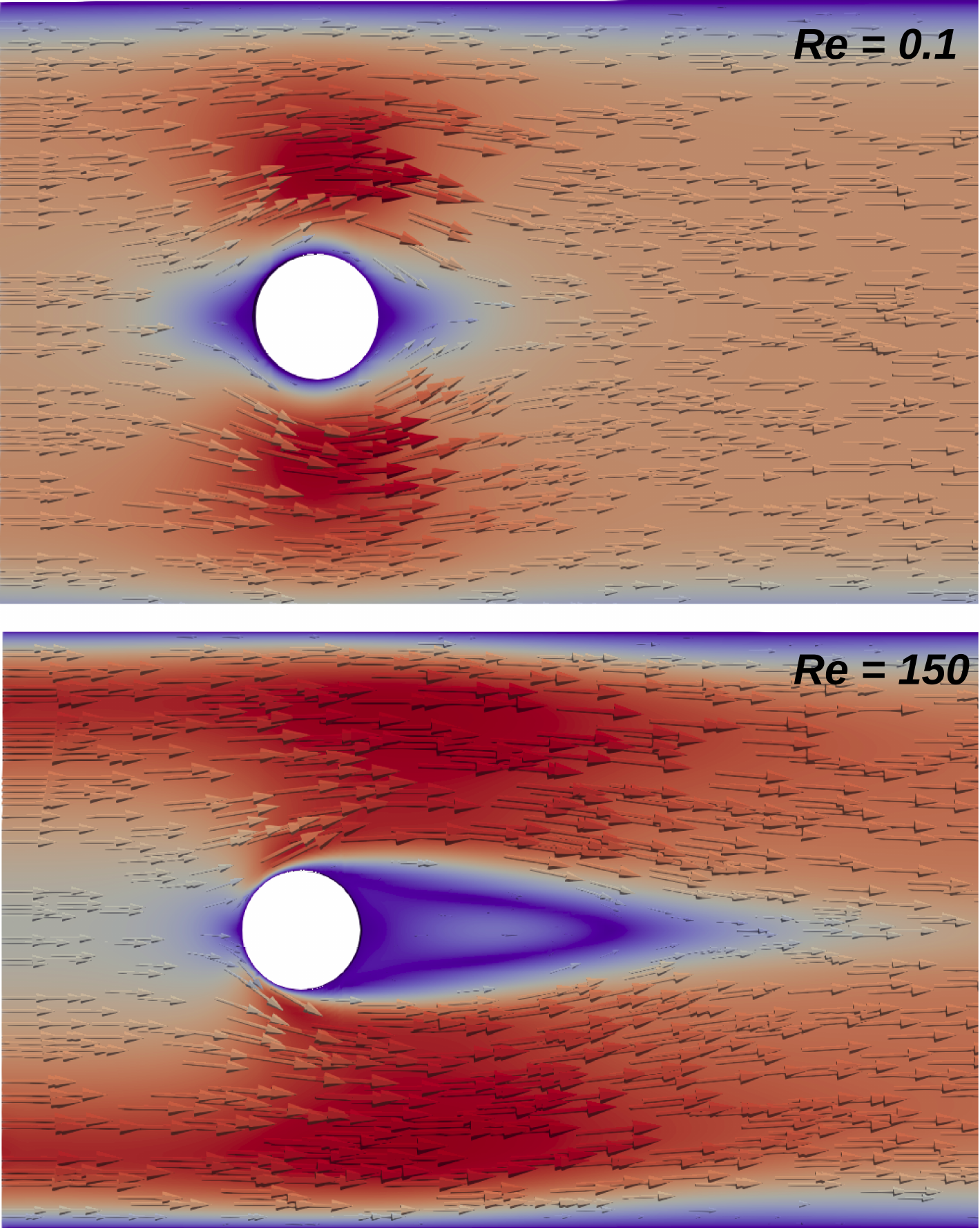}
\caption{Numerical simulation of flow around of cylinder (sphere in 2D) under laminar regime ($Re = 0.1$) and transitional regime ($Re = 150$). Flow at high Reynolds number is not stable and instabilities, such as a Karman vortex street, and turbulence might develop if the simulation is continued.}
\label{ch5_fig1}
\end{figure}

For particles finer than sand, sieving does not work adequately, first because it might not be feasible to create sieves with apertures small enough for the particles, which might have diameters on the micrometer range, second because it might be virtually impossible to disaggregate these particles in the dried state, and third because even if these particles could be disaggregated and precision micrometer aperture sized sieves created, they might cause the clogging of the apertures of the sieve which at that size range would act analogous to pores. For smaller particles, most methods of particle size determination rely on settling of the dispersed particles in a fluid, usually water. Particles are assumed as perfectly spherical, non-interacting, slowly settling in an infinitely dispersed solution. Under these conditions the Navier-Stokes equation solution for a sphere settling in a viscous fluid is the solution to the problem. The fact that the settling is slow under low Reynolds number guarantees that the flow around the sphere is laminar (Figure \ref{ch5_fig1}). Under turbulent flows, analytical solutions to the Navier-Stokes equations, such as the one presented below, might not be feasible due to the complexity of the chaotic fluctuations of the flow regime. In any case, even before turbulence developed, the settling velocity might be large enough that the inertial component of the Navier-Stokes equation is not negligible, and analytic solutions might not be achievable.

The solution of the sphere settling problem was provided by G.G. Stokes \cite{stokes}, many authors present interpretations of Stokes solution which are more or less similar to the original and to each-other \cite{lamb, landau, batchelor, acheson}. The solution given by Acheson \cite{acheson} is presented here for reasons of clarity and conciseness. For slow, steady-state settling conditions the mathematical problems is to find an analytical solution to the Stokes and conservation equations presented in Chapter \ref{ch4}.   
\begin{equation*}
 \nabla p =  \mu \nabla^2 \mathbf{u} 
\end{equation*}
\begin{equation*}
\nabla \cdot \mathbf{u} = 0 
\end{equation*}
Not surprisingly, the problem of flow past a sphere is solved by using spherical coordinates. Thus the vector 
\begin{equation*}
 \mathbf{u} = \langle u_x, u_y, u_z \rangle = u_x \mathbf{i} + u_y \mathbf{j} + u_z \mathbf{k}  
\end{equation*}
is written in spherical coordinates as 
\begin{equation}
 \mathbf{u} = \langle u_r(r, \theta), u_{\theta}(r, \theta), 0 \rangle 
\end{equation}
The conservation equation $ \nabla \cdot \mathbf{u}$ is satisfied by a Stokes stream function $\Psi(r , \theta)$ which satisfies the conditions
\begin{equation}
u_r = \frac{1}{r^2 \sin{\theta}} \frac{\partial \Psi}{\partial \theta}, ~ u_{\theta} = \frac{1}{r \sin{\theta}} \frac{\partial \Psi}{\partial r}
\end{equation}
Now, the curl of $\mathbf{u}$ in spherical coordinates (consult Chapter \ref{ch1}) is  
\begin{equation}
\nabla \times \mathbf{u} =  \langle 0, 0,  \frac{1}{r \sin{\theta}} E^2 \Psi \rangle
\end{equation}
With $E^2$ representing the differential operator
\begin{equation}
E^2 = \frac{\partial^2}{\partial r^2} + \frac{\sin{\theta}}{r^2} \frac{\partial}{\partial \theta} (\frac{1}{\sin{\theta}} \frac{\partial}{\partial \theta} )
\end{equation}
By virtue of vector identities, Stokes equation can be written as 
\begin{equation}
\nabla p = -\mu \nabla \times (\nabla \times \mathbf{u}) 
\end{equation}
in which the gradient operators in $p$ and after $\mu$ can be written in spherical coordinates as 
\begin{equation}
\nabla = \langle \frac{\partial}{\partial r}, \frac{1}{r} \frac{\partial}{\partial \theta}, \frac{1}{r \sin{\theta}} \frac{\partial}{\partial \phi}  \rangle  
\end{equation}
from  Equations 5.3, 5.5. and 5.6 the components in $\mathbf{r}$ and $\boldsymbol{\theta}$ can be obtained
\begin{equation}
\frac{\partial p}{\partial r} = \frac{\mu}{r^2 \sin{\theta}} \frac{\partial}{\partial \theta} E^2 \Psi, 
~
\frac{1}{r} \frac{\partial p}{\partial \theta} = \frac{-\mu}{r \sin{\theta}} \frac{\partial}{\partial r} E^2 \Psi
\end{equation}
The detailed procedure for cross differentiation considering a stream function $\Psi$ is given by Lamb \cite{lamb}. Acheson \cite{acheson} explains it in terms of the pressure being eliminated by cross-differentiation which implies that $E(E^2\Psi) = 0$ such that\footnote{Obviously there is much more to this, the user is encouraged to consult Stokes (1901)\cite{stokes} and Lamb (1932)\cite{lamb}, for an in depth discussion. Stokes (1901) is a reprint of the original and Lamb's discussion draws directly from the source and the level of detail provided is unmatched in modern literature (Horace Lamb was a student of Sir George Stokes at Trinity College, University of Cambridge). There is a very elegant derivation in Landau and Lifshitz \cite{landau} based largely on vector calculus operators. Although precise, Landau and Lifshitz  and Batchelor \cite{batchelor} derivations rely heavily in implicit arguments. We tried to find a compromise between conciseness and clarity and most likely have not been entirely successful to say the least.}  

\begin{equation}
[\frac{\partial^2}{\partial r^2} + \frac{\sin{\theta}}{r^2} \frac{\partial}{\partial \theta} (\frac{1}{\sin{\theta}} \frac{\partial}{\partial \theta} ) ]^2 \Psi= 0
\end{equation}
The boundary conditions on the settling sphere are
\begin{equation}
u_r = u_{\theta} = 0 ~\text{at} ~r = a
\end{equation}
\begin{equation}
u_r = U\cos{\theta},~u_{\theta} = -U\sin{\theta}   ~\text{as} ~r \rightarrow a
\end{equation}
Because the decomposition of the velocity components in $\mathbf{r}$ and $\boldsymbol{\theta}$. The latter conditions along with Equation 5.2, results, by integration in
\begin{equation}
\Psi = \frac{U}{2} r^2 \sin^2{\theta}   ~\text{as} ~r \rightarrow a
\end{equation} 
Now we search for a solution to the partial differential equation 5.8. The theory of solutions to partial differential equations plays a fundamental role in analytical methods in fluid mechanics and soil physics, readers unfamiliar with it are encouraged to consult Boas \cite{boas} for a concise introduction and Asmar \cite{asmar} for a more detailed discussion. Equation 5.8 is satisfied by a solution of the form 
\begin{equation}
\Psi =  \sin^2{\theta}   f(r)
\end{equation} 
provided that 
\begin{equation}
[\frac{d^2}{dr^2} - \frac{2}{r^2}]   f(r) = 0 
\end{equation} 
with solution of the form
\begin{equation}
 f(r) = \frac{A}{r} + Br + Cr^2 + Dr^4  
\end{equation} 
with A, B, C and D constants. Substituting Equation 5.14 into 5.12 results
\begin{equation}
\Psi =  \sin^2{\theta} [ \frac{A}{r} + Br + Cr^2 + Dr^4]
\end{equation} 
The values of the constant can be obtained by applying the boundary conditions previously defined to Equation 5.15, resulting in
\begin{equation}
\Psi =  \frac{U}{2}[ r^2 + \frac{a^3}{2r} - \frac{3ar}{2}]  \sin^2{\theta} 
\end{equation} 
By combining the $\mathbf{r}$ component of Equation 5.7 , Equation 5.4 and Equation 5.16 and integrating from $r = 0$ from $r = \infty$  a pressure function is obtained
\begin{equation}
p =  p_{\infty} - \frac{3}{2} \frac{\mu U a}{r^2} \cos{\theta} 
\end{equation} 
The stress components on a sphere settling in a viscous fluid are 
\begin{equation}
\tau_r = -p + 2\mu \frac{\partial u_r}{\partial r}, ~ \tau_{\theta} = \mu r \frac{\partial}{\partial r}(\frac{u_{\theta}}{r}) + \frac{\mu}{r} \frac{\partial u_r}{\partial \theta}, ~\tau_{\phi} = 0 
\end{equation} 
From Equations 5.2, 5.16 and 5.18 we have
\begin{equation}
\tau_r = -p_{\infty} + \frac{3}{2} \frac{\mu U}{a}\cos{\theta} 
\end{equation} 
\begin{equation}
\tau_{\theta} = - \frac{3}{2} \frac{\mu U}{a}\sin{\theta} 
\end{equation} 
The resulting stress vector on the sphere is 
\begin{equation}
\tau_{net} = \tau_r  \cos{\theta} - \tau_{\theta}\sin{\theta} = -p_{\infty}\cos{\theta} + \frac{3}{2} \frac{\mu U}{a} 
\end{equation} 
and the total drag force on the sphere can be calculated by integrating the net force in spherical coordinates
\begin{equation}
F_{D} =  \int_0^{2\pi} \int_0^{\pi} a^2~\tau_{net}~\sin{\theta}~d \theta d \phi
\end{equation} 

\begin{align*}
F_{D} =  \int_0^{2\pi} \int_0^{\pi} a^2~(-p_{\infty}\cos{\theta} + \frac{3}{2} \frac{\mu U}{a})~\sin{\theta}~d \theta d \phi \\
F_{D} =  \int_0^{2\pi} \int_0^{\pi} a^2~(-p_{\infty} ~\sin{\theta}~\cos{\theta})~d \theta d \phi + \int_0^{2\pi} \int_0^{\pi} (a~\frac{3}{2} \mu U~\sin{\theta})~d \theta d \phi \\
F_{D} =  \int_0^{2\pi}  a^2~(-p_{\infty} ~\frac{\sin^2{\theta}}{2})\bigg|_0^{\pi}~d \phi + \int_0^{2\pi} -(a~\frac{3}{2} \mu U~\cos{\theta})\bigg|_0^{\pi} ~ d \phi \\
\end{align*} 
Remembering that $\sin{0}$ = 0 and $\sin{\pi} = 0$ and that $\cos{0}$ = 1 and $\cos{\pi} = -1$, the first term on the right side is equal to zero
\begin{align*}
F_{D} =   \int_0^{2\pi} (2 a~\frac{3}{2} \mu U) ~ d \phi \\
F_{D} =   \int_0^{2\pi} (a 3 \mu U) ~ d \phi \\
F_{D} =   (a 3 \mu U) \phi \bigg|_0^{2\pi} \\
\end{align*} 
\begin{equation}
\boxed{
F_{D} =  6 \pi \mu U a
}
\end{equation} 
This is Stokes formula for the drag on a sphere settling at low velocities in a viscous fluid. It is valid for creeping flow, often called Stokes flow, in which the inertial forces are much smaller than the viscous forces. Flow is laminar, meaning that fluid particles follow smooth trajectories, without turbulent mixing. It is often conceptualized as continuous flow lines around the object which is moving through the fluid or around which the fluid is moving. 
The terminal settling velocity can be found by the force balance in the $z$ direction since the sphere is settling in direction of the gravitational force, against viscosity and buoyancy in a stationary fluid  
\begin{equation}
\sum F_z = F_D + F_b + F_g = ma
\end{equation} 
Buoyancy force is proportional to the mass of fluid displaced by the settling object 
\begin{equation}
F_b = m_{fluid} g
\end{equation} 
While the gravitational force is 
\begin{equation}
F_g = m_{sphere} g
\end{equation} 
But the volume of displaced fluid and the volume of the sphere are the same and equal to 
\begin{equation}
V_{sphere} = \frac{4}{3} \pi R^3
\end{equation} 
If the density of the fluid is 
\begin{equation}
\rho_{fluid} = \frac{m_{fluid}}{V_{fluid}} 
\end{equation} 
and that of the sphere is 
\begin{equation}
\rho_{sphere} = \frac{m_{sphere}}{V_{sphere}} 
\end{equation} 
the buoyancy force can be written as 
\begin{equation}
F_b = \rho_{fluid} \frac{4}{3} \pi R^3 g
\end{equation} 
while the gravitational force is 
\begin{equation}
F_g = \rho_{sphere} \frac{4}{3} \pi R^3 g
\end{equation} 
Now in the drag force equation $a$ is R, the radius of the sphere and the terminal translational velocity can be written as $v_s$ such that
\begin{equation}
\boxed{
F_{D} =  6 \pi \mu v_s R
}
\end{equation} 
Replacing all the forces in Equation 5.24 and remembering that the terminal velocity is constant, i.e., there is no acceleration, $ma = 0$, and 
\begin{equation}
\sum F_z = 6 \pi \mu v_s R + \rho_{fluid} \frac{4}{3} \pi R^3 g - \rho_{sphere} \frac{4}{3} \pi R^3 g = 0
\end{equation}  
Where the coordinates systems establishes that downward forces have negative sign. Solving for $v_s$
\begin{align*}
6 \pi \mu v_s R = \rho_{sphere} \frac{4}{3} \pi R^3 g - \rho_{fluid} \frac{4}{3} \pi R^3 g \\
6 \pi \mu v_s R = \frac{4}{3}(\rho_{sphere} - \rho_{fluid}) \pi R^3 g \\
v_s = \frac{2}{9} \frac{ (\rho_{sphere} - \rho_{fluid})  g}{\mu}   R^2 
\end{align*}  
Since the sphere in this case is the solid particles settling, or solid minerals, we can rewrite using notation from previous chapters as 
\begin{equation}
\boxed{
v_s = \frac{2}{9} \frac{ (\rho_{p} - \rho_{f})  g}{\mu}   R^2 
}
\end{equation}
This equation is know as Stokes law of settling and it is used to model the settling behavior of infinitely dispersed spherical solid particles settling in laminar regime. Terminal velocity is a function of the squared radius of the particle, therefore, everything else constant in a viscous fluid, larger particles settle faster. Settling velocity is also controlled by the densities of the fluid and the solid, viscosity and gravity. Fluid density and viscosity are controlled by temperature and empirical laws are needed to estimate these quantities.  

Stokes law of settling is valid for laminar flow, where $Re \ll 1$ \footnote{The limits are determined empirically so the critical value of the Reynolds number might depend on the nature of the problem being studied. More specifically, there might be transitional flow regimes between laminar and fully turbulent.}. The Reynolds number for a settling spherical particle is 
\begin{equation}
Re = \frac{2\rho_f R v_s}{\mu}
\end{equation}
Considering a maximum Reynolds number for laminar flow in settling of $Re \approx 1$, the maximum radius of a smooth spherical  particle  falling under laminar flow regime is 

\begin{align*}
\frac{\mu}{2\rho_{f} R} = \frac{2}{9} \frac{ (\rho_{p} - \rho_{f})  g}{\mu}   R^2 \\
\end{align*}
\begin{equation}
R = \sqrt[3]{ \frac{9 \mu^2}{4 \rho_f(\rho_s-\rho_f)g }}
\end{equation}

For water density of $\rho \approx 1000 $ kg m$^{-3}$, and viscosity at 25 $^{o}$C of $\mu \approx 0.00089 $ Pa s$^{-1}$, the maximum radius of a smooth spherical quartz particle with $\rho_s \approx 2650 $ kg m$^{-3}$ and $g = 9.80665 m~s^{-2}$ falling under laminar flow regime is approximately $R = 4.793 \times 10^{-5}~m $ or $R = 0.048~mm$, which translates into a diameter of approximately $0.1~mm$. 

On the other hand of the spectrum, for colloidal particles, Stokes law is no longer valid because under real circumstances, the solution is not infinitely dispersed and the particles are subjected to thermal motion (also known as Brownian motion). As smaller particles settle, a concentration gradient is established along the length of the settling column and a concentration gradient will keep particles in suspension as long as there is no coagulation or aggregation in suspension \cite{leao2013, piazza}.  In other words, the thermal motion is able to overcome gravitational forces\footnote{Albert Einstein's Ph.D. dissertation was on Brownian motion, before he was made famous by his studies in relativity and the photoelectric effect}. The Péclet number is another dimensionless number in hydrodynamics which expresses the ratio between advective and diffusive transport rates \cite{piazza}. If $Pe \ll 1$, diffusive transport predominates and settling is perturbed by thermal motion.  Considering $Pe = 1$ the radius of a sphere settling is given by 
\begin{equation}
R = [\frac{3 k T}{4 \pi (\rho_s - \rho_f)g}]^{\frac{1}{4}}
\end{equation}
In which $k = 1.380649 \times 10^{-23}~J~K^{-1}$ is the Boltzmann constant and all other variables were previously defined. Using the values for a spherical silicate mineral settling in water presented for the previous calculations, the radius above which diffusive transports can be consider to start affecting settling is $ r = 4.96 \times 10^{-7} ~m$ or $r = 0.4965~\mu m$, or approximately $1~\mu m$ in diameter. The conclusion is that Stokes law for an average spherical mineral particle settling in water at room temperature is valid for diameters above that of colloidal particles and below that of fine sand particles. Below the colloidal limit settling is hindered by thermal motion and above the fine sand limit settling is under transitional or turbulent flow regime. Stokes law only applies to creeping (laminar) flow. In reality, the dynamics of particles on the clay size range is much more complicated because of surface forces among the interacting particles. These forces can cause aggregation of particles in suspension to which Stokes law applies to some degree and settling via coagulation and other mechanisms. Stokes law in particle size characterization of soils and fine sediments assumes that aggregation and coagulation do not apply and the analysis should be disregarded if these processes are observed. 

We saw in Chapter \ref{ch3} that when charged particles are put in contact with an electrolyte solution an Electric Double Layer (EDL) develops with charged ions adsorbed onto the surface by electrostatic forces. The concentration of ions is larger at the surface and at certain concentrations the EDL extends far enough that repulsive forces are dominant in the interaction between charged particles and the colloidal suspension remains dispersed. However, this is not the entire story. Between particles, atoms and molecules, \emph{dispersion forces}\footnote{The term dispersion forces here refers to intermolecular and interatomic forces and not to particles in dispersion} originating from quantum fluctuations arise. These dispersion forces are grouped as what is called van der Waals forces (vdW) and for the systems we are concerned are attractive in nature. If the electrolyte concentration is increased, the EDL is compressed and the vdW forces which are stronger than the repulsive forces in the short range predominate and the colloidal particles can aggregate and coagulate. This can happen with dispersed soil solutions depending on the valence and concentration of the electrolyte as clays are charged particles of small diameter. The theory of balance of these attractive and repulsive forces in colloidal suspensions is called DLVO after Boris Derjaguin, Lev Landau, Evert Verwey and Theodoor Overbeek who developed it, the first two working together and the last two independently. Considering the interaction of two charged spheres of radius R in a 1:1 electrolyte solution, the energy of interaction due to electrostatic double layer forces is given by \cite{israelachvili} 
\begin{equation}
	W_{EDL} = \frac{64 \pi k T R \rho_{\infty}}{\kappa} \tanh^2 (\frac{z e \psi_0  }{4kT}) \exp{(-\kappa D')} 
\end{equation}
in which the eletrochemical parameters are the same as in Chapter 3, i.e. $\rho_{\infty}$ is the electrolyte concentration in bulk solution, $\psi_0$ is the surface potential, $z$ is the valence of the ions, $k$ is the Boltzmann constant, $T$ is the absolute temperature, $\kappa$ is a parameter related to the thickness of the double layer, $e$ is the electronic charge, and $R$ is the spheres radius.  The attractive vdW force is given for two spheres of the same radius as \cite{israelachvili}
\begin{equation}
W_{vdW} = \frac{A_HR}{12 D'} 
\end{equation}
in which $D'$ is the separation distance usually in $nm$ or angstrons (\AA) and $A_H$ is the Hamaker constant
\begin{equation}
A_H = \pi^2 C \rho_1 \rho_2 
\end{equation}
In which $\rho_1$ and $\rho_1$ are the number densities of the interacting spheres and $C$ is the London constant\footnote{For the time being we will refrain from exploring the derivations of these relationships as they might require exploring quantum mechanical perturbation theory}. In the case of equal particles $\rho_1 = \rho_2$ and $\rho_1\rho_2 = \rho^2$. The Hamaker constant has an important role in the adsorption of water films in surfaces and is fundamental for modeling dry water adsorption in soils. The pair interaction energy considering the two forces is 
\begin{equation}
W(D') = W_{EDL} - W_{vdW} 
\end{equation}
Negative values of the pair interaction energy (usually given in J) indicate potential for van der Waals attraction and coagulation. However the relationship is complex and depending on several of the factors seen in the preceding equations including temperature, which affects the kinetic energy of the colloidal particles, surface potential, which might change with factors such as pH, as we discussed in Chapter \ref{ch3}, and others. The proper analysis must be done in terms of the energy barrier that needs to be overcome for coagulation to occur. The electrolyte concentration at which coagulation occurs is the critical coagulation concentration (ccc) which plays a critical role in grain size analysis of soils, especially in the grain size analysis of oxidic soils which are variable charged and in which the analysis is almost always performed under dispersion using sodium hydroxide or other sodium salts \cite{leao2013}. 

If the constraints of settling velocity, lower limit of settling and dispersing agent concentration are observed, then grain size analysis by settling methods can be performed normally.

%% *** See Lamb p.236 and 602

\section{Particle size distribution}

For the determination of particle size distribution and the proportion of the particle fractions we need some more or less arbitrary scale for classification of grain sizes (Table \ref{table:ch5_table1}). For particle size analysis, soils must be first dispersed chemically and or physically using salt solutions such as sodium hydroxide and/or sodium hexametaphoshate at concentrations high enough to create repulsive EDL forces, but not above the ccc. Mechanical energy is applied to the samples by stirring, ultrasonic waves, gas bubbles or shaking. Several combinations of chemical and physical methods are possible, some more efficient than others \cite{geeor}.   

\begin{table}[h]
\centering
\caption{Classification of soil particles based on diameter \cite{soilsurvey}.}
\begin{center}
\begin{tabular}{ll}
\hline
Fraction Name & Particle Size Limit (mm)  \\ \hline
Fine Clay &    0.0002          \\ 
Coarse Clay &    0.002          \\ 
Fine Silt &       0.02        \\ 
Coarse Silt &       0.05        \\ 
Very Fine Sand & 0.1               \\ 
Fine Sand & 0.25                \\ 
Medium Sand &  0.5               \\ 
Coarse Sand & 1.0                 \\ 
Very Coarse Sand & 2.0                 \\ 
Fine Gravel & 5                \\ 
Medium Gravel & 20                \\ 
Coarse Gravel & 76                \\ 
Cobbles         &      250             \\ 
Stones        &      600		\\
Boulders &  $>$ 600  \\
\hline
\end{tabular}
\end{center}
\label{table:ch5_table1}
\end{table}

After dispersion, the sample is again vigorously stirred and the amount of particles in suspension is measured at prescribed times directly, by collecting a volumetric sample of the dispersion, or indirectly, by using a hydrometer or by transmission/diffraction of electromagnetic radiation methods as in laser and x-ray methods. The most common direct method is the pipette method in which a solution sample is taken and oven dried and the mass in suspension measured using a precision scale. The mass in suspension is then extrapolated for the entire volume. As for the indirect methods the most commonly used is the hydrometer method. Hydrometers measure the amount or particles in suspension by buoyancy, the greater the mass in suspension the greater the buoyancy force, as particles settle, the hydrometer will tend gradually sink in response to settling of particles until settling stops, when the hydrometer will be reading the amount of particles in suspension under equilibrium, or the density of the dispersing solution if the material is composed only of coarser particles. Most hydrometers are nothing more than a weighed glass tube with a bulbous shape and a scale. The scale is calibrated at the solution level and will read zero for pure water at a given temperature and as the mass of particles in suspension increases, the reading will increase in the downward direction indicating the mass of particles in suspension, usually in grams, per unit volume, usually liters. For any settling based method, the settling velocity is given by Stokes law and the mass of particles greater than a given radius (or diameter) in suspension will depend on the sampling depth and on the calculated settling time. From basic physics the velocity of a particle is given by  
\begin{equation}
v_s = \frac{\Delta S}{\Delta t}  
\end{equation}
Because we are concerned with a spherical particle settling under Stokes regime, and therefore at a constant terminal velocity, $v_s$ is the settling velocity, $S$ is the settling depth and $t$ is the time for the particle to reach $S$.  Replacing the velocity in Stokes law
\begin{equation*}
\frac{\Delta S}{\Delta t} = \frac{2}{9} \frac{ (\rho_{p} - \rho_{f})  g}{\mu}   R^2 
\end{equation*}
\begin{equation*}
\Delta t = \frac{9 \mu \Delta S}{2 (\rho_{p} - \rho_{f})  g  R^2}    
\end{equation*}
and because it is convenient to set $t_0$ and $S_0$ to zero we have
\begin{equation}
\boxed{
t = \frac{9 \mu S}{2 (\rho_{p} - \rho_{f})  g  R^2}    
}
\end{equation}
This equation implies that at time $t$, all particles with radius greater than $R$ will have settled to a depth greater than $S$. In addition to the settling depth and the particle radius, the time for a reading for determining particle size distribution using a given method is dependent on the fluid viscosity, the settling mineral particles density, the density of the fluid and gravitation acceleration.  Particle density, discussed in Chapter \ref{ch2}, can be measured with relatively simple methods, while fluid viscosity and density will require more specialized high precision techniques. It is much more convenient to perform the measurements at a standard temperature to which water viscosity and density have tabulated values or to use published empirical relationships for water viscosity and density, for example  
\begin{equation}
\mu =  1 \times 10^{-6} [280.68 (\frac{T}{300})^{-1.9} + 511.45 (\frac{T}{300})^{-7.7} + 61.131 (\frac{T}{300})^{-19.6} + 0.45903 (\frac{T}{300})^{-40.0} ]
\end{equation}
in which $\mu $ is water dynamic viscosity in Pa s and T is the temperature in Kelvin in the range of  253.15 K to 383.15 K \cite{huber}, and 
\begin{equation}
\rho_f =  999.974950 [1 -  \frac{(Tc - 3.983035)^2(Tc + 301.797)}{522528.9(Tc + 69.34881)}] 
\end{equation}
in which $\rho_f $ is the density of water in kg m$^{-3}$ and Tc is the temperature in $^{o}$C in the range of 0 $^{o}$C to 40 $^{o}$C \cite{tanaka}. For example, using the equations above at 25 $^{o}$C = 298.15 K, $\mu = 0.000889997~ Pa~ s$, $\rho_f = 997.047~ kg~ m^{-3}$, and  using $\rho_s $ and $g$ as in previous calculations, for the silt $D = 0.05~mm$ and clay $D = 0.002~mm$ limits, the time for all the particles with diameter greater than $D$ to settle 0.1 m (10 cm) will be $t_{clay} = 6.86~h$  and $ t_{silt} = 39.53~s $. Thus, a reading at $t_{silt}$ will indicate the amount of particles finer than 0.05 mm in suspension and the reading at $t_{clay}$ will indicate the amount of particles finer than 0.002 mm in diameter in suspension.  

The particle size distribution curve is the measured cumulative mass fraction of particles in suspension (usually in \%) at a given time plotted as a function of the logarithm of particle diameter corresponding to each time, given by 
\begin{equation}
D = \sqrt{\frac{18 \mu S}{ (\rho_{p} - \rho_{f})  g  t}}    
\end{equation}

\begin{figure}[h!]
\centering
 \includegraphics[width=0.7\textwidth]{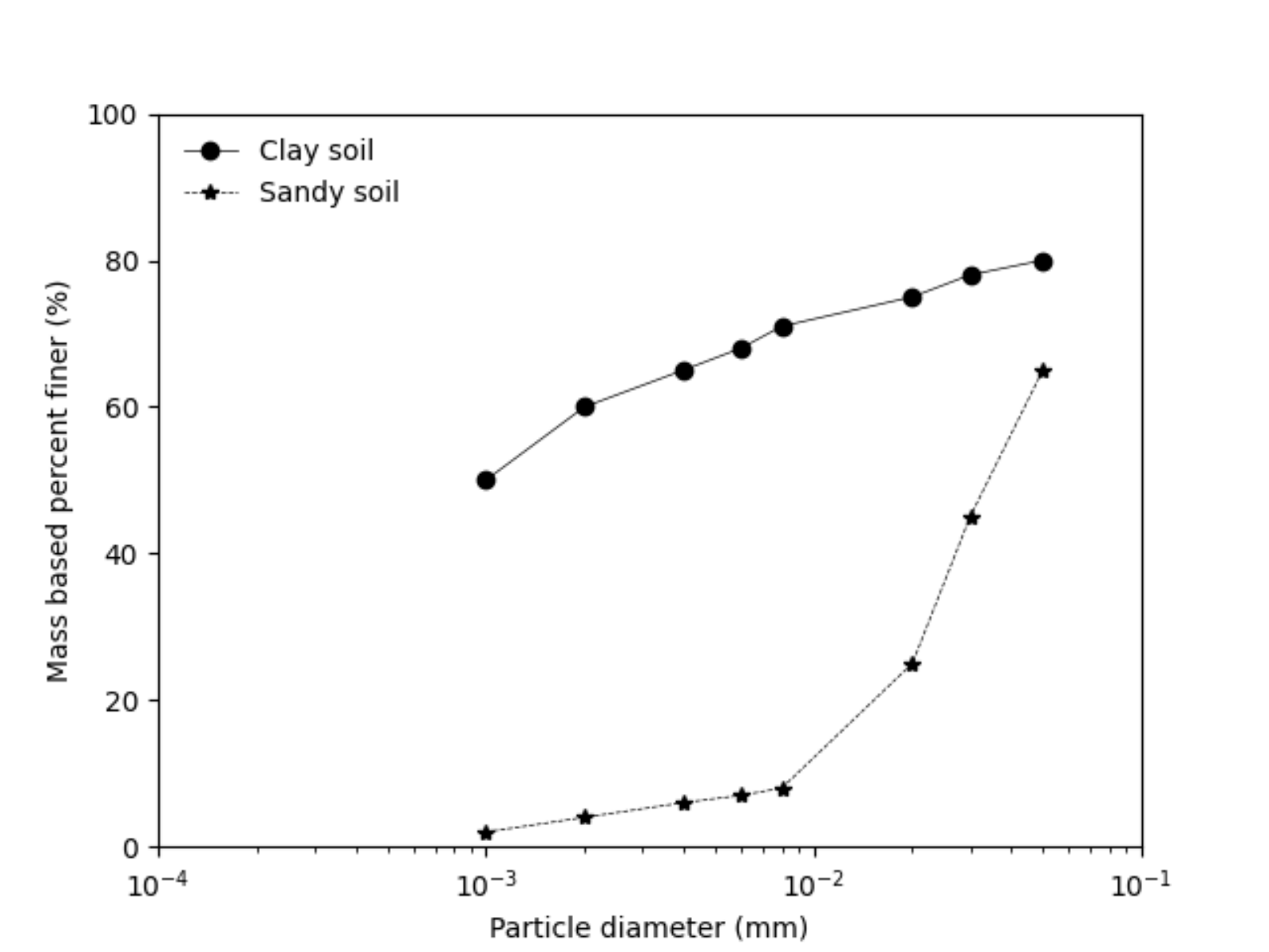}
\caption{Grain size distribution curves for hypothetical clay and sandy soils.}
\label{ch5_fig2}
\end{figure}

Examples of hypothetical granulometric curves for clay and sandy soils are illustrated in Figure \ref{ch5_fig2}

\begin{figure}[h!]
\centering
 \includegraphics[width=0.8\textwidth]{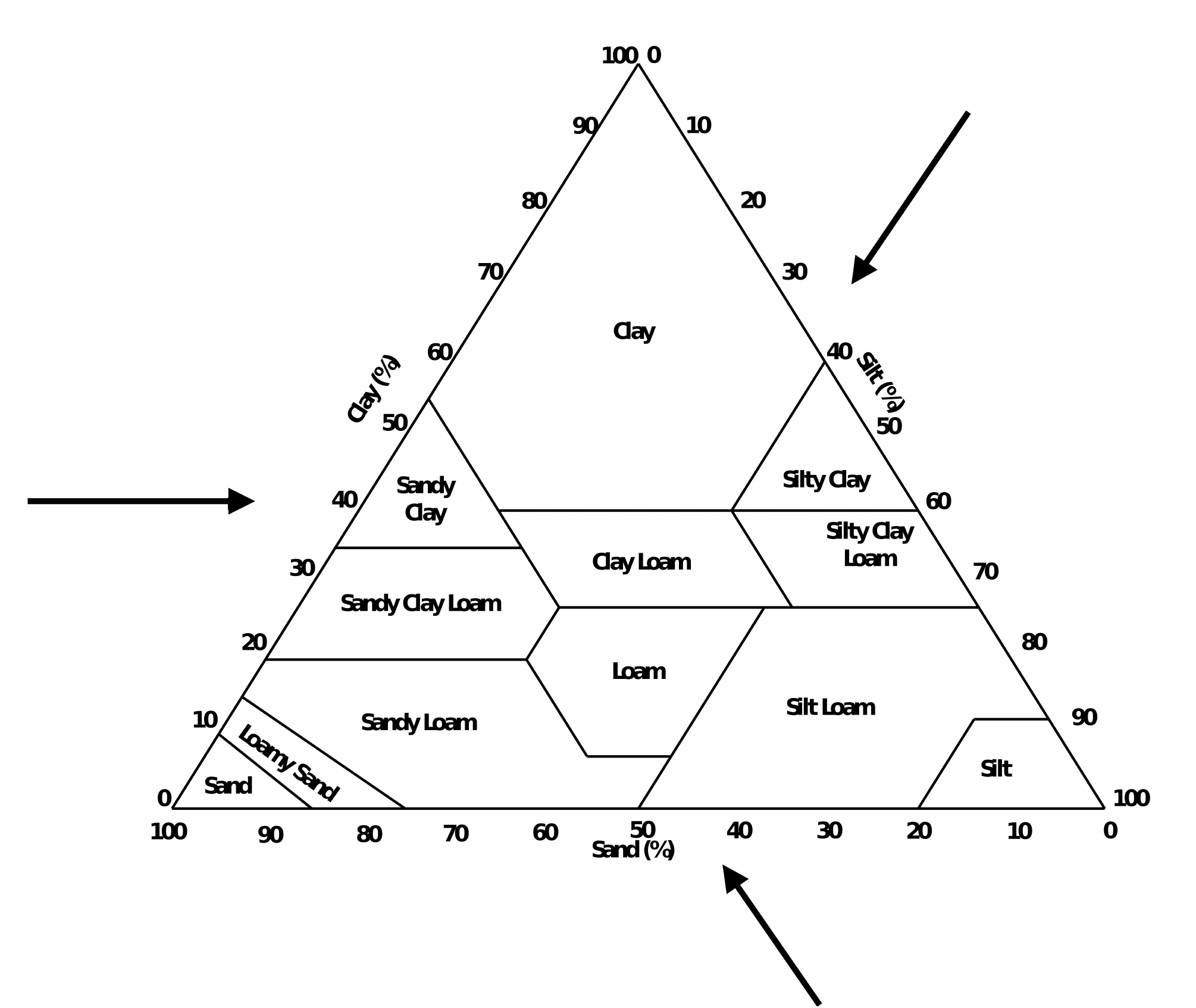}
\caption{Textural triangle for determination of soil texture class. The direction of data entry for each class is indicated by the arrows.}
\label{ch5_fig3}
\end{figure}

In soil science, and soil physics to some degree, once the mass based fraction of particles in the general classes clay ($D< 0.002~mm$), silt ($0.002~mm \le D < 0.05~mm$) and sand ($0.05~ mm \le D < 2.00~ mm$) is determined, the data can be entered into a ternary plot generally referred to as \emph{textural triangle} and the nominal soil texture class determined (Figure \ref{ch5_fig3}). The nominal texture classes are a relic from soil pedology and are an attempt to classify soils using the ``feeling'' method in the field. The feeling method can be useful to discern the general texture classes after extensive training of the field technician. In broader terms, the soil sample is disagregatted, moistened to provide some degree of dispersion by manipulating it using the fingers and rubbed between the index and thumb so that the soil texture is literally felt by the person. It can be useful as an expedite field method for estimating soil texture and for soil classification where laboratory methods might not be available.

Grain size distribution and soil texture are important properties of soils because they will directly or indirectly affect almost all soil physical and chemical phenomena. Aggregation and pore size distribution will affect water transport and storage as well as aeration and susceptibility to compaction. The amount of clay will affect surface area the reactivity of soils as we discussed previously. For a given mineralogy, increasing clay content, will tend to increase plasticity and susceptibility to compaction and compression under certain conditions. 

\section{List of symbols for this chapter}

\begin{longtable}{ll}
       	$ p $ & Pressure  \\
        $ \mu $ & Viscosity  \\
    	$ \mathbf{u}  $ & Vector velocity function  \\
        $ u_x, u_y, u_z $ & Components of the velocity vector in Cartesian coordinates  \\
    	$ \mathbf{i}, \mathbf{j}, \mathbf{k}$ & Unit vectors in Cartesian coordinates \\    
        $ \mathbf{r}, \boldsymbol{\theta}, \boldsymbol{\phi}$ & Unit vectors in spherical coordinates \\
      	$ u_r, u_{\theta}, u_{\phi} $ & Components of the velocity vector in spherical coordinates  \\
    	$ r, \theta, phi $ & Length and angle components in spherical coordinates  \\
    	$ \Psi $ & Stokes stream function  \\
    	$ E^2 $ & Differential operator  \\
    	$ a $ & Radius  \\
    	$ A, B, C, D $ & Constants  \\
    	$ \tau $ & Stress components  \\
    	$ F_D $ & Drag force  \\
    	$ U $  & Velocity  \\
 	$ F_b $ & Buoyancy force  \\
    	$ F_g $ & Gravitational force  \\
    	$ m $ & Mass  \\
    	$ \rho $ & Density  \\
    	$ R $ & Radius  \\
    	$ v_s $ & Settling velocity  \\
    	$ Re $ & Reynolds number  \\
    	$ Pe $ & P\'{e}clet number  \\
    	$ T $ & Temperature  \\
    	$ k $ & Boltzmann constant  \\
    	$ W_{EDL} $ & Energy of interaction due to double layer forces  \\
    	$ \rho_{\infty} $ & Electrolyte concentration in bulk solution  \\
    	$ \kappa $ & Parameter related to the thickness of the double layer  \\
    	$ z $ & Valence of the ion  \\
    	$ e $ & Electronic charge  \\
    	$ W_{vdW} $ & Attractive van der Waals force  \\
    	$ A_H $ & Hamaker constant  \\
    	$ D' $ & Separation distance  \\
    	$ \rho_1, \rho_2  $ & Number densities of interacting spheres  \\
    	$ C  $ & London constant  \\
    	$ W(D') $ & Pair interaction energy  \\
    	$ S $ & Settling distance  \\
    	$ T_c $ & Temperature in Celsius  \\   
        $ D $ & Particle diameter     \\
      	
\end{longtable}

% !TEX TS-program = pdflatex
% !TEX encoding = UTF-8 Unicode

% Example of the Memoir class, an alternative to the default LaTeX classes such as article and book, with many added features built into the class itself.

\chapter{Structure and aggregation}
\label{ch6}

\section{Sphere packing models}

Because soils and sedimentary deposits are composed of solid particles and porous space, forming an unconsolidated medium, mechanical stresses caused by foundations, plant growth, traffic of machines, animals and people, overburden of confining layers, water in pores, clay expansion and contraction and many other natural and man-made processes can cause reorganization of solid particles and, depending on the circumstances, compaction, consolidation or compression or increase in pore volume by processes such as loosening.  Before we consider aggregation of soil primary particles and organic matter, it is convenient to treat porous media as an idealized stacking of spheres. This is directly applied to round, homogeneous, coarse grained particles in a few soils and sedimentary deposits and can be an important tool in modeling water retention and transport phenomena, but it is less representative to heterogeneous particle size distributions and as the particle shapes deviate from an ideal sphere.  

Treating porous media as a set of equally sized spherical particles, the porosity or pore fraction of the material will depend on how the spheres are organized in space. There are more and less efficient ways of packing spheres, a pile of oranges in a supermarket tray is an example of a packing system. The packing models are organized in \emph{lattices} which can be defined as a repetitive arrangement in space. The lattices discussed in this chapter are directly applied to the organization of atoms in space seen in Chapter \ref{ch3} and the structure of minerals is organized in lattices, often following the systems below, but usually with more than one type of chemical element, each element represented by different spheres \cite{hammond}.  The systems described below can be applied to soil aggregates to some extent. Soil aggregates are usually less resistant to soil stresses than most primary particles and will be discussed in this chapter.  The most common sphere packing arrangements are described below. The packing density is derived from geometric arguments. \\

\noindent
\textbf{\underline{I. Cubic Lattice}}

\begin{figure}[h!]
\centering
 \includegraphics[width=0.75\textwidth]{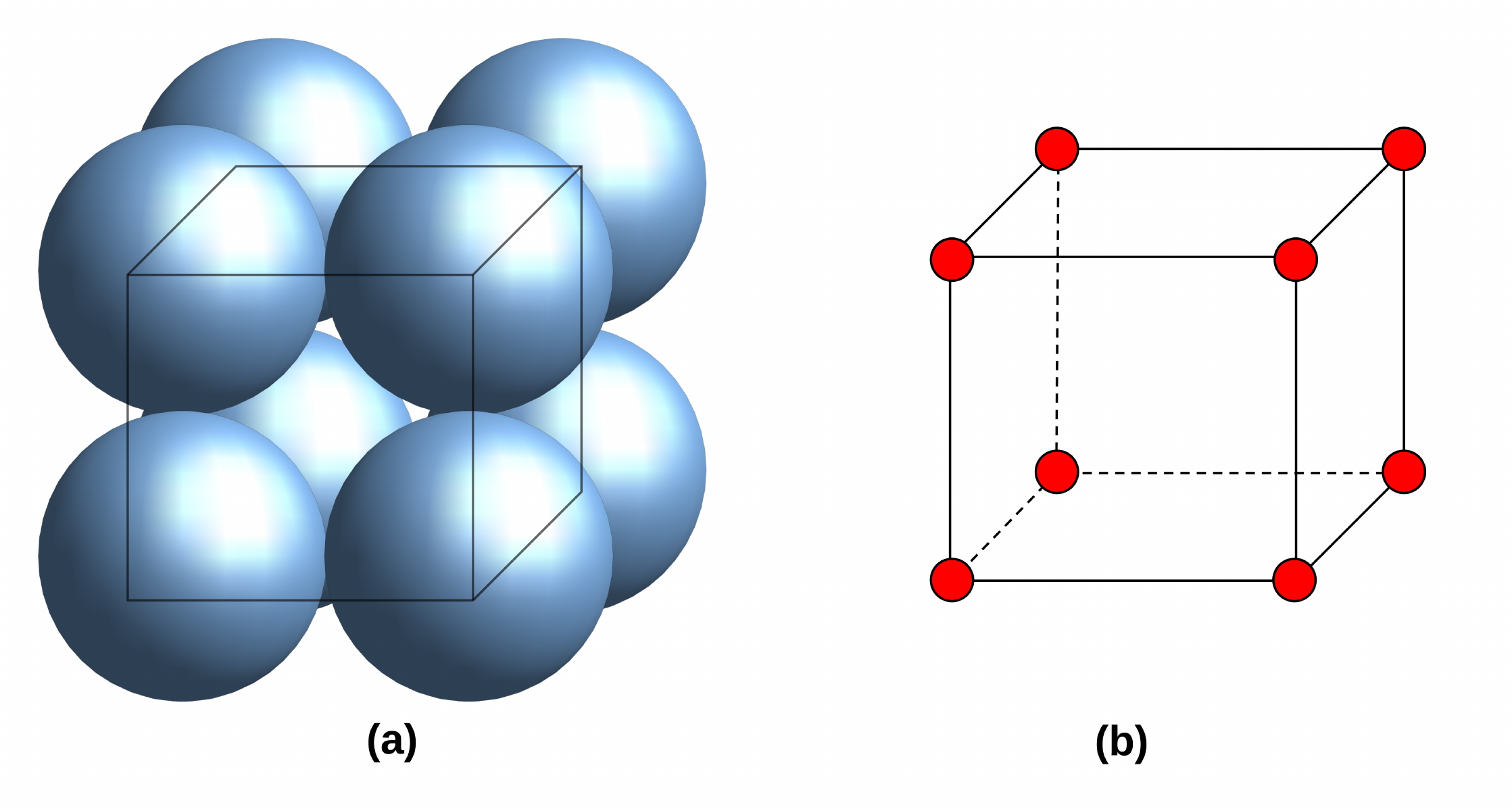}
\caption{Cubic lattice.}
\label{ch6_fig1}
\end{figure}

\noindent
In the case of cubic lattice, or \emph{simple cubic lattice}, as the spheres directly touch, the distance between the spheres centers is two times the radius ($R$). The volume of the unit cell considered is the cube that unites the points at the centers of the eight spheres (Figure \ref{ch6_fig1}). Considering the side of the cube as $A = 2R$, the volume of the unit cell, the volume occupied by spheres and the packing density, given by the volume occupied by the spheres over the unit cell volume are given by    

\begin{align*}
V_{cell} = A^3 = (2R)^3 = 8R^3 \\
V_{spheres} = 8 \cdot \frac{1}{8} \cdot \frac{4}{3} \pi R^3 = \frac{4}{3} \pi R^3 \\ 
\eta = \frac{V_{sphere}}{V_{cell}} = \frac{\frac{4}{3} \pi R^3}{8R^3}
\end{align*}
\begin{equation}
\eta = \frac{\pi}{6} = 0.52359 ...
\end{equation}
In which $V_{cell}$ is the volume of the unit cell, $V_{sphere}$ is the corresponding volume of spheres located within the unit cell and $\eta$ is dimensionless packing density.
Notice that within the unit cell there are eight $1/8$s of spheres such that the total volume of spheres within the unit cell is $8 \cdot 1/8 = 1$ sphere. Similar logic will be used in the following packing schemes. \\

\noindent
\textbf{\underline{II. Face-Centered Cubic}}

\begin{figure}[h!]
\centering
 \includegraphics[width=0.50\textwidth]{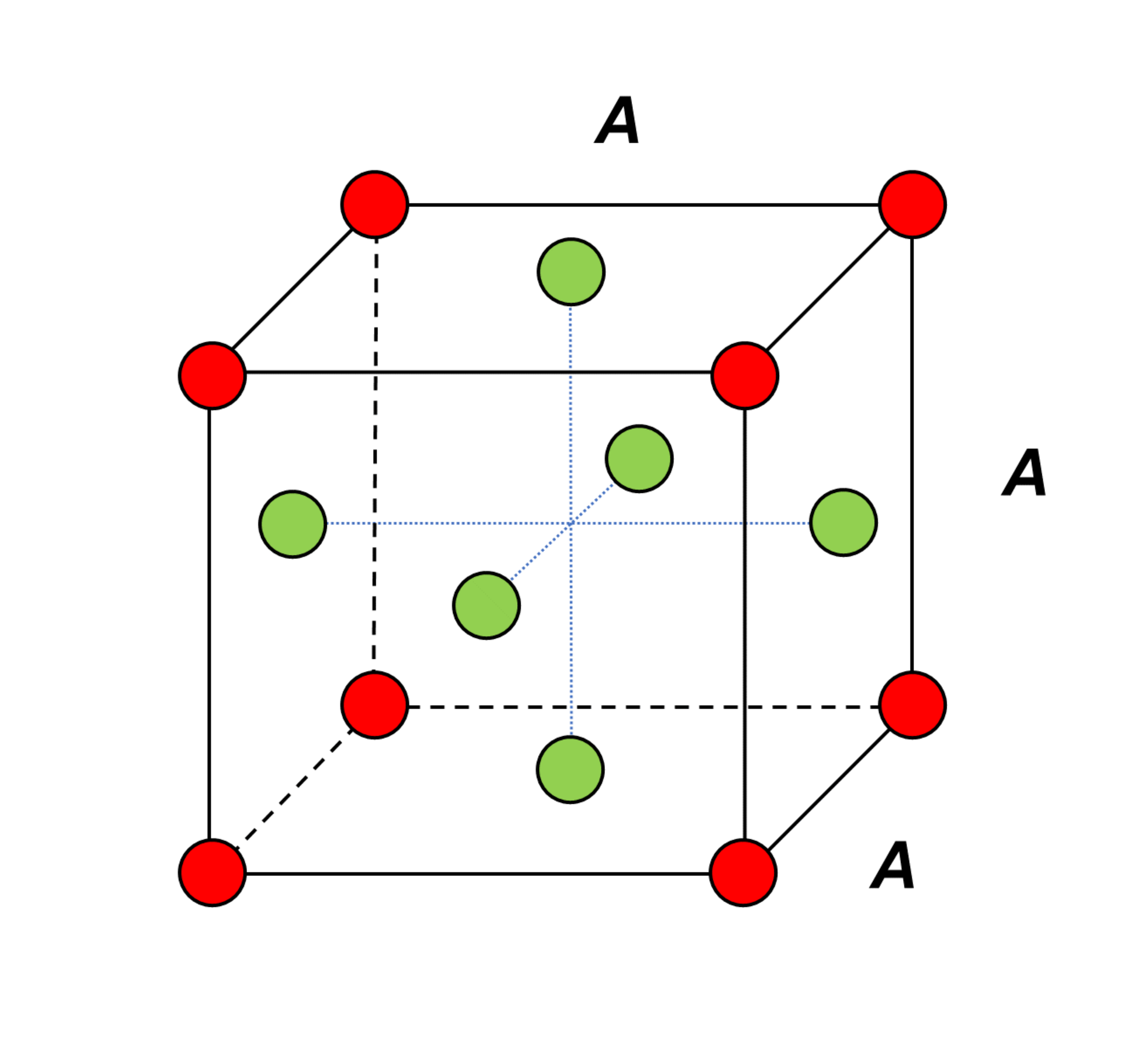}
\caption{Face-centered cubic packing.}
\label{ch6_fig2}
\end{figure}

\noindent
In this arrangement, an additional sphere is located on the center of each face (Figure \ref{ch6_fig2}). Now the side $A$ is no longer $2R$ as additional space is needed between the spheres in the corners to accommodate the spheres on the faces. Notice that the simplified ball-and-stick representation is used, but each corner and face is occupied by a large sphere of radius $R$. The color red in indicates the corner spheres and the light green indicates the face spheres (Figure \ref{ch6_fig2}).

Because of the additional space, the side of the unit cell is given by $A = 2 \sqrt{2} R$ and the volumes and  packing density can be calculated as 

\begin{align*}
V_{cell} =  A \cdot A \cdot A =  (2 \sqrt{2} R)^3 = 16 \sqrt{2} R^3\\
V_{spheres} = (8 \cdot \frac{1}{8} + 6 \cdot \frac{1}{2}) \cdot \frac{4}{3} \pi R^3 = \frac{16}{3} \pi R^3 \\ 
\eta = \frac{V_{sphere}}{V_{cell}} = \frac{\frac{16}{3} \pi R^3}{16 \sqrt{2} R^3}
\end{align*}
\begin{equation}
\eta = \frac{\pi}{3 \sqrt{2}} = 0.74048 ...
\end{equation}
Notice that the packing density is considerably higher and the porosity, given by $ 1 - \eta$ is considerably lower, around 25\%, when compared to the simple cubic lattice. \\ 

\noindent
\textbf{\underline{III. Body-Centered Cubic}}

\begin{figure}[h!]
\centering
 \includegraphics[width=0.50\textwidth]{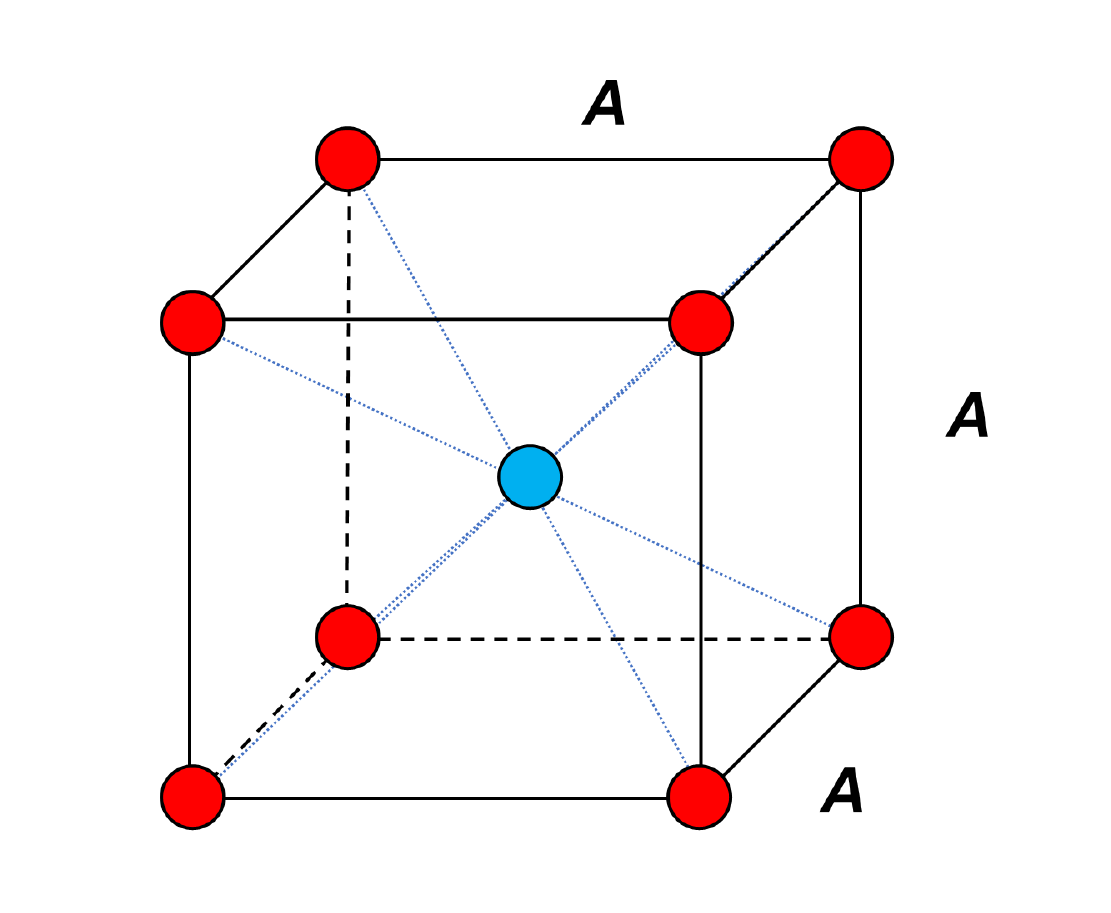}
\caption{Body-centered cubic packing.}
\label{ch6_fig3}
\end{figure}

\noindent
In this case there is one sphere in each corner and one sphere (represented in blue) in the center of the structure (Figure \ref{ch6_fig3}). Notice that the sides $A$ are still equal on the cube, but the diagonals are formed by the radius of the spheres in the transverse corners plus the diameter of the sphere in the center ($2R$) totaling $4R$. From the triangle formed by the height $A$, the diagonal $4R$, the diagonals in each face can be calculated from simple geometric arguments resulting in $d = \sqrt{16 R^2 - A^2}$. Again the relationship between $A$ and $R$ can be calculated from the diagonal of each face and the sides $A$ using simple geometric arguments resulting in $ A = 4R/\sqrt{3}$. The volumes and packing density can then be calculated as before

\begin{align*}
V_{cell} = (\frac{4R}{\sqrt{3}})^3 = \frac{64}{3\sqrt{3}} R^3 \\
V_{spheres} =  (8 \cdot \frac{1}{8} + 1) \cdot \frac{4}{3} \pi R^3 = \frac{8}{3} \pi R^3\\ 
\eta = \frac{V_{sphere}}{V_{cell}} = \frac{\frac{8}{3} \pi R^3}{\frac{64}{3\sqrt{3}} R^3} 
\end{align*}
\begin{equation}
\eta = \frac{\pi \sqrt{3}}{8} = 0.68017 ...
\end{equation}
\\

\noindent
\textbf{\underline{IV. Hexagonal Lattice}}

\begin{figure}[h!]
\centering
 \includegraphics[width=0.55\textwidth]{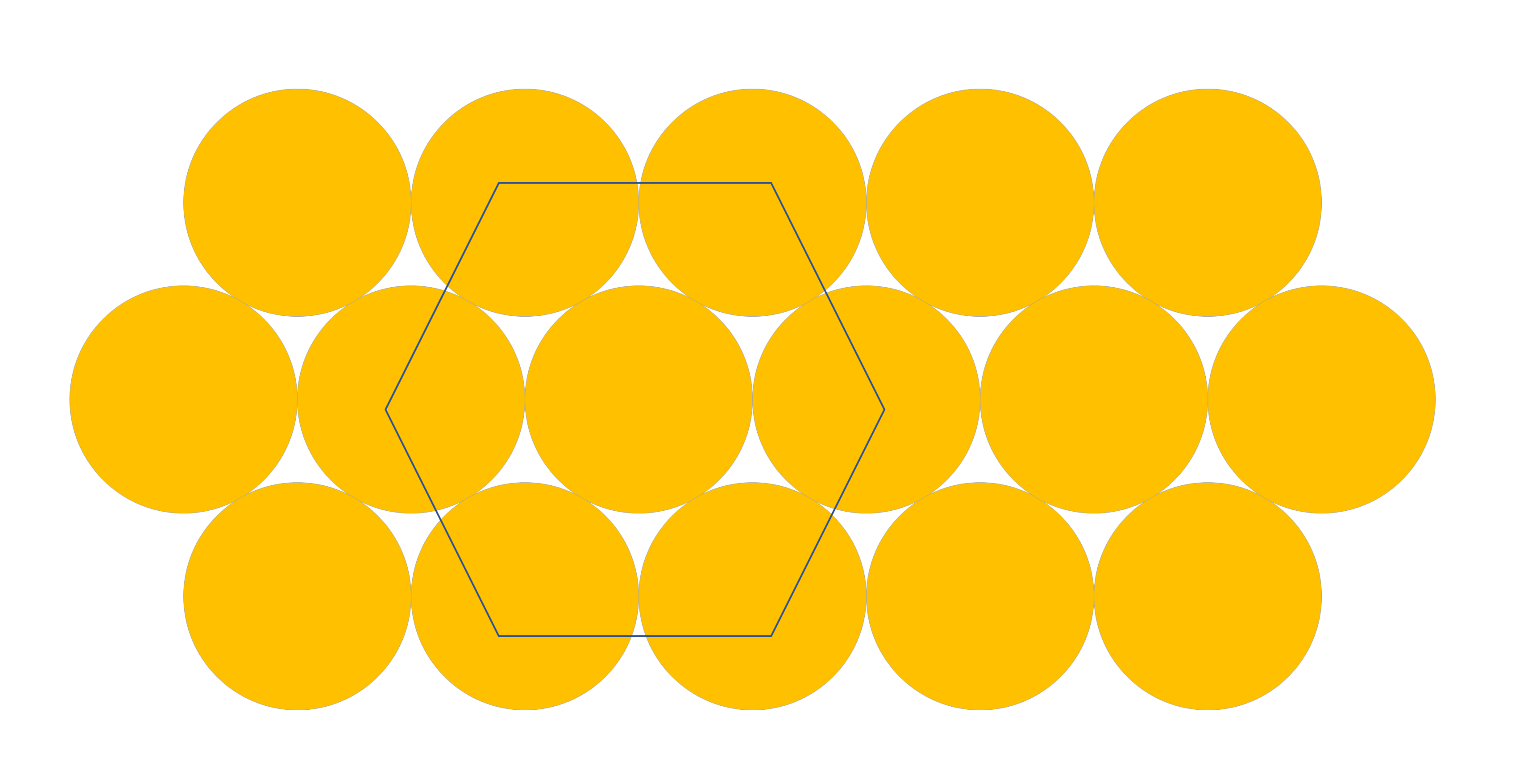}
\caption{Spheres plane in an hexagonal arrangement.}
\label{ch6_fig4}
\end{figure}

\begin{figure}[h!]
\centering
 \includegraphics[width=0.35\textwidth]{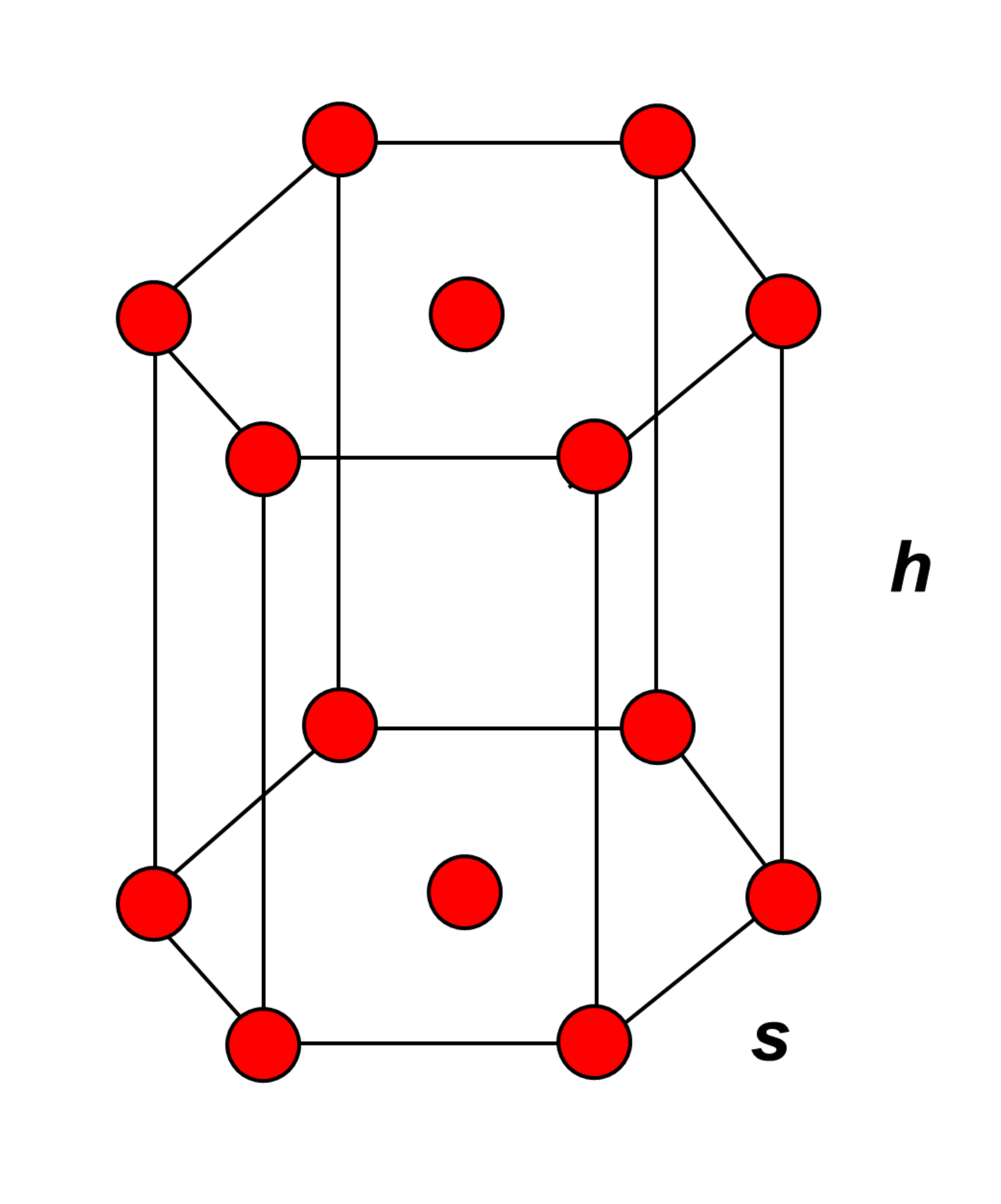}
\caption{Hexagonal lattice.}
\label{ch6_fig5}
\end{figure}

\noindent
Now on the two hexagonal packing schemes presented here, the geometric polyhedron considered is no longer a cube but an hexagonal prism. Looking at one layer of packed spheres from above it is clear that the center of the spheres can be linked to form an imaginary hexagon  (Figure \ref{ch6_fig4}). The height of the hexagonal prism will depend of how the additional layers of spheres are stacked. If the additional layer of spheres is staked directly on top of the first layer, an hexagonal packing lattice is formed (Figure \ref{ch6_fig5}). Notice that because the spheres on the base touch, the base edge, $s$, is $2R$ and because the next layer is stacked directly above the first layer, the height of the hexagonal prism linking the center of the group of spheres is also $2R$. Recalling that the volume of an hexagonal prism is given by $ (3 \sqrt{3}/2) s^2 h $, the volume of the unit cell is the volume of the hexagonal prism and the packing density can be calculated as  

\begin{align*}
V_{cell} = \frac{3}{2} \cdot \sqrt{3} \cdot  s^2 \cdot  h = \frac{3}{2} \cdot \sqrt{3} \cdot (2R)^2 \cdot (2R)  = 12 \sqrt{3} R^3\\
V_{spheres} = 3 \cdot  \frac{4}{3} \pi R^3 = 4 \pi R^3 \\ 
\eta = \frac{V_{sphere}}{V_{cell}} = \frac{4 \pi R^3}{12 \sqrt{3} R^3}
\end{align*}
\begin{equation}
\eta = \frac{\pi}{3 \sqrt{3}} = 0.64599 ...
\end{equation}
Notice that there are 12 spheres in the corners of the hexagonal prism and each one contributes to 1/6th of the volume within the unit cell, plus two half spheres in the upper and lower faces, thus we have $ 12 \cdot 1/6 + 2 \cdot 1/2 = 3$ spheres.  
\\

\noindent
\textbf{\underline{V. Hexagonal Close Packing}}

\begin{figure}[h!]
\centering
 \includegraphics[width=0.8\textwidth]{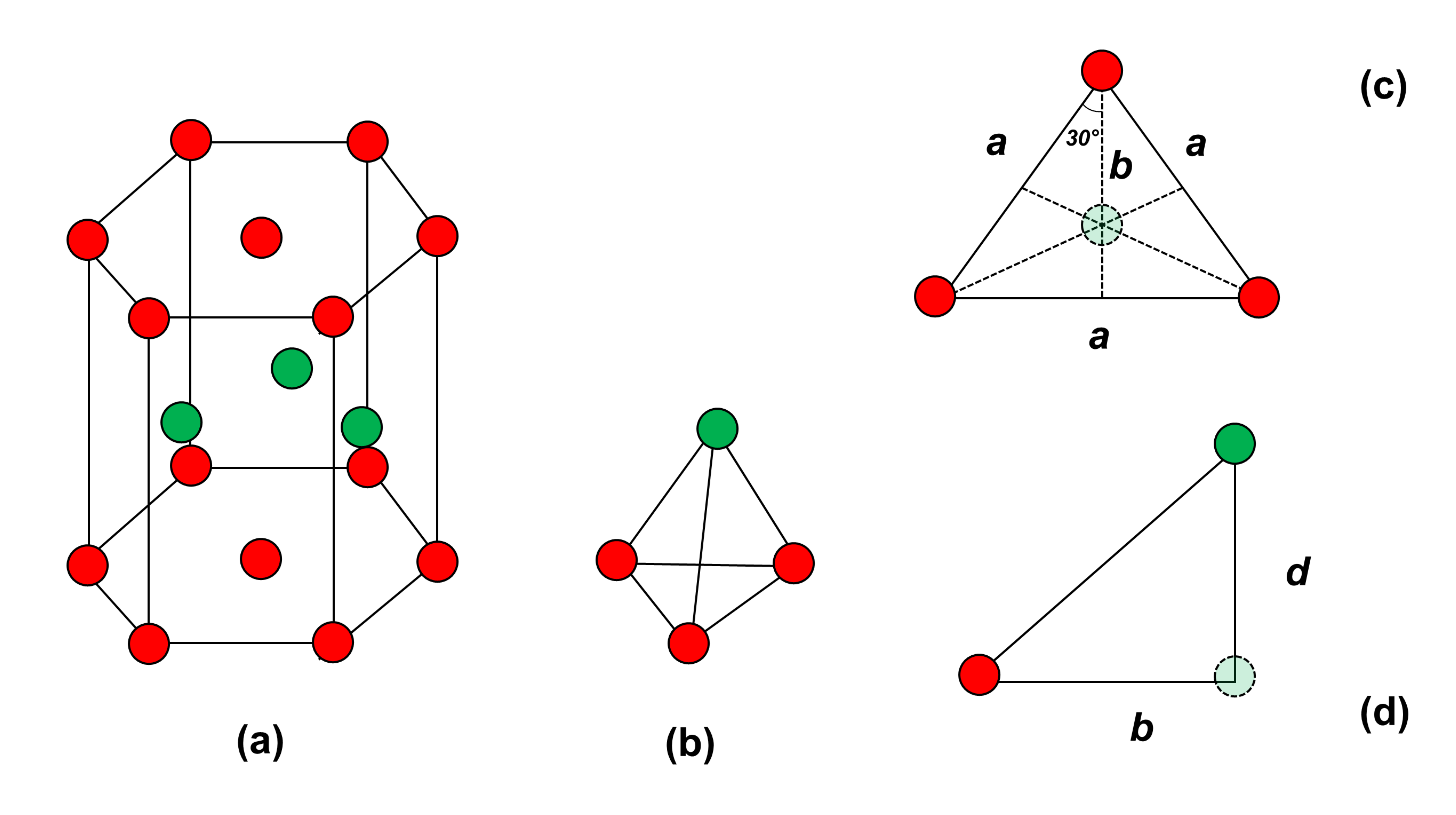}
\caption{Hexagonal close packing.}
\label{ch6_fig6}
\end{figure}

\noindent
The hexagonal close packing is similar to hexagonal packing but there are three spheres located in between the two staked planes  (Figure \ref{ch6_fig6}a). It is important to note that these three spheres will also form a plane of spheres when multiple layers are considered in three dimensional space, in addition to the two planes of spheres represented in red (Figure \ref{ch6_fig6}a). The calculation of the volume of the unit cell is more complicated than in the hexagonal packing lattice because the distance between the centers of the spheres in the two planes is no longer $2R$. 

Consider a tetrahedron formed by the central sphere in the base plus two spheres on the edges forming the base of the tetrahedron and one of the spheres in the middle plane as the upper vertex (Figure \ref{ch6_fig6}b). Considering the sides of the base of the tetrahedron as $a$, the distance to the point of projection of the vertex sphere $b$ can be calculated from the internal angle and the position at the middle point of the side of the base $a/2$ (Figure \ref{ch6_fig6}c)

\begin{align*}
\cos{30^{o}} &= \frac{a/2}{b} 
\\
b &= \frac{a}{\sqrt{3}} 
\end{align*}
The height $d$ to the plane of spheres in the middle of the unit cell is then given by (Figure \ref{ch6_fig6}d)

\begin{align*}
a^2 &= d^2 + b^2 = d^2 + (\frac{a}{\sqrt{3}})^2 = d^2 + (\frac{a^2}{3})
\\
d &= a \frac{\sqrt{2}}{\sqrt{3}}
\end{align*}
Because the sides of edge of the base of the hexagonal prism is $a = 2R$  and the total distance between the  two planes in red in Figure \ref{ch6_fig6}a is $2d = 2 \cdot 2R \frac{\sqrt{2}}{\sqrt{3}}$, the total height of the hexagonal prism are 

\begin{equation}
h'_T = 4R \frac{\sqrt{2}}{\sqrt{3}}
\end{equation}
and the volumes and packing density of the hexagonal close packing lattice are given by 

\begin{align*}
V_{cell} = \frac{3}{2} \cdot \sqrt{3} \cdot (2R)^2 \cdot (4R\frac{\sqrt{2}}{\sqrt{3}}) = 24R^3\sqrt{2} \\
V_{spheres} = 6 \cdot  \frac{4}{3} \pi R^3 = 8 \pi R^3 \\ 
\eta = \frac{V_{sphere}}{V_{cell}} = \frac{8 \pi R^3}{24R^3\sqrt{2}}
\end{align*}
\begin{equation}
\eta = \frac{\pi}{3 \sqrt{2}} = 0.74048 ...
\end{equation}

Notice that the packing density is the same as the Face Centered Cubic. In essence the difference between these two systems is geometrical, as in the way the layers of spheres are conceptualized and organized. One system can be transformed into the other by rotation of coordinates.  

\section{Aggregation}

In soil science, aggregation refers to the union of primary\footnote{The meaning here is not as in \emph{primary minerals} but in the basic building blocks of soil aggregates} mineral particles clay, silt and sand with organic compounds and various amorphous materials forming clusters with shapes and sizes that are more or less constant in certain naturally occurring soil types and depths, in response to soil forming processes. Soil structure generally refers to the shape, size and structural stability of these aggregates and larger domains formed by the set of aggregates in a given soil. In natural soils, structure is a direct product of soil forming processes, namely climate, organisms, relief, parent material and time. In fine textured soils, close to surface, the structure tends to be composed of larger aggregates and with higher porosity, especially in humid tropical climates where there is higher biological activity of plants, animals and other organisms. Aggregates in natural soils also tend to be more stable than in areas disturbed by human activity. Human activity, especially agriculture, tends to disturb soil structure, destroying large size aggregates due to tillage and chemical modification on soil upper layers. 

The destruction of aggregates and soil structure is usually a physical and chemical process. Traffic of animals, humans and machinery tends to compress and break soil aggregates, causing increase in bulk density and decrease in porosity, creating an environment that is less adequate to biological activity which further causes reduction in creation of new aggregates and pores, and loss of organic matter. Agricultural practices also employ a variety of tillage implements which in some cases are directly designed to destroy soil structure for sowing operations, while others tend to destroy soil structure indirectly. Chemically, the input of ionic chemicals such as fertilizers and other contaminants can cause dispersion via the surface chemistry mechanisms discussed in Chapter \ref{ch3} which decreases soil structural stability on one end while some compounds can also kill soil organisms, contributing to loss of organic matter and biological activity, further driving soil structural degradation.         

We will not discuss tillage systems and soil quality in this book. There are a lot of sociological, ecological, philosophical and political issues related to soil quality, tillage and agriculture that rely on personal opinion of different groups none of which we have any interest in supporting.  

Water retention due to capillary phenomena tends to be controlled by the pore size distribution and pore shape in soils and these are to a larger extent controlled by the size and shape of soil aggregates. In most soils, aggregation creates at least two pore domains, the intra-aggregate pores and inter-aggregate pores, intra-aggregate pores tends to be larger and usually hold water by capillary phenomena at lower energies of retention when compared to inter-aggregate smaller pores. In soils with high amounts of clay, these inter-aggregate pores tend to be very small and as such, water retention in them can have contribution of both capillary and adsorption phenomena.  Water retention can be modeled in terms of aggregation and pore size distribution in self-similar and fractal models for example \cite{perrier}. 

Formation of aggregates is a slow and complex process dependent on several factors. There are several theories of soil aggregate formation (e.g. \cite{taylorashcroft}), but overall it is thought that aggregation in most soils is dependent on biological activity and organic matter which creates physical and chemical cementing agents contributing to uniting primary particles in larger structural domains. Biological activity can directly contribute to formation of aggregates as some organisms such as termites and earthworms produce aggregate-like structures by ingesting or manipulating soil organic matter and mineral particles. From the chemical point of view, charged surfaces in soil particles, colloidal organic matter, and ions can create aggregating and dispersing forces during aggregate formation. Higher valence cations tend to favor aggregation while monovalent high hydrated radius such as sodium tend to favor dispersion\footnote{As we seen, this depends also on the concentration in solution.}. Physically, stresses caused by roots and organisms and mechanical stresses caused by natural processes such as soil drying, expansion and shrinking of soil minerals due to loss of interlayer water as well as capillary forces related to drying can also contribute to aggregation of primary particles. 

Soil structure, shapes, sizes and stability of aggregates, total porosity and pore size distribution are all affected by stresses applied to soils by human activities. The modifications caused by these activities will in turn affect soil functioning. In many engineering applications such as in the construction of earth dams and foundations, the increase in bulk density and reduction of porosity to a minimum are usually desired while in agriculture, where the soil is viewed as a medium to plant growth,  porosity, and pore size distribution must be at an ideal state for root growth, water storage, water and air circulation and biological activity. 

\section{Compaction, consolidation and compression}
Much of the theory behind soil porosity reduction and increase in bulk density in response to applied stresses comes form geotechnical engineering, specific from the discipline of \emph{Soil Mechanics}. In soil mechanics, \emph{compaction}  is the increase in density of a soil by reduction of the pore volume occupied by air,    \emph{consolidation} is a reduction in volume of a fully saturated, low permeability soil due to drainage of soil water on the pores, and  \emph{compression}  is a reduction of volume of a soil under compressive stress \cite{craig, das}. Although we have not discussed soil permeability to water, it can be understood as a coefficient which measures how easy it is for water to flow out of the soil under a (hydraulic) pressure gradient. On the geotechnical engineering literature, consolidation applies to low permeability soils, such as soils with high phylossilicate clays content, such that there is a time lag between the application of the load and the extrusion of water from pores and soil settlement \cite{das}.

\begin{figure}[ht]
\centering
 \includegraphics[width=0.8\textwidth]{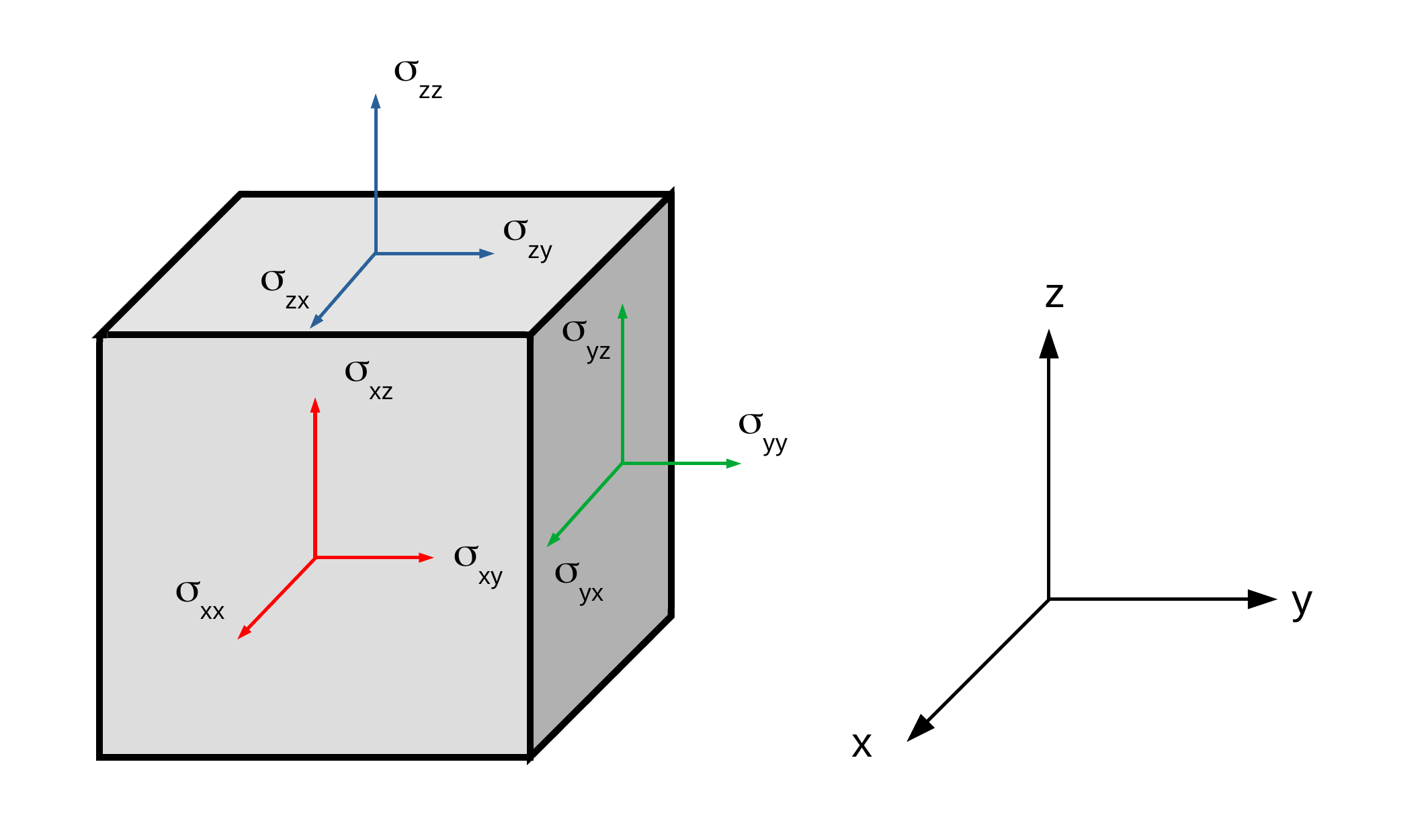}
\caption{Components of stress in a tridimensional solid.}
\label{ch6_fig7}
\end{figure}

We have studied the concept of stress tensor applied to fluids in porous media in Chapter \ref{ch4} and now we need to apply the same concept to stresses in a solid (Figure \ref{ch6_fig7}). Imagining a volume of soil shaped like a cube, forces applied to each surface can be parallel or perpendicular to each face. The force component divided by the area over which it is acting acting defines the stress $\sigma$  (Figure \ref{ch6_fig7}). As with the stress tensor in a fluid, the stress components in each direction, $x$, $y$ and $z$ can be further decomposed into three components, one for each original direction. Therefore, the stress components acting on the face perpendicular to the direction $x$ are $\sigma_{xx}$, $\sigma_{xy}$ and $\sigma_{xz}$ and so forth for the other three directions. The full stress tensor is, as before  

 \begin{equation}  
 \sigma_{ij}  =
\begin{bmatrix}
\sigma_{xx} & \sigma_{xy}  & \sigma_{xz} \\
\sigma_{yx} & \sigma_{yy}  & \sigma_{yz} \\
\sigma_{zx} & \sigma_{zy}  & \sigma_{zz} \\
\end{bmatrix}
 \end{equation}

The perpendicular (or normal) stresses $ \sigma_{ii} = \sigma_{xx} = \sigma_{yy} = \sigma_{zz} $ are equivalent to pressures, or force over area ($F/A$) acting over each surfaces while the stresses parallel to each face $ \sigma_{ij}, ~ i \ne j$ are called the shearing stresses, also $F/A$, in units analogous to pressure units.  Stresses applied by foundations, traffic of animals and machinery, tillage and other agricultural operations are usually multi-directional and have normal and shearing components, especially in non-rigid, granular porous materials such as soils and sedimentary deposits. In some cases the stresses can be modeled as unidimensional pressures as in modeling compaction by agricultural machinery and animal traffic.

If we imagine a foundation, a tire or an animal foot applying a vertical force to a soil, the response of the soil will depend on the magnitude of the applied force and on the mechanical strength of the soil.  The mechanical strength of the soil will depend directly and indirectly on a series of factors such as soil structure, grain and aggregate strength, grain size distribution, porosity, water content, mineralogy, story of previous applied loads and many other internal mechanisms. If the applied stress is greater than the mechanical strength of the soil, deformation will occur. Considering only unidimensional compressive stresses on a soil, deformation will correspond to a decrease in porosity. There are different mechanisms and regimes of deformation in soils, a soil can suffer \emph{elastic deformation} when the soil returns to its original state after the force is removed or \emph{plastic deformation} when the soils does not returns to its original state after the force is removed. On a macro scale, in soils subjected to unequal stresses in multiple directions, \emph{rupture} or \emph{compressive failure} can also occur, although these concepts are more applied to rigid solids. However, internally, rupture of aggregates due to compressive stresses is an important mechanism of soil compaction, even under uniaxial loads. As larger aggregates break the overall strength of the soil decreases and the volume of pores also decreases, as the structural units are destroyed and the pores filled with smaller aggregates or loose particles.  

Dry soils are more subjected to elastic deformation and rupture while wet soils tend to become plastic above a given water content depending on specific characteristics of the soil, like grain size distribution and mineralogy. Compaction in dry soils can be caused by reorganization of aggregates and individual particles, as seen in the packing models at the beginning of this chapter, rupture of aggregates and in some cases of individual mineral particles under load, and elastic deformation of aggregates and individual particles (usually sand and coarser fractions) under elevated loads. Usually aggregates in dry soil tend to be more \emph{brittle}, i.e. they can fracture with little deformation, while as the water content increases, water molecules act as lubricants within particles and aggregates, contributing to plastic deformation. 

\section{Atterberg limits}

The set of critical water contents at which a fine soil transition to different consistency states is called \emph{Atterberg limits}. The Atterberg limits are used in engineering and in agricultural sciences to predict soil behavior in the field for construction, traffic and tillage. In simple terms the principal soil conceptual consistency limits considered in soil physics and soil mechanics are \cite{das} 

\begin{enumerate}
\item Shrinkage limit - Most fine texture soils, especially soils rich in smectites, shrink as the soil dries. This shrinking can be thought to be related to two main mechanisms, first in phylossilicate clays, water is lost from the interlayer planes causing shrinkage at the mineral structure level which reflects on the macrostructure, second, drying tends to cause internal stresses between particles and aggregates due to forces related to menisci in pore water water. These forces can pull the particles together as the soil dries and the radius of the meniscus becomes smaller increasing the capillary forces. The shrinkage limit is the water content below which decrease in water content will not cause further volume reduction of the soil (i.e. shrinkage). A soil in a water content below the shrinkage limit is said to be in a \emph{solid state} and it is said to be in a \emph{semisolid state} at water contents above the shrinkage limit and below the plastic limit.   
\item Plastic limit - The water content below which the soil ceases to display plastic behavior. In practical terms the soil will start to crumble or fracture if manipulated when below the plastic limit. Above the plastic limit the soil is subjected to plastic deformation and is in a \emph{plastic state}.  
\item Liquid limit - It is the water content above which the soil changes from a plastic state to a \emph{liquid sate}. In practical terms, above the liquid limit the water content is such that the soil will begin to flow similar to a liquid when under stress.  
\end{enumerate}

Atterberg limits are strongly dependent on mineralogy and soil texture. The main goal in agricultural soils is that traffic and other agricultural operations are to be performed below the plastic limit. Plastic deformations are related to structural degradation and compaction that is usually expensive and labor intensive to recover. Above the liquid limit machinery can sink and become stuck depending on the soil and other conditions. In practice however, as the soil dries the mechanical resistance can become such that it is impossible or not economically viable to perform tillage or sowing operations. Thus, the limits for \emph{trafficability} and \emph{workability} in soils can be narrow, either due to excess water content, turning the soil plastic and subjected to compaction or consolidation, or due to water content that is too low for efficient tillage operations. Soils that are too plastic, such as those rich in smectites, are often too sticky, requiring high amounts of energy for operations or even being impossible to till under certain conditions

\section{Soil penetrability}

In many practical applications, either agricultural, military or geotechnical engineering, a simple and easy approach to estimate soil mechanical resistance is the use of \emph{cone penetrometers}. The principle is very simple, a rod with a cone at its end is inserted into the soil and the mechanical resistance in terms of stress or force required is measured. Devices based on this principle are called \emph{penetrometers}. There are many different designs of penetrometers, either analogical or with digital sensors, for measuring force and data collecting. Soil mechanical resistance measured by a penetrometer can give an idea of the mechanical resistance of the soil to withstand loads and can be used to estimate trafficability. In agricultural sciences, the \emph{resistance to penetration} or \emph{cone index} is used as a proxy to the mechanical resistance encountered by plant roots. 

\begin{figure}[h!]
\centering
 \includegraphics[width=0.8\textwidth]{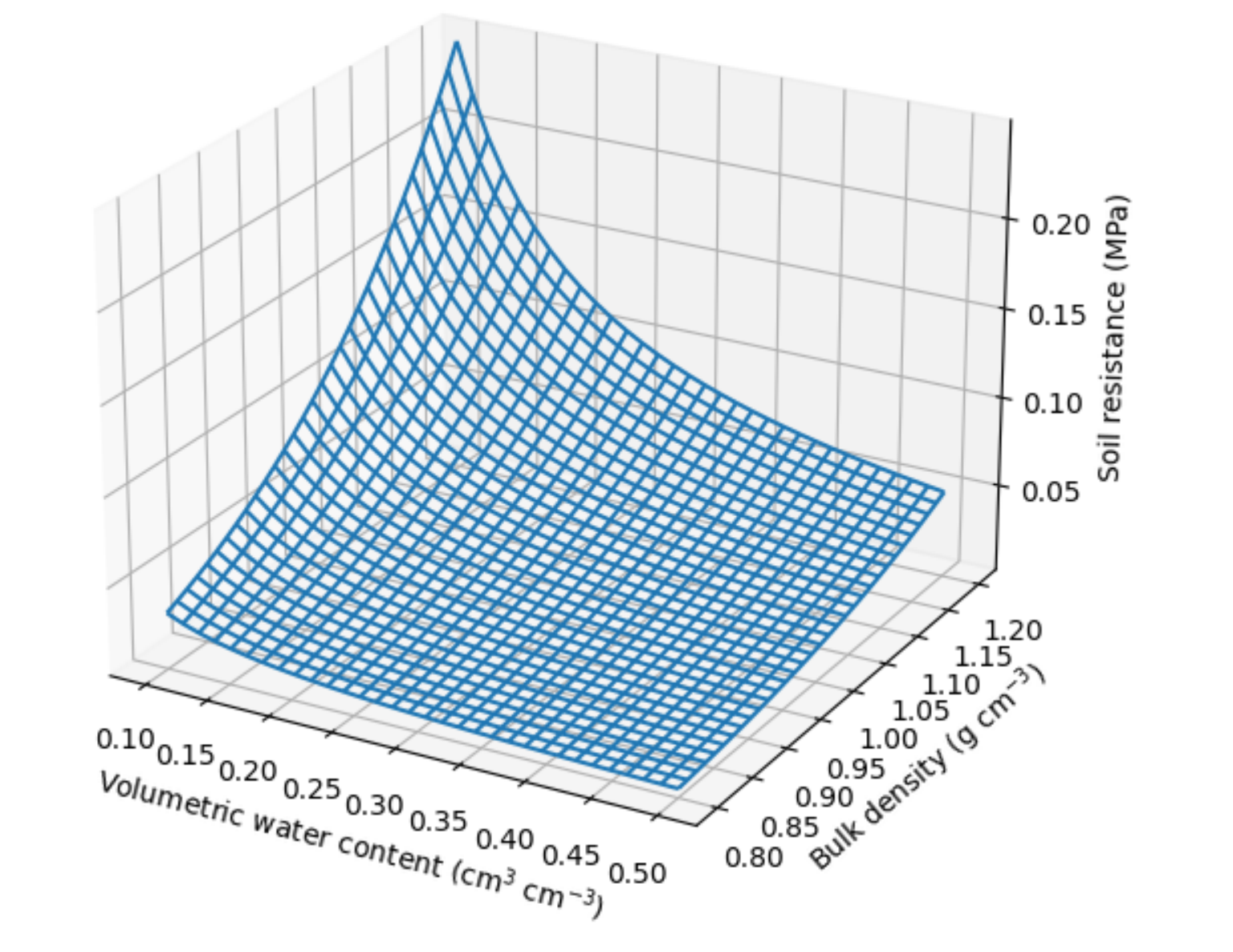}
\caption{Soil penetration resistance as a function of bulk density and volumetric water content based on empirical data. The parameters of Equation 6.8 used are within the ranges for real soils, $a_{0} = 0.01$, $a_{1} = 5.0$ and $a_{2} = -1.0$. }
\label{ch6_fig8}
\end{figure}

The mechanisms of root growth are complex and multiple and it is wrong to attribute critical limits of soil resistance to root growth based on penetrometer readings. In practical terms, a penetrometer can be used as one parameter, among many, for assessing soil compaction and decision making in terms of compaction remediation measures. Penetrometer resistance measurements should never be used without checking soil water content and bulk density as it is an exponential function of these two soil variables (Figure \ref{ch6_fig8}). An empirical relationship between soil penetration resistance,  bulk density and water content usually encountered on the literature is \cite{leao2019}

\begin{equation}
SR = a_0 \rho^{a_1} \theta^{a_2}
\end{equation}
In which $SR$ is the soil mechanical resistance, usually in stress (analogous to pressure) units, $\rho$ is soil bulk density, $\theta$ is the volumetric water content, and $a0$, $a1$ and $a2$ are empirical parameters commonly fitted using regression techniques such as least squares regression. 

This equation has useful practical applications for adjusting $SR$ to a given water content and bulk density so that comparisons between different areas and conditions can be made, and also has pedagogical applications, illustrating the exponential relationship between $SR$ and $\theta$ and $\rho$. However, you must be aware that it is an \emph{empirical relationship} and has no \emph{physical basis} and was not derived based on \emph{first principles} and thus does not explain the physical basis of the relationship between resistance, density and water content. As such, it cannot be extrapolated to different soils, or to conditions not observed on the fitting data. First principles derivations of the $SR = f(\rho, \theta)$ relationship should be possible accounting for internal stresses of the soil provided by cohesive forces between particles and within aggregates, and adhesive forces generated by water in contact with solid surfaces. These forces will also depend on soil texture, mineralogy, organic matter content among other soil properties.    

\section{List of symbols for this chapter}

\begin{longtable}{ll}
    	$ V_{cell} $ & Volume of the unit cell \\
    	$ V_{spheres} $ & Volume of spheres inside the unit cell \\
    	$ A $ & Side length in cubic unit cells \\
    	$ R $ & Radius of spheres \\
    	$ \eta $ & Packing density \\
    	$ s $ & Side length in an hexagonal prism \\
    	$ h $ & Height of the hexagonal prism \\
    	$ a, b, c $ & Parameters for calculation in an hexagonal close packing system \\
    	$ h'_T $ & Height of the unit cell in an hexagonal close packing system \\
    	$ \sigma_{ij} $ & Stress tensor \\
   	$ SR $ & Soil penetration resistance \\
   	$ \rho $ & Soil bulk density \\
   	$ \theta $ & Volumetric water content \\
   	$ a_0, a_1, a_2 $ & Empirical parameters on the soil penetration resistance function \\
\end{longtable}

% !TEX TS-program = pdflatex
% !TEX encoding = UTF-8 Unicode

% Example of the Memoir class, an alternative to the default LaTeX classes such as article and book, with many added features built into the class itself.

\chapter{Hydrostatics}
\label{ch7}

\section{Surface tension}
In a pure, static liquid in thermal equilibrium, away from surfaces, and ignoring any external body forces, all molecules are subjected to the same forces. The molecules interact with neighbor molecules through intermolecular surface forces which, when averaged over time, are constant in all directions and thus, the resultant of these forces is zero. This is not true when there is a discontinuity between the fluid and a different fluid, liquid, solid or gas. On an interface with a gas, the balance of forces is altered such that a resulting force on the direction of the interior of the fluid appears. Because the molecules at the interface are surrounded by other liquid molecules at the liquid end, but not at the gas end, resulting \emph{cohesion} forces pulling the molecules in direction of the liquid appear. Intermolecular forces also generate the \emph{adhesion} and \emph{adsorption} forces between the liquid and solid surfaces. 

The surface tension at the interface can be conceptualized using the thermodynamics or the mechanics frameworks. The mechanics framework states that, because of these interfacial forces, the external work at constant volume and temperature, $\delta W$, required to increase the surface area by an infinitesimal amount $dA$ is directly proportional to the increase in area with the proportionality coefficient being called \emph{surface tension}\footnote{The symbol $\delta$ with respect to W instead of d is because of thermodynamic considerations, $d\hspace*{-0.08em}\bar{}\hspace*{0.1em}$ is also often used. Heat and work in thermodynamics are imperfect differentials. See Callen (2005). } 
\begin{equation}
\delta W = \gamma dA
\end{equation}
In which if the energy $W$ is in $J$ and the area in $m^2$, and the surface tension coefficient needs to have units of $J~m^{-2}$, which is consistent with the definition of energy per unit surface area.  The surface tension is very often conceptualized using the wire and soap film analogy and experiment. Imagine that you have a square wire with one of the sides movable, now imagine a soap bubble is created within the square, if the square has a fixed dimension $l$ and a movable dimension $x$, the area can be increased by increasing or decreasing the length of $x$, such that 
\begin{equation*}
dA = ldx 
\end{equation*}
this implies that 
\begin{equation}
\delta W = \gamma l dx
\end{equation}
In this case, the surface tension coefficient is usually conceptualized as a force per unit length in $N~m^{-1}$. Note that $1~J = 1~N~m$ such that $1~N~m^{-1}$ = $1~J~m^{-2}$ 

Now lets consider the liquid, the gas and the interface as a thermodynamical system. Because the variables of interest in this system are temperature and volume, the thermodynamic potential of interest here is the Helmholtz free energy, $\mathcal{F}$ \cite{callen, pellicer}
\begin{equation}
d\mathcal{F} = -SdT - pdV + \mu_i dN_i 
\end{equation}
Since the temperature and number of moles are kept constant, $dT$ and $dN_i$ are zero, and the equation reduces to
\begin{equation}
d\mathcal{F} = - pdV 
\end{equation}
Since we know from thermodynamics that
\begin{equation}
\delta W = - pdV 
\end{equation}
we have, for an interface
\begin{equation}
d\mathcal{F} = d\mathcal{F}_1 + d\mathcal{F}_2 + d\mathcal{F}_3  = - \delta W_1 - \delta W_2 - \delta W_3
\end{equation}
in which 1 and 2 refer to the solid and liquid phases, respectively, and 3 to the interface phase. Considering that Equation 7.2 is known experimentally, and that the work in that case is negative because it is carried by the interface instead of by an external agent we have, after some reorganizing
\begin{align*}
- \delta W_1 - \delta W_2 - \delta W_3 =  - p_1dV_1 - p_2dV_2 + \gamma dA \\
- p_1dV_1 - p_2dV_2 + d F_3 =  - p_1dV_1 - p_2dV_2 + \gamma dA  \\
d \mathcal{F}_3 = \gamma dA \\
\gamma = \frac{d \mathcal{F}_3}{dA}
\end{align*}
Rewriting in a more elegant symbology
\begin{equation}
\boxed{
\gamma = \bigg(\frac{\partial \mathcal{F}_{int}}{\partial A}\bigg)_T
}
\end{equation}
The surface tension coefficient is the change in Helmholtz free energy of the interface ($\mathcal{F}_{int}$) per unit area at constant temperature \cite{pellicer}. Note that we have not yet discussed the concept of \emph{thermodynamic potential} because it would take us to a long detour which is not strictly necessary at this point. This long detour will be necessary when we discuss the concept of soil water potentials later in this chapter. 

%%% Not clear at all how to get to the F = 2 gamma L

\section{Young-Laplace equation}
Capillary theory is based on developments by Thomas Young\footnote{Young, T. 1805. An essay on the cohesion of fluids. Transactions of the Royal Society of London. 95:65-87.} who laid the descriptive foundations of the process, Pierre Laplace\footnote{Laplace, M. 1805. Trait\'{e} de m\'{e}canique c\'{e}leste. Tome Quatri\`{e}me. Courcier, Paris.} who provided a rigorous initial mathematization and Carl Fredrich Gauss\footnote{Gavss, C.F. 1830. Principia generalia theoriae figvrae flvidorvm in statv aeqvilibrii. Gottingae.} who unified both descriptions in terms of the virtual work principle, based on the principle of least action. The works are based on a rigorous mathematical description and occupy several pages of works in French for Laplace and Latin for Gauss. Like most of what is now known as the core of soil physics, simplifications and ad-hoc derivations came later in order to provide more palatable descriptions. The Young-Laplace equation is a differential equation of the form
\begin{equation}
\Delta p = \gamma \nabla \cdot \mathbf{n}
\end{equation}
in which $\nabla \cdot \mathbf{n}$ represents the mean curvature and can be calculated using differential geometry for different shapes of the interfacial surface.
 
\begin{figure}[ht]
\centering
 \includegraphics[width=0.8\textwidth]{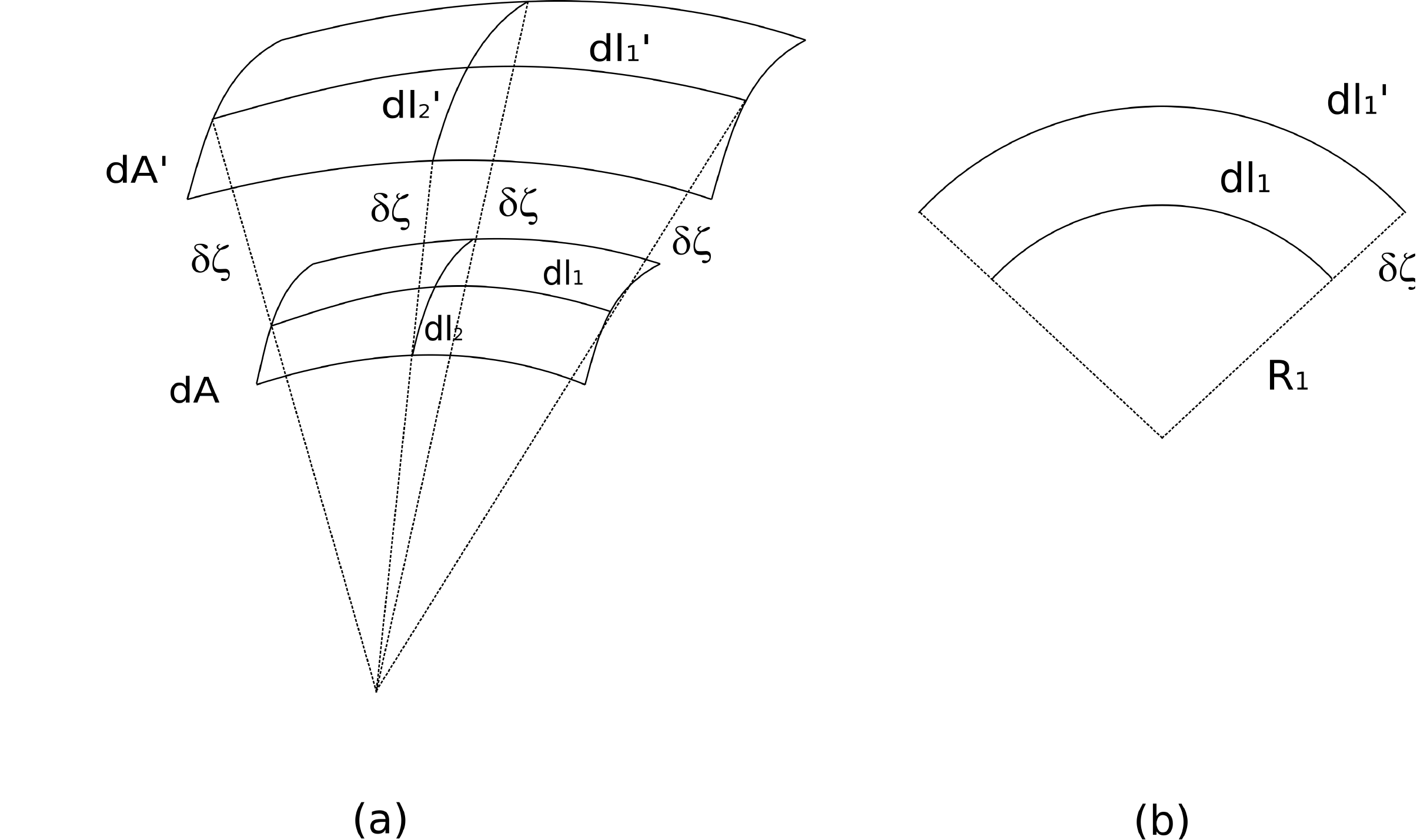}
\caption{Schematics for the energy derivation of the Young-Laplace equation.}
\label{ch7_fig1}
\end{figure}

For reasons of simplicity and elegance we will present here the derivations of Landau and Lifshitz \cite{landau} followed by that of Defay and Prigogine \cite{defay}.
Suppose you have a curved surface with two principal radius of curvature $R_1$ and $R_2$ on an interface between two media. A limiting case would be a sphere in which $R_1 \equiv R_2$. An element of surface area $dA$ is defined in the interface between the two mediums (Figure \ref{ch7_fig1}a). The area element can be calculated as the product of two elements of circumference of a circle, defined in terms of the two radii $R_1$ and $R_2$ as $dl_1$ and $dl_2$. Suppose now that the surface area is increased by displacing the surface by an amount $\delta \zeta$ in the direction of the normal of the surface $dA$ (Figure \ref{ch7_fig1}). The two elements of circumference are now  $dl_1'$ and $dl_2'$ and the surface area is increased from $dA = dl_1dl_2$ to $dA' = dl_1'dl_2'$. An element of volume between the surfaces is $dV = \delta \zeta dA$. If the pressures on the two media are $p_1$ and $p_2$, the work necessary for the volume increase is related to the pressure difference between the two surfaces by

\begin{equation}
\delta W_v = \int p dV = \int (-p_1 + p_2) dV = \int (-p_1 + p_2) \delta \zeta dA 
\end{equation}
and the total work is the work necessary for the volume increase added to the work necessary for surface expansion, as described on the previous section, thus
\begin{equation}
\delta W = \delta W_v + \delta W_s  = \int p dV + \gamma \delta A =  -\int (p_1 - p_2) \delta \zeta dA + \gamma \delta A
\end{equation}
in which $\delta A$ is a change in the area of surface. As we have two areas $dA$ and $dA'$, the change in surface area after displacement, $\delta A$, is needed. Suppose that the length element $dl_1$ is increased to $dl_1'$ after the $\delta \zeta$ displacement in the direction of the normal. Considering an element of angle $\theta$, and using the circumference of a circle formula (Figure \ref{ch7_fig1}b) we have
\begin{equation} 
dl_1 = \theta R_1 
\end{equation} 
and  
\begin{equation} 
dl_1' = \theta (R_1 + \delta \zeta)    
\end{equation} 
Isolating $\theta$ in the first equation and replacing into the second we have
\begin{equation} 
dl_1' = \frac{dl_1}{R_1} (R_1 + \delta \zeta) = (1 + \frac{\delta \zeta}{R_1}) dl_1     
\end{equation} 
Using the same argument in $dl_2$ 
\begin{equation} 
dl_2' = \frac{dl_2}{R_2} (R_2 + \delta \zeta) = (1 + \frac{\delta \zeta}{R_2}) dl_2     
\end{equation} 
Now $dA'$ can be calculated in terms of $dl_1$ and $dl_2$

\begin{align*}
dA' & =  dl_1'dl_2' \\
& =  (1 + \frac{\delta \zeta}{R_1}) dl_1  (1 + \frac{\delta \zeta}{R_2}) dl_2 \\
& =  dl_1dl_2 + dl_1dl_2 \frac{\delta \zeta}{R_2} + dl_1dl_2 \frac{\delta \zeta}{R_1} +dl_1dl_2 \frac{\delta \zeta}{R_1} \frac{\delta \zeta}{R_2}  
\end{align*}
Neglecting the much smaller second order term $\delta \zeta \delta \zeta = \delta \zeta^2$
\begin{equation}
dA' = (1 + \frac{\delta \zeta}{R_1} + \frac{\delta \zeta}{R_2}) dl_1dl_2 \\
\end{equation}
Now the element change in area as the surface is displaced is  
\begin{align*}
da & = dA' - dA  \\
& = (1 + \frac{\delta \zeta}{R_1} + \frac{\delta \zeta}{R_2}) dl_1dl_2 - dl_1dl_2 \\
& = (\frac{\delta \zeta}{R_1} + \frac{\delta \zeta}{R_2}) dl_1dl_2
\end{align*}
The total change in area of the surface of separation is
\begin{equation}
\delta A = \int da = \int (\frac{1}{R_1} + \frac{1}{R_2}) \delta \zeta dA
\end{equation}
and replacing into Equation 7.10
\begin{equation}
\delta W =   -\int (p_1 - p_2) \delta \zeta dA + \gamma \int (\frac{1}{R_1} + \frac{1}{R_2}) \delta \zeta dA
\end{equation}
Because the displacement is required to occur in thermodynamic equilibrium, $\delta W = 0$ and we can write
\begin{equation}
\int [ (p_1 - p_2) - \gamma (\frac{1}{R_1} + \frac{1}{R_2}) ] \delta \zeta dA = 0 
\end{equation}
In which, as in other cases seen before, the integral vanishes identically, thus
\begin{equation}
\boxed{
\Delta p = p_1 - p_2 = \gamma (\frac{1}{R_1} + \frac{1}{R_2}) 
}
\end{equation}
This is the Young-Laplace equation for a curved interface with constant radii $R_1$ and $R_2$ between two media. For a sphere in which $R \equiv R_1 \equiv R_2$ it reduces to
\begin{equation}
\boxed{
\Delta p = p_1 - p_2 =  \frac{2 \gamma}{R} 
}
\end{equation}

\begin{figure}[ht]
\centering
 \includegraphics[width=0.6\textwidth]{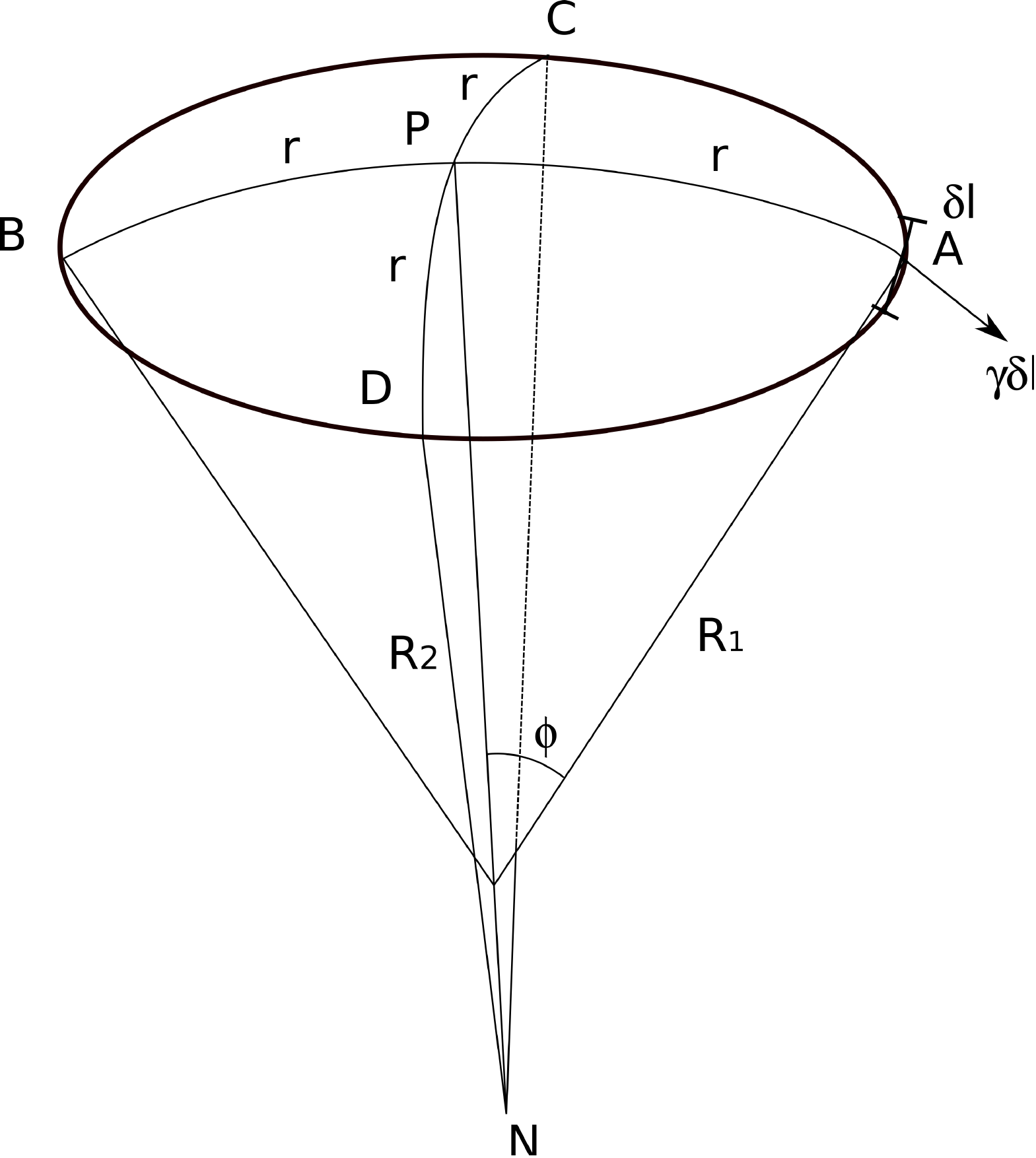}
\caption{Schematics for the mechanics derivation of the Young-Laplace equation (Adapted from \cite{defay}).}
\label{ch7_fig2}
\end{figure}

The derivation from Defay and Prigogine \cite{defay} is simpler and relies on mechanical and geometrical arguments. Consider a non-spherical cap with principal radii of curvature $R_1$ and $R_2$ (Figure \ref{ch7_fig2}). Because of surface tension, a line element $\delta l$ at A is subjected to a force $F_A = \gamma \delta l$. The projection of $F_A$ on the normal P-N is then 
\begin{align*}
\sin{\phi} = \frac{proj_{F_A}}{\gamma \delta l} \\ 
proj_{F_A} = \gamma \delta l \sin{\phi}  
\end{align*}
Using the small angle approximation on the triangle A-P-N it is possible to define
\begin{equation}
\sin{\phi} \approx \tan{\phi} \approx \phi \approx \frac{r}{R_1}
\end{equation}
and
\begin{equation}
proj_{F_A} = \gamma  \frac{r}{R_1} \delta l 
\end{equation}
The same argument for points B, C and D results in  
\begin{equation}
proj_{F_B} = \gamma  \frac{r}{R_1} \delta l 
\end{equation}
\begin{equation}
proj_{F_C} = \gamma  \frac{r}{R_2} \delta l 
\end{equation}
\begin{equation}
proj_{F_D} = \gamma  \frac{r}{R_2} \delta l 
\end{equation}
The resulting force on the N direction on the cap is the sum of the force components of each point in the N direction
\begin{align*}
F =  proj_{F_A} + proj_{F_B} + proj_{F_C} + proj_{F_D} = \gamma  \frac{r}{R_1} \delta l + \gamma  \frac{r}{R_1} \delta l + \gamma  \frac{r}{R_2} \delta l + \gamma  \frac{r}{R_2} \delta l     
\end{align*}
\begin{equation}
F =  2 r \gamma (\frac{1}{R_1} + \frac{1}{R_2}) \delta l    
\end{equation}

The total downward ($F_{Do}$) force around the  border of the cap can be calculated by integrating in $\delta l$. Since there are four quadrants and the perimeter is treated as that of a circle, the integration is made in the $0$ to $\pi/2$ interval. The element of length needs to be transformed into an element of angle using the arc length formula $\delta l = r \delta \theta$.
\begin{equation}
F_{Do} =  \int_0^{\frac{\pi}{2}}  2r\gamma (\frac{1}{R_1} + \frac{1}{R_2}) r \delta \theta = \gamma \pi r^2 (\frac{1}{R_1} + \frac{1}{R_2})   
\end{equation}

For the cap to be in equilibrium, the force downward caused by surface tension has to be in equilibrium with the upward force which in this case is caused by the pressure gradient between the two sides of the interface
\begin{equation}
F_U =  (p_1 - p_2) \pi r^2  
\end{equation}
Note that the pressure difference is force over area and $\pi r^2$ is an area such that we have the resulting force. On equilibrium $F_U = F_{Do}$   
\begin{equation*}
F_U = F_{Do} = (p_1 - p_2) \pi r^2 = \gamma \pi r^2 (\frac{1}{R_1} + \frac{1}{R_2})   
\end{equation*}
And we again arrive at the Young-Laplace equation
\begin{equation*}
p_1 - p_2 = \gamma (\frac{1}{R_1} + \frac{1}{R_2})   
\end{equation*}

\section{Young-Dupré equation}
\begin{figure}[ht]
\centering
 \includegraphics[width=0.6\textwidth]{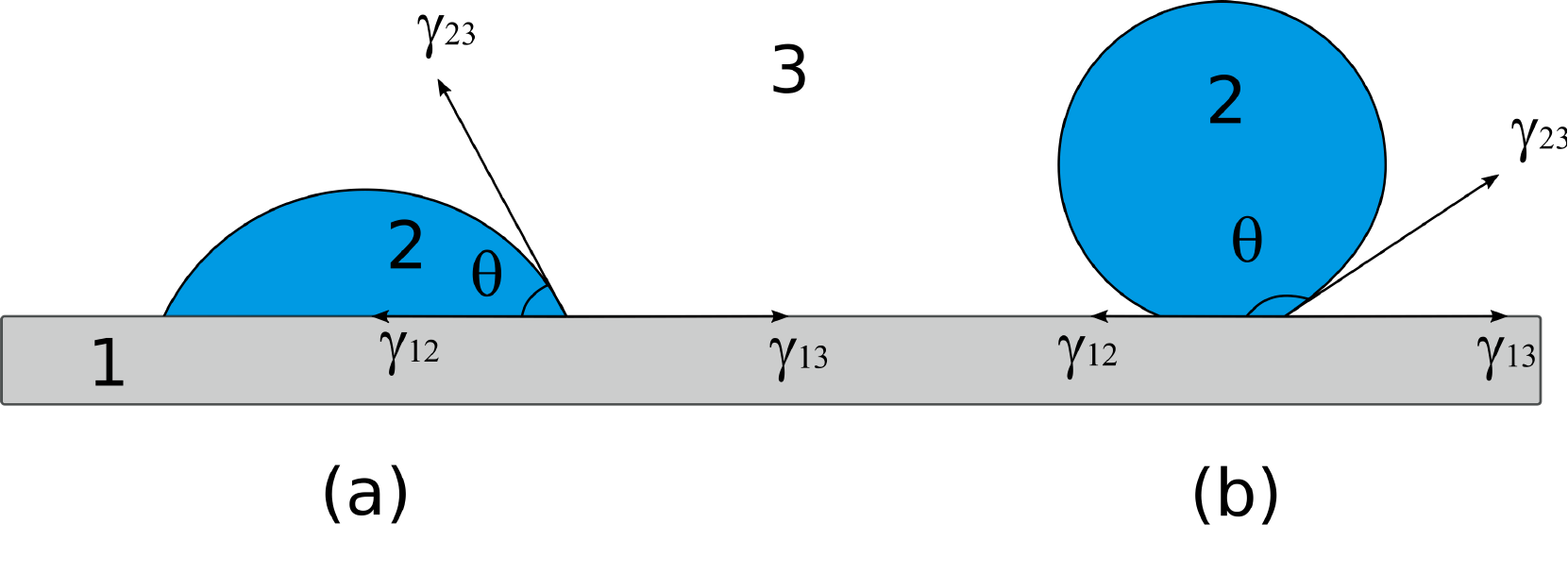}
\caption{Angle of contact on an interface solid (1) - liquid (2) - gas (3). (a) Is a wetting liquid $\theta < 90 ^{o}$ and (b) is a nonwetting liquid $\theta > 90 ^{o}$.}
\label{ch7_fig3}
\end{figure}

The Young-Dupré equation defines the equilibrium condition, more specifically the contact angle, when three homogeneous phases are placed in contact and are in thermodynamic equilibrium. The condition applies to an interface solid-liquid-gas where the shape of a liquid drop is defined by the balance of the surface tensions on the three interfaces, liquid-solid, liquid-gas and gas solid. In general, liquids in contact with solids can behave as \emph{wetting} or \emph{nonwetting} phases. If the fluid behaves as a wetting phase it will tend to maximize the surface area of contact with the solid (Figure \ref{ch7_fig3}a) while nonwetting phases will tend to minimize the contact area with the solid (Figure \ref{ch7_fig3}b). Water over a hydrophilic surface is an example of a wetting phase while it is a nonwetting phase when in contact with hydrophobic surfaces such as surfaces covered with wax. The simplest derivation of the force balance is based on the force equilibrium on the $x$ plane (Figure \ref{ch7_fig3}). For a wetting phase, for example, the surface tension on the direction of the solid-gas interface $\gamma_{13}$ is balanced by summation of the forces on the opposite direction, the surface tension on the solid-liquid interface $\gamma_{12}$ and the $x$ component of the surface tension on the liquid-gas interface, $\gamma_{23} \cos{\theta}$ 
\begin{equation}
\boxed{
\gamma_{13} =  \gamma_{12} +  \gamma_{23} \cos{\theta} 
}
\end{equation}
$\theta$ in this case is the contact angle between the liquid-gas and solid-liquid interfaces.

\section{Capillary rise}
\begin{figure}[ht]
\centering
 \includegraphics[width=0.5\textwidth]{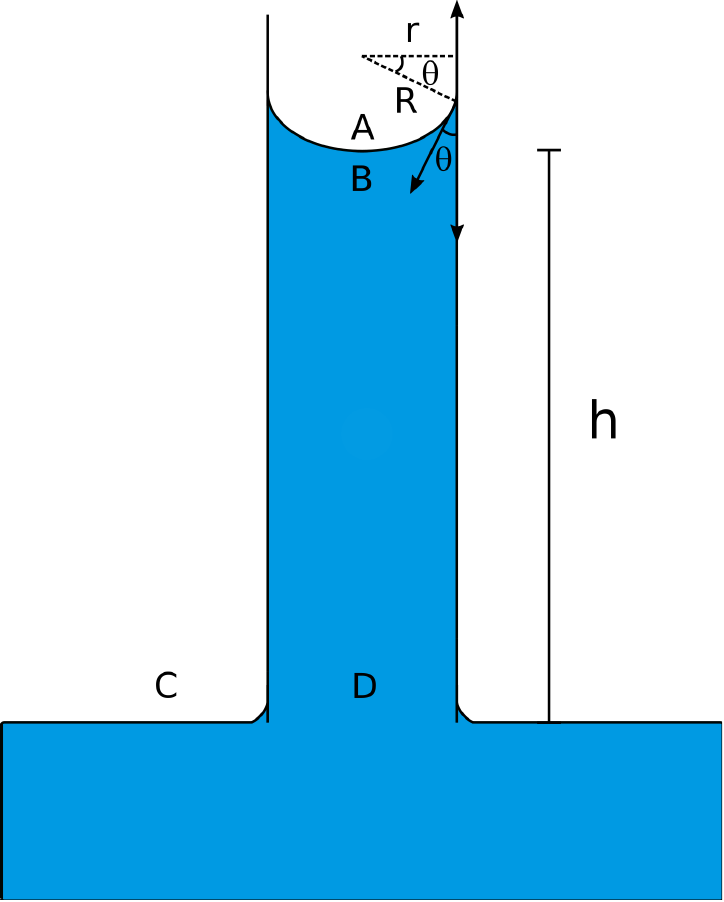}
\caption{Capillary rise.}
\label{ch7_fig4}
\end{figure}

The physical interpretation of the capillary rise phenomenon is a controversial issue. The force balance interpretation given in many textbooks is criticized in terms of accuracy, alternative mechanical and thermodynamic interpretations being preferred. The force balance interpretation due to pressure difference will be presented here due to its simplicity. Consider water rising to the level $h$ in a fine glass tube (Figure \ref{ch7_fig4}), the pressures at points A and C are equal to atmospheric pressure being exerted at the surface of the liquid, $P_A = P_C = P_{atm}$,  and following Pascal's principle, $P_C = P_D = P_{atm}$. Capillary rise and the Young-Laplace equation indicate that there is a pressure difference between the two sides of the curved surface represented by the meniscus between A and B. Because there is capillary rise, the pressure is lower below the surface of the meniscus, in B, than above the meniscus, in A, such that $P_B < P_A$. The capillary rise also indicates that the pressure in D is greater than in B, such that a pressure gradient exists producing an upwards force. The condition of equilibrium in terms of force is
\begin{equation}
(P_D  - P_B - \rho g h) \pi r^2 = 0
\end{equation}
In other words, the upwards force $P_D \pi r^2 $ = force/area $\cdot$  area = force, is balanced by the weight of the water column $\rho g h \pi r^2$ = P $\cdot$ area = force/area $\cdot$ area = force, and the force exerted at B. Because of the equilibrium condition, $P_D - P_B = P_A - P_B  $ and
\begin{equation}
P_A  - P_B = \rho g h
\end{equation}
substituting the Young-Laplace equation for the pressure gradient on the interface
\begin{equation*}
\frac{2 \gamma}{R} = \rho g h
\end{equation*}
This equation can be written in terms of the radius of the capillary tube instead of the radius of the spherical meniscus by considering the geometrical relation in Figure \ref{ch7_fig4} 
\begin{equation*}
R = \frac{r}{\cos{\theta}}
\end{equation*}
The capillary rise equation results from replacing $R$ into the Young-Laplace equation
\begin{equation}
\boxed{h = \frac{2 \gamma \cos{\theta}}{\rho g r}
}
\end{equation}
The rise considered is up to the lowest point of the meniscus, the volume of water above that point being neglected from the weight of the column because it is usually much less than the weight of the rest of the column.

%%\section{Capillary pressure}

\section{The theory of soil water potentials}
We will leave capillary rise and curved interfaces aside for a little while in order to discuss the theory of soil water potentials. Potential theory of soil water is nothing more than an application of one of the most fundamental principles in physics - and in nature for that matter -, the \emph{principle of conservation of energy}, to soil systems. We already discussed several phenomena in terms of energy throughout the book, now we need to formalize how energy plays a role in transport and storage of water in porous media.  Energy is a fundamental concept on the ontology of physics and it is often easier to be understood intuitively than to be given an actual definition. The concept of energy in classical mechanics is a good starting point for an intuitive discussion.  We all know that a body will fall under the force of gravity if it is released close to Earth's surface. The nature of gravitational force which accelerates the body in direction of the center of Earth is complex and the most accepted explanation is given in terms of Einstein's general relativity and curvature of space-time. The theory is beyond the scope of this book and for the sake of simplicity we can accept that such force exists as has been experimentally verified.  

One of the most important ideas in the concept of energy is that it can only be defined in relation some reference state or condition. When we lift a body 10 m into the air close to Earth's surface we are defining Earth's surface itself as the reference level. We could lift a body 10 m above the surface in Mars, on the Moon or above some point in an inertial space station millions of kilometers away from the nearest star or planet and use any of these points as a reference level and calculate stored energy. The gravitational force near Earth is defined experimentally as

\begin{equation}
\mathbf{F} = -m \mathbf{g} 
\end{equation}

The negative sign indicates that the force is downwards in a coordinate system where upwards is defined as positive. If we defined Earth's surface at a particular point and we wish to increase the elevation of an object with mass, work needs to be done against this force. If the object is then released, work is done by the gravitational field as the object falls to a new equilibrium condition. The work done against gravitational force by elevating an object with mass \emph{m} from the surface reference point where \emph{h} is set to zero to a height \emph{h} is 

\begin{equation}
W = \int_{0}^{h} \mathbf{F} dh = \int_{0}^{h}  -m \mathbf{g}  dh =  -m \mathbf{g} \bigg\rvert_{0}^{h} = -m \mathbf{g} h - (-m \mathbf{g}0)   = -m \mathbf{g} h 
\end{equation}
 If the force is conservative, or path independent, the work done is directly related to an energy gradient as 
\begin{equation}
W = -\Delta U 
\end{equation}
Thus the gravitational potential energy of a body lifted to a height \emph{h} above a reference level where height is arbitrarily set as zero is 
\begin{equation}
U = mgh
\end{equation}

A body falls from height because it has mechanical energy stored. Water in rivers, streams, pipes, and pores in soils and rocks also moves because of an energy gradient. In pipes, rivers and streams, the flow can be fast enough that there is a contribution from kinetic energy to the overall energy conservation calculations, as evidenced by the use of the Bernoulli equation, but this is usually not the case in small pores where the flow is slow enough that the kinetic energy contribution can be ignored.  In water in soils and rocks, many different forms of energy can be involved. In Chapter \ref{ch4}, where we discussed the fundamental equations of fluid mechanics, we saw that pressure gradients are involved in the flow of fluids, and we also saw that gravitational forces can be involved along with pressure gradients. In porous media, especially in unsaturated porous media, things can be a little more complicated, in both statics and dynamics a variety of other forces are involved, including complex intermolecular and surface forces. Because of the irregularity and often random distribution of pores and solids, other governing laws in addition to the Navier-Stokes equations might be necessary for modeling such systems. 

Although in this section we discuss the role of potentials in statics, a truly static condition is virtually impossible in natural porous media. Even a block of wet soil in laboratory conditions might be constantly subjected to potential gradients with the surrounding air and thermal fluctuations within the soil mass, which will cause energy gradients and water movement in response to these gradients. However, the assumption of a transitory static condition is very important for the use of  methods of determination  of potential and water content in porous materials.   

To discuss the state of energy of water in contact with solids in rocks and soils we need to address a different form of energy, other than the mechanical forms gravitational potential energy and kinetic energy, the internal energy.  The internal energy, contrary to forms of energy which are related to external fields and motion of the system as whole, is the energy within the system as defined by a particular state in comparison to a reference state in which the energy is arbitrarily taken as zero. In these equilibrium states, the system can be completely characterized macroscopically by the internal energy, \emph{U}, the volume, \emph{V}, and the number of moles of its chemical components, \emph{$N_{1}$, $N_{2}$, ..., $N_{i}$ }. Another extensive function that can be posited for equilibrium states of a system is the entropy, \emph{S}\footnote{Interpretations of the concept of entropy in popular media and introductory courses are most often inadequate, if not incorrect. Entropy must be defined on a physical and mathematical basis, and different definitions can be given based on different subfields of physics, we will address the issue in later editions. For now refer to \cite{guggenheim, callen}.}. The entropy is a function of the other extensive properties        
\begin{equation}
S = S(U, V, N_i) 
\end{equation}
and because the entropy function is continuous, differentiable and is a monotonically increasing function of energy, it can be solved for energy resulting in  
\begin{equation}
U = U(S, V, N_i) 
\end{equation}

Now the internal energy and entropy can be stated for a multiphased porous media as long as it is in a state of equilibrium\footnote{Non-equilibrium thermodynamics plays an important role in transport phenomena, but for the sake of simplicity it is necessary to assume that states of equilibrium are attained for a reasonable simple treatment of the problem.}, and in such condition they contain all information necessary to characterize the system. By computing the differential of the energy form we have \cite{callen}

\begin{equation}
dU = \left(\frac{\partial U}{\partial S}\right)_{V, N_i} dS + \left(\frac{\partial U}{\partial V}\right)_{S, N_i} dV + \sum_i \left(\frac{\partial U}{\partial N_i}\right)_{V, N_j} dN_i
\end{equation}
From this equation the well known intensive properties temperature, pressure and chemical potential of the $i^{th}$ chemical species can be defined by the following partial derivatives \cite{callen}  
\begin{equation}
T = \left(\frac{\partial U}{\partial S}\right)_{V, N_i} 
\end{equation}
\begin{equation}
-P = \left(\frac{\partial U}{\partial V}\right)_{S, N_i} 
\end{equation}
\begin{equation}
\mu_i =  \left(\frac{\partial U}{\partial N_i}\right)_{S, V, N_j}
\end{equation}
The internal energy relationship can then be written as 
\begin{equation}
dU = T dS -P dV + \mu_i dN_i
\end{equation}
For a partially saturated porous media composed of a solid, liquid and a gas, the chemical potential can be defined for each phase, assuming that there are no different species within each phase, such that
\begin{equation}
dU = T dS -P dV + \mu_{solid} dN_{solid} + \mu_{liquid} dN_{liquid} + \mu_{gas} dN_{gas}
\end{equation}
\emph{T}, \emph{P} and \emph{$\mu_i$} are partial derivatives of functions of \emph{S}, \emph{V} and \emph{$N_i$} and because of that they are themselves functions of these quantities. These functional relationships are called \emph{equations of state}. For a natural porous media system we are interested in the equation of state which describes the chemical potential of pore water
\begin{equation}
\mu_w = \mu_w(S, V, N_1, N_2, ...) 
\end{equation}
However, $S$, $V$ and $N_i$ are inconvenient variables to control in experimental studies in partially saturated soils and rocks. Since the chemical potential of water in this case was derived from the energy function, these variables also appear on the equation state. One alternative is to change the independent variables on the water chemical potential equation of state by performing a \emph{Legendre transformation} in order to obtain new functions called \emph{thermodynamic potentials} which can be a function of one or more more easily measured and controlled intensive parameters under experimental conditions.  Examples of thermodynamic potentials are the internal energy itself, Helmholtz free energy, enthalpy and Gibbs free energy, although others can be derived. A convenient potential for porous media studies is the Gibbs (free) energy $G$.   In the case of the Gibbs energy we want to replace both S and V as independent variables and in this case G is defined by differential of $U$ minus the two conjugate pairs one wishes to replace, $TS$ and $-PV$ \cite{guggenheim}
\begin{equation*}
dG = d(U - TS - (-PV)) = d(U -TS + PV) = dU - TdS - SdT + PdV + VdP
\end{equation*}
Recalling \emph{dU = TdS - pdV}
\begin{equation*}
dG = TdS - PdV + \sum_i \mu_i dN_i - TdS - SdT + PdV + VdP 
\end{equation*}
arriving at
\begin{equation}
dG =  - SdT +  VdP + \sum_i \mu_i dN_i    
\end{equation}
In which the Gibbs free energy is a function of $T$, $P$ and $N_i$
\begin{equation}
G = G(T, P, N_i) 
\end{equation}
As with the internal energy function, we can also define intensive parameters and equations of state for the Gibbs free energy 
\begin{equation}
-S = \left(\frac{\partial G}{\partial T}\right)_{P, N_i} 
\end{equation}
\begin{equation}
V = \left(\frac{\partial G}{\partial P}\right)_{T, N_i}
\end{equation}
\begin{equation}
\mu_i =  \left(\frac{\partial G}{\partial N_i}\right)_{T, P, N_j}
\end{equation}
This last term is the partial molar Gibbs energy of component $i$ and is defined at constant T, P and number of moles of any other constituents, $N_j$ \cite{boltfrissel}.  Thus, for the component $i$, the equation of state is a function of the form
\begin{equation}
\mu_i = \mu_i(T, P,  N_j) 
\end{equation}
with differential
\begin{equation}
d\mu_i = \left(\frac{\partial \mu_i}{\partial T}\right)_{P, N_j} dT + \left(\frac{\partial \mu_i}{\partial P}\right)_{T, N_j} dP + \sum_j \left(\frac{\partial \mu_i}{\partial N_j}\right)_{T, P} dN_j
\end{equation}

There are many ways of using these equations in thermodynamic systems, the user having freedom within the constraints of the theory to choose the independent variables. In soils and other porous systems we are usually concerned with chemical potential of the water in the pores and surfaces $\mu_w$. It is convenient to modify the chemical potential to include the mas based or gravimetric water content $\theta_g$, which is essentially the mass based water fraction. Under these conditions, the chemical potential of water can be defined as a function of $T$, $P$, $\theta_g$ and the number of moles of each dissolved solute fraction $N_j$ \cite{babcockoverstreet, boltfrissel, taylorashcroft}\footnote{An alternative derivation can be found in \cite{sposito81}.}

\begin{equation}
\mu_w = \mu_w(T, P,  \theta_g, N_j) 
\end{equation}
thus,
\begin{equation}
d\mu_w = \left(\frac{\partial \mu_w}{\partial T}\right)_{P, \theta_g, N_j} dT + \left(\frac{\partial \mu_w}{\partial P}\right)_{T, \theta_g, N_j} dP + \left(\frac{\partial \mu_w}{\partial \theta_g}\right)_{T, P, N_j} d\theta_g + \sum_j \left(\frac{\partial \mu_w}{\partial N_j}\right)_{T, P, \theta_g} dN_j
\end{equation}
In the soil physics literature what is defined as water potential ($\psi_w$) is the chemical potential of water in soil in reference to a \emph{Standard State}, usually pure water at 293.15 K and 101325 Pa ($\mu_w^0$)   
\begin{equation}
\boxed{\psi_w \equiv d\mu_w \equiv \mu_w - \mu_w^0}
\end{equation}

Make sure to understand this concept, it is one of the most fundamental principles in applied soils physics and, along with other forms energy and potential gradients, defines how water is stored and moves in soils, and the principles of operation of various field and laboratory devices such as tensiometers, pressure plate apparatus, psychrometers and what the water retention curve is. Most soil physicists and vadose zone hydrologists take this basic knowledge for granted and it is impossible to have a firm grasp on anything related to soil water phenomena without understanding these concepts.  Since $\mu_w$ is an intensive property in units of energy/amount (usually $J~kg^{-1}$ or $J~mol^{-1}$) the water potential has the same units.  

Having defined what the water potential $\psi_w$ is we now have to define its components as usually encountered on the soil physics literature\footnote{We mention soil physics because these concepts evolved within and along with early soil physics. It is interesting to follow the evolution of the concepts and discussions on journal articles published by N.E. Edlefsen and A.B.C. Anderson, the Gardner brothers, G.H. Bolt, K.L. Babcock, R.D. Miller, S. Takagi and many of their collaborators and students during the 1940s, 50s and 60s.}. Because\footnote{$j$ here has the same function as $i$ from when this function was first presented.} 
\begin{equation}
\mu_w = \left(\frac{\partial G}{\partial N_j}\right)
\end{equation}
the coefficients in $dP$ and in $dT$ in Equation 7.54 can be transformed by maintaining the appropriate quantities constant and changing the order of differentiation as \cite{boltfrissel, callen}
\begin{equation}
\frac{\partial \mu_w}{\partial T} = \frac{\partial }{\partial T}(\frac{\partial G}{\partial N_j}) = \frac{\partial }{\partial N_j} (\frac{\partial G}{\partial T}) = \frac{\partial}{\partial N_j}(-S) = -\frac{\partial S}{\partial N_j} =  -S_w
\end{equation}
\begin{equation}
\frac{\partial \mu_w}{\partial P} = \frac{\partial }{\partial P} (\frac{\partial G}{\partial N_j}) = \frac{\partial }{\partial N_j} (\frac{\partial G}{\partial P}) = \frac{\partial }{\partial N_j}(V) = \frac{\partial V}{\partial N_j} = V_w
\end{equation}
in which $V_w$ is the partial molar volume of water and $S_w$ is the partial molar entropy of pore water. Replacing into the original formula we arrive at 
\begin{equation}
\boxed{d\mu_w = -S_w dT + V_w dP + \left(\frac{\partial \mu_w}{\partial \theta_g}\right)_{T, P, N_j} d\theta_g + \sum_j \left(\frac{\partial \mu_w}{\partial N_j}\right)_{T, P, \theta_g} dN_j
}
\end{equation}
This is the thermodynamic equation that defines the potentials of soil water, it is possible to arrive at this equation using different thermodynamic potentials although Gibbs free energy and the Groenevelt-Parlange \cite{groeneveltparlange} potentials are most often employed in the soil physics literature \cite{sposito81}.  Using the soil physics $\psi$ notation, this equation is usually written as  \cite{taylorashcroft}
\begin{equation}
\boxed{\psi_w = -S_w dT + \psi_p + \psi_m + \psi_o}
\end{equation}
Considering isothermal conditions, $dT = 0$, and the equation reduces to  
\begin{equation}
\boxed{\psi_w =\psi_p + \psi_m + \psi_o}
\end{equation}
In reality, many field and laboratory applications of these equations must consider isothermal conditions, not doing so would require some entropy function for the temperature potential, $\psi_T$, to be considered, which would cause great mathematical complications.  The $\psi$ terms in this equation are the most commonly defined components of soil water potential $\psi_w$, the \emph{pressure potential}
\begin{equation}
\psi_p = V_w dP 
\end{equation}
the \emph{matric potential}
\begin{equation}
\psi_m =  \left(\frac{\partial \mu_w}{\partial \theta_g}\right)_{T, P, N_j} d\theta_g
\end{equation}
and the \emph{osmotic potential}
\begin{equation}
\psi_o =  \sum_j \left(\frac{\partial \mu_w}{\partial N_j}\right)_{T, P, \theta_g} dN_j
\end{equation}

The pressure potential is usually associated with hydraulic pressure in saturated soils below the water table, however it can be positive or negative\footnote{A body of water can be under tension, and not only under compressive stresses.}. Under the water table the pressure potential corresponds to the hydraulic pressure plus the atmospheric pressure, at the water table the hydraulic pressure is zero and the second component of pressure is the atmospheric pressure. Because we define the standard state at atmospheric pressure, the pressure potential is zero at the water table. %In unsaturated soils the liquid pressure is zero and the soil water is at atmospheric pressure. 

The soil matric potential is negative and is related to the capillary and adsorptive forces exerted by the soil solid phase on soil water. Everything else constant, the water potential in an unsaturated soil is lower than that of free water at the same height and elevation, so that the state of energy of soil water is lower than that of free water. In simpler terms, water molecules in contact with soil solid particles are ``less free to move'' because of the forces exerted by these surfaces. At higher water contents, the adsortion forces of water molecules with surfaces play a small role when compared to capillary forces. 
As the soil dries and the water is restricted to thin films around the particles and very small pores, the adsorption component cannot be neglected and the matric potential will be some function of the capillary potential and the adsorption potential. Detailed discussion of the components of capillary and adsorptive forces can be found in \cite{nitaobear, tulleretal99}. 

The osmotic (or solute) potential is the reduction of water potential in contact with dissolved solids, or more specifically, solutes. Water molecules in contact with dissolved solids have lower energy, thus, as with cells and in other semi-permeable membranes, a larger solute concentration within the porous membrane or the cell will tend to draw water from the exterior until a pressure equilibrium is achieved in response to the osmotic forces. 

Different authors report different soil water potentials. This is a complex issue and as we mentioned before, the potentials will depend on thermodynamic theoretical considerations such as the thermodynamic potential used to derive the water potentials, see for example the discussions on pneumatic, submergence and overburden potentials, among others in \cite{taylorashcroft, sposito81}. 

Because we initially expressed the water potential as a partial molar Gibbs energy $\partial G/\partial N_j$, the units of the water potential are in $J~mol^{-1}$, and this water potential can be defined as the \emph{molar water potential}. However, this unit is not often used in soil physics and groundwater hydrology, the most used units are derived from this basic unit. Considering the molar mass of water $M_w$ in $kg~mol^{-1}$, we can divide the soil molar water potential by it, defining the water potential in units of  $J~kg^{-1}$
\begin{equation}
\frac{\psi_w}{M_w} \rightarrow \frac{J~mol^{-1}}{kg~mol^{-1}} = \frac{J}{kg} 
\end{equation}
Perhaps the most common form encountered in soil physics is in terms of pressure units, or the \emph{soil water pressure}. The water potential in joules per mass is converted to pressure units by multiplying by water density 
\begin{equation}
\psi_w \rho_w \rightarrow \frac{J }{kg} \frac{kg}{m^{-3}} = \frac{J}{m^{3}} = \frac{N~m}{m^{3}} = \frac{N}{m^{2}} = Pa  
\end{equation}
Pascal ($Pa$) and its equivalents $kPa$, $hPa$ and $MPa$ are frequently used in soil physics, especially when dealing with matric potentials measured using tensiometers and pressure plates. The third most commonly used conversion of water potential is in terms of head units or length, or the \emph{soil water head}. To convert to length units, the water potential in joules per mass is divided by the gravitational acceleration, $g$
\begin{equation}
\frac{\psi_w}{g} \rightarrow \frac{J~kg^{-1} }{m~s^{-2}} = \frac{N~m~kg^{-1} }{m~s^{-2}} = \frac{N~kg^{-1} }{ s^{-2}} = \frac{kg~m~s^{-2}~kg^{-1} }{ s^{-2}} = m  
\end{equation}
Recall for the above derivations that $1~J = 1~N~m$ and that $1~N = 1~kg~m~s^{-2}$. The head units are common in hydrogeology and to some extent in soil physics, especially for the pressure potential. Sposito \cite{sposito81} warned that the different units should not obscure the physical interpretation of the problem. No matter in what unit it is expressed, the soil water potential represents an gradient of energy per unit amount of water which drives movement of soil water.   

Because water close to Earth's surface is subjected to a gravitational field and gravitational forces, the description of the potentials is incomplete without considering the effect of gravity. For a system under the effect of gravity the gravitational potential $ \phi = gz$ must be added to the chemical potential of water to completely define its state. The complete treatment of the inclusion of the gravitational potential into the thermodynamic treatment of soil water is complex and a more detailed analysis of the methods can be found in Babcock \cite{babcock}. The general theoretical account of including the gravitational potential with the chemical potential can be found in Guggenheim \cite{guggenheim}.  For now\footnote{A complete treatment is planned to be included in later editions.} and for simplicity we will follow the approach of \cite{taylorashcroft} and simply add the soil water gravitational potential to the other potentials as\footnote{Note that this relationship can be easily demonstrated to hold by considering the gravitational potential energy $ U = mgh$ and solving the density of water $\rho_w = m_w/V$ for $m_w$ and replacing into the gravitational potential energy. The soil water gravitational potential then is simply $U/V \rightarrow J~m^{-3} = N~m^{-2} = Pa $.}   
\begin{equation}
\psi_g =  \rho_w g z
\end{equation}
In which $z$ is the vertical elevation. Considering now the potentials in pressure units, the total potential of soil water is the sum of the components of water potential and the gravitational potential, plus any other potentials that might be relevant under specific conditions 
 \begin{equation}
\boxed{\psi_t = \psi_p + \psi_m + \psi_o + \psi_g + ....}
\end{equation}

\begin{figure}[ht]
\centering
 \includegraphics[width=1.0\textwidth]{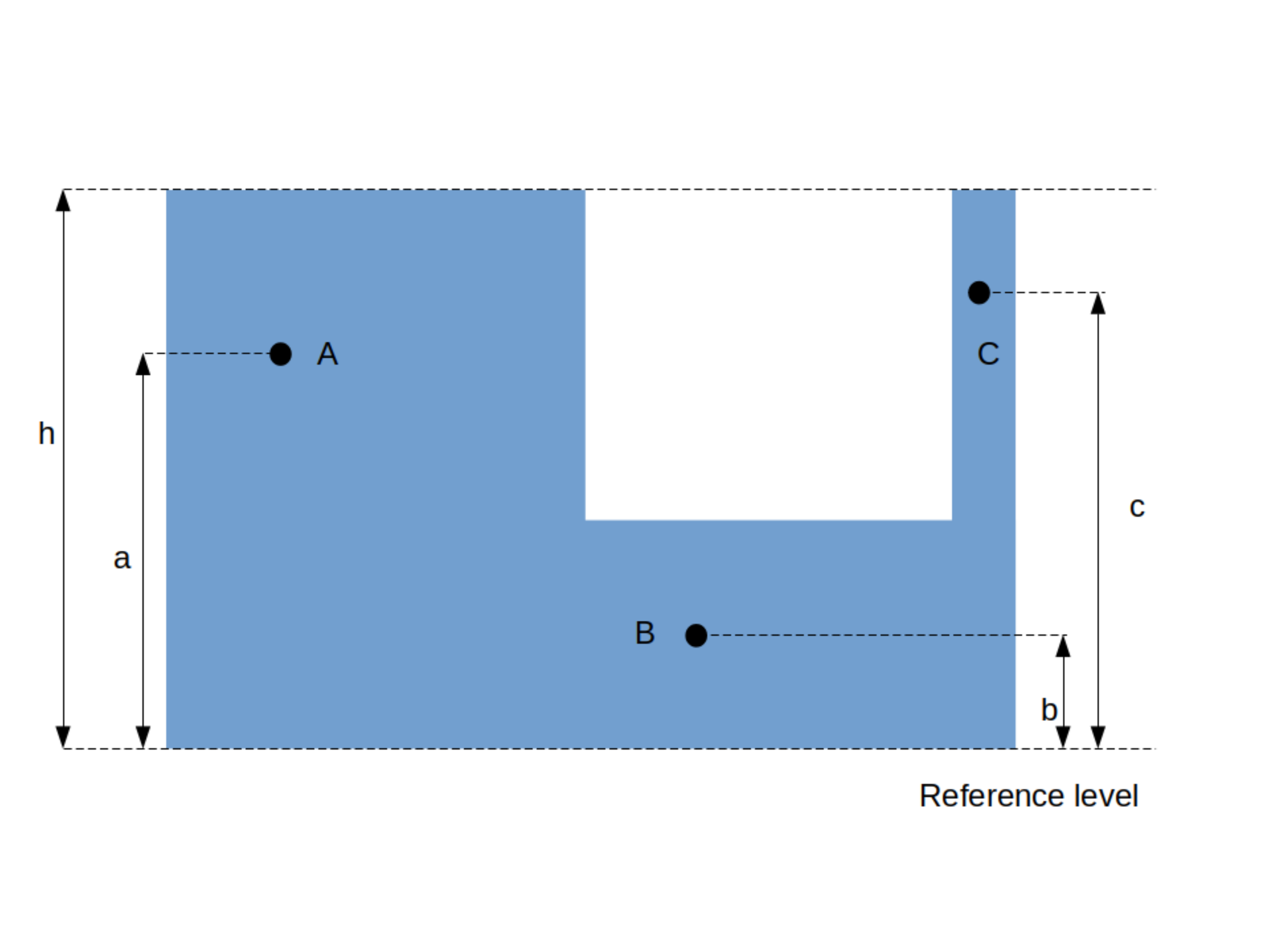}
\caption{Total potential in a vessel.}
\label{ch7_fig5}
\end{figure}

Remember that a potential gradient can be understood as a potential to perform work, movement of water will occur as a response to a gradient of total potential. In a water body under static conditions the total potential must be the same everywhere. Notice that the individual components might not be equal, but the sum of the components must be the same everywhere, otherwise there is a potential gradient and the water will flow in response to such gradient. Consider the points $A$, $B$ and $C$ at heights $a$, $b$ and $c$ on the vessel in Figure \ref{ch7_fig5}. If the water level is at $h$ and we consider the gravitational reference at the dotted line at the bottom, the gravitational potential, in head units, at each point is

\begin{align*}
\psi_{gA} =  a \\
\psi_{gB} =  b \\
\psi_{gC} =  c 
\end{align*}
and the pressure potential is 
\begin{align*}
\psi_{pA} =  h - a \\
\psi_{pB} =  h - b \\
\psi_{pC} =  h - c 
\end{align*}
Since only these two potentials are being considered, the total potential $\psi_t = \psi_p + \psi_g$ is 
\begin{align*}
\psi_{tA} =  (h - a) + a = h \\
\psi_{tB} =  (h - b) + b = h\\
\psi_{tC} =  (h - c) + c  = h
\end{align*}
Thus the total potential is constant and equal to $h$ everywhere. The same principle is valid for saturated soils, porous rocks and sedimentary deposits, if the total potential is constant everywhere the fluid is static. 

\begin{figure}[ht]
\centering
 \includegraphics[width=1.0\textwidth]{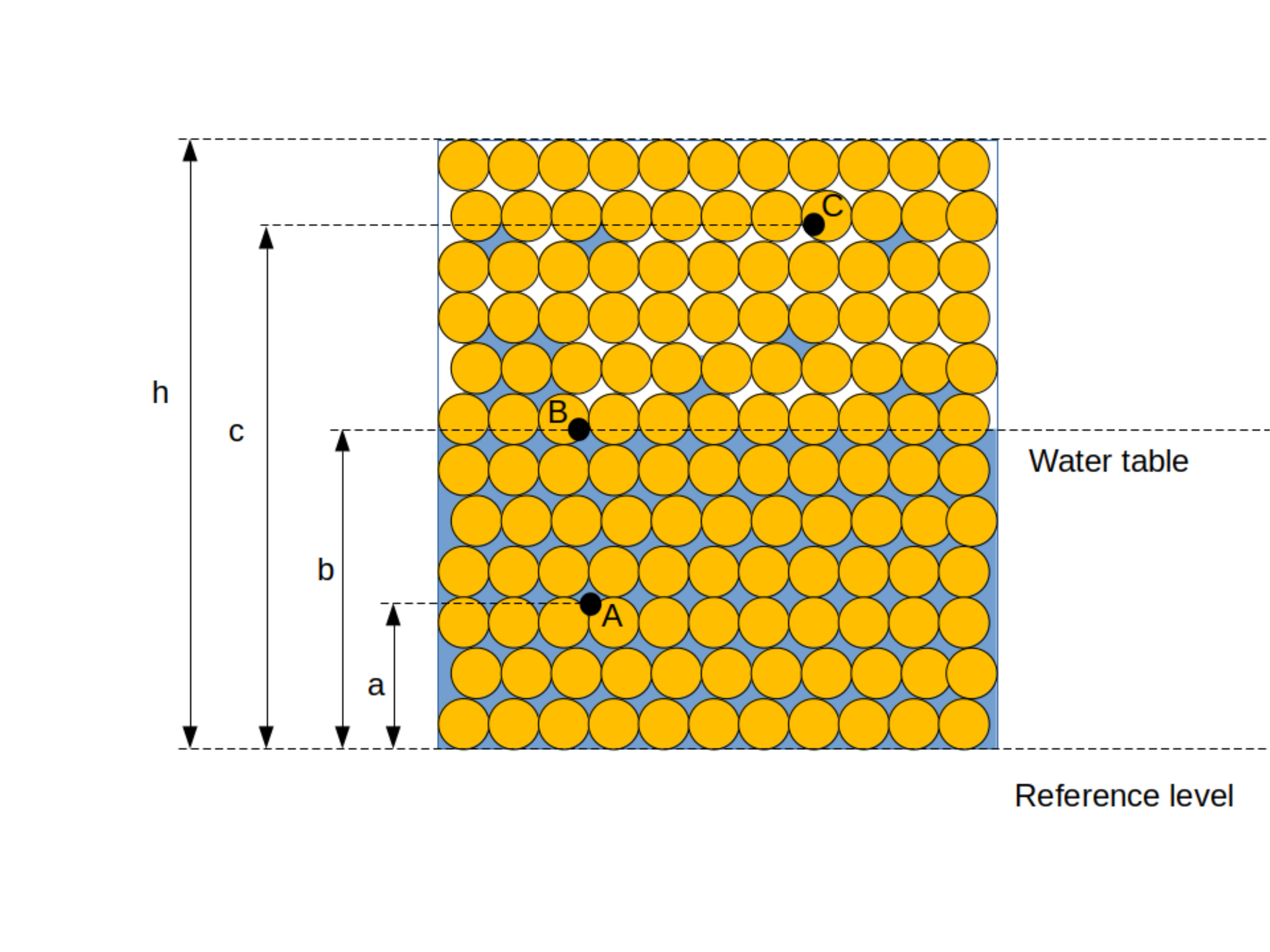}
\caption{Total potential in a partially saturated porous medium.}
\label{ch7_fig6}
\end{figure}

Consider now a porous medium which is half saturated and half unsaturated (Figure \ref{ch7_fig6}). Above the water table the water column is under tension following the mechanisms described for a capillary tube. We showed that the height of capillary rise was given by  
\begin{align*}
h = \frac{2 \gamma \cos{\theta}}{\rho g r}
\end{align*}

Because water at each point above the water table is under tension as demonstrated for the capillary rise equation derivation, the capillary potential in head units at any point above the water table can be shown to be equivalent to the height above the water table\footnote{This offers no mathematical or physical proof of any kind and is presented as a statement. A mathematical proof is being pursued.}
\begin{equation}
\psi_c = -h
\end{equation}

Considering that the matric potential can be approximated by the capillary potential $\psi_m \approx \psi_c$, the matric potential, in head units, will correspond to the negative of the height of water column above the water table. Thus, for a porous medium that is half saturated and half unsaturated (Figure  \ref{ch7_fig6}) we can calculate the gravitational, pressure, matric and total potentials at the points $A$, $B$ and $C$. The gravitational potential is
\begin{align*}
\psi_{gA} =  a \\
\psi_{gB} =  b \\
\psi_{gC} =  c 
\end{align*}
the pressure potential is 
\begin{align*}
\psi_{pA} =  b - a \\
\psi_{pB} =  0 \\
\psi_{pC} =  0 
\end{align*}
the matric potential is 
\begin{align*}
\psi_{mA} =  0 \\
\psi_{mB} =  0 \\
\psi_{mC} =  -(c - b) 
\end{align*}
such that the total potential, $\psi_t = \psi_g + \psi_g + \psi_m $, is 
\begin{align*}
\psi_{tA} =  a + (b - a) + 0 = b\\
\psi_{tB} =  b + 0 + 0 = b\\
\psi_{tC} =  c + 0 -(c - b) = b  
\end{align*}
Thus the total potential is constant and equal to $b$ on the whole system, as expected.

\section{Measuring water potential}

Most devices for measuring soil water potential fall into four categories: 1) suction devices, such as Haines apparatus or tension tables, 2) tensiometers, 3) pressure plates or pressure chamber apparatus and 4) psychrometers \cite{danetopp}.

 Methods 1, 2 and 3 rely on similar physical principles, the soil sample is put into contact with a porous membrane which is contact with a water reservoir. In 1 the pressure in the water reservoir is then decreased by decreasing the elevation of the opening of a water column in contact with the porous membrane which in turn is in contact with the soil sample. As the pressure decreases, water in an initially saturated soil sample will flow to the reservoir, and from there to the outflow tube, until equilibrium is reached. In head units, the capillary potential, thus the matric potential, are effectively the negative of the decrease in elevation of the opening of the water column with respect to the porous membrane. 
\begin{equation}
\psi_m \approx  \psi_c = -h
\end{equation}
If the soil sample is relatively small in height, less than one to two cm \cite{danetopp}, at equilibrium the matric potential of soil water is the same as the porous membrane. Different porous membranes can be used, depending on the application of the method and the desired matric potential range. Fine sand, kaolin clay, paper, porous stone and others can be found on the literature. Often in the literature the term \emph{tension} is used instead or matric potential, tension ($\tau$) being simply the negative value of the matric potential
\begin{equation}
\tau = -\psi_m
\end{equation}    
Tensiometers (category 2) function in a similar way, except that the pressure on the water reservoir is not controlled by the user as in the case of hanging water column devices. The device is usually composed of a tube full of water with a porous cup on one end. The porous cup is put into contact with soil and if the soil is unsaturated, water will flow from the device and into the soil until equilibrium is reached. When equilibrium is reached, the water within the tube is in a state of tension and the tension or pressure can be measured using a variety of methods. Historically, the tension was measured by the decrease in height of the water level in a tube connected to the porous medium, however, because this would require a very long column, to avoid this issue, a liquid with a much greater density, mercury (Hg) was used in contact with a water reservoir, resulting in the need of much smaller columns and in devices that could be conveniently deployed to the field. Later, due to the extreme environmental toxicity or Hg, and for convenience, analog pressure gauges were used. Although analog gauges are still used, with the decrease in cost of electronics in the last decades, digital pressure and tension sensors connected to data logging devices are now common in research and in agricultural and environmental monitoring applications. 

At equilibrim, and considering the water reservoir and soil at the same elevation, the sum of the pressure and matric potentials are going to be the same on the soil and on the water reservoir \cite{taylorashcroft}
\begin{align*}
\psi_{ps} + \psi_{ms} = \psi_{pr} + \psi_{mr}  
\end{align*}
Where the subscripts $s$ and $r$ stand for soil and water reservoir, respectively. However, the pressure potential is zero if the soil is unsaturated and the matric potential is zero in the water reservoir as there is no soil matrix such that
\begin{align*}
\psi_{ms} = \psi_{pr}  
\end{align*}
Therefore, the (negative) pressure reading corresponding to the water on the reservoir is the soil matric potential. Notice that atmospheric pressure still acts on the unsaturated soil, but since the pressure reading on the device is with respect to the atmospheric pressure (by definition the gauge pressure), the two terms cancel out and there is no need to consider atmospheric pressure. Because the boiling point of water at pressures below atmospheric pressures decreases as the pressure difference increases at constant temperature, the maximum practical tension possible at the sea level and at average surface temperature is around $8~m$, although in laboratory conditions, using de-areated water, higher tensions are possible \cite{youngsisson}. 

In order to increase the range of pressures (or tensions), the soil sample and porous membrane can be put inside a pressurized chamber (category 3). As the pressure inside the chamber increases, the capillary pressure of the water in the porous membrane decreases as the pores will gradually drain according to their radius. At higher pressures, water exists in smaller pores of small radius, such that the radius of curvature of the water meniscus is also very small, corresponding to very small capillary pressures (higher numbers in absolute value, or in terms of tension). The soil samples in hydraulic contact with the porous membrane will lose water until equilibrium is reached\footnote{True equilibrium is never reached in reality, but fluxes are microscopic only or gradients small enough so that they can be disregarded.}. If the height of the soil sample is small, the gravitational potential can be neglected and the tension on the soil sample is the gauge pressure within the chamber.  The porous membrane, usually a porous ceramic plate, is in contact with a water reservoir which is open to the atmosphere such that, as before 
\begin{align*}
\psi_{ps} + \psi_{ms} = \psi_{pr} + \psi_{mr}  
\end{align*}
and the matric and pressure potential are now zero on the water reservoir such that 
\begin{align*}
\psi_{ms} = -\psi_{ps}  
\end{align*}
In which $\psi_{ps}$ is effectively the gauge pressure within the chamber. Pressure chamber devices can reach tensions well beyond those of tension devices. The limit usually considered is more or less arbitrarily set at $15~m$\footnote{In part because of the idea that there is some universal \emph{permanent wilting point}.}. At pressures above this value, adsorption forces start to play a more prominent role and the pressure gradient alone is not enough to further remove water from the soil. The water that cannot be removed by pressure plate apparatus is often referred to \emph{residual water content} and can often be removed using other methods.

Psychromters, hygrometes and potentiameters (category 4) are, generally speaking, devices that measure the water potential based on the Kelvin equation 
\begin{equation}
\psi_{w} = \frac{R_g T}{V_w} \ln{\frac{p}{p_0}}  
\end{equation}
in which the water potential measured is the sum of the matric and osmotic potential, $R_g$ is the universal gas constant, T is the absolute temperature, $V_w$ is the molar volume of water, $p$ is the air actual vapor pressure and $p_0$ is the air saturation vapor pressure, and $p/p_0$ is the relative humidity of the air in contact with the sample. The water activity is also used in some devices and is, by definition, $a_w \equiv  p/p_0$. The water potential measured by these devices will not necessarily correspond to the water potential measured by pressure and suction devices, since the nature of the components measured are different. One of the main advantages of psychrometric and other similar devices is that they can measure potentials at extremely low water contents, well beyond what is possible with the pressure chamber, well below -1.5 MPa. Details on these devices can be found in \cite{andraskiscanlon, scanlonetal}

\section{Measuring water content}

The gravimetric method, based on weighing a soil or porous material sample, oven drying at 105 °C for at least 24 h and reweighing the oven dried sample and calculating the mass of water divided the mass of oven dried material is the standard method for measuring water content and it is used to calibrate \emph{indirect} water content estimation methods. The gravimetric method is a \emph{direct} method because the mass of water is directly measured by weighing. Because it is a direct method, if conducted properly using scales with high enough accuracy and precision, results in errors, both systematic and random, much smaller than indirect methods. In indirect methods, the water content is estimated indirectly and they are usually calibrated for a particular soil or condition using the gravimetric method. Errors related to calibration, or lack of thereof, and instrumental errors, both on the sensors and electronics usually result in lower precision and/or accuracy when compared to direct methods. There are several different types of indirect methods for measuring water content, the two most common types in use today being methods using radioactive sources, such as neutrons and gamma radiation, and general electromagnetic methods including time domain reflectometry (TDR), capacitance and other devices \cite{danetopp}.    

The neutron thermalization method historically has been an important method for water content measurement and monitoring in the field. Once calibrated it provides fast and accurate measurements of soil water content. Physically, the method relies on a radioctive source that emits fast neutrons, with kinetic energy on the MeV\footnote{eV = electron-volt; 1 $eV = 1.60217646 \cdot 10^{-19}~J$.} range and speed on the $10^4~km~s^{-1}$ order of magnitude, and a detector of slow neutrons, with energies less than about 1 eV \cite{hignettevett}. The fast neutrons emitted by the source collide with hydrogen atoms (H) in water molecules. Because H has a high scattering cross section, it can absorb or scatter fast neutrons, causing attenuation of the energies and correspondingly the velocity of the neutrons. As the water content increases, more hydrogen nuclei are available for collisions, attenuating fast neutrons and producing slow neutrons which are then counted by the detector. Of course other atoms can attenuate neutrons, but because H present in water is particularly effective in doing so in comparison to other atoms commonly found in soils, it can be used for water content estimation in most soils. The calibration is performed by direct measurement of volumetric water content around the sensing volume and corresponding slow neutron count ratios \cite{hignettevett}
 \begin{equation}
C_R = \frac{x}{x_s} 
\end{equation}
 in which $x$ is the count in the measured material and $x_s$ is a standard count. The count ratio can then be related to the volumetric water content by a calibration equation of the form 
\begin{equation}
\theta_v = b_0 + b_1 C_R
\end{equation}
in which $b_0$ and $b_1$ are empirical parameters obtained from statistical or mathematical fitting  procedures and $\theta_v$ is the volumetric water content \cite{hignettevett}. Other forms of calibration equations can be found on the literature for different materials and water content ranges. 
        
The principle of operation of most electromagnetic methods including TDR, frequency domain reflectometry (FDR), ground penetrating radar (GPR), capacitance, impedance and others are based on the principle that the relative permittivity\footnote{Referred to as the dielectric constant in older literature} of water is much greater than that of other soil components, including air, minerals and organic matter. The relative permittivity of a material is a ratio of the absolute permittivity of the material to that of vacuum
\begin{equation}
\epsilon_r = \frac{\epsilon}{\epsilon_0}
\end{equation}
Because the absolute permittivity of the material, $\epsilon$, and that of vacuum have both the same units, usually  $F~m^{-1}$, and the permittivity of vacuum is a physical constant $\epsilon_0 = 8.8541878128(13)\cdot10^{-12} ~ F~m^{-1}$, $\epsilon_r$ is a dimensionless normalized quantity, with values around 80 for water, 1 for air and around  5 for quartz rich soils \cite{robinsonfriedman}. The relative permittivity of the porous medium as a whole will be some form of average of the individual permittivities of each phase, the term \emph{apparent permittivity} ($\epsilon_{r(app)}$) being used on the literature \cite{leao2020}. Because the solid phase is usually constant in fractional volume for most porous media applications and because the relativity permittivity of one of the dynamic phases, water, is much greater than that of the other, air, under unsaturated conditions the apparent permittivity will be a function of the volumetric water fraction, i.e. the volumetric water content. Thus, for most practical applications, water content measured using direct methods, usually gravimetric, can be related to apparent relativity permittivity measurements obtained using electromagnetic measuring devices, generating \emph{calibration equations}. The most common empirical calibration equations found in the literature are third degree polynomials, square root and logarithmic \cite{leaoetal2020}
\begin{equation}
\theta_v = a_0 + a_1 \epsilon_{r(app)} + a_2 \epsilon_{r(app)}^2 + a_3 \epsilon_{r(app)}^3
\end{equation}
\begin{equation}
\theta_v = a_0 + a_1 \sqrt{\epsilon_{r(app)}}
\end{equation}
\begin{equation}
\theta_v = a_0 + a_1 \log{(\epsilon_{r(app)})}
\end{equation}
in which $a_0$ to $a_3$ are empirical fitting parameters, fitted for each individual equation and $\log$ is the natural logarithm. Most commercial devices come with ``universal'' calibration equations from the manufacturers, but it is almost always recommended that the device should be calibrated for the individual soils or porous materials in which is going to be used.  Water content can also be estimated using more physically based models based on \emph{effective medium approximations} theory \cite{landaulifshitz1960, leao2020}.

There are several different designs and electronics for electromagnetic water content measuring sensors. The relative permittivity is in reality an imaginary number and is highly dependent on the frequency of operation of the device in terms of the frequency of the electromagnetic radiation emitted. For most soils, but not all, the relative apparent permittivity can be approximated by the real component of the complex permittivity. A detailed discussion of measuring water content with electromagnetic devices is beyond the scope of this book. The user is highly encouraged to study electromagnetic theory before using any electromagnetic device for soil and porous media measurements, there are many misconceptions related to these devices. For more information the reader is referred to \cite{vonhippel, danetopp, leao2020}.

There are many other methods used for estimating water content, including methods based on remote sensors, thermal methods and other ionizing radiation methods. These methods will not be discussed in this book, as they are not as broadly employed as the methods discussed above. The user interested in these specific methods should consult \cite{danetopp} and the references listed therein.

\section{The water retention curve}

For a set of tubes of different radiuses $r_i$, there is an exact relationship between the percent saturation and the capillary pressure.  At the capillary pressure $Pc_i$, all tubes of radius greater than $r_i$ will have drained, following the Young-Laplace equation
\begin{equation}
Pc_{i} =\frac{ 2\gamma }{r_i} 
\end{equation}
thus the volume of water retained will correspond to the sum of the volumes of water retained in all tubes of radius smaller or equal than $r_i$. But the volume will also depend on the capillary rise for each individual tube which is related to the radius by the capillary rise equation \footnote{For the angle $\theta = 0$ and $\cos \theta = 1$.}.
\begin{equation}
h_{i} =\frac{ 2\gamma }{\rho g r_i} 
\end{equation}
Considering the volume of each tube as that of a cylinder
\begin{equation}
V_{i} =\pi r_i^2 h_i 
\end{equation} 
and replacing the capillary rise equation
\begin{equation}
V_{i} = \pi r_i^2 \frac{ 2\gamma }{\rho g r_i} =  \frac{ 2\gamma \pi r_i}{\rho g } 
\end{equation} 
Now the volume of water is the sum of the volumes of water in all $i$ pores with radius equal or smaller than $r_i$
\begin{equation}
V_{water} =\sum_i \pi r_i^2 \frac{ 2\gamma }{\rho g r_i} = \sum_i \frac{ 2\gamma \pi r_i}{\rho g } 
\end{equation} 
and the fractional water content is simply the volume of water divided by the total volume of all tubes in the system
\begin{equation}
\theta_{f} =\frac{ V_{water}}{V_{t} } 
\end{equation} 
Thus, for each capillary pressure $Pc_{j}$ corresponding to a radius $r_j$, the  fractional water content can be directly calculated from the sum of the volumes of water in all pores with radius less or equal than $r_j$ \footnote{An explicit mathematical relationship can be derived by manipulating summation terms, this will be explored in following editions.}.  This is what is called the \emph{capillary bundle model} for water retention. It is also an important model for hydraulic conductivity, as we will discuss in later chapters. The capillary bundle model is a useful tool to model and understand transport and retention of water in porous media. In this approach, the pore space is simplified as a set of continuous tubes of constant radius in each pore. Because pores are irregular, do not have constant radius, are usually not cylindrical, and because forces related to other than simple capillary phenomena act in complex natural porous systems, there is no direct way of calculating water saturation from matric potential and empirical and semi-empirical models are necessary. Empirical models rely on fitting to observed data of water content and matric potential, in other words, there is no analytical expression to calculate the volumetric water content by knowing only the matric potential, water content data is necessary. 

The \emph{water retention curve} is nothing more than the empirical relationship between water content and matric potential obtained from laboratory and field data. The relationship has been observed to follow a characteristic sigmoid function (S-shaped) when the matric potential is considered in log scale.  To avoid mathematical and graphing difficulties it is convenient to consider the absolute values of matric potential ($|\psi_m|$), which as we have seen is a negative quantity, when dealing with water retention equations. There has been several efforts to derive analytical expressions for the water retention curve. These have important practical applications in irrigation and environmental modeling and because, in most cases, an unsaturated hydraulic conductivity function was derived from the water retention curve (e.g. \cite{brookscorey, campbell, vangenuchten, kosugui}). Some of the models most often found on the literature are presented below. We will leave the related unsaturated hydraulic conductivity functions for later chapters.  

One of the earliest models that saw wide application is that of Brooks and Corey (BC) \cite{brookscorey} \footnote{For simplicity and to be consistent we will refer to soil water potential as $\psi$ on the water retention equations.}
 \begin{equation}
\theta_v = 
 \begin{cases}
    \theta_r + (\theta_s - \theta_r) (\frac{\psi_b}{|\psi|})^\lambda, & \text{for } |\psi| > \psi_b \\
    \theta_s, & \text{for } |\psi| \le \psi_b 
  \end{cases}
\end{equation}
The BC equation, like many water retention models, is a piecewise equation, meaning that it is segmented into two parts. Its parameters are the saturation water content $\theta_s$, the residual water content $\theta_r$, the air entry value, often called bubbling pressure $\psi_b$, and the pore size distribution parameter $\lambda$. Most water retention models can be defined in terms of saturation and residual water contents. The saturation water content corresponds to the water content at a matric potential of zero, whenever it is defined, where in theory the pore space is completely water saturated. The residual water content is the remaining water content at very high values of matric tension\footnote{Or very low values if you are thinking in potential, remember that tension is defined as the negative of the potential.}, where further increase in tension does not cause decrease in water content. In most models, the residual water content is an asymptotic parameter reached by taking tension to infinity. The bubbling pressure is the matric tension in which air starts to enter the pore system, usually at relatively low matric tensions for most soils. Below the bubbling pressure, increase in tension does not cause a corresponding decrease in water content. Most water retention models will have parameters related to pore size distribution since, as we saw, water retention is controlled by the distribution of pores. In the BC model the piecewise condition is defined by the $\psi_b$ parameter. At tensions below or equal to $\psi_b$ the water content is constant and equal to saturation water content, while at tensions above $\psi_b$ water content decreases following a power law of $\lambda$.     

Most water retention equations can be expressed in terms of the effective saturation, given by
 \begin{equation}
S_e =   \frac{\theta_v - \theta_r}{\theta_s - \theta_r}    
 \end{equation}
The BC equation, for example, can be expressed as
 \begin{equation}
S_e = 
 \begin{cases}
     (\frac{\psi_b}{|\psi|})^\lambda, & \text{for } |\psi| > \psi_b \\
    1, & \text{for } |\psi| \le \psi_b 
  \end{cases}
\end{equation}
Using the effective saturation normalizes the water content between zero and one, making it possible to compare the shapes of the curves for different materials. Using the effective saturation can, in some cases, make fitting easier using computer methods by reducing the number of fitting parameters. 
The Campbell equation is in essence the BC equation, but defining the effective saturation as $\theta/\theta_s$ \cite{campbell}
 \begin{equation}
\theta = 
 \begin{cases}
    \theta_s (\frac{\psi_b}{|\psi|})^b, & \text{for } |\psi| > \psi_b \\
    \theta_s, & \text{for } |\psi| \le \psi_b 
  \end{cases}
\end{equation}
In which the physical meaning is the same as in the BC equation, and $b$ is essentially $\lambda$.

Brutsaert \cite{brutsaert} proposed an equation of the form
 \begin{equation*}
S_e =    \frac{a}{a + \psi^b}
\end{equation*}
where $a$ and $b$ are empirical parameters. Brutsaert's equation was later improved upon by van Genuchten, resulting in one of the most frequently used water retention equations in soil physics and hydrology  \cite{vangenuchten}
 \begin{equation*}
\theta_v = \theta_r + \frac{(\theta_s - \theta_r)}{[1 + (\alpha \psi)^n]^m}
\end{equation*}
In which $m$ and $n$ are shape parameters related to the pore size distribution. It is often convenient to define the Mualem \cite{mualem} restriction as
 \begin{equation*}
m = 1 - \frac{1}{n}
\end{equation*}
Reducing the number of empirical parameters to four.

\begin{figure}[ht]
\centering
 \includegraphics[width=1.0\textwidth]{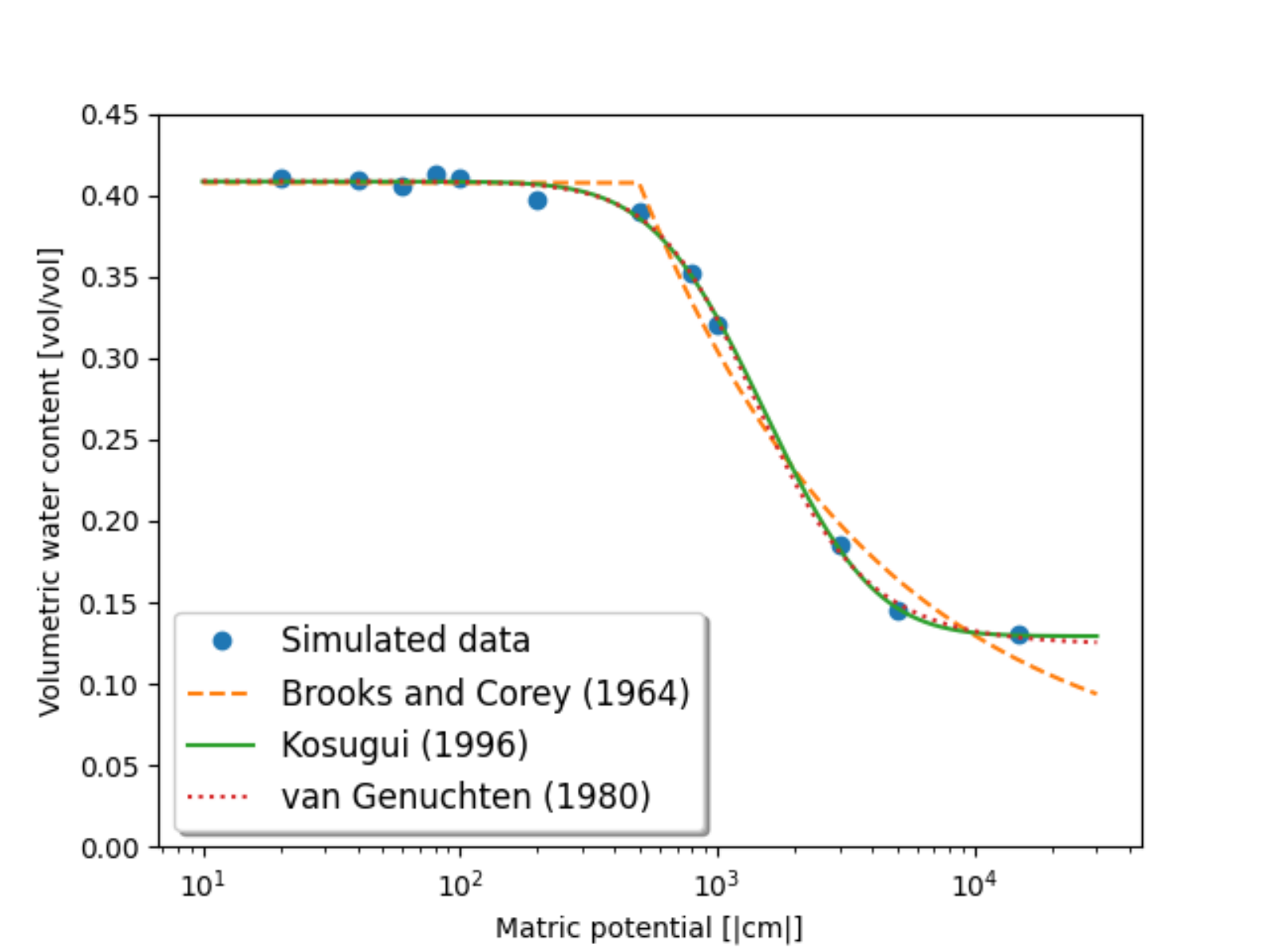}
\caption{Commonly used water retention models fitted to simulated oxisol data.}
\label{ch7_fig7}
\end{figure}

Kosugui \cite{kosugui} derived an equation based on a lognormal pore size distribution 

 \begin{equation*}
\theta_v = 
 \begin{cases}
    \theta_r + \frac{(\theta_s - \theta_r)}{2}  \text{erfc}[ \frac{\log(\psi_b-\psi)/(\psi_b - \psi_0) - \sigma^2}{\sqrt{2} \sigma}]  , & \text{for } |\psi| > \psi_b                                                    \\
    \theta_s, & \text{for } |\psi| \le \psi_b 
  \end{cases}
\end{equation*}
with two new parameters $\sigma$, the standard deviation of the transformed pore capillary pressure distribution, and $\psi_0$, the capillary pressure at the inflection point. Kosugui's model uses the complementary error function (erfc), defined as \cite{boas}

\begin{equation}
\text{erf}(x) = \frac{2}{\sqrt{\pi}}\int_0^x e^{-t^2} dt 
\end{equation}

\begin{equation}
\text{erfc}(x) = 1 - \text{erf}(x) =  \frac{2}{\sqrt{\pi}}\int_x^\infty e^{-t^2} dt 
\end{equation}
In which erf is the error function. However, the erfc is calculated numerically in almost all practical applications and is available as a function in most programming languages. Thus, the user does not need to use the integral form when fitting water retention data. The BC, Brutsaert-van Genuchten and Kosugui equation are fitted to simulated oxisol data and shown in Figure \ref{ch7_fig7} 

%%% discuss dry water range models in later editions
%%% discuss dry range water retention in later editions

%\section{Stress tensor**}

%\section{Hydrostatic pressure}

%\section{Potentials, potential density, energy density, ...}

%\section{Gradient of a potential, ...}

%\section{Hydraulic head, potential head, pressure head, ...}

\section{List of symbols for this chapter}

\begin{longtable}{ll}
    	$ W $ & Thermodynamical work \\
    	$ \gamma $ & Surface tension \\
    	$ A, A', a $ & Area \\
    	$ l $ & length \\
    	$ dx $ & Infinitesimal increase in $x$  \\
    	$ \mathcal{F} $ & Helmholtz (free) energy \\
    	$ S $ & Entropy \\
    	$ p, P $ & Pressure, except when specified otherwise \\
    	$ \mu $ & Chemical potential of a substance \\
    	$ N_{i} $ & Number of moles of a substance \\
   	$ T $ & Temperature \\
   	$ \mathbf{n} $ & Unit normal vector \\
   	$ R, r $ & Radius \\
   	$ l_1, l_2, l'_1, l'_2 $ & Elements of arc \\
	$ \zeta $ & Unspecified displacement \\
	$ \theta, \phi $ & Angles \\
	$ proj_F $ & Component of a force  \\
   	$ F $ & Force  \\
   	$ \rho $ & Generalized density  \\
   	$ \rho, \rho_w $ & Water density  \\
   	$ g $ & Earth's gravitational acceleration  \\
   	$ U $ & Potential of a force \\
   	$ m $ & Mass \\
   	$ h $ & Height \\
   	$ G $ & Gibbs (free) energy \\
   	$ \theta_g $ & Gravimetric water content \\
   	$ \theta_v $ & Volumetric water content \\
   	$ V_w $ & Partial molar volume of water \\
   	$ S_w $ & Partial molar entropy of water \\
   	$ \psi $ & Soil potential \\
   	$ \psi_w $ & Soil water potential \\
   	$ \psi_p $ & Soil pressure potential \\
   	$ \psi_m $ & Soil matric potential \\
   	$ \psi_o $ & Soil osmotic potential \\
   	$ \psi_c, P_c $ & Soil capillary potential \\
   	$ \psi_g $ & Soil water gravitational potential \\
   	$ \psi_t $ & Total soil potential  \\
   	$ z $ & Elevation  \\
   	$ \tau $ & Soil water tension  \\
   	$ p, p0 $ & Actual air vapor pressure and air saturation vapor pressure on Kelvin's equation  \\
   	$ R_g $ & Universal gas constant on Kelvin's equation  \\
   	$ a w $ & Water activity on Kelvin's equation  \\
   	$ C_R $ & Count rate  \\
   	$ a, b, a_0, a_1, _2, A_3, b_0, b_1, m, n $ & Empirical parameters  \\
   	$ \epsilon_r  $ & Relative permittivity  \\
   	$ \epsilon  $ & Absolute permittivity of a material  \\
   	$ \epsilon_0  $ & Permittivity of free space  \\
	$ \theta_f $ & Fractional water content \\
	$ V_{water} $ & Volume of water \\
	$ V_t $ & Total volume \\
	$ \theta_r $ & Residual volumetric water content \\
	$ \theta_s $ & Saturation volumetric water content \\
	$ \psi_b $ & Air entry value \\
	$ \lambda $ & pore size distribution parameter \\
	$ S_e $ & Effective saturation \\
	$ \alpha $ & Inverse air entry value \\
	$ \psi_0 $ & Capillary pressure at the inflection point of Kosugi's equation \\
	$ \sigma $ & Standard deviation of transformed capillary pressure distribution in Kosugi's equation \\
	$ \text{erf} $ & Error function \\
	$ t $ & Generalized variable in the definition of erf \\
	$ \text{erfc} $ & Complementary error function \\
\end{longtable}

% !TEX TS-program = pdflatex
% !TEX encoding = UTF-8 Unicode

% Example of the Memoir class, an alternative to the default LaTeX classes such as article and book, with many added features built into the class itself.

\chapter{Saturated porous media hydrodynamics}
\label{ch8}

\section{A fundamental problem in transport in porous media}
As you will see in the following sections, although there are analytical solutions for the Navier-Stokes equations to simple cases of laminar flow in open conduits, for flow in porous media, the statement given by Morris Muskat (\cite{muskat46} Muskat, 1946, p.55) summarizes the problem

\begin{quotation}
\noindent ``Unfortunately, however, in spite of the justifiable simplification of neglecting the inertial forces - due to the small velocities generally characteristic of flow through porous media - the mathematical difficulties of applying these equations to porous media are for practical purposes entirely unsurmountable."
\end{quotation}

Although theoretical derivations of Darcy's law from the Navier-Stokes equations have been proposed over the last decades, a solution for random porous media will invariable depend on some stochastic mathematical method, such as representative volume averaging or homogenization theory, a direct analytical derivation for a random and irregular porous medium remains a problem of difficulty on par with that of modelling turbulence. In simpler words, the random nature of the behaviour of fluid flows cannot be directly solved analytically in all but the simplest cases.

We will be mentioning Muskat's work on several occasion throughout this chapter. He was a trained theoretical physicist who earned his Ph.D. at California Institute of Technology in 1929 with the dissertation ``The continuous spectra of hydrogen like atoms'' \cite{muskat29}. After graduating, he joined the petroleum industry were he applied his training to problems of multi-phase flow in porous media, making important contributions to the mathematics of flow in porous media in the process, especially by modifying and generalizing the original form of Darcy's equation. His original contributions remain an important topic not only in engineering and applied mathematics, but also in pure mathematics (e.g. \cite{alazard}). His point is that there are no analytical solutions of the Navier-Stokes equation for true random porous media. On top of that, there is no guarantee that a solution exists for many fluid mechanics problems. This remains an open problem in mathematics in the form of the \emph{Navier-Stokes existence and smoothness problem} one of the so called \emph{millennium problems} for which you can win a large sum of money you manage to solve it. 

Getting back to Darcy's law, it is an empirical law, initially based on experimental data. There has been many \emph{a posteriori} attempts to derive Darcy's law from first principles, including averaging the Navier-Stokes equations, but the fact remain that most of these make several assumptions along the way and are for the most part largely heuristic, there are no first principles derivation from a direct analytical solution of the Navier-Stokes equations and the hydraulic conductivity remains an empirical constant. Before we get too much ahead of ourselves, as you haven't been formally introduced to Darcy's law, let's look at a few cases of flow through free conduits where there is nothing but the fluid going through the conduit. In porous media, as we have discussed, there is a solid porous matrix inside a volume through which the fluid is moving. Under these conditions, the fluid does not occupy the entire inner volume and is forced to flow through an often very complex geometry of pores. This is one of the main factors which makes flow through porous media so mathematically complex.     

\section{Navier-Stokes solutions for flow in open conduits}

In this section, two analytical solutions of the Navier-Stokes will be discussed. The first is that for flow through two infinite and flat surfaces parallel to each-other and the second is for flow through a cylindrical conduit. The second is often seen in soil physics and transport of fluid through porous media books as an introduction to Darcy's law, given that the pore network of a porous material can be generalized to a network of tubes, from which an approximation to Darcy's law can be obtained \footnote{Which as an approximation, does not adequatelly describes real porous media}.  \\ 

\noindent
\textbf{Case I. Flow between two infinite plates} \\
Let us start be recalling the Navier-Stokes equations in vector form
\begin{equation*} 
\frac{\partial  \mathbf{u}}{ \partial t} + (\mathbf{u} \cdot \nabla) \mathbf{u}   = \mathbf{F} -\frac{1}{\rho} \nabla p + \frac{\mu}{\rho}  \nabla^{2} \mathbf{u} 
\end{equation*}    

For steady state flow, the velocity field does not vary with time, so $\mathbf{u}$ is constant with time everywhere such that  $\frac{\partial  \mathbf{u}}{ \partial t} = 0$ and
\begin{equation*} 
(\mathbf{u} \cdot \nabla) \mathbf{u}   = \mathbf{F} -\frac{1}{\rho} \nabla p + \frac{\mu}{\rho}  \nabla^{2} \mathbf{u} 
\end{equation*}    
There are many conditions in which body forces can be neglected, such as in purely horizontal flow of a fluid in which the horizontal dimension is much larger than the vertical direction in which a gravitational field acts or if the contact forces are much larger than the body forces. Ignoring the body forces results in

\begin{equation*} 
(\mathbf{u} \cdot \nabla) \mathbf{u}   = -\frac{1}{\rho} \nabla p + \frac{\mu}{\rho}  \nabla^{2} \mathbf{u} 
\end{equation*}    

Now we have to take a look at each term, the term $ (\mathbf{u} \cdot \nabla) \mathbf{u} $ was discussed earlier in the context of the material derivative, based on that discussion, it is possible to show that it represents the set of equations
\begin{equation*} 
u_x \frac{\partial u_x}{\partial x} + u_y \frac{\partial u_x}{\partial y} + u_z \frac{\partial u_x}{\partial z} 
\end{equation*}    
\begin{equation*} 
u_x \frac{\partial u_y}{\partial x} + u_y \frac{\partial u_y}{\partial y} + u_z \frac{\partial u_y}{\partial z} 
\end{equation*}    
\begin{equation*} 
u_x \frac{\partial u_z}{\partial x} + u_y \frac{\partial u_z}{\partial y} + u_z \frac{\partial u_z}{\partial z} 
\end{equation*}    
and it is called the inertial or convective term, due to change of velocity as the fluid element is transported to different positions in the velocity field. The first term on the right side is the pressure gradient and represent the contact forces which cause movement of the fluid if the gradient is not equal to zero. Recalling the gradient operator, $ \nabla p$ can be represented as   
\begin{equation*}
\nabla p = \frac{\partial p}{\partial x} \mathbf{i} + \frac{\partial p}{\partial y} \mathbf{j} + \frac{\partial p}{\partial x} \mathbf{k} 
\end{equation*}
Do not worry about the density $\rho$, it is assumed as constant and thus can be manipulated to either side of the equation. To have an intuitive understanding of the pressure gradient, think about a garden hose connected to a reservoir a few meters above ground, it is the pressure gradient between both ends of the hose that causes the water to flow out of the end where the overall pressure is lower (assuming the hose is perpendicular in relation to the gravitational field), but in this case the gradient is unidimensional, say $\partial p/\partial x$. The second term on the right is the viscosity or diffusion term, recalling the Laplacian operator, it can be written as   
\begin{equation*}
\mu \nabla^2 \mathbf{u} = \mu [\nabla \cdot \nabla (u_x \mathbf{i} + u_y \mathbf{j} + u_z \mathbf{k})] 
\end{equation*}
which in Cartesian coordinates resolves in three components on the directions $\mathbf{i}$, $\mathbf{j}$ and $\mathbf{k}$ 
\begin{equation*} 
\mu (\frac{\partial^2 u_x}{\partial x^2} + \frac{\partial^2 u_x}{\partial y^2} + \frac{\partial^2 u_x}{\partial z^2} )
\end{equation*}    
\begin{equation*} 
\mu (\frac{\partial^2 u_y}{\partial x^2} + \frac{\partial^2 u_y}{\partial y^2} + \frac{\partial^2 u_y}{\partial z^2} ) 
\end{equation*}    
\begin{equation*} 
\mu (\frac{\partial^2 u_z}{\partial x^2} + \frac{\partial^2 u_z}{\partial y^2} + \frac{\partial^2 u_z}{\partial z^2} )
\end{equation*}
or, simply
\begin{equation*}
\mu \nabla^2 \mathbf{u} = \mu (\nabla^2  u_x  \mathbf{i} + \nabla^2  u_y  \mathbf{j} + \nabla^2  u_x  \mathbf{k}) 
\end{equation*}

For a fluid moving slowly, the inertial forces are much smaller than the viscous forces, such that $ \mu \nabla^2 \mathbf{u} \gg \mathbf{u} \cdot \nabla \mathbf{u} $ and the latter can be ignored resulting in a much simpler form of the Navier-Stokes equations, the Stokes equations  
\begin{equation} 
\nabla p  =  \mu  \nabla^{2} \mathbf{u} 
\end{equation}    
complemented with the incompressibility condition, as before
\begin{equation*} 
\nabla  \cdot  \mathbf{u} = 0
\end{equation*}    
The Stokes equations can be written in Cartesian coordinates as
\begin{equation} 
\frac{\partial p}{\partial x} \mathbf{i} + \frac{\partial p}{\partial y} \mathbf{j} + \frac{\partial p}{\partial z} \mathbf{k}  =  \mu [ \nabla \cdot \nabla (u_x \mathbf{i} + u_y \mathbf{j} + u_z \mathbf{k}) ] 
\end{equation}    
Many analytical solutions of the Navier-Stokes equations can be derived from this simplification, for a broad range of experimental conditions. The important point here is that at low flow velocities, when the viscous forces are much greater than the inertial forces the flow is often laminar (this is true for many fluids found in daily life, the most important being water). If the flow velocity is too high and/or viscosity too low a phenomenon known as turbulence might occur and it might become impossible to solve the differential equations analytically.  

We can finally state our problem as the ``laminar flow of water between two smooth parallel plates in which the pressure gradient is in the $x$ direction, the plates are infinitely long in the $y$ direction and the distance $d$ between the plates, through which the water flows, is oriented in the $z$ direction''. The infinite plates simplification is a common occurrence in electromagnetism and other areas of physics and engineering as it becomes much easier to solve the differential equations that arise from the geometry of the problem. The simple fact of assuming flow through a square tube results in a partial differential equation with a fairly more complicated \emph{series} solution.   
First if the pressure gradient is in the $x$ direction, there are no pressure gradients on the $y$ or $z$ directions and the equation further simplifies to
\begin{equation} 
\frac{\partial p}{\partial x} \mathbf{i}  =  \mu [ \nabla^2 (u_x \mathbf{i} + u_y \mathbf{j} + u_z \mathbf{k}) ] 
\end{equation}
Second, if the flow is in the $x$ direction, and is steady-state, the velocity does not vary on that direction, and if the plates are infinite in $y$, as stated it also does not vary in $y$ such that    
\begin{equation*} 
\frac{\partial p}{\partial x} \mathbf{i}  =  \mu \frac{\partial^2 u_x}{\partial z^2} \mathbf{i}  
\end{equation*}
Showing that $ \mu \nabla^2 \mathbf{u} = \mu \frac{\partial^2 u_z}{\partial z^2} \mathbf{i} $ when the flow velocity does not vary in $x$ or $y$ is left as an exercise to the reader. A simple solution can be found if the pressure gradient in the $x$ direction is constant. Omitting the unit vectors the equation to be solved is  
\begin{equation} 
\frac{\partial p}{\partial x}  =  \mu \frac{\partial^2 u_x}{\partial z^2}   
\end{equation}
Whose solution can be found by separating the variables and integrating
\begin{equation*} 
\int \partial^2 u_z = \int \frac{1}{\mu} \frac{\partial p}{\partial x}  \partial z^2   
\end{equation*}
\begin{equation*} 
\int \partial u_z = \int (\frac{1}{\mu} \frac{\partial p}{\partial x}  z + C_1) \partial z   
\end{equation*}
\begin{equation} 
 u_z = \frac{1}{\mu} \frac{\partial p}{\partial x} \frac{z^2}{2} + C_1 z + C_2   
\end{equation}
As always, we need the boundary conditions to solve the equation, in this case, because of the viscosity condition, the velocity is zero at the surface of the plates  
\begin{equation} 
 u_z = 0 ~\text{at} ~z = 0, ~u_z = 0 ~\text{at} ~z = h{'}   
\end{equation}
The first condition gives $C_2 = 0$, while applying the second  
\begin{align*} 
 0 = \frac{1}{\mu} \frac{\partial p}{\partial x} \frac{h{'}^2}{2} + C_1 h{'} + 0 \\
C_1 h = -\frac{1}{\mu} \frac{\partial p}{\partial x} \frac{h{'}^2}{2}
\end{align*}
\begin{equation} 
C_1 = -\frac{1}{\mu} \frac{\partial p}{\partial x} \frac{h{'}}{2}
\end{equation}
Replacing into Equation 8.5 results
\begin{equation*} 
 u_z = \frac{1}{\mu} \frac{\partial p}{\partial x} \frac{z^2}{2} -\frac{1}{\mu} \frac{\partial p}{\partial x} \frac{h{'}}{2} z    
\end{equation*}
\begin{equation} 
\boxed{ 
u_z = \frac{1}{2\mu} \frac{\partial p}{\partial x} z(z - h{'})    
}
\end{equation}
The discharge (or volumetric flux) passing through a section can be calculated by integrating the flow velocity over the area (or length in this case)
\begin{align*} 
Q = \int_0^{h{'}} u dz \\
Q = \frac{1}{2\mu} \frac{\partial p}{\partial x} \int_0^{h{'}}  z(z - h{'}) dz \\
Q = \frac{1}{2\mu} \frac{\partial p}{\partial x} (\int_0^{h{'}}  z^2 dz - \int_0^{h{'}}  zh{'} dz) \\
Q = \frac{1}{2\mu} \frac{\partial p}{\partial x} (  \frac{z^3}{3}\bigg|_0^{h{'}} -   h' \frac{z^2}{2}\bigg|_0^{h{'}}) \\
Q = \frac{1}{2\mu} \frac{\partial p}{\partial x} (  \frac{h{'}^3}{3} -   \frac{h{'}^3}{2}) \\
\end{align*}
\begin{equation} 
\boxed{
Q = - \frac{1}{12\mu} \frac{\partial p}{\partial x} h{'}^3 \\
}\end{equation}
Dimensional analysis shows that in this case the formula results in an ``areal discharge'' in [L$^2$ T$^{-1}$]. Note that this has to be the case because of the infinite plates assumption. 
\\
\\
\noindent
\textbf{Case II. Flow in a cylindrical conduit - Poiseuille flow}  \\
We can start with Equation 8.5, the preliminary analysis is the same as discussed in the previous section. The pressure gradient driving the flow is still in the $x$ direction but now the circular geometry in the cylider cross section requires both $y$ and $z$ components
\begin{equation} 
\frac{\partial p}{\partial x} \mathbf{i}  =  \mu (\frac{\partial^2 u_y}{\partial y^2}\mathbf{j} + \frac{\partial^2 u_z}{\partial z^2} \mathbf{k})  
\end{equation}
This equation can be much more easily solved in cylindrical coordinates in which the Laplacian operator can be written as 
\begin{equation} 
\nabla^2 = \frac{1}{r} \frac{\partial }{\partial  r}( r \frac{\partial }{\partial  r}) + \frac{1}{r^2} \frac{\partial^2 }{\partial \theta^2} + \frac{\partial^2 }{\partial z^2}  
\end{equation}
However, the flow velocity components do not vary in $z$ nor with the polar angle $\theta$ along the cylindrical conduit, varying only with the radius of the tube $r$, such that   
\begin{equation} 
\nabla^2 = \frac{1}{r} \frac{\partial }{\partial  r}( r \frac{\partial }{\partial  r}) 
\end{equation}
In this case, Equation 8.10 can be written in cylindrical coordinates as 
\begin{equation} 
\frac{\partial p}{\partial x}   =  \mu \frac{1}{r} \frac{\partial }{\partial  r}( r \frac{\partial u_r}{\partial  r})
\end{equation}
Now integration can be performed in a single variable $r$ instead of on $y$ and $z$ which simplifies the process.
\begin{align*} 
\int \partial ( r \frac{\partial u_r}{\partial  r}) = \frac{1}{\mu}\frac{\partial p}{\partial x} \int r  \partial r\\
 r \frac{\partial u_r}{\partial  r} = \frac{1}{\mu}\frac{\partial p}{\partial x} \frac{r^2}{2} + C_1\\
\int \partial u_r = \int (\frac{1}{\mu}\frac{\partial p}{\partial x}  \frac{r}{2} + \frac{C_1}{r}) \partial r \\
\end{align*}
\begin{equation}
 u_r = \frac{1}{\mu}\frac{\partial p}{\partial x} \frac{r^2}{4} + C_1 \log{r} + C_2 \\
\end{equation}
The boundary conditions in this case are that the velocity is zero at the walls of the tube, as before, because of the viscosity condition, the velocity is maximum at center of the tube, at $r = 0$. 
\begin{equation} 
	u_r = 0 ~\text{at} ~r = a, ~u_r = u_{max} ~\text{at} ~ r = 0   
\end{equation}
Another important condition is that the velocity is finite at the center of the tube. This implies that $C_1$ has to be zero, otherwise $C_1 \log{0}$ results in a singularity. Applying the first boundary condition and with $C_1 = 0 $   
\begin{align*}
 0 = \frac{1}{\mu}\frac{\partial p}{\partial x} \frac{a^2}{4} + 0 + C_2 \\
\end{align*}
\begin{equation}
C_2 = - \frac{1}{\mu}\frac{\partial p}{\partial x} \frac{a^2}{4}
\end{equation}
Replacing into Equation 8.14
\begin{equation}
 u_r = \frac{1}{4\mu}\frac{\partial p}{\partial x} (r^2 - a^2)
\end{equation}
The discharge (volumetric flux) can now be calculated by integrating the velocity function in $y$ and $z$
\begin{equation}
Q = \int_0^y \int_0^z u dydz 
\end{equation}
But again, because of the circular geometry of the tube's cross section, this is much more easily done by converting to polar coordinates, taking into account that $dydz$ is an area element $dA$, which in polar coordinates is $ r dr d\phi$. There should be no mystery here, if you take a circle of radius r and rotate this radius around 360° you will draw an area. Rewriting the volumetric flux equation and solving
\begin{align*}
Q =  \frac{1}{4\mu}\frac{\partial p}{\partial x} \int_0^{2\pi} \int_0^a  (r^2 - a^2) r dr d\phi \\
Q =  \frac{1}{4\mu}\frac{\partial p}{\partial x} \int_0^{2\pi} \int_0^a  (r^3 - r a^2) dr d\phi \\
Q =  \frac{1}{4\mu}\frac{\partial p}{\partial x} \int_0^{2\pi}  (\frac{r^4}{4} - \frac{r^2}{2} a^2)\bigg|_0^a d\phi \\
Q =  \frac{1}{4\mu}\frac{\partial p}{\partial x} \int_0^{2\pi}   (-\frac{a^4}{4}) d\phi \\
Q =  \frac{1}{4\mu}\frac{\partial p}{\partial x} \int_0^{2\pi} (-\frac{a^4}{4})   d\phi \\
Q =   \frac{1}{4\mu}\frac{\partial p}{\partial x} (-\frac{a^4}{4}) \phi\bigg|_0^{2\pi} \\
Q =   \frac{1}{4\mu}\frac{\partial p}{\partial x} (-\frac{a^4}{4}) 2\pi \\
\end{align*}  

\begin{equation}
\boxed{
Q =   -\frac{\pi}{8\mu}\frac{\partial p}{\partial x} a^4 \\
}
\end{equation}  
 in which $Q$ is the discharge in [L$^3$ T$^{-1}$] as expected.
 
\section{Darcy's equation in its original form}

\begin{figure}[ht]
\centering
 \includegraphics[width=0.6\textwidth]{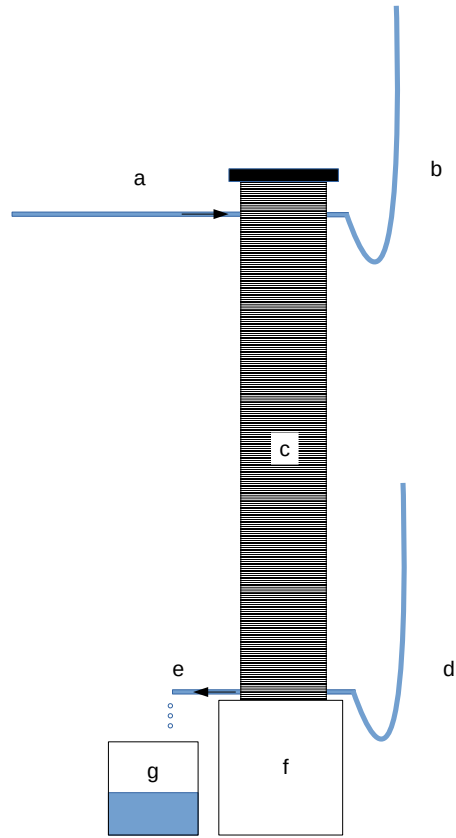}
\caption{Simplified schematics of sand column used in the original Darcy experiment (adapted from \cite{darcy}). (a) is the water inlet, (b) and (d) are manometers, (c) is space to be filled with sand, note that the height of sand within the column varied from experiment to experiment\cite{darcy,brown}, being below the inlet, (e) is the water outlet, (f) is the support, and (g) is a reservoir for water collecting and measuring outflow volume.}
\label{ch8_fig1}
\end{figure}

Henry Darcy was a french engineer who, while working on water distribution system for the city of Dijon in the 1800s, wrote a monumental report concerning various aspects of water supply and distribution \cite{darcy}. In one of the appendixes of the document he presents an equation that describes the flow of water through a column filled with sand under a hydraulic gradient (Figure \ref{ch8_fig1})

\begin{equation}
q = k \frac{s}{e} [h + e \mp h_0]
\end{equation}  
In which $q$ was the water flux, $e$ is the height of the sand column, $s$ is the cross sectional area of the column, $h$ is the water height in the upper part of the column (on the inlet), $h_0$ is the water height on the inferior part of the column (on the outlet).   

If the water height is zero at the outlet, $h_0 = 0$ 
\begin{equation}
q = k \frac{s}{e} [h + e]
\end{equation}  
He realized that the water flux through the sand column was proportional to the pressure difference between the inlet and the outlet and he defined the constant of proportionality as a coefficient k, dependent on the permeability of the sand (\emph{``dependant de la permeabilité de la couche''}). So $k$ in this context was defined as a permeability, dependent on the physical characteristics of the solid phase. In contemporary terms, the height of water at the inlet and outlet is effectively the pressure potential in terms of height of water. Considering the potentials at the inlet and outlet, and if we establish the gravitational reference at the bottom of the column, we have at the inlet (top of the column): \\
\begin{align*}
\psi_{gi} =  e \\
\psi_{pi} =  h + P_0\\
\psi_{ti} =  e +  h + P_0
\end{align*}
At the outlet (bottom of the column):\\
\begin{align*}
\psi_{go} =  0 \\
\psi_{po} =  h_0 + P_0\\
\psi_{to} =  h_0 + P_0
\end{align*}
in which $P_0$ is the atmospheric pressure. Therefore, Darcy's law can be expressed as 
\begin{align*}
q = k A\frac{\psi_{ti} - \psi_{to}}{L} = k s \frac{(e +  h + P_0) - (h_0 + P_0)}{e} \\
= k s\frac{(e +  h + P_0) - (h_0 + P_0)}{e} = k s\frac{(e +  h)}{e}   
\end{align*}  
for $L = e$, the height of the column, $A = s$ the cross sectional area of the column and $h_0 = 0$. The unidimensional form of Darcy's law, as applied to a column filled with saturated porous media is represented in modern notation in introductory soil physics and groundwater hydrology books as 
\begin{equation}
\boxed{
Q = K_s A\frac{\psi_{ti} - \psi_{to}}{L} = Q = K_s A\frac{\Delta \psi_{t}}{L}    
}
\end{equation}
for total potential expressed in terms of height of water.  A more precise general representation is in terms of differentials, thus for a one dimensional column oriented in some arbitrary $x$ direction, Darcy's law is   
\begin{equation}
\boxed{
Q =  -K_s A\frac{d \psi_{t}}{dx}    
}
\end{equation}  
in which $Q$ is the water discharge in L$^3$~T$^{-1}$,   $K_s$ is now defined as the saturated hydraulic conductivity in L~T$^{-1}$ and the term $d \psi_{t}/dx$ is called the \emph{hydraulic gradient} or the potential difference over the length between the two points in which potential is being measured, in L~L$^{-1}$. The negative sign, source of much confusion in soil physics, means that the flow is opposite to the direction of the gradient. In other words, the flow is always from the high total potential to the low total potential. As with the original Darcy formulation, these equations can be expressed in terms of height of water on the inlet and outlet of a saturated column filled with some porous material. If the column is oriented on the $z$ direction, here defined as in the direction of the gravitational field, we have   
\begin{align*}
Q =  -K_s A \frac{d \psi_{t}}{dz} = -K_s A \frac{ (\psi_{pi} + \psi_{gi}) - (\psi_{po} + \psi_{go})    }{dz}     
\end{align*}  
If the length of the column is $z$ and establishing the gravitational potential at the bottom of the column at $z = 0$, we have  
\begin{align*}
Q =   -K_s A \frac{ (\psi_{pi} + z) - (\psi_{po} + 0) }{z}     
\end{align*}  
Representing the pressure potential in terms of height $h_i$
\begin{equation}
Q =   -K_s A \frac{ (h_i-h_o) + z}{z} = -K_s A (\frac{ h_i-h_o}{z} + 1)       
\end{equation} 
If the pressure potential is zero on the outlet
\begin{equation}
Q =   -K_s A (\frac{ h_i}{z} + 1)       
\end{equation} 
It is easy to show that if the column is oriented perpendicular to the gravitational field in some arbitrary $x$ (or $y$) direction, Darcy's law in terms of the height of water on the inlet and  outlet will be 
\begin{equation}
Q =     -K_s A (\frac{ h_i - h_o}{x})       
\end{equation}  
For $h_o$ = 0 this reduces to
\begin{equation}
Q =     -K_s A \frac{ h_i}{x}       
\end{equation}  
where $x$ (or $y$) represents the horizontal length of the column and $z$ the vertical length.  It is clear that Darcy's law is a linear relationship between discharge and gradient with slope equal to the saturated hydraulic conductivity. Darcy's law, along with Fourier's law of heat  conduction, Ohm's law in electromagnetism, Fick's law of molecular diffusion and Poiseuille's law is one of the linear gradient transport laws discovered in the 1800s \cite{simmons}. Although Darcy only mentions Poiseuille's law in his report, it is very likely that he was aware and was influenced by the physics of the time in creating his empirical law \cite{simmons}. Darcy's law can be written in terms of a flux density $q$, similar to the other laws of physics mentioned, by dividing the discharge $Q$ by  the area $A$    
\begin{equation}
q = \frac{Q}{A} =  -K_s A \frac{d \psi_t}{dx}       
\end{equation}  
Darcy obviously knew about and investigated Poiseuille's, as he was an engineer concerned with water distribution, which would obviously require studying water flow in pipes. Analyzing Poiseuille's solution for flow in a cylindrical conduit, 
\begin{align*}
Q =   -\frac{\pi}{8\mu}\frac{\partial p}{\partial x} a^4 
\end{align*}  
we can rewrite this equation by reorganizing it
\begin{align*}
Q =   -\frac{\pi a^2 a^2}{8\mu}\frac{\partial p}{\partial x}  
\end{align*}  
and introducing $A = \pi a^2$ as the area of the tube
\begin{align*}
Q =   - A\frac{\pi a^2}{8\mu}\frac{\partial p}{\partial x}  
\end{align*}  
Rewriting the term $ \pi a^2/(8\mu)$ as some proportionality constant $k^*$
\begin{align*}
Q =   - A k^*\frac{\partial p}{\partial x}  
\end{align*}  
Poiseulle's law can be interpreted as some limiting case of Darcy's law for when there is no porous media within the pipe. The presence of a porous medium causes additional stresses in the fluid that decrease the discharge under the same potential gradient. Because natural porous media usually have complex grain size distribution, grain shape and rugosity, resulting in complex pore size distribution and geometry, the empirical constant in Darcy's law, $K_s$, has to be determined experimentally, there is no analytical solution of Darcy's law from the Navier-Stokes equation. There are a few studies that use stochastic considerations to derive Darcy's law from the Navier-Stokes equations, but as discussed previously, they must rely in some averaging mechanism, unless the porous media model is extremelly simplistic.    

\begin{figure}[ht]
\centering
 \includegraphics[width=1.0\textwidth]{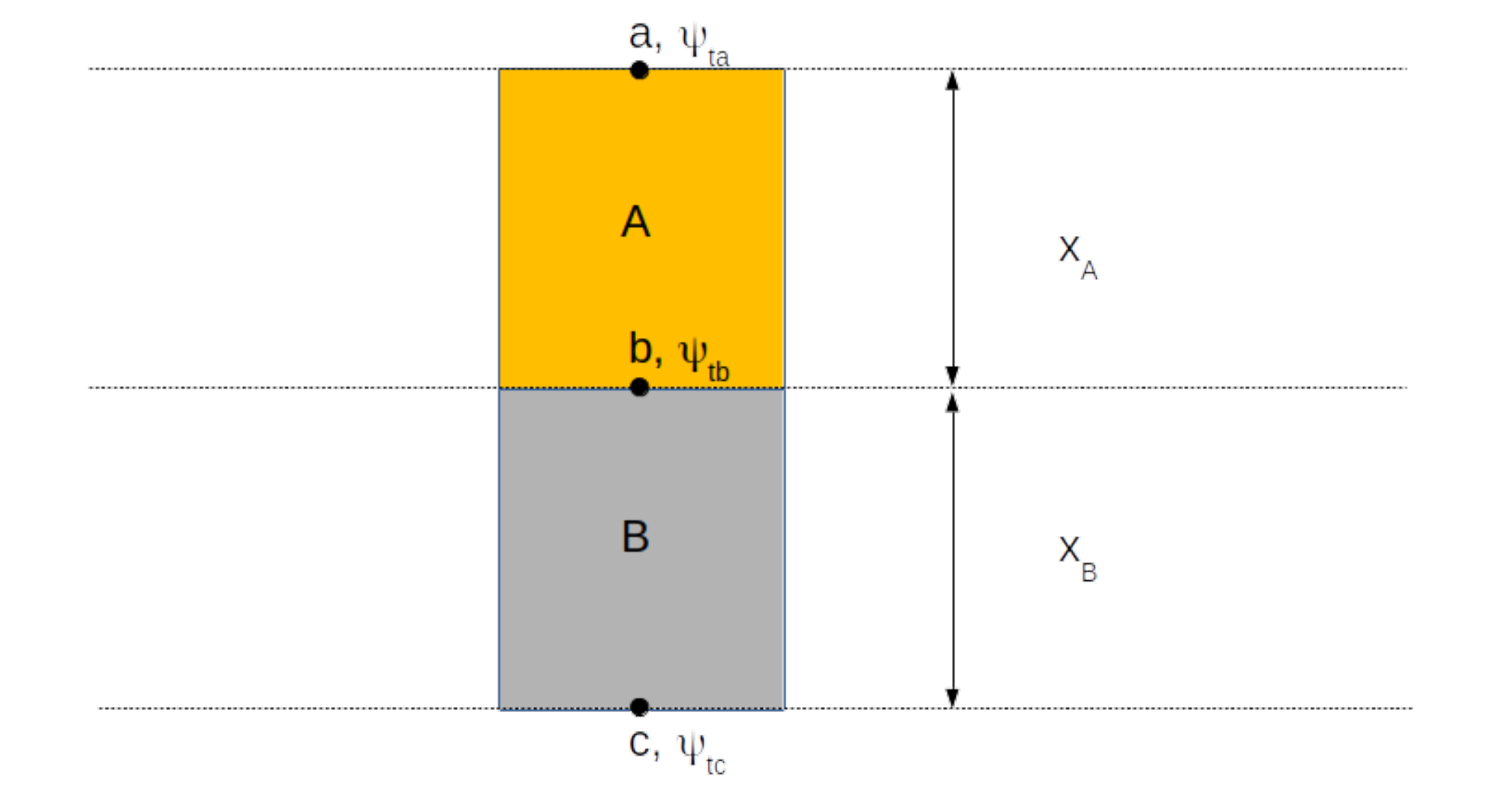}
\caption{Darcy's law in a composite column with individually homogeneous but different materials $A$ and $B$.}
\label{ch8_fig2}
\end{figure}

In stratified media, such as a soil profile, sedimentary rock or deposit, or laboratory column, where the direction of flux is perpendicular to the layering, the flow will be determined by some form of averaging of the individual hydraulic conductivity of the layers\footnote{Here as a direct average, in a much more simplistic way than what was discussed previously for averaging of the Navier-Stokes equation at the pore scale}.  Consider a composite column, composed of two independently homogeneous but different materials $A$ and $B$ with heights $x_A$ and $x_B$ (Figure \ref{ch8_fig2}). The flux is in the $z$ direction and the total potential is measured on the points $a$, $b$ and $c$ as $\psi_{ta}$, $\psi_{tb}$ and $\psi_{tc}$. Darcy's law is defined for material $A$ in terms of the flux between points $a$ and $b$

\begin{align*}
q_{ab} =   -K_{sA} \frac{\psi_{tA} - \psi_{tB}}{x_A}       
\end{align*}  
and for material $B$ in terms of the flux between points $b$ and $c$
\begin{align*}
q_{bc} =   -K_{sB} \frac{\psi_{tB} - \psi_{tC}}{x_B}       
\end{align*}  

As point $b$ is located in the interface between the two materials, the total potential is equal and we can solve both equations for $\psi_{tB}$.
\begin{align*}
\psi_{tB} = \psi_{tA} + \frac{q_{ab}x_A}{ K_{sA}}        
\end{align*}  
\begin{align*}
\psi_{tB} = \psi_{tC} - \frac{q_{bc}x_B}{ K_{sB}}        
\end{align*}  
Replacing $\psi_{tB}$ in any of the equations and solving for the total potential difference between the column's inlet in $a$ and the outlet in $c$ 
\begin{align*}
\psi_{tA} - \psi_{tC} = - (\frac{q_{bc}x_B}{ K_{sB}}   + \frac{q_{ab}x_A}{ K_{sA}})
\end{align*}
For constant (steady-state) saturated flow, for which Darcy's law is being considered, the flux in any point within the length of the composite column is constant, and thus $q_{ab} = q_{bc} = q_{ac} = q$ , and 
\begin{align*}
\psi_{tA} - \psi_{tC} = - q(\frac{x_B}{ K_{sB}}   + \frac{x_A}{ K_{sA}})
\end{align*}  
Finally solving for $q$ we arrive at an expression for a composite material 
\begin{equation}
q = -\frac{ (\psi_{tA} - \psi_{tC})}{(\frac{x_B}{ K_{sB}}   + \frac{x_A}{ K_{sA}})}  
\end{equation}  
The term  
\begin{equation}
\frac{1}{(\frac{x_B}{ K_{sB}}   + \frac{x_A}{ K_{sA}})} = \frac{K_{sA}K_{sB}}{x_BK_{sA} + x_AK_{sB}}
\end{equation}  
is a form of weighed mean of the hydraulic conductivity of the layers with the height of each layer acting as a weighing term. 

In practical terms, especially in laboratory, saturated hydraulic conductivity can be determined using constant head and falling head permeameters\footnote{As we will see, hydraulic conductivity and permeability are different things, however, these devices historically have been called permeameters. As they can also be used to measure the permeability of the medium, we will maintain the use of the terminology }. In constant head permeameters, the total potential or hydraulic head is kept constant in both ends of the column. The hydraulic head must be equal or greater than zero on both ends. If the total potential is negative the water will be under tension, the water content might be below saturation and hydraulic conductivity is no longer constant. Saturated hydraulic conductivity is calculated by measuring discharge or flux density as  
\begin{equation}
K_s = -\frac{QL}{A(h_i - h_o)} 
\end{equation}  
for a horizontal column, and
\begin{equation}
K_s = -\frac{Q}{A[(h_i - h_o)/L +1]} 
\end{equation}  
for a vertical column with the gravitational reference chosen at the outlet. Note that the flow can be in any direction including upwards, the direction being indicated by the sign of the hydraulic conductivity.

\begin{figure}[ht]
\centering
 \includegraphics[width=0.6\textwidth]{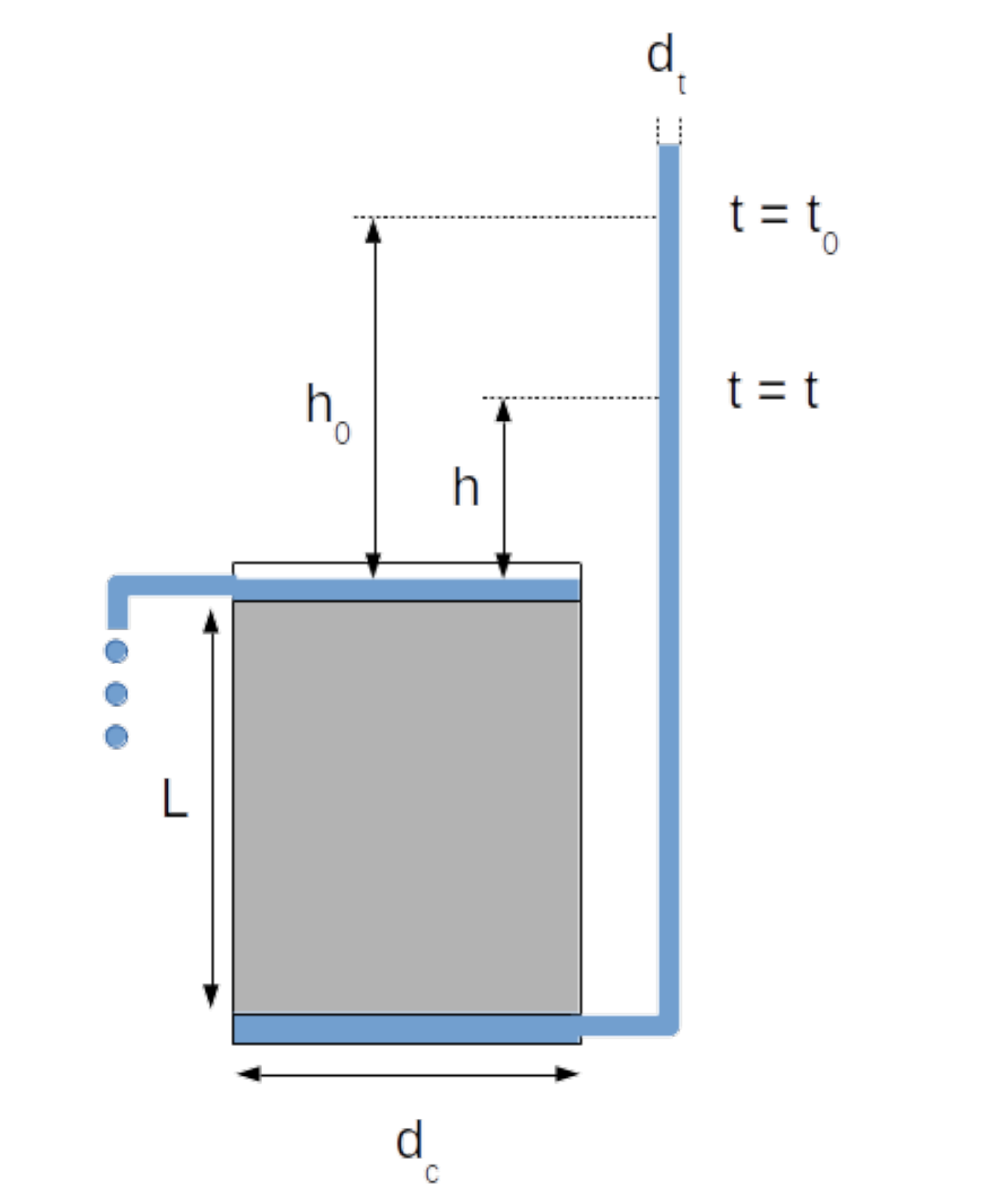}
\caption{Diagram illustrating a generic the falling head permeameter.}
\label{ch8_fig3}
\end{figure}

The second common type of laboratory method for measuring saturated hydraulic conductivity are falling head permeameters. In the falling head method the hydraulic head is not constant but falls from a height $h_0$ to a height $h$ in time. Falling head permeameters are convenient when high hydraulic heads are required \cite{bear}. To derive the expression for measuring saturated hydraulic conductivity, consider a permeameter connected to a tube filled with water on its base (Figure \ref{ch8_fig3}). The flow is upwards and the discharge is measured at the outlet on the water level on the upper part of the column. The tube has diameter $d_t$, radius $r_t$ and area $A_t$ and the column filled with porous media has diameter $d_c$, radius $r_c$ and area $A_c$. The inflow on the smaller tube is given by \cite{fetter}

\begin{align*}
Q_{in} = -A_t \frac{dh}{dt}
\end{align*}  
while on the column the flow obeys Darcy's law 
\begin{align*}
Q_{out} =  \frac{K A_c h}{L}
\end{align*}  
Under saturated flow conditions the discharge into the column has to be equal to the discharge out of the column
\begin{align*}
Q_{in} = Q_{out}
\end{align*}  

\begin{align*}
-A_t \frac{dh}{dt} = \frac{K A_c h}{L}
\end{align*}  
Reorganizing the terms and integrating
\begin{align*}
 \frac{dh}{h} = - \frac{A_c K}{A_t L} dt
\end{align*}  

\begin{align*}
\int  \frac{dh}{h} = - \int  \frac{A_c K}{A_t L} dt
\end{align*}  

\begin{align*}
\log{|h|} \bigg |_{h}^{h_{0}} = - \frac{A_c K}{A_t L} t \bigg |_{t_{0}}^t
\end{align*}  
By applying the boundary conditions $t = 0$ at $h = h_0$, and considering $h$ as always positive we arrive at 

\begin{align*}
\log{h}  - \log{h_{0}} = - \frac{A_c K}{A_t L} t
\end{align*}  
Solving for $K$ and considering the area of the small tube as $A_t = \pi r_t^2 $ and the area of the soil column as $A_t = \pi r_t^2$ \footnote{Remember that $\log x - \log y = \log{x/y}$ and notice that we multiplied both sides by $-1$  to eliminate the negative sign in $K$.} 

\begin{align*}
K = \frac{r_t^2 L}{r_c^2 t} \log \frac{h_0}{h} 
\end{align*}  
or, in terms of diameter
\begin{equation}
K  = \frac{d_t^2 L}{d_c^2 t} \log \frac{h_0}{h}
\end{equation}  
For alternative methods and equipment for measuring saturated hydraulic conductivity on the field and on the laboratory, the reader is referred to \cite{danetopp}.

\section{Generalization of Darcy's equation}

Up until now we have avoided dealing with vectors by working with flow in one dimension. A column filled with a porous medium saturated with water, where the diameter of the column is small and where the walls are impermeable can be treated as an unidimensional system and the flux density vector is oriented in the general direction of water flow. In soils and sedimentary deposits, aquifers, porous rocks and many other natural materials, flow is often not constrained to one dimension. This is especially true on the field, where water and contaminant transport occur in large scales and in response to total potential gradients, but also in laboratory, for large enough samples and when heterogeneity cannot be disregarded. Under such conditions, not only the flux density is a vector but the gradient is also a vector, and often hydraulic conductivity is a second order tensor.  Recalling Chapter \ref{ch1}, Darcy's law can be generalized as

\begin{equation}
\mathbf{q}  = \mathbf{K}   \cdot \nabla \psi_t
\end{equation}  
In three-dimensional Cartesian coordinates, the flux density is decomposed as   
\begin{equation}
\mathbf{q}  = q_x \mathbf{i} + q_y \mathbf{j} + q_x \mathbf{k} 
\end{equation}  
while the three-dimensional gradient is 
\begin{equation}
\nabla  = \frac{\partial \psi_t}{\partial x} \mathbf{i} + \frac{\partial \psi_t}{\partial y} \mathbf{j} + \frac{\partial \psi_t}{\partial z} \mathbf{k} 
\end{equation}  
Under conditions of anisotropy, where hydraulic conductivity varies with direction in the material, saturated hydraulic conductivity is a second rank tensor with $3^2 = 9$ elements   
\begin{equation}
\mathbf{K} = K_{ij} =  
\begin{pmatrix}
K_{xx} & K_{xy} & K_{xz}\\
K_{yx} & K_{yy} & K_{yz}\\
K_{zx} & K_{zy} & K_{zz}
\end{pmatrix}
\end{equation} 
With the conductivity tensor, Darcy's law can be written as 
\begin{equation}
\begin{pmatrix}
q_{x}\\
q_{y} \\
q_{z} 
\end{pmatrix}
 =  
\begin{pmatrix}
K_{xx} & K_{xy} & K_{xz}\\
K_{yx} & K_{yy} & K_{yz}\\
K_{zx} & K_{zy} & K_{zz}
\end{pmatrix}
\cdot
\begin{pmatrix}
\partial \psi_t/\partial x\\ 
\partial \psi_t/\partial y \\
\partial \psi_t/\partial z 
\end{pmatrix}
\end{equation} 
Therefore, the components of flux density in $x$, $y$ and $z$ are
\begin{equation}
q_x  = K_{xx}  \frac{\partial \psi_t}{\partial x}  + K_{xy} \frac{\partial \psi_t}{\partial y} + K_{xz}\frac{\partial \psi_t}{\partial z}
\end{equation}  
\begin{equation}
q_y  = K_{yx}  \frac{\partial \psi_t}{\partial x}  + K_{yy} \frac{\partial \psi_t}{\partial y} + K_{yz}\frac{\partial \psi_t}{\partial z}
\end{equation}  
\begin{equation}
q_z  = K_{zx}  \frac{\partial \psi_t}{\partial x}  + K_{zy} \frac{\partial \psi_t}{\partial y} + K_{zz}\frac{\partial \psi_t}{\partial z}
\end{equation}  
The hydraulic conductivity tensor is proposed as a symmetrical tensor\cite{bear}, meaning that $K_{ij} = K_{ji}$ such that the tensor reduces to six different components. There are profound implications related to these equations. The fact that the conductivity tensor is symmetrical is by no means a trivial issue and is a property of most of the transport equations and related to the theory know as the Onsager reciprocal relations, in the realm of non-equilibrium thermodynamics.   

Another important property of the conductivity tensor is that it is assumed as orthotropic, meaning that there are three principal directions for the conductivity vectors which are orthogonal to eachother. If this condition is valid, it is possible to orientate the axis in such a way that the conductivity tensor is reduced to three principal directions, such that $K_{ij} = 0$ for $i \ne j$ and $K_{ii}$ for $i = j$
\begin{equation}
\mathbf{K} = K_{ij} =  
\begin{pmatrix}
K_{xx} & 0 & 0\\
0 & K_{yy} & 0\\
0 & 0 & K_{zz}
\end{pmatrix}
\end{equation} 
In practice, this means that it is possible to orientate a sample of an anisotropic media in laboratory conditions such that only the three principal permeability directions need to be measured. If the medium is isotropic, there is no difference between $K_{ii}$ measured in any direction and $K$ is reduced to the constant hydraulic conductivity scalar 
\begin{equation}
\mathbf{K} = K_{ij} = K 
\begin{pmatrix}
1 & 0 & 0\\
0 & 1 & 0\\
0 & 0 & 1
\end{pmatrix}
\end{equation} 
Suppose now that the saturated hydraulic conductivity does not vary with direction at each point in space, but the value, even being equal in all directions, is different in different points in space. In this case, the saturated hydraulic conductivity is a scalar function in space $K(x,y,z)$ and Darcy's law is  
\begin{equation}
\mathbf{q}  = -K(x,y,z)   \nabla \psi_t
\end{equation}  
A medium where the hydraulic conductivity does not vary with direction at a given point in space is said to be isotropic and if the scalar hydraulic conductivity does vary in space is called heterogeneous. If the saturated hydraulic conductivity does not vary with direction and is the same in all points in space the medium is said to be homogeneous and isotropic and Darcy's law simplifies to
\begin{equation}
\mathbf{q}  = -K    \nabla \psi_t
\end{equation}
A consistent representation of Darcy's law requires that the potentials are expressed in terms of hydraulic height [L]. This guarantees that the gradient $\partial \psi_t/\partial x$ (or in $\partial y$ or $\partial z$) is nondimentional [L L$^{-1}]$ and that the flux density has the same units of the saturated hydraulic conductivity [L T$^{-1}$]. It is convenient now to fully introduce the notation used before for the total potential in  height units as 
\begin{equation}
h_t  = h_p + h_g
\end{equation}
in which $h_t$ is the total potential, $h_p$ is the pressure potential and $h_g$ is the gravitational potential, all in [L] units. Using this notation and considering the $z$ direction as the direction of the gravitational force, the components of Darcy's law in $x$, $y$ and $z$ are 
\begin{equation}
q_x  = -K    \frac{\partial h_t}{\partial x} = -K \frac{h_{ti} - h_{to}}{x - x_0} = -K \frac{(h_{pi} + h_{gi}) - (h_{po} + h_{go})}{x - x_0}   
\end{equation}
\begin{equation}
q_y  = -K    \frac{\partial h_t}{\partial y} = -K \frac{h_{ti} - h_{to}}{y - y_0} = -K \frac{(h_{pi} + h_{gi}) - (h_{po} + h_{go})}{y - y_0}   
\end{equation}
\begin{equation}
q_z  = -K    \frac{\partial h_t}{\partial y} = -K \frac{h_{ti} - h_{to}}{z - z_0} = -K \frac{(h_{pi} + h_{gi}) - (h_{po} + h_{go})}{z - z_0}   
\end{equation}
where the subscripts $i$ and $o$ stand for inlet and outlet. Setting the gravitational reference at $x_0$, $y_0$ and $z_0$, on the outlet, the gravitational potential will be $z_0$ in both the inlet and outlet in $x$ and $y$ and in the outlet only in the direction $z$. If the column height is $z$, and the inlet is located in at the distance $z$, this will be the gravitational potential in the inlet for the vertical column. Using these conditions, the components of Darcy's law in three dimensions are
\begin{equation}
q_x  = -K \frac{(h_{pi} + z_0) - (h_{po} + z_0)}{x - x_0} = \frac{\partial h_{p}}{\partial x}    
\end{equation}
\begin{equation}
q_y  = -K \frac{(h_{pi} + z_0) - (h_{po} + z_0)}{y - y_0} = \frac{\partial h_{p}}{\partial y}    
\end{equation}
\begin{align*}
q_z  = -K \frac{(h_{pi} + z) - (h_{po} + z_0)}{z - z_0} = -K \frac{(h_{pi} - h_{po}) + (z - z_0)}{z - z_0} = -K (\frac{\partial h_{p} + \partial z}{\partial z})      
\end{align*}
\begin{equation}
q_z  = -K (\frac{\partial h_{p}}{\partial z} + 1)   
\end{equation}

Suppose now the total potential is expressed in  pressure units. Using the $z$ coordinate for convenience
\begin{align*}
q_z  = -K    \frac{\partial \psi_t}{\partial z}   
\end{align*}
it becomes clear that the the gradient is no longer dimensionless and the equation is not consistent in terms of units. Expressing the total potential in terms of the sum of the potentials
\begin{align*}
q_z  = -K    \frac{\partial (\psi_p + \psi_g)}{\partial z}   = -K \frac{\partial (\rho_w g h + \rho_w g z)}{\partial z} 
\end{align*}
and using the definitions from Chapter \ref{ch7}, for consistence of units to be achieved, the saturated hydraulic conductivity would have to be divided by $\rho_w g$  
\begin{align*}
q_z  = -\frac{K}{\rho_w g}     \frac{\partial (\rho_w g h + \rho_w g z)}{\partial z} 
\end{align*}
Therefore, for a liquid of constant density, Darcy's law in its original form is retrieved by placing the constant terms outside of the derivative
\begin{align*}
q_z  = -\frac{K \rho_w g}{\rho_w g}     \frac{\partial (h + z)}{\partial z} = -K    \frac{\partial (h + z)}{\partial z} 
\end{align*}

Although stated otherwise in his report, the saturated hydraulic conductivity as defined by Darcy in his original experiment is a composite property of the solid and liquid phases \cite{muskat46}. Whereas this is convenient in soil physics where the phases concerned are usually soil and water, it might not be convenient in groundwater hydrology and petroleum engineering, where different and often multiple phases are concerned and where the characteristics of the porous material might be of concern. Suppose you define a saturated conductivity term for water using a column experiment. You could repeat the same experiment with other fluids, for example oil, salt solution, gasoline, and for each fluid you would obtain one saturated conductivity. To circumvent this issue, and using dimensional analysis, the saturated conductivity term was  rewritten in terms of the properties of the fluid being used, namely viscosity and density and a new coefficient, called \emph{intrinsic permeability} or simply \emph{permeability} $\kappa$, which depends on the solid phase alone, thus \cite{nutting, wyckoffetal, muskat46}
\begin{equation}
K = \frac{\kappa \rho_f g}{\mu}
\end{equation}
In which $\rho_f$ and $\mu$ are the density and viscosity of a fluid moving through the porous material\footnote{These laws should be valid for viscous, incompressible newtonian fluids in general}. Thus, Darcy's law can be expressed in terms of pressure using the permeability as
\begin{align*}
q_z  = -\frac{\kappa \rho_f g}{\mu} \frac{1}{\rho_f g}     \frac{\partial (\rho_f g h + \rho_f g z)}{\partial z}  =  -\frac{\kappa}{\mu}   \frac{\partial (\rho_f g h + \rho_f g z)}{\partial z} 
\end{align*}
and in terms of length for the potential as
\begin{align*}
q_z  = -\frac{\kappa \rho_f g}{\mu}     \frac{\partial ( h +  z)}{\partial z}  
\end{align*}
For an arbitrary $x$ direction perpendicular to Earth's gravitational field these reduce to
\begin{equation}
q_x  = - \frac{\kappa}{\mu}   \frac{\partial (\rho_f g h)}{\partial z} = -  \frac{\kappa}{\mu}   \frac{\partial P}{\partial z}  
\end{equation}
and
\begin{equation}
q_x  = -\frac{\kappa \rho_f g}{\mu}     \frac{\partial h}{\partial x}  
\end{equation}
Note that the similarity between Darcy's law and Poiseuille's equation is much more evident when the pressure form is considered. 

In general, the permeability is a tensor with the same properties of the saturated hydraulic conductivity, i.e. for an anisotropic medium
\begin{equation}
\mathbf{k} = k_{ij} =  
\begin{pmatrix}
k_{xx} & k_{xy} & k_{xz}\\
k_{yx} & k_{yy} & k_{yz}\\
k_{zx} & k_{zy} & k_{zz}
\end{pmatrix}
\end{equation} 
which can be orientated in the direction of the principal axes
\begin{equation}
\mathbf{k} = k_{ij} =  
\begin{pmatrix}
k_{xx} & 0 & 0\\
0 & k_{yy} & 0\\
0 & 0 & k_{zz}
\end{pmatrix}
\end{equation} 
and for an isotropic medium
\begin{equation}
\mathbf{k} = k_{ij} = k  
\begin{pmatrix}
1 & 0 & 0\\
0 & 1 & 0\\
0 & 0 & 1
\end{pmatrix}
\end{equation}

\section{Validity of Darcy's law}

We have seen that the flux density vector in Darcy's law is the discharge over area. Because the pore geometry is complex, a variety of shapes and sizes of pores are available for flow in porous materials. The flux density in Darcy's law is not the velocity of flow of fluid in the pores. Because the flow only takes place within the pores, the average flow velocity is the flux density divided by the porosity. 
\begin{equation}
v  = \frac{Q}{An}  =   \frac{q}{n}  
\end{equation}
in which $n$ is the porosity. This relationship is valid because for an homogeneous porous material, area porosity is equivalent to the volumetric porosity \cite{bear}. In other words, considering a cross section of the column, only part of the area is available for flow and for a principle equivalent to Bernoulli's law, the flux velocity at each point will be higher for smaller pores, within the limitations of viscosity and wall effects. The difference in velocities in a porous material will be an important mechanism in contaminant transport. 

Like many other equations and laws in fluid mechanics, Darcy's law is valid for laminar flow. Laminar flow here meaning the same as we discussed when analyzing the problem of a sphere settling under gravity. The flow is slow enough that there is no turbulence or vortexes. Poiseuille's law also has a restriction of laminar flow. As we seen for the sphere problem, one way of identifying an approximate limit between laminar and turbulent flow is by using the Reynolds number. For flow in a cylindrical conduit, the Reynolds number is  
\begin{equation}
Re  =  \frac{\rho_f u d_h}{\mu} 
\end{equation}
In which $d_h$ is the hydraulic diameter of the pipe, simple the inner diameter if the tube is circular. For Darcy's law, this can be approximated by 
\begin{equation}
Re  =  \frac{\rho_f u d'}{\mu} 
\end{equation}
In which $d'$ is some representative length dimension of the porous matrix representing the elementary channels of the porous medium, usually a parameter related to average grain diameter or other representative grain diameters \cite{bear} such as the diameter with 10$\%$, 30$\%$ or or 50$\%$  of the material passing through a sieve of mesh equal to the diameter of interest.

Darcy's law implies a linear relationship between flux density and hydraulic gradient. The experimentally determined $Re$ values for the transition zone from laminar to turbulent flow for Darcy's law ranges from 1 to 10, with laminar flow for values less than 1, turbulent flow for values above 10 \cite{bear}. For values above 10 the relationship between flux density and gradient is no longer linear and Darcy's law cannot be used.  As the flow velocity increases, inertia terms are no longer negligible.  At the transition zone, the Darcy-Forchheimer equation can be used, presented here in one dimension as \cite{sobieskitrykozko}.

\begin{equation}
-\frac{\partial p}{\partial x} = \frac{\mu}{\kappa}q + \beta \rho_f q^2  
\end{equation}
In which $\beta$ is the Forchheimer coefficient or the inertial permeability. Note the quadratic term on the flux density. 

If the flow is too slow, for very small gradients and for some very fine soils and sediments, surface forces (Chapter \ref{ch3}) can become important as well as viscous effects. It has been speculated that under such conditions water can display non-newtonian behavior and non-darcyan flow is observed \cite{kirkhampowers, bear}.

%\section{Darcy-Muskat equation}

%\section{Muskat generalization for multi-phase flow}

%% Saturated hydraulic conductivity values

Saturated hydraulic conductivity is a function of total porosity and pore size distribution. As such it is a function of bulk density and grain size distribution. Compaction and consolidations will decrease the total porosity and will tend to reduce saturated hydraulic conductivity. In soil science saturated hydraulic conductivity can be used as an indicator of soil degradation. Grain size distribution controls not only pore size distribution but also influences bulk density. Coarse grained materials can have lower total porosity but the average pore diameter will be greater, resulting in larger values of saturated hydraulic conductivity. For soils rich in 2:1 layer silicates, high clay content will result in lower saturated hydraulic conductivity, because of the higher average pore radius and in some cases because of expansion of the silicate clays. For soils rich in iron and aluminum oxides, the saturated hydraulic conductivity is often very high, due to the microaggregate structure with two pore domains, micropores and mesopores, and macropores. Under saturated conditions the bulk of the water transport will take part in macropores. In sediments and rocks, grain size distribution and degree of consolidation will control saturated hydraulic conductivity. In highly consolidated igneous and metamorphic rocks, water transport can occur in fractures and hydraulic conductivity can be highly directional. Saturated hydraulic conductivity in the field is not a normally distributed variable and this should be accounted for when analyzing data. 

Bear \cite{bear} reports saturated hydraulic conductivity values ranging from $10^{2}~cm~s^{-1}$ for clean gravel to $10^{-11}~cm~s^{-1}$ for impervious rocks. The same source reports values ranging from   $10^{-3}~cm~s^{-1}$ for sandy soils to $10^{-11}~cm~s^{-1}$ for dispersive clay soils. Because of the relationship of saturated hydraulic conductivity with porous media parameters, various attempts have been made to derive equations to calculate the saturated hydraulic conductivity from physical properties of the material. These models vary from physically based but approximated models based on Poiuseuille's equation to purely empirical models based on statistical regression. 
One model commonly found in engineering and hydrogeology is the Kozeny-Carman equation\footnote{Derivations based on the original sources should be provided in later editions} 
\begin{equation}
k = \frac{n^3}{(1-n)^2} \frac{c_0}{M_s^2}  
\end{equation}
in which $\kappa$ is the permeability, $M_s$ the the specific surface area with respect to unit volume of porous medium, $n$ is the porosity and $c_0$ is the Kozeny's constant, dependent on the geometrical form of the channels on the model  \cite{bear}.  The term
\begin{equation}
f(n) = \frac{n^3}{(1-n)^2}   
\end{equation}
is the porosity factor. It has been found that  $c_0 = 1/5$  resulted good agreement with measured data, resulting in  
\begin{equation}
k = \frac{n^3}{(1-n)^2} \frac{1}{5M_s^2}  
\end{equation}
Defining the mean particle size as $d_m = M_s/6$, the Kozeny-Carman equation can be written in terms mean of particle diameter as
\begin{equation}
k = \frac{n^3}{(1-n)^2} \frac{d_m^2}{180}  
\end{equation}

\section{Laplace's equation}

In Chapter \ref{ch4} we derived a mass conservation equation (continuity equation) of the form
\begin{align*}
\frac{\partial \rho}{\partial t} = - \nabla \cdot (\rho \mathbf{u}) 
\end{align*}
Suppose that water is being accumulated in the pore space of the representative volume of interest. If the media is completely saturated and the density of water is considered constant, 
\begin{align*}
\frac{\partial \rho}{\partial t} = 0 
\end{align*}
and 
\begin{align*}
\nabla \cdot (\rho \mathbf{u}) = \rho \nabla \cdot \mathbf{u} 
\end{align*}
and, as before, we arrive at
\begin{align*}
\nabla \cdot \mathbf{u} = 0
\end{align*}
For flow in porous media, the velocity term is replaced by the flux density
\begin{align*}
\nabla \cdot \mathbf{q} = 0
\end{align*}
We already know that the flux density is represented by Darcy's law, thus, expanding the components of flux and replacing Darcy's law

\begin{align*}
(\frac{\partial }{\partial x} \mathbf{i} + \frac{\partial }{\partial y} \mathbf{j} + \frac{\partial }{\partial y} \mathbf{k}) \cdot (q_x \mathbf{i} +  q_y \mathbf{j} + q_z \mathbf{k}) = 0\\
(\frac{\partial q_x}{\partial x} + \frac{\partial q_y}{\partial y} + \frac{\partial q_z}{\partial z}) = 0 \\
(\frac{\partial }{\partial x}(-K_x \frac{\partial h}{\partial x}) + \frac{\partial }{\partial y}(-K_y \frac{\partial h}{\partial y}) + \frac{\partial }{\partial z} (-K_z \frac{\partial h}{\partial z}) ) = 0\\
\end{align*}
dropping the parentheses and multiplying both sides by $-1$ results in  
\begin{equation}
\frac{\partial }{\partial x}(K_x \frac{\partial h}{\partial x}) + \frac{\partial }{\partial y}(K_y \frac{\partial h}{\partial y}) + \frac{\partial }{\partial z} (K_z \frac{\partial h}{\partial z}) ) = 0 
\end{equation}
This equation is valid for a porous media where the unsaturated hydraulic conductivity cannot be assumed as constant. Because for a truly anisotropic media, the saturated hydraulic conductivity is a second rank tensor, a more general representation would be 
\begin{equation}
\nabla \cdot  (\mathbf{K} \cdot \nabla{h}) = 0
\end{equation}
Back to more general cases, if the saturated hydraulic conductivities in each direction does not vary with the position in space, but the conductivities in each direction are different,  we have $ K_x \ne K_y \ne K_z $ all constant, and
\begin{equation}
K_x  \frac{\partial h^2}{\partial^2 x} + K_y  \frac{\partial^2 h}{\partial y^2} + K_z  \frac{\partial^2 h}{\partial z^2} = 0 
\end{equation}

Finally, if the saturated hydraulic conductivity is constant everywhere, independent of direction, such that $ K_x \ne K_y \ne K_z = K $ we have 
\begin{equation*}
K  \frac{\partial h^2}{\partial^2 x} + K  \frac{\partial^2 h}{\partial y^2} + K  \frac{\partial^2 h}{\partial z^2} = 0 
\end{equation*}
and dividing both sides by $K$ we arrive at
\begin{equation}
\boxed{
\frac{\partial h^2}{\partial^2 x} + \frac{\partial^2 h}{\partial y^2} + \frac{\partial^2 h}{\partial z^2} = 0 
}
\end{equation}
This is a partial differential equation know as Laplace equation. It is ubiquitous in physics and mathematics and can be represented using the aptly named Laplace operator ($\nabla^2$), of which we have made extensive usage in this text  
\begin{equation}
\boxed{
\nabla^2 h = 0 
}
\end{equation}
Laplace's equation is a second-order partial differential equation that appears in electrostatics, gravitation, head conduction, transport phenomena and many other problems in applied physics, mathematics and engineering. As discussed previously, there is vast amount of literature covering methods of solution of partial differential equations, the reader being encouraged to consult the references. Analytical solutions in three and two dimensions can be found in the literature. As an example we will show how Laplace's equation can be solved in one-dimension to find the distribution of $h$ along a length of saturated porous media. Let us consider flow in the $x$ direction\footnote{To avoid any complication with implicit gravitational field}, such that h is constant in $y$ and $z$ and the partial derivatives are zero  
\begin{align*}
\frac{\partial h^2}{\partial^2 x}  = 0 \\
\int \partial h^2 = \int 0 \partial^2 x \\ 
\int \partial h = \int C_1 \partial x \\ 
\end{align*}
\begin{equation}
h =  C_1  x + C_2 \\ 
\end{equation}
 To find the values of the constants one just needs to repeat the same procedure of applying the boundary conditions. Supposing, for example that $h = h_1$ at $x = 0$ and $h = h_2$ at $x = h$, we have   
\begin{equation}
h =  \frac{(h_2 - h_1)}{h}  x + h_1 \\ 
\end{equation}
This shows that $h$ varies linearly along the length of the $x$ gradient.
Laplace's equation solutions in 2D and 3D are highly dependent on the geometry of the problem, and can be challenging. For example, a general solution of Laplace's equation in a square geometry with boundary conditions with $h = 0$ in all sides except the right vertical wall will have the form 
\begin{equation}
h(x, y)  = \sum_{n = 1}^\infty c_n \sinh(\frac{n \pi x}{b}) sin(\frac{n \pi y}{b})  
\end{equation}
with
\begin{equation}
c_n  = \frac{2}{b \sinh (\frac{n \pi a}{b})} \int_0^b g_2(y) sin(\frac{n \pi y}{b}) dy  
\end{equation}
in which $g_2(y)$ is a function that defines the nonzero boundary condition\footnote{\url{http://www.math.ubc.ca/~peirce/M257_316_2012_Lecture_24.pdf}}.

\section{List of symbols for this chapter}

\begin{longtable}{ll}
       	$ p $ & Pressure  \\
        $ \mu $ & Viscosity  \\
    	$ \mathbf{u}  $ & Vector velocity function  \\
        $ u_x, u_y, u_z $ & Components of the velocity vector in Cartesian coordinates  \\
        $ \mathbf{i}, \mathbf{j}, \mathbf{k} $ & Unit vectors in Cartesian coordinates  \\  
        $ x, y, z$ & Generalized Cartesian coordinates  \\  
    	$ C_1, C_2 $ & Constants  \\
        $ h $ & Distance between plates  \\
        $ Q $ & Discharge  \\
    
        $ r $ & Radial dimension  \\
        $ \phi $ & Angle in polar coordinates  \\
        $ q $ & Water flux  \\

        $ h' $ & Height of water above a given point (hydraulic head)  \\
        
        $ A, s $ & Cross section area  \\
        $ L, e $ & Length  \\
    
        $ \psi $ & Water potential  \\
        $ P $ & Pressure  \\
    
        $ k , K_s $ & Saturated hydraulic conductivity  \\
    
        $ d_t $ & Tube diameter  \\
        $ r_t $ & Tube radius  \\
        $ A_t $ & Tube cross sectional area  \\
    
        $ d_c $ & Column diameter  \\
        $ r_c $ & Column radius  \\
        $ A_c $ & Column cross sectional area  \\
    
        $ t $ & Time  \\
    
        $ \mathbf{q} $ & Flux density vector  \\
        $ \mathbf{K} $ & Saturated hydraulic conductivity tensor, vector form  \\
        $ q_x, q_y, q_z $ & Components of the flux density vector in Cartesian coordinates  \\
        $ K_{ij} $ & Saturated hydraulic conductivity tensor  \\
        $ \rho_w $ & Water density  \\
        $ \rho_f $ & Fluid density  \\
        $ g $ & Earth's gravitational acceleration  \\
        $ k $ & Intrinsic permeability  \\
        $ \mathbf{k} $ & Permeability tensor, vector form  \\
        $ k_{ij} $ & Permeability tensor  \\
        $ n $ & Porosity  \\

    	$ Re $ & Reynolds number  \\

        $ \beta $ & Forchheimer coefficient  \\
        $ M_s $ & Specific surface area  \\
        $ c_0 $ & Kozeny's constant  \\
        $ d_m $ & Mean particle size  \\
        $ \rho $ & Generalized density  \\

\end{longtable}

% !TEX TS-program = pdflatex
% !TEX encoding = UTF-8 Unicode

% Example of the Memoir class, an alternative to the default LaTeX classes such as article and book, with many added features built into the class itself.

\chapter{Unsaturated porous media hydrodynamics}
\label{ch9}

\section{Darcy-Buckingham equation for unsaturated flow}

Edgar Buckingham, one of the forefathers of soil physics, wrote a 61 page report to the U.S. Department of agriculture which set the foundations of what today is the theory of unsaturated flow in soils \cite{buckingham}.  Using an analogy to other physical transport laws, he postulated that the capillary current density (equivalent to $q$ in modern notation) is proportional to the capillary potential gradient. Adapting to the notation used in this book and in one dimension. \footnote{His $\lambda$ was replaced with $\beta$ as we will need the former letter later in this chapter.}

\begin{equation}
q = \beta \frac{\partial \psi_c}{\partial x}
\end{equation}
Based on his observations and theoretical analysis he concluded that the capillary conductivity, $\beta$, was not constant and was a function of water content $\beta = f(\theta)$. His conductivity law can be understood as a generalization of Darcy's law to unsaturated porous media, thus, writing the total potential in terms of hydraulic head, $h$, and $\beta$ as an unsaturated hydraulic conductivity, which is no longer constant but dependent on water content, Darcy's law can be written, for the one dimensional case as 
\begin{equation}
\boxed{
q = K(\theta) \frac{\partial h}{\partial x}
}
\end{equation}
This is the one dimensional Darcy-Buckingham equation. Contrary to the saturated case, the unsaturated hydraulic conductivity is an empirical function of water content, and not a constant which can be determined experimentally. There has been a lot of effort in developing these empirical functions, with a lot of the fundamental equations being derived in the XX century. Many of the water retention functions presented in Chapter \ref{ch7} were essentially derived to arrive at an unsaturated hydraulic conductivity function. 

\begin{figure}[ht]
\centering
 \includegraphics[width=1.0\textwidth]{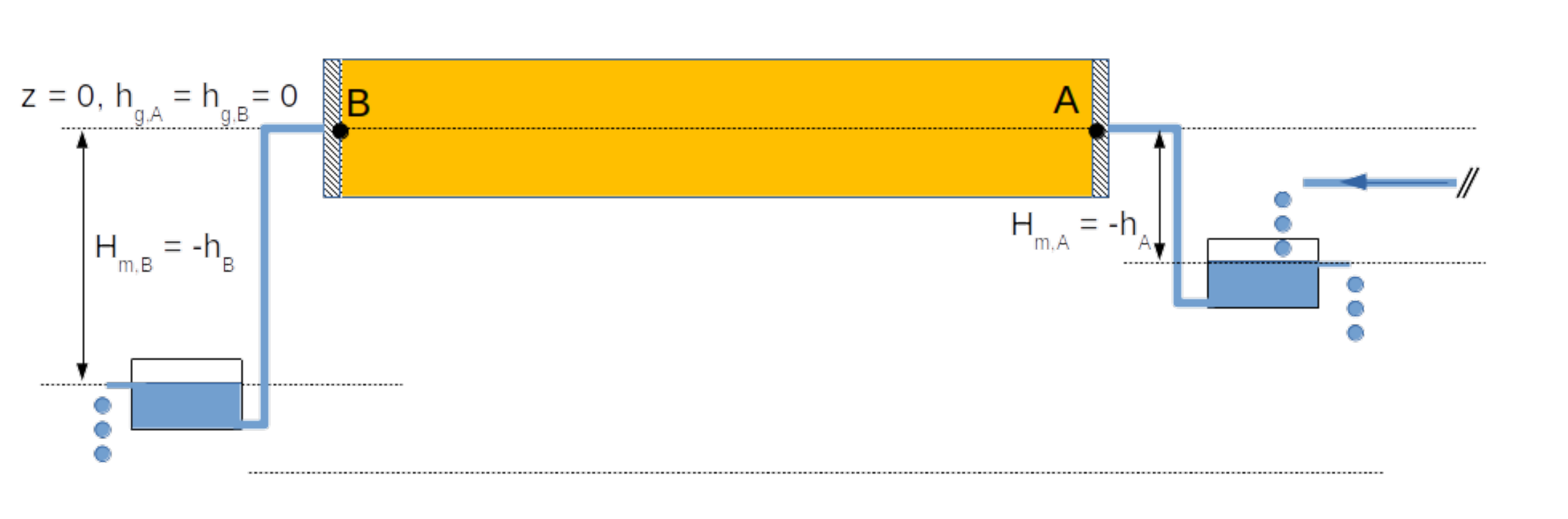}
	\caption{Unsaturated horizontal column.}
\label{ch9_fig1}
\end{figure}

One of the fundamental differences between Darcy's law and the Darcy-Buckingham equation is that in all, or at least in some part of the porous media, the water content is below saturation. This implies that the pressure potential is zero and the matric potential is the component of interest along with the gravitational potential in the total potential. Instead of applying pressure, represented by a height of water column at the inlet and outlet of the column filled with soil or other porous media of interest, the outlet and inlet can be under tension or suction, by placing the water level below the height at the point of interest. Generally speaking, the pressure potential can be positive at the inlet and negative at the outlet which causes the column to be partially saturated in part of its extension and saturated on the rest. The most common example, given in many soil physics textbooks is a column which is under tension on both ends (Figure \ref{ch9_fig1}). Usually, a porous plate is used on both ends of the column, similar to the system on a tensiometer. 

Flow is from point A from point B (right to left). By fixing the gravitational reference ($z = 0$) at the center of the column, where the inlet and outlet tubes are located, the matric potential is equal to the distance between the center of the column and the water level on the recipients connected to the inlet and outlet. The recipient connected to the outlet is fed from a water source to guarantee that the level of water is constant. Another way of achieving this, which is commonly used in laboratory, is to use a Mariotte's bottle. The scheme illustrated in Figure \ref{ch9_fig1} is valid as long as the column's height is small enough not to create significant gravitational potentials on the $z$ direction, this principle also being valid for saturated columns. The problem is critical in unsaturated soils because gravitational potential gradients can induce total potential gradients on the $z$ direction which induce hydraulic conductivity and velocity gradients. The gradient on the $x$ direction follows the decrease in water content along the column. For simple illustrative examples, if the gradient small and the column can be assumed as small enough on the $x$ length, the unsaturated hydraulic conductivity can be approximated as the average between the inlet and outlet. In real world applications an unsaturated hydraulic conductivity function is necessary. 

Because unsaturated hydraulic conductivity was theorized to be a function of water content and because water content has been experimentally shown to be a function of matric potential, hydraulic conductivity can be expressed as a function of matric potential
\begin{equation}
\boxed{
q = -K(h_m) \frac{\partial h}{\partial x}
}
\end{equation}
in which $h_m = \psi_m$ is, as before, the matric potential expressed in height units. Expanding the total potential, we arrive at similar equations as for the saturated hydraulic conductivity in a general horizontal $x$ or $y$ direction
\begin{equation}
\boxed{
q = -K(h_m) \frac{\partial h_m}{\partial x}
}
\end{equation}
\begin{equation}
\boxed{
q = -K(h_m) \frac{\partial h_m}{\partial y}
}
\end{equation}
and on the $z$ direction, parallel to the gravitational field
\begin{equation}
\boxed{
	q = -K(h_m) (\frac{\partial h_m}{\partial z} + 1)
}
\end{equation}

Applying the Darcy-Buckingham equation to Figure \ref{ch9_fig1}, and expressing the total potential as $h_t$ for clarity
\begin{align*}
	q = -K(h_m) (\frac{h_{t,A} - h_{t,B}}{x_A - x_B}) =  -K(h_m) (\frac{(h_{m,A}-h_{g,A}) - (h_{m,B}-h_{g,B})}{x_A - x_B}) 
\end{align*}
Because $h_{g,A} = h_{g,B} = 0$ for a horizontal column and making $x_A - x_B = L$ the length of the column
\begin{align*}
	q = -K(h_m) (\frac{(h_{A}-0) - (h_{B}-0)}{L}) =  -K(h_m) (\frac{h_{A} - h_{B}}{L}) 
\end{align*}
where $h_A$ and $h_B$ are the distances between the center of the column, where the inlet and outlet are located, and the respective water levels at the reservoirs.

We now know the transport equation for unsaturated porous media. The relationship indicated by the Darcy-Buckingham equation is physically and mathematically more complex than Darcy's law because introduces a hitherto unknown relationship between unsaturated hydraulic conductivity and water content or matric potential. Early experimental investigations indicated a nonlinear relationship between unsaturated hydraulic conductivity and water content or matric potential. As we have already indicated, unsaturated hydraulic conductivity functions have been derived from the water retention curve function.

\section{Unsaturated hydraulic conductivity functions}

\begin{figure}[ht]
\centering
 \includegraphics[width=0.8\textwidth]{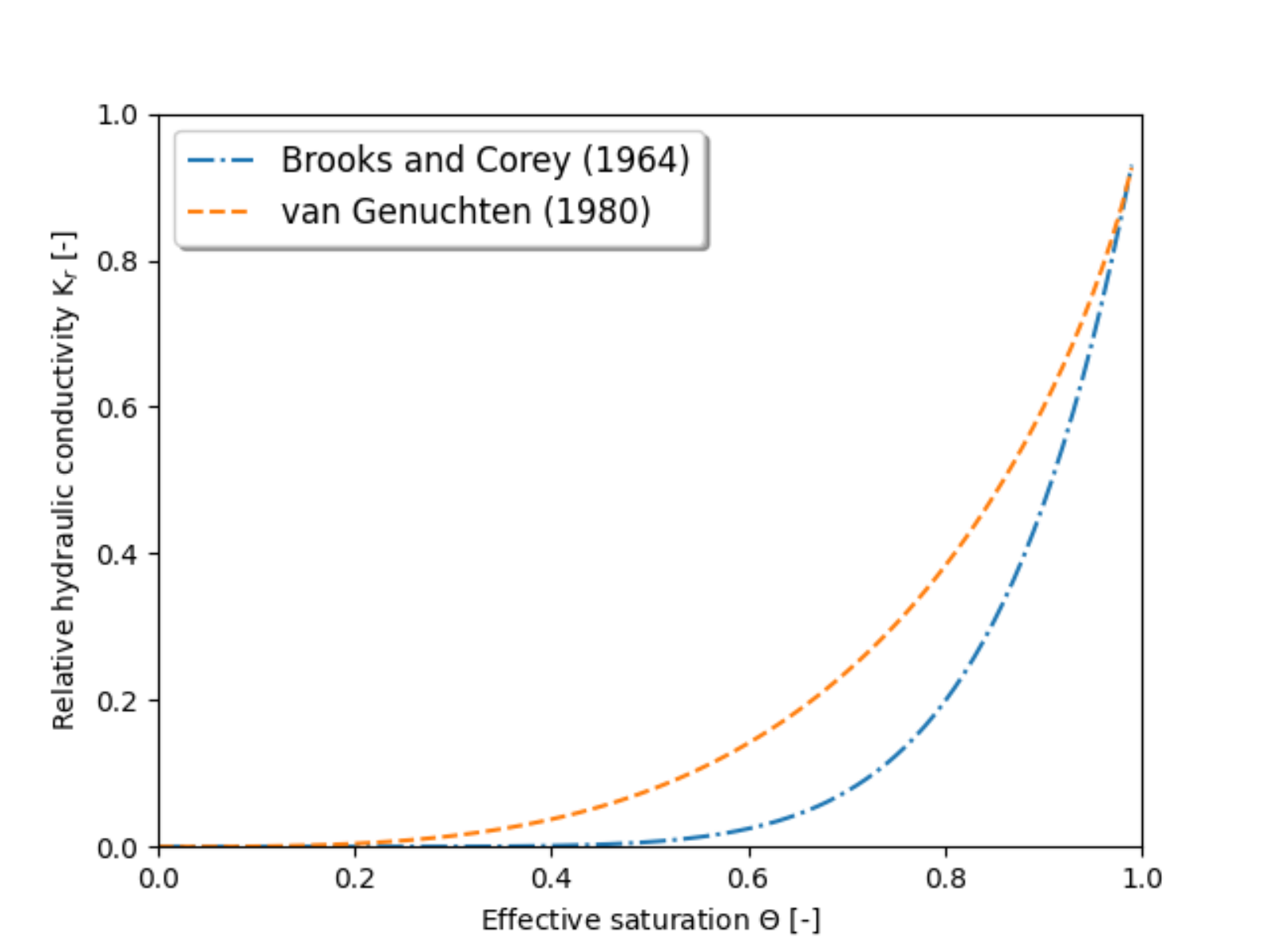}
	\caption{Relative unsaturated hydraulic conductivity as a function of effective saturation for the van Genuchten \cite{vangenuchten} equation with $m = 0.77$ and the Brooks and Corey \cite{brookscorey} equation with $\lambda = 0.47$.}
\label{ch9_fig2}
\end{figure}

Unsaturated hydraulic conductivity functions can be derived from the water retention function using the frameworks proposed by Burdine \cite{burdine, brookscorey, vangenuchten}\footnote{Remember from the previous chapter that $\Theta$ is the relative saturation equal to  $(\theta - \theta_r)/(\theta_s - \theta_r)$.}
\begin{equation}
	K_r(\Theta) = \Theta^2 \int_0^{\Theta} \frac{d \Theta}{\psi_m^2(\Theta)}\bigg / \int_0^1 \frac{d \Theta}{\psi_m^2(\Theta)} 
\end{equation}
and Mualem \cite{mualem, vangenuchten}
\begin{equation}
	K_r(\Theta) = \Theta^{1/2} [\int_0^{\Theta} \frac{d \Theta}{\psi_m(\Theta)}\bigg / \int_0^1 \frac{d \Theta}{\psi_m(\Theta)}]^2
\end{equation}
Details about the derivation of these formulas can be found in \cite{burdine, brookscorey, mualem}. They are essentially based on pore size distribution considerations. The symbology used in these equations is the same as the one used in the previous chapter. Note that we are using $\psi_m$ here to avoid being constrained by units, since these equations do not necessarily require matric potential in height units. The new term $K_r$ is the relative unsaturated hydraulic conductivity
\begin{equation}
	K_r(\Theta) = \frac{K(\Theta)}{K_s}
\end{equation}
varying from 0 to 1, and in which $K_s $ is the saturated hydraulic conductivity. To obtain the unsaturated hydraulic conductivity function it is necessary to solve the water retention function for $\psi_m$, insert it on the Burdine or Mualem equations and solve the integrals. For the Brooks and Corey equation \cite{brookscorey}, the process is straightforward. Expressing the Brooks and Corey in terms of the relative saturation, $\Theta$ 
\begin{align*}
	\Theta = (\frac{\psi_b}{\psi_m})^{\lambda}
\end{align*}
we first solve for the matric potential
\begin{align*}
	\psi_m = (\frac{\psi_b}{\Theta^{1/\lambda}})
\end{align*}
then plug into the Burdine equation
\begin{align*}
	K_r(\Theta) = \Theta^2 \int_0^{\Theta} \frac{d \Theta}{(\frac{\psi_b}{\Theta^{1/\lambda}})^2}\bigg / \int_0^1 \frac{d \Theta}{(\frac{\psi_b}{\Theta^{1/\lambda}})^2}\\
	= \Theta^2 \int_0^{\Theta} \frac{\Theta^{2/\lambda} d \Theta}{\psi_b^2} \bigg / \int_0^1 \frac{\Theta^{2/\lambda} d \Theta}{\psi_b^2}
\end{align*}
because $\psi_b$ is constant
\begin{align*}
	K_r(\Theta) =  \Theta^2 \int_0^{\Theta} \Theta^{2/\lambda} d \Theta \bigg / \int_0^1 \Theta^{2/\lambda} d \Theta
\end{align*}
The solution of the integral form in the numerator and denominator is trivial, i.e.
\begin{align*}
	\int_a^b x^{2/c} dx = \frac{x^{2/c+1}}{2/c+1} \bigg |_a^b
\end{align*}
thus 
\begin{align*}
	K_r(\Theta) =  \Theta^2 \frac{[\Theta^{2/\lambda +1}]_0^{\Theta}}{[\Theta^{2/\lambda +1}]_0^1}
\end{align*}
Applying the integration limits
\begin{align*}
	K_r(\Theta) =  \Theta^2 \frac{\Theta^{2/\lambda +1}}{1} =  \Theta^2 \Theta^{2/\lambda +1}  
\end{align*}
or 
\begin{equation}
K_r(\Theta) =   \Theta^{2/\lambda +3}  
\end{equation}
Recalling that $\Theta$ is the effective saturation and $K_r$ is the relative unsaturated conductivity, the equation can be written in terms of volumetric water content and unsaturated hydraulic conductivity
\begin{equation}
K(\theta) =  K_s (\frac{\theta - \theta_r}{\theta_s - \theta_r})^{2/\lambda +3}  
\end{equation}
To retrieve the matric potential form of unsaturated hydraulic conductivity just replace the original Brooks and Corey function for $\Theta$
\begin{equation}
K(\psi_m) =  K_s (\frac{\psi_b}{\psi_m})^{2 +3 \lambda}  
\end{equation}
These last two equations are the Brooks and Corey unsaturated hydraulic conductivity equations derived using Burdine's approach. The formulas can also be derived using Mualem's formulas by the same process. For  Brutsaert-van Genuchten type equations the process is not as straightforward. Following \cite{vangenuchten}, we can solve the effective saturation form of the water retention function resulting in 
\begin{align*}
	\psi_m = [\frac{1}{\alpha} (\frac{1}{\Theta^{1/m}} - 1)^{1/n}]
\end{align*}
Replacing into the Mualem function 
\begin{align*}
	K_r(\Theta) = \Theta^{1/2} [\int_0^{\Theta} \frac{d \Theta}{[\frac{1}{\alpha} (\frac{1}{\Theta^{1/m}} - 1)^{1/n}]}\bigg / \int_0^1 \frac{d \Theta}{[\frac{1}{\alpha} (\frac{1}{\Theta^{1/m}} - 1)^{1/n}]}]^2
\end{align*}
Doing some simple algebraic manipulations and placing the constant $1/\alpha$ term outside of the integral
\begin{equation}
	K_r(\Theta) = \Theta^{1/2} [\int_0^{\Theta} (\frac{\Theta^{1/m} }{1 - \Theta^{1/m}})^{1/n} d \Theta \bigg / \int_0^1 (\frac{\Theta^{1/m} }{1 - \Theta^{1/m}})^{1/n} d \Theta]^2
\label{eq:ch9_eq12}
\end{equation}
The issue here is the solution of the integral
\begin{equation}
\int (\frac{\Theta^{1/m} }{1 - \Theta^{1/m}})^{1/n} d \Theta
\label{eq:ch9_eq13}
\end{equation}
The integral on the denominator is the same equation with different bounds. This equations does not have a closed form analytical solution. This means that the solution is in form of special functions, there is no simple expression in which you can plug the constants and obtain results without using numerical methods or approximations to the special functions. In the last chapter we encountered the complementary error function when dealing with water retention functions. Other important special functions are the gamma and beta functions and their incomplete counterparts \cite{boas}. To find an analytical solution to this equation van Genuchten \cite{vangenuchten} used manipulations of the incomplete beta function \cite{boas}.  The first step is apply the transformation 
\begin{align*}
\Theta = y^m \\
d\Theta = m y^{m-1} dy
\end{align*}
which after replaced into the Equation \ref{eq:ch9_eq13} results
\begin{align*}
\int_0^{y^m} (\frac{(y^m)^{1/m} }{1 - (y^m)^{1/m}})^{1/n} m y^{m-1} dy
\end{align*}
simplifying and rearranging
\begin{equation}
m\int_0^{y^m} y^{1/n + m -1} (1-y)^{-1/n} dy 
\end{equation}
This equation is a form of the incomplete beta function \cite{boas}. These functions have a very important role in physics and mathematics and have been studied for hundreds of years, dating at least to the time of Euler and Legendre. There is no closed form analytical solution for the general case, but it is possible to introduce simplifications such that a solution can be found. The next simplification was to propose that for all integers  
\begin{align*}
k = m-1+1/n
\end{align*}
the integration could be performed directly. For the particular case $k=0$
\begin{align*}
m = 1-1/n
\end{align*}
We already seen this term in the last chapter when studying the Brutsaert-van Genuchten equation. It was then called the Mualem restriction, the reason why being apparent now. Applying this restriction to the incomplete beta function representation allows a straightforward solution of the integral. For simplicity we will replace  $n$ rewriting the equation in terms of $m$, thus
\begin{equation}
m\int_0^{y^m} y^{m -m} (1-y)^{m-1} dy = m\int_0^{y^m} (1-y)^{m-1} dy
\end{equation}
The solution to this equation is very simple and can be found using the $u$ substitution technique studied in introductory calculus books. Let 
\begin{align*}
u = 1-y \\
du = -dy
\end{align*}
we have
\begin{align*}
-m\int_0^{y^m} u^{m-1} du
\end{align*}
Integrating in $u$
\begin{align*}
-m \frac{u^{m-1+1}}{m-1+1} |_0^{y^m} = u^{m} |_0^{y^m} 
\end{align*}
Substituting $y$ back
\begin{align*}
-(1-y)^{m} |_0^{y^m} 
\end{align*}
At the beginning of the derivation we established that $\Theta = y^m$, replacing the original variable
\begin{align*}
-(1-\Theta^{1/m})^{m} |_0^\Theta = 1-(1-\Theta^{1/m})^{m} 
\end{align*}
The exact same steps apply to the denominator, however the limits of integration result in 
\begin{align*}
-(1-\Theta^{1/m})^{m} |_0^1 = 1
\end{align*}
Replacing the solutions of the integrals into the Mualem expression (Equation \ref{eq:ch9_eq12})
\begin{align*}
K_r(\Theta) = \Theta^{1/2} [\frac{1-(1-\Theta^{1/m})^{m} }{1}]^2
\end{align*}
and finally 
\begin{equation}
K_r(\Theta) = \Theta^{1/2} [1-(1-\Theta^{1/m})^{m} ]^2
\end{equation}
with
\begin{align*}
m = 1-1/n \\
0  < m < 1
\end{align*}
The van Genuchten-Mualem hydraulic conductivity function can be written in terms of $\theta$ and $K(\theta)$ as 
\begin{equation}
K(\theta) = K_s (\frac{\theta-\theta_r}{\theta_s-\theta_r})^{1/2} [1-(1-(\frac{\theta-\theta_r}{\theta_s-\theta_r})^{1/m})^{m} ]^2
\end{equation}
The matric potential form can be retrieved by replacing the Brutsaert-van Genuchten in terms of $\Theta$ \footnote{Be careful not to mix the the subscript ``m'' on the matric potential symbol $\psi_m$ with the empirical parameter ``m''}
\begin{equation}
K_r(\psi_m) = \frac{[1-(\alpha \psi_m)^{nm}((1+(\alpha \psi_m)^n)^{-m}]^2}{[1+(\alpha \psi_m)^n]^{m/2}}
\end{equation}
the derivation being left to the reader as an exercise.  The van Genuchten unsaturated hydraulic conductivity functions can also be derived from the Burdine expression and the Brooks and Corey equations can be derived from the Mualem expression. In fact many new equations have been derived using these expressions. There are other approaches to derive unsaturated hydraulic conductivity functions.  A comparison of the Brooks and Corey \cite{brookscorey} and van Genuchten \cite{vangenuchten} relative unsaturated conductivity functions is shown in Figure \ref{ch9_fig2} for $\lambda = 0.47$ and $m = 0.77$.  

\section{Conservation equation for unsaturated porous media}

In the previous chapter we derived the Laplace equation for a saturated porous media from the mass conservation equation (continuity equation). To derive the conservation equation for unsaturated porous media we again start from the generalized mass conservation equation presented in Chapter \ref{ch4}.

\begin{align*}
\frac{\partial \rho}{\partial t} = - \nabla \cdot (\rho \mathbf{u}) 
\end{align*}
The original derivation of this equation considered a control volume full of water or other Newtonian viscous fluid. We retrieved Laplace's equation on the last chapter by replacing the density of the fluid as the density of water $\rho_w$ and by considering water as an incompressible fluid such that its density is constant resulting in it vanishing from the conservation equation. Note that for a saturated porous medium, the volume available for water storage is less than the total volume of interest and is equal to the porosity of the medium. For saturated porous media, assuming the porosity and water density as constant has the same implications as if the media was simply a volume filled with water, the rate of variation of water content with time is always zero. For unsaturated porous media these assumptions might not be valid as the volume (or mass) of water within the pore space might vary with time. In this case we are not immediately concerned with the variation of the density of fluid within the volume of interest, but we are concerned with the variation of the concentration of the fluid within the porous media, thus we can rewrite the conservation equation as
\begin{align*}
\frac{\partial c_w}{\partial t} = - \nabla \cdot (\rho_w \mathbf{u}) 
\end{align*}
in which $c_w = m_w/V_t$ is now the water concentration in mass $m_w$ per unit of total volume of the porous medium $V_t$
\begin{align*}
	\frac{\partial \frac{m_w}{V_t}}{\partial t} = - \nabla \cdot (\rho_w \mathbf{u}) 
\end{align*}
We know that the density of water is $\rho_w = m_w/V_w$, thus solving for $m_w$ and replacing into the conservation equation 
\begin{align*}
	\frac{\partial \frac{\rho_w V_w}{V_t}}{\partial t} = - \nabla \cdot (\rho_w \mathbf{u}) 
\end{align*}
Assuming that the density of water is constant, it can be placed outside the derivative and divergence terms
\begin{align*}
	\rho_w \frac{\partial \frac{V_w}{V_t}}{\partial t} = - \rho_w \nabla \cdot \mathbf{u} 
\end{align*}
Recalling that $V_w/V_t$ is the volumetric water content $\theta$, and assuming that for a porous medium the velocity term $\mathbf{u}$ can be represented by the average velocity over a surface $\langle \mathbf{u} \rangle $, which by definition is flux density $\langle \mathbf{u} \rangle \equiv \mathbf{q}$ 
\begin{equation}
	 \frac{\partial \theta}{\partial t} = -  \nabla \cdot \mathbf{q} 
\end{equation}
With volumetric water content varying with time, this is a partial differential equation that is difficult to solve analytically without certain assumptions and simplifications. If the water content does not vary with time $\partial \theta / \partial t = 0$ and 
\begin{equation}
	 \nabla \cdot \mathbf{q} = 0  
\end{equation}
Note, however, that many of the simplifying assumptions used for saturated flow are not valid because $K$ and $\theta$ are no longer constant in space, even under steady state conditions for a partially saturated porous medium.

\section{Richardson-Richards equation}

Before deriving the general equation that governs unsaturated flow in porous media we need to find a general form of the Darcy-Buckingham's equation. Theoretically, the most general form of Darcy-Buckingham's equation for an anisotropic, heterogeneous porous media would include a vector valued function in space and matric potential for $K$ \footnote{The tensor form for anisotropic and heterogeneous porous media is speculative, the mathematical implications and the calculations have not being investigated in depth by the author.}
\begin{equation}
	\mathbf{q} = -\mathbf{K}(\mathbf{r}, \psi_m) \cdot \nabla \psi_t   
\end{equation}
%with unsaturated hydraulic conductivity tensor
%\begin{equation}
%\mathbf{K}(\mathbf{r}, \psi_m) = \mathbf{K}(\mathbf{r}, \psi_m)_{ij} =  
%\begin{pmatrix}
%\mathbf{K}(\mathbf{r}, \psi_m)_{xx} & \mathbf{K}(\mathbf{r}, \psi_m)_{xy} & \mathbf{K}(\mathbf{r}, \psi_m)_{xz}\\
%\mathbf{K}(\mathbf{r}, \psi_m)_{yx} & \mathbf{K}(\mathbf{r}, \psi_m)_{yy} & \mathbf{K}(\mathbf{r}, \psi_m)_{yz}\\
%\mathbf{K}(\mathbf{r}, \psi_m)_{zx} & \mathbf{K}(\mathbf{r}, \psi_m)_{zy} & \mathbf{K}(\mathbf{r}, \psi_m)_{zz}
%\end{pmatrix}
%\end{equation} 
in which 
\begin{equation}
	\mathbf{r} = x~ \mathbf{i} + y~ \mathbf{j}+ z~\mathbf{k}  
\end{equation}
For an homogeneous and anisotropic medium, unsaturated hydraulic conductivity does not vary in space, only with direction at a given point. In this case the components of the conductivity tensor are scalar functions of matric potential alone
\begin{equation}
\mathbf{K}(\psi_m) = -K(\psi_m)_{ij} =  
\begin{pmatrix}
K(\psi)_{xx} & K(\psi_m)_{xy} & K(\psi_m)_{xz}\\
K(\psi)_{yx} & K(\psi_m)_{yy} & K(\psi_m)_{yz}\\
K(\psi)_{zx} & K(\psi_m)_{zy} & K(\psi_m)_{zz}
\end{pmatrix}
\end{equation} 
For an heterogeneous and isotropic system, the unsaturated hydraulic conductivity is no longer a tensor but a scalar function of space and matric potential, such that $K(x,y,z,\psi_m)$ and 
\begin{equation}
	\mathbf{q} = -K(x, y, z, \psi_m)  \nabla \psi_t   
\end{equation}
The simplest case is when hydraulic conductivity is a function of matric potential only everywhere
\begin{equation}
	\mathbf{q} = -K(\psi_m)  \nabla \psi_t   
\end{equation}

Richards \cite{richards31} derived an equation governing the fluids in porous media by replacing the Darcy-Buckingham equation into the mass conservation equation for water
\begin{align*}
	\boxed{
	\frac{\partial \theta}{\partial t} =  \nabla \cdot ( K(\psi_m)  \nabla \psi_t  )
}
\end{align*}
This is known as the Richards or Richardson-Richards equation. Richardson derived an analogous equations and published it in a small section of a book dealing with numerical methods for weather prediction \cite{richardson}. The equation derived by Richardson was not derived from Darcy's law but included a ``conductance of the channels connecting opposite faces'' in a volume of soil. Whether Richards was aware of Richardson's work is unknown, although some authors believe he wasn't \cite{knightraats}\footnote{Richards did not cite Richardson in his original work, but it is entirely possible that his ideas could have reached Richards, ideas seem to float through several means within a scientific epoch.}. 

Expansion of the divergence and gradient terms results in 

\begin{align*}
	\frac{\partial \theta}{\partial t} =  \frac{\partial}{\partial x}  [K(\psi_m) \frac{\partial \psi_t}{\partial x}] +  \frac{\partial}{\partial y}  [K(\psi_m) \frac{\partial \psi_t}{\partial y}] +  \frac{\partial}{\partial z}  [K(\psi_m) \frac{\partial \psi_t}{\partial z}]   
\end{align*}
The unsaturated hydraulic conductivity is not constant and cannot be placed outside the derivative. Writing in terms of the total potential in height of water $h = h_m + h_g$

\begin{align*}
	\frac{\partial \theta}{\partial t} =  \frac{\partial}{\partial x}  [K(h_m) \frac{\partial h}{\partial x}] +  \frac{\partial}{\partial y}  [K(h_m) \frac{\partial h}{\partial y}] +  \frac{\partial}{\partial z}  [K(h_m) \frac{\partial h}{\partial z}]   
\end{align*}

\begin{align*}
	\frac{\partial \theta}{\partial t} =  \frac{\partial}{\partial x}  [K(h_m) \frac{\partial (h_m + h_g)}{\partial x}] +  \frac{\partial}{\partial y}  [K(h_m) \frac{\partial (h_m + h_g)}{\partial y}] +  \frac{\partial}{\partial z}  [K(h_m) \frac{\partial (h_m + h_g)}{\partial z}]   
\end{align*}
\begin{align*}
	\frac{\partial \theta}{\partial t} =  \frac{\partial}{\partial x}  [K(h_m) (\frac{\partial h_m}{\partial x} + \frac{\partial h_g}{\partial x})] +  \frac{\partial}{\partial y}  [K(h_m) (\frac{\partial h_m}{\partial y} + \frac{\partial h_g}{\partial y})] +  \frac{\partial}{\partial z}  [K(h_m) (\frac{\partial h_m}{\partial z} + \frac{\partial h_g}{\partial z})]   
\end{align*}
Considering that the gradient of the gravitational potential is zero in $x$ and $y$, i.e. the gravitational potential is constant such that  $\partial h_g/\partial x = 0 $ and $\partial h_g/\partial y = 0$ and that in length units the gravitational potential difference in $z$ is equal to $\partial z$, Richardson-Richards equations in $x$, $y$ and $z$ are
\begin{equation}
	\frac{\partial \theta}{\partial t} =  \frac{\partial}{\partial x}  [K(h_m) \frac{\partial h_m}{\partial x} ]
	\end{equation}
\begin{equation}
	\frac{\partial \theta}{\partial t} =   \frac{\partial}{\partial y}  [K(h_m) \frac{\partial h_m}{\partial y}]   
\end{equation}
\begin{equation}
	\frac{\partial \theta}{\partial t} =   \frac{\partial}{\partial z}  [K(h_m) (\frac{\partial h_m}{\partial z} + \frac{\partial z}{\partial z})] =  \frac{\partial}{\partial z}  [K(h_m) (\frac{\partial h_m}{\partial z} + 1)] 
\end{equation}
Richardson-Richards equation is a nonlinear partial differential equation which has analytical solutions only in very specific cases which might be physically realistic. One idea to put the equation in a form which could be solved more easily is to include a diffusivity term. 

Diffusion type equations have been studied for centuries and before high processing power computers were available, any simplification would help immensely in terms of simplifying an equation to find analytical solutions or to solve it numerically. Klute \cite{klute52} noted an artifice used by Buckingham \cite{buckingham} and later by \cite{childscollisgeorge48, childscollisgeorge50} in order to write the Darcy-Buckingham and Richardson-Richards equations in terms of a \emph{hydraulic diffusivity} term.
First let us rewrite Richardson-Richards equation in the $z$ direction by simple using the distributive property of multiplication and recalling that the unsaturated hydraulic conductivity can be written as a function of volumetric water content 
\begin{equation}
	\frac{\partial \theta}{\partial t} =  \frac{\partial}{\partial z}  [K(\theta) \frac{\partial h_m}{\partial z}] + \frac{\partial K(\theta) }{\partial z}  
\end{equation}
Note that in the gradient term
\begin{equation}
	\frac{\partial h_m}{\partial z}  
\end{equation}
if we can assume that the matric potential is a unique function of water content \cite{nielsenetal72}, then, by the chain rule for an implicit function \cite{boas} 
\begin{align*}
	\frac{\partial h_m}{\partial z} =\frac{\partial h_m}{\partial \theta}  \frac{\partial \theta}{\partial z}
\end{align*}
Replacing the expanded differential into the Richardson-Richards equation

\begin{align*}
	\frac{\partial \theta}{\partial t} =  \frac{\partial}{\partial z} [K(\theta) \frac{\partial h_m}{\partial \theta} \frac{\partial \theta}{\partial z}]
  + \frac{\partial K(\theta) }{\partial z}  
\end{align*}
The term
\begin{equation}
	D(\theta) =   K(\theta) \frac{\partial h_m}{\partial \theta}
\end{equation}
is now called the \emph{hydraulic diffusivity} or \emph{soil water diffusivity} \cite{childscollisgeorge50, klute52} and the Richardson-Richards equation can be written in water content form as  
\begin{equation}
	\frac{\partial \theta}{\partial t} =  \frac{\partial}{\partial z} [D(\theta) \frac{\partial \theta}{\partial x}] + \frac{\partial K(\theta) }{\partial z}  
\end{equation}
Conversely, Richardson-Richards equation can be also written in the matric potential form by defining the implicit differential for $\partial \theta/\partial t$ and replacing into the left side of the equation in terms of matric potential \cite{haverkamp77}

\begin{align*}
	\frac{\partial \theta}{\partial h_m}\frac{\partial h_m}{\partial t} =  \frac{\partial}{\partial z} [K(h_m) \frac{\partial h_m}{\partial z}] + \frac{\partial K(h_m) }{\partial z}  
\end{align*}
By defining a new function 
\begin{equation}
	C(\theta) =   \frac{\partial \theta}{\partial h_m}
\end{equation}
called the \emph{hydraulic capacity}, we can write the Richardson-Richards equation in terms of matric potential only as  
\begin{align*}
	C(h_m)\frac{\partial h_m}{\partial t} =  \frac{\partial}{\partial z} [K(h_m) \frac{\partial h_m}{\partial z}] + \frac{\partial K(h_m) }{\partial z}  
\end{align*}
The hydraulic capacity is the slope of the water retention function. The reasoning for reducing the number of variables in these equations is that it should simplify the process of solution either analytically, when an analytical solution is possible, or numerically. Because of the functional relationship between water content and matric potential, and unsaturated hydraulic conductivity, the functions discussed in Chapter \ref{ch7} and in this chapter can be used when solving these equations. Because these functions exist, there is also a functional relationship between hydraulic diffusivity and water content. Notice that based on the discussion on the previous paragraphs, the hydraulic diffusivity can be written as  \cite{haverkamp77}
\begin{equation}
	D(\theta) = \frac{K(\theta)}{C(\theta)}
\end{equation}
It is also worth noticing that the Darcy-Buckingham equation can be written in terms of diffusivity. In a generalized horizontal $x$ direction 
\begin{equation}
q =  D(\theta) \frac{\partial \theta}{\partial x}   
\end{equation}
Beware, however, of the interpretation of the hydraulic diffusivity \cite{kirkhampowers, nielsenetal72}. The introduction of the diffusivity form made it easier to solve the equation in certain cases, but the hydraulic diffusivity does not have the same meaning as a diffusion term in heat and solute dispersion processes. Keep in mind that many of these equations require a unique relationship between the variables unsaturated hydraulic conductivity, diffusivity and water content and/or matric potential, and between water content and matric potential. This assumption is not physically realistic because of \emph{hysteresis}, the relationship between water content and matric potential is not unique in real natural porous media and dependents on the history of the processes, either if a drying or wetting process is being considering. In addition, for wetting-drying cycles within a given water content range, there might by multiple possible paths (hysteresis loops). In other words, there is a non-unique relationship between water content and matric potential for real porous media, for a given matric potential there might be at least two corresponding water content values depending on the wetting history.   

\section{Boltzmann transformation and sorptivity}

Bruce and Klute \cite{bruceklute56} described a procedure do find the diffusivity function following a procedure that Ludwig Boltzmann \cite{boltzmann1894} used to solve diffusion type equations, now known as the Boltzmann transformation.  Recall the diffusivity form of the Richardson-Richards equation in a generalized horizontal $x$ direction  
\begin{align*}
	\frac{\partial \theta}{\partial t} =  \frac{\partial}{\partial x} [D(\theta) \frac{\partial \theta}{\partial x} ] 
\end{align*}
Boltzmann \cite{boltzmann1894} showed that diffusion type equations can be solved by introducing a new variable 
\begin{equation}
	\Lambda =  x t^{-1/2} 
\end{equation}
For the diffusivity equation of concern in soil and porous media transport, the volumetric water content is now a function of the new variable \cite{kirkhampowers}
\begin{equation}
	\theta = f(\Lambda) 
\end{equation}
The idea now is to replace $\Lambda$ into the Richardson-Richards equation in order to reduce it from a partial differential equation in $\theta$, $t$ and $x$ to a ordinary differential equation in $\Lambda$ and $\theta$. The first step is to calculate the partial derivatives of $\Lambda$ in $x$ and $t$
\begin{equation}
	\frac{\partial \Lambda }{\partial x} = \frac{1}{\sqrt{t}}  
\end{equation}
\begin{equation}
	\frac{\partial \Lambda }{\partial t} = -\frac{x}{2t^{3/2}}  = -\frac{\Lambda}{2t}
\end{equation}
Next, the water content partial derivatives can be expanded by the chain rule. Recall that, by the definition of $\Lambda$, the unsaturated hydraulic conductivity is an implicit function of it such that 
\begin{align*}
	\frac{\partial \theta }{\partial x} = \frac{\partial \theta }{\partial \Lambda} \frac{\partial \Lambda }{\partial x}   
\end{align*}
\begin{align*}
	\frac{\partial \theta }{\partial t} = \frac{\partial \theta }{\partial \Lambda} \frac{\partial \Lambda }{\partial t}   
\end{align*}
Since we have already calculated the partial derivatives of $\Lambda$ it is just a matter of replacing them into the two equations above
\begin{equation}
	\frac{\partial \theta }{\partial x} = \frac{\partial \theta }{\partial \Lambda} \frac{1}{\sqrt{t}}  
\end{equation}
\begin{equation}
	\frac{\partial \theta }{\partial t} = -\frac{\partial \theta }{\partial \Lambda}  \frac{\Lambda}{2t}
\end{equation}
Replacing these into the Richardson-Richards equation results in 
\begin{align*}
	-\frac{\partial \theta }{\partial \Lambda}  \frac{\Lambda}{2t} =  \frac{\partial}{\partial x} [D(\theta) \frac{\partial \theta }{\partial \Lambda} \frac{1}{\sqrt{t}}] 
\end{align*}
Because the partial derivatives on the right side no longer depend on $t$ they can be placed outside of the differentials
\begin{align*}
	-\frac{\partial \theta }{\partial \Lambda}  \frac{\Lambda}{2t} =  \frac{1}{\sqrt{t}} \frac{\partial}{\partial x} [D(\theta) \frac{\partial \theta }{\partial \Lambda}  ]
\end{align*}
We can now expand the partial derivative in $x$ by the chain rule
\begin{align*}
	-\frac{\partial \theta }{\partial \Lambda}  \frac{\Lambda}{2t} =  \frac{1}{\sqrt{t}} \frac{\partial}{\partial \Lambda} [D(\theta) \frac{\partial \theta }{\partial \Lambda}  ] \frac{\Lambda}{\partial x}
\end{align*}
and, since we already know $\partial \Lambda/ \partial x$
\begin{align*}
	-\frac{\partial \theta }{\partial \Lambda}  \frac{\Lambda}{2t} =  \frac{1}{\sqrt{t}} \frac{\partial}{\partial \Lambda} [D(\theta) \frac{\partial \theta }{\partial \Lambda}  ]\frac{1}{\sqrt{t}} 
\end{align*}
and finally \cite{bruceklute56}
\begin{equation}
	-\frac{\Lambda}{2} \frac{\partial \theta }{\partial \Lambda}  =   \frac{\partial}{\partial \Lambda} [D(\theta) \frac{\partial \theta }{\partial \Lambda}  ] 
\end{equation}
An experimentally based solution to this ordinary differential equation was proposed by \cite{bruceklute56} for the boundary conditions
\begin{align}
	\theta = \theta_i~ \text{for}~ x > 0, t = 0 \\
	\theta = \theta_s~ \text{for}~ x = 0, t \ge 0
\end{align}
in which $\theta_i$ is the initial water content of the system and $\theta_s$ is the saturated water content. These conditions are valid for an infinitely long column, such that the wetting front will not reach the other end. In practical terms, a long column filled with porous material at constant initial water content is placed in contact with a water source, such that the water content will increase with time. After an elapsed time $t$, the $\Lambda$ function can be evaluated along the length of the column. The $\Lambda$ function would later be called the \emph{sorptivity} of the soil or porous media. Under these conditions, the $\Lambda$ form of Richardson-Richards equation can be integrated in the limits of initial water content to water content at a position $x$ and from $\Lambda$ at the initial condition to $\Lambda$ evaluated at $x$
\begin{align*}
	-\int_{\theta_i}^{\theta_x}\frac{\Lambda}{2} \frac{\partial \theta }{\partial \Lambda}  =  \int_{\Lambda_i}^{\Lambda_x} \frac{\partial}{\partial \Lambda} [D(\theta) \frac{\partial \theta }{\partial \Lambda}  ] 
\end{align*}
which immediately simplifies to
\begin{align*}
	-\frac{1}{2}\int_{\theta_i}^{\theta_x} \Lambda \partial \theta  =  \int_{\Lambda_i}^{\Lambda_x} \partial [D(\theta) \frac{\partial \theta }{\partial \Lambda}  ] 
\end{align*}
Noticing that $\Lambda_i = \Lambda_\infty$ corresponds to $\theta_i$ and that $\Lambda_x$ corresponds to $\theta_x$, a change of variables can be performed on the right integral
\begin{align*}
	-\frac{1}{2}\int_{\theta_i}^{\theta_x} \Lambda \partial \theta  =  \int_{\theta_i}^{\theta_x} \partial [D(\theta) \frac{\partial \theta }{\partial \Lambda}  ] 
\end{align*}
or
\begin{align*}
	-\frac{1}{2}\int_{\theta_i}^{\theta_x} \Lambda \partial \theta  =  [D(\theta) \frac{\partial \theta }{\partial \Lambda}  ]_{\theta_i}^{\theta_x} =  [D(\theta_x) \frac{\partial \theta }{\partial \Lambda}  ]_{\theta_x} - [D(\theta_i) \frac{\partial \theta }{\partial \Lambda}  ]_{\theta_i}  
\end{align*}
Because $\theta_i$ is constant, the partial derivative $\partial \theta / \partial \Lambda$ evaluated at $\theta_i$ is zero
\begin{align*}
	\frac{1}{2}\int_{\theta_i}^{\theta_x} \Lambda \partial \theta  =  - [D(\theta_x) \frac{\partial \theta }{\partial \Lambda}  ]_{\theta_x}  
\end{align*}
Thus, the diffusivity can be calculated numerically and experimentally by \cite{bruceklute56}
\begin{equation}
	D(\theta_x) = -\frac{1}{2}   (\frac{\partial \Lambda }{\partial \theta}  )_{\theta_x}  \int_{\theta_i}^{\theta_x} \Lambda \partial \theta    
\end{equation}
The procedure is described in \cite{kirkhampowers} and on the original publication, among others \cite{danetopp}. 

\section{A comment on analytical solutions}
You have seen that the mathematical complexity of the models for describing unsaturated flow is substantially greater than that of saturated flow equations. The partial differential equations used includes the variables $K$, $D$, $\theta$ and $h_m$ in addition to spatial coordinates. These variables are usually related to each other by nonlinear functions, which must be considered when searching for the solution. In many real world applications, in transient flow conditions and in dimensions greater that one, these equations might not have analytical solutions or can only be solved analytically using unrealistic boundary conditions. One example of simplification was described on the previous section. When analytical solutions are not possible, numerical solutions are used. In numerical methods the equations are solved numerically through approximation methods. In fact, much of fluid mechanics and many areas of applied physics and groundwater hydrology now rely heavily on numerical methods. A whole subarea of fluid mechanics called \emph{computational fluid mechanics} is now on the frontier of knowledge in the field. Simple numerical solutions can be computed by hand but today and for real world scenarios, most applications will use a computer. Knowledge of informatics and programming languages is essential to applied numerical methods in soil physics and groundwater hydrology. 

\section{List of symbols for this chapter}

\begin{longtable}{ll}
    $\beta$ & Capillary conductivity defined by Buckingham \cite{buckingham} \\
    $\psi_c$ & Capillary potential \\
    $x$ & Generalized horizontal direction or coordinate \\
    $q$ & Flux density \\
    $K(\theta)$ & Unsaturated hydraulic conductivity as a function of volumetric water content \\
    $h$ & Total potential in length units \\
    $h_m$ & Matric potential in length units \\
    $K(h_m)$ & Unsaturated hydraulic conductivity as a function of matric potential \\
    $z$ & Generalized vertical direction or coordinate \\
    $h_t$ & Total potential in length units  \\
    $h_g$ & Gravitational potential in length units  \\
    $K_r(\Theta) $ & Unsaturated hydraulic conductivity as a function of effective saturation \\
    $\Theta$ & Effective saturation $= (\theta - \theta_r)/(\theta_s - \theta_r)$ \\
    $\psi_m$ & Matric potential in general units \\
    $\lambda$ & Pore distribution parameter of Brooks and Corey \cite{brookscorey}  \\
    $a, b, c$ & Arbitrary constants \\
    $\theta$ & Volumetric water content  \\
    $\theta_r$ & Residual water content \\
    $K_s$ & Saturated hydraulic conductivity \\
    $y$ & Parameter for transformation on the van Genuchten \cite{vangenuchten} solution or arbitrary horizontal direction or coordinate elsewhere \\
    $k$ & Generic parameter for solving the Mualem equation in the van Genuchten framework \\
    $m$ &  Shape parameter for the van Genuchten equation \\
    $n$ & Shape parameter for the van Genuchten equation \\
    $\alpha$ & Inverse of air entry pressure for the van Genuchten equation \\
    $\rho$ & Arbitrary density of a material for the generalized conservation equation  \\
    $\mathbf{u}$ & Velocity vector for fluid flow in general \\
    $c_w$ & Water concentration within a porous medium \\
    $\rho_w$ & Water density \\
    $t$ & Time \\
    $m_w$ & Water mass \\
    $V_w$ & Water volume within a porous medium \\
    $\mathbf{q}$ & Flux density in a porous medium \\
    $\mathbf{K}(\mathbf{r}, \psi_m) $ & Unsaturated hydraulic conductivity tensor \\
    $\mathbf{r}$ & Position vector \\
    $K(x,y,z, \psi_m) $ &  Unsaturated hydraulic conductivity scalar function\\
    $D(\theta)$ & Diffusivity function \\
    $C(\theta)$ & Water (hydraulic) capacity function $ = \partial \theta/\partial \psi_m$  \\
    $\Lambda$ & Variable introduced for the Boltzmann transformation
\end{longtable}

% !TEX TS-program = pdflatex
% !TEX encoding = UTF-8 Unicode

% Example of the Memoir class, an alternative to the default LaTeX classes such as article and book, with many added features built into the class itself.

\chapter{Other transport phenomena}
\label{ch10}

%%%This chapter does not specify the framework used for modeling transport (see Bear, Dullien or Scheidegger). It appears that unsaturated transport is not as consolidated theoretically in terms of framewords, as it has been developed by soil physicists

\section{Transport of dissolved chemicals}

In the last two chapters we studied the equation of conservation of water for porous media under saturated and unsaturated conditions. To study conservation of other types of mass in porous media, similar principles can be applied. Soluble solids are of immense interest in groundwater hydrology and soil physics because a large part  of the groundwater and soil contaminants are soluble solids. Examples include pesticides, organic molecules, viruses, bacteria, radionuclides, ammonium, nitrates, phosphates and many others.   In agriculture, fertilizers and other plant nutrients can be also transported through soil with water fluxes and by diffusion. Many compounds can act as either plant nutrients or contaminants, depending on where they are and on the amount and on intrinsic factors of soils and aquifers. The process can be also complicated as many solids are volatile and/or can be adsorbed onto the solid phase, suffer transformations, decay and many other processes. 

Suppose you have a volume of a porous media either soil, an aquifer or other geologic material. Suppose now that we are concerned with the dynamics of a single soluble solid. The concentration of the solid within the porous media is $C$. One of the mechanisms by which the contaminant can be transported in and out of the volume of interest is by \emph{advection}. Advective transport is the mass transport with the flow of the fluid in which it is dissolved. When the conservation equation was derived, we saw that the flux density was given by $\mathbf{q}$. The mass flux of the dissolved solid being transported in an unsaturated medium is given by 

\begin{equation}
	\mathbf{J}_{\text{adv}} = \theta \mathbf{v} C
\end{equation}
in which $\theta$ is the volumetric water content [L$^3$~L$^{-3}$], $\mathbf{v}$ is the water velocity within the pores [L T$^{-1}$], as discussed in the previous chapter and $C$ is the concentration of the dissolved solid in the liquid phase [M L$^{-3}$]. Dimensional analysis indicates that the vector $\mathbf{J}_{adv}$ is the mass of the dissolved solid entering (or exiting) per unit area per unit time [M L$^{-2}$ T$^{-1}$]. If there was no diffusion or hydrodynamic dispersion, an unrealistic scenario for dissolved solids in porous media, the conservation equation would be of the form  

\begin{equation}
	\frac{\partial (\theta C)}{\partial t} = -\nabla \cdot (\theta \mathbf{v} C) %%Verify this negative signal here
\end{equation}

However, dissolved solids are subject to diffusion due to concentration gradients. As we discussed, regions of high concentration of dissolved solids will result in lower water potential, in this scenario diffusion tends to create equilibrium of concentration throughout the fluid, subjected that there are no other constraints regarding other potentials e.g. thermal and electrostatic.  Considering an element of static, incompressible fluid, not within a porous medium, the diffusion of a soluble solid in the fluid can be modeled using Fick's law 

\begin{equation}
	\mathbf{J}_{\text{diff}} = - D_d \nabla C
\end{equation}
In this equation $\mathbf{J}_{\text{diff}}$ is the diffusive flux [M L$^{-2}$ T$^{-1}$], $D_d$ is the diffusion coefficient in [L$^{2}$ T$^{-1}$] and $\nabla C$ is the concentration gradient. Within a porous medium, not all the space is available for diffusion. In saturated porous media, the space occupied by solid particles is not available for diffusion, only the water saturated pore space. Even within a saturated porous medium, it might be that not all of the pore space is available for transport due to the fact that not all pores might be connected, and also because of dead end pores \cite{fetteretal18}. In unsaturated porous media, only the fraction of space saturated with water is available for transport \cite{bearcheng}. To account for the geometry of the pore space available for fluid flow and its modification with changing water content, a second rank \emph{tortuosity} tensor is included into Fick's law of diffusion \cite{bearbachmat, bearcheng}

\begin{equation}
	\mathbf{J}_{\text{diff}} = - D_d \mathbf{T}(\theta) \cdot \nabla C
\end{equation}
The product of the diffusion coefficient by the tortuosity tensor can be written as a tensor, defined as the coefficient of diffusion for a a given phase in a partially saturated porous medium \cite{bearcheng}

\begin{equation}
	\mathbf{J}_{\text{diff}} = - \mathbf{D}^*(\theta) \cdot \nabla C
\end{equation}
where $  \mathbf{D}^*(\theta) = D_d\mathbf{T}(\theta)$ \footnote{An analysis of the diffusion and tortuosity tensors will be left for later editions, the interested reader should consult \cite{bearbachmat, bearcheng}.}. In essence this equation shows that the concentration of soluble solids will still tend to equilibrium due to concentration gradients in partially saturated porous media, but the volume available for diffusive movement will be limited by the volume occupied by water, and the path for movement will be constrained by the geometry of the porous space.

%----> note that hydrodynamic dispersion seems to lump diffusion and dispersion
Because of the geometry of the solid and pore space, the velocity and paths of water flux in porous media tends to be highly heterogeneous. In general, the more heterogeneous the porous medium is in terms of grain size distribution and solid particles shapes and surface rugosity, the more heterogeneous the flux behavior will be, especially if the shapes and rugosity distributions are more or less random. Grain and pore size distributions are still random but tend to follow certain statistical distributions and this has been taken advantage for modeling flow and transport using stochastic techniques. Because of the irregular distribution of velocities within the pores, the irregularity of the pore paths, a third transport mechanism called \emph{mechanical dispersion} is also important. Even within a single pore, because of viscosity and surface forces, the velocity will be maximum at the center of the pore and zero at the surface if the pore is modeled as two infinite, parallel plates or as a cylinder, as we have seen when deriving Hagen-Poiseuille's equation in Chapter \ref{ch8}.  All these sub-mechanisms cause mixing of the dissolved solids when the water in which it is dissolved is flowing through a porous medium. Dispersion has been shown experimentally to follow a Fick's diffusion type law \cite{bearcheng} 
\begin{equation}
	\mathbf{J}_{\text{disp}} = - \mathbf{D} \cdot \nabla C
\end{equation}
in which $\mathbf{D}$ is the dispersion coefficient, also a second rank tensor.  Considering the three forms of transport, the total flux of a dissolved solid in a partially saturated porous medium is 
\begin{equation}
	\mathbf{J}_{\text{tot}} =\mathbf{J}_{\text{adv}} + \mathbf{J}_{\text{diff}} + \mathbf{J}_{\text{disp}} = \theta \mathbf{v} C - (\mathbf{D}^*(\theta) \cdot \nabla C  + \mathbf{D} \cdot \nabla C)
\end{equation}
The diffusion and mechanical dispersion components can be lumped into a single tensor called the \emph{coefficient of hydrodynamic dispersion} $\mathbf{D}_h$
\begin{equation}
	\mathbf{D}_h =  \mathbf{D} + \mathbf{D}^*(\theta)
\end{equation}
such that the total flux can be written as 
\begin{equation}
	\mathbf{J}_{\text{tot}} =\mathbf{J}_{\text{adv}} + \mathbf{J}_{\text{diff}} + \mathbf{J}_{\text{disp}} = \theta \mathbf{v} C - \mathbf{D}_h \cdot \nabla C
\end{equation}
In theory our work is almost be done, considering the conservation of mass and the total flux of the dissolved solid, the conservation equation should be something like

\begin{equation}
	\frac{\partial (\theta C)}{\partial t} = -\nabla \cdot (\theta \mathbf{v} C - \mathbf{D}^*(\theta) \cdot \nabla C  - \mathbf{D} \cdot \nabla C)
\end{equation}

In reality, the behavior of chemicals in porous media is much more complicated than that. Many chemicals can transition from one phase to another, for example, several compounds exist as a solid or as a liquid or as both depending on chemical equilibrium conditions, other chemicals can be adsorbed onto solid surfaces by several different mechanisms, including those discussed in Chapter \ref{ch3}, while some are subjected to both adsorption and volatilization.  In addition to that, some chemicals can be created or destroyed while being transported, for example, radionuclides can decay generating different isotopes or elements, organic matter can be transformed or destroyed by organisms and other biodegradation processes, and both mechanisms can be effectively creating new products in the environment. Because of that, a general equation of transport of chemicals in porous media must include terms for exchanges among phases, and \emph{sources} and \emph{sinks} of compounds, a source meaning a term for creation of a substance and a sink a term for the elimination of a given substance.\footnote{Obviously conservation laws still apply, a substance is eliminated by being transformed into other forms of matter or energy and created from other forms of matter or energy.} 

Following the language of \cite{bearcheng} the mass balance equation for a single chemical species in a porous media including sources and sinks is 

\begin{equation}
	\frac{\partial (\theta C + \rho F)}{\partial t} = -\nabla \cdot (\theta \mathbf{v} C - \mathbf{D}^*(\theta) \cdot \nabla C  - \mathbf{D} \cdot \nabla C) + \theta \rho_{f} \Gamma_{f} + \rho \Gamma_s
\end{equation}
in which the total mass balance term in time 
\begin{align*}
\frac{\partial (\theta C + \rho F)}{\partial t}
\end{align*}
now accounts for the rate of variation of the amount adsorbed onto the solid phase in the form of the component
\begin{align*}
\frac{\partial \rho F}{\partial t}
\end{align*}
where $F$ is the mass of a substance adsorbed into the solid phase per unit mass of solid phase and $\rho$ is the bulk density of the porous medium. The source (or sink) term of the chemical in the liquid phase is given by 
\begin{align*}
\theta  \rho_f \Gamma_f 
\end{align*}
in which $\theta$ is, as before, the volumetric water content, $\rho_f$ is the density of water, or of the liquid phase in general, and $\Gamma_f$ is the rate of creation (or destruction if the signal is negative) of the chemical per unit mass of water [M T$^{-1}$ M$^{-1}$]. The source (or sink) term for the chemical in the solid phase is given by  
\begin{align*}
\rho \Gamma_s 
\end{align*}
in which  $\Gamma_s$ is the rate of creation (or destruction if the signal is negative) per unit mass of solid phase [M T$^{-1}$ M$^{-1}$]. 

In theory, the transport equation can be further expanded to include transport in the gas phase, by including a source term in the gas phase and the vapor diffusive flux \cite{juryhorton04}. The form without the gas transport is commonly used in soil physics and hydrology and has been called the  \emph{advection-dispersion} or \emph{advective-dispersive} equation when the lumped hydrodynamic dispersion term is considered

\begin{equation}
	\frac{\partial (\theta C + \rho F)}{\partial t} = -\nabla \cdot (\theta \mathbf{v} C - \mathbf{D}_h \cdot \nabla C) + \theta \rho_{f} \Gamma_f + \rho \Gamma_s
\end{equation}

The full equation with the tensor components in three dimensions is obviously too complex for any analytical treatment. Solutions for several cases are available on the literature, in general \cite{bearcheng, fetteretal18} and for particular cases \cite{vangenuchtenalves82}.\footnote{Another fundamental reference for the mathematics of diffusion and numerical methods is the book by John Crank \cite{crank75}, Crank along with Phyllis Nicolson also created an important finite differences scheme for solving differential equations.}. One obvious simplification, especially useful for laboratory column experiments and field studies of soil profiles or geological strata is to consider a single dimension. Thus, in a generalized horizontal $x$ direction 

\begin{equation}
	\frac{\partial (\theta C + \rho F)}{\partial t} = - \frac{\partial }{\partial x}  (\theta \mathbf{v} C) + \frac{\partial }{\partial x} (\mathbf{D}_h \frac{\partial C}{\partial x}) +  \theta \rho_{f} \Gamma_f + \rho \Gamma_s 
\end{equation}
For one dimensional flow of a non-adsorbing, chemical which suffers no degradation and is not created within the porous media the sources the terms can be discarded. An example would be dissolved sodium chloride (NaCl) being transported in water through a column composed of quartz (essentially silica) sand. A further simplification can be applied if the hydrodynamic dispersion coefficient is constant for the fluid and can be placed outside the partial derivative. With all these simplifications, the conservation equation is 

\begin{equation}
	\frac{\partial (\theta C)}{\partial t} = - \frac{\partial }{\partial x}  (\theta v_x C) + D_h  \frac{\partial^2 C}{\partial x^2} 
\end{equation}
If the water content and velocity are constant they can be placed outside the derivative terms and 
\begin{equation}
	\frac{\partial C}{\partial t} = - v_x \frac{\partial C}{\partial x}     + D  \frac{\partial^2 C}{\partial x^2} 
\end{equation}
where $D = D_h/\theta$.
This second order partial differential equation can be solved analitically for different boundary conditions. For an infinite unidimensional domain saturated with water with initial chemical concentration $C = 0$ in which a solution containing a chemical (or a tracer) is introduced in the inlet with concentration $C = C_0$ subjected to the following the initial condition 
\begin{align}
	C(x,0) = 0  ~ \text{at}~ x \ge 0
\end{align}
and boundary conditions
\begin{align}
	C(0,t) = C_0~ \text{for}~ t \ge 0 \\
	C(\infty,t) = 0~ \text{for}~ t \ge 0 
\end{align}
the solution is provided by Ogata and Banks \cite{ogatabanks61}\footnote{The solution spans at least three pages in the original and might be included in further editions.}
\begin{equation}
	\frac{C(x,t)}{C_0} =  \frac{1}{2} [\text{erfc}(\frac{x-v_xt}{2\sqrt{Dt}}) + \exp{(\frac{v_x x}{D})}\text{erfc}(\frac{x+v_xt}{2\sqrt{Dt}})]
\label{eq:ch10_eq19}
\end{equation}
where erfc is the complementary error function introduced in Chapter \ref{ch8}. 

For a cylinder filled with a static fluid in which a pulse of solution with concentration $C_0$ is injected at $x=0$, the concentration will spread symmetrically from the point of application with time following \cite{crank75}
\begin{equation}
	C(t) =  \frac{M}{2\sqrt{\pi D t}} \exp{(-\frac{x^2}{4Dt})}  
\end{equation}
in which $M$ is the mass of substance deposited at time $t= 0$ at $x = 0$.

\begin{figure}[h!]
\centering
 \includegraphics[width=0.75\textwidth]{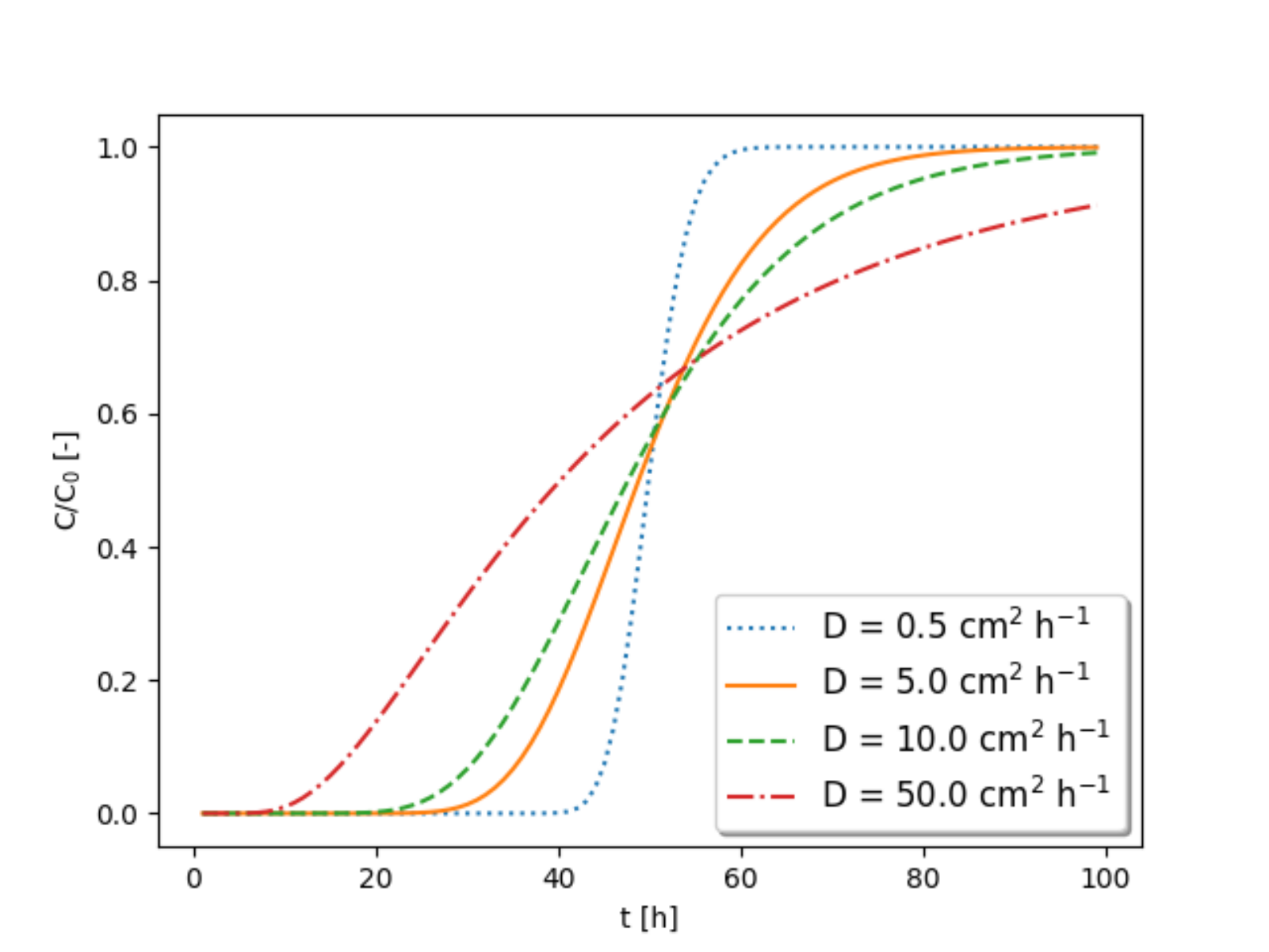}
\caption{Breakthrough curves for a $100.0~cm$ long saturated horizontal column with flow velocity $v = 2.0~cm~ h^{-1}$ and different hydrodynamic dispersion parameter values.}
\label{fig:ch10_fig1}
\end{figure}

This solution has been adapted to a cylinder filled with a porous medium in which the fluid is moving with velocity $v_x$ by considering a relative displacement term $x - v_x t$ \cite{bear, sauty80}
\begin{equation}
	\frac{C(x,t)}{C_0} =  \frac{1}{2\sqrt{\pi D t}} \exp{[-\frac{(x-v_x t)^2}{4Dt}]}  
\label{eq:ch10_eq21}
\end{equation}
Jury and Sposito \cite{jurysposito85} working within a different framework arrived at a solution of similar form with the exception of a cubic time term and an additional $x$ length term 
\begin{equation}
	\frac{C(x,t)}{C_0} =  \frac{x}{2\sqrt{\pi D t^3}} \exp{[-\frac{(x-v_x t)^2}{4Dt}]}  
\label{eq:ch10_eq22}
\end{equation}
Equations \ref{eq:ch10_eq19}, \ref{eq:ch10_eq21} and \ref{eq:ch10_eq22} can be applied to a column of finite length $L$ to predict the relative concentration $C/C_0$ at $x = L$ with time $t$. Simulated values of $C/C_0$ as a function of time for a $100~cm$ saturated horizontal column with fluid moving with velocity $v_x = 2.0~cm~h^{-1}$ for the hydrodynamic dispersion coefficients $D = D_h/\theta $ of $0.5~cm^2~h^{-1}$, $5.0~cm^2~h^{-1}$, $10.0~cm^2~h^{-1}$ and $50.0~cm^2~h^{-1}$ are presented for each of the analytical solutions. Equation \ref{eq:ch10_eq19} predicts an increase of the time for solute to first appear on the outlet, and the time for the relative concentration at the outlet to reach one, depending on the diffusivity (Figure \ref{fig:ch10_fig1}). These types curves are called \emph{breakthrough curves} because the model predicts the time and behavior for the solute to break through the column. With increasing hydrodynamic dispersion coefficient, the contribution of dispersion and diffusion to solute transport increases, ultimately resulting in some of the solute breaking through earlier than the average velocity of the fluid alone would predict. Dispersion and diffusion also cause a dilution of the front with respect to the concentration of the applied pulse. Small values of $D$ indicate little contribution of diffusion and dispersion as can be observed from the increase in the slope of the curve at lower values of $D$. At a theoretical $D=0$ the solute pulse would move exactly at the velocity of the fluid flow and would appear at once at the end of the column without change in concentration. This type of theoretical flow is called \emph{piston flow}, because it would mimic the behavior of a fluid being pushed by a piston, as in a syringe for example.

\begin{figure}[h!]
\centering
 \includegraphics[width=0.75\textwidth]{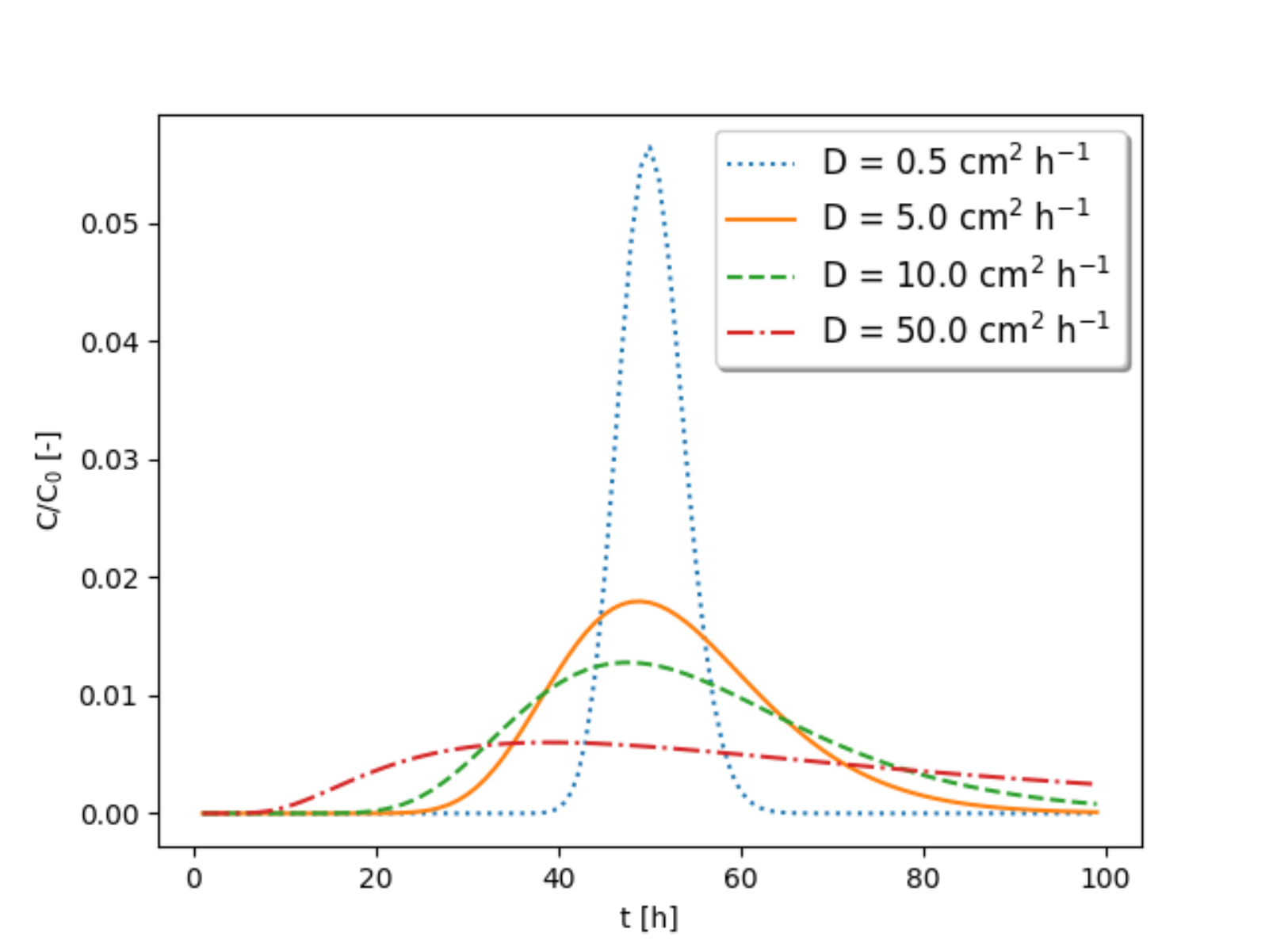}
\caption{Behaviour of a pulse of a chemical for a $100.0~cm$ long saturated horizontal column with flow velocity $v = 2.0~cm ~h^{-1}$ and different hydrodynamic dispersion parameter values \ref{eq:ch10_eq21}.}
\label{fig:ch10_fig2}
\end{figure}

\begin{figure}[h!]
\centering
 \includegraphics[width=0.75\textwidth]{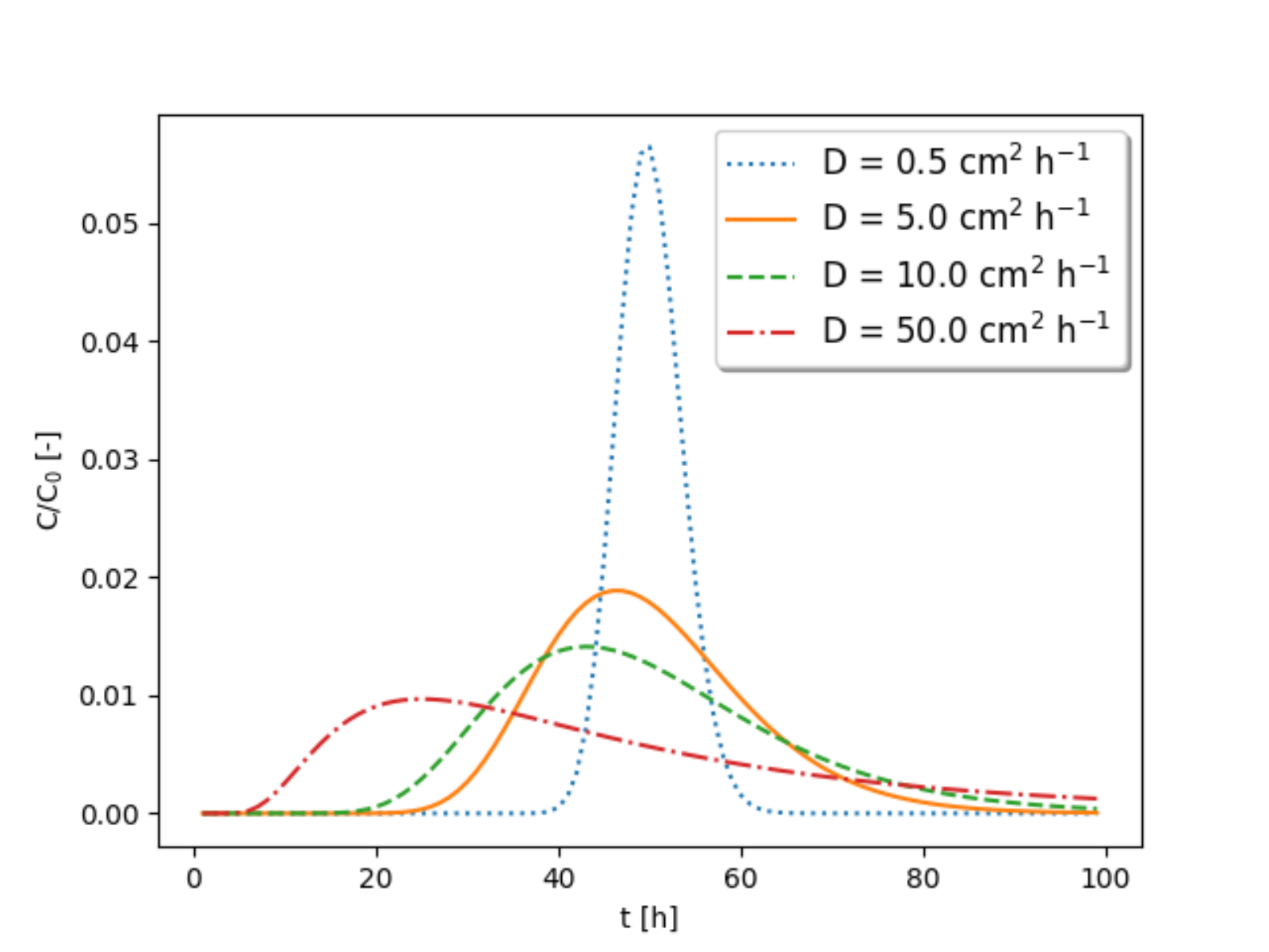}
\caption{Behavior of a pulse of a chemical for a $100.0~cm$ long saturated horizontal column with flow velocity $v = 2.0~cm~h^{-1}$ and different hydrodynamic dispersion parameter values \ref{eq:ch10_eq22}.}
\label{fig:ch10_fig3}
\end{figure}

For the same simulation parameters used for Equation \ref{eq:ch10_eq19}, Equations \ref{eq:ch10_eq21} and \ref{eq:ch10_eq22} predict the behavior of a pulse of a solution of concentration $C_0$ with time (Figures \ref{fig:ch10_fig2}, \ref{fig:ch10_fig3}). The two equations predict similar results as $D$ decreases. Equation \ref{eq:ch10_eq22} tends to predict higher concentrations as $D$ is increased when compared to Equation \ref{eq:ch10_eq21}. For a theoretical $D = 0$ both equations would predict a single instantaneous pulse, sometimes modeled as a \emph{dirac delta function}.  As $D$ increases, an increase in the spreading of the pulse is observed, the outlet concentration reaching smaller maximum output concentrations.   
 
Suppose now you have a chemical that does not have any sources of sinks, but can be adsorbed onto the solid phase, as it is the case with ions in solution and charged compounds in soils. In one dimension, the mass balance equation for a single species can be written as \cite{juryhorton04, bearcheng}
\begin{equation*}
	\frac{\partial (\theta C)}{\partial t} + \frac{\partial \rho F}{\partial t} = \frac{\partial}{\partial x} D_h \frac{\partial C}{\partial x} - \frac{\partial}{\partial x} (\theta v_x C) 
\end{equation*}
For a saturated medium, the volumetric water content $\theta$ is constant and equal to the total porosity $\phi$ and can be placed outside of the partial derivatives. Assuming also the hydrodynamic dispersion coefficient and flow velocity as constant, the equation simplifies to 
\begin{equation}
	\phi \frac{\partial C}{\partial t} + \frac{\partial \rho F}{\partial t} =  D_h \frac{\partial^2 C}{\partial x^2} - \phi v_x  \frac{\partial C}{\partial x}  
\end{equation}

If the mass adsorbed onto the solid phase is a linear function of the concentration in the liquid phase, we can define the following relationship 
\begin{equation}
	F =  K_d  C 
\end{equation}
This equation describes a linear \emph{adsorption isotherm} with $K_d$ being the \emph{partition coefficient} between the solid and liquid phases, and it is the slope of the relationship between $F$ and $C$. Replacing it into the mass balance equation 
\begin{equation*}
	\phi \frac{\partial C}{\partial t} + \frac{\partial \rho K_d  C }{\partial t} =  D_h \frac{\partial^2 C}{\partial x^2} - \phi v_x  \frac{\partial C}{\partial x}  
\end{equation*}
Because $K_d$ does not depend on the concentration and the bulk density of the porous medium it is assumed as constant
\begin{equation*}
	\phi \frac{\partial C}{\partial t} + \rho K_d\frac{\partial   C }{\partial t} =  D_h \frac{\partial^2 C}{\partial x^2} - \phi v_x  \frac{\partial C}{\partial x}  
\end{equation*}
Reorganizing and rearranging
\begin{equation*}
	(1 + \frac{\rho K_d}{\phi})\frac{\partial   C }{\partial t} =  D \frac{\partial^2 C}{\partial x^2} -  v_x  \frac{\partial C}{\partial x}  
\end{equation*}
In which $D_h = D/\phi$. Defining a  coefficient of retardation $R_d$ \cite{bearcheng}
\begin{equation*}
R_d \equiv 1 + \frac{\rho K_d}{\phi}
\end{equation*}
the mass balance equation is
\begin{equation}
	R_d\frac{\partial   C }{\partial t} =  D \frac{\partial^2 C}{\partial x^2} -  v_x  \frac{\partial C}{\partial x}  
\end{equation} 
The retardation coefficient is a number equal to or greater than zero. There are different adsorption isotherms, including nonlinear relationships between $F$ and $C$ and the mass balance equation needs to be derived for each specific case. For linear adsorption isotherms, the retardation coefficient appears in the breakthrough and pulse solutions as  \cite{skaggsleij02}
\begin{equation}
	\frac{C(x,t)}{C_0} =  \frac{1}{2} [\text{erfc}(\frac{R_d x-v_xt}{2\sqrt{D R_d t}}) + \exp{(\frac{v_x x}{D})}\text{erfc}(\frac{R_d x+v_xt}{2\sqrt{DR_d t}})]
\end{equation}
\begin{equation}
	\frac{C(x,t)}{C_0} =  \frac{x}{2\sqrt{\pi D t^3 / R_d }} \exp{(-\frac{(R_d x-v_x t)^2}{4DR_d t})}  
\label{eq:ch10_eq27}
\end{equation}

\begin{figure}[h!]
\centering
 \includegraphics[width=0.75\textwidth]{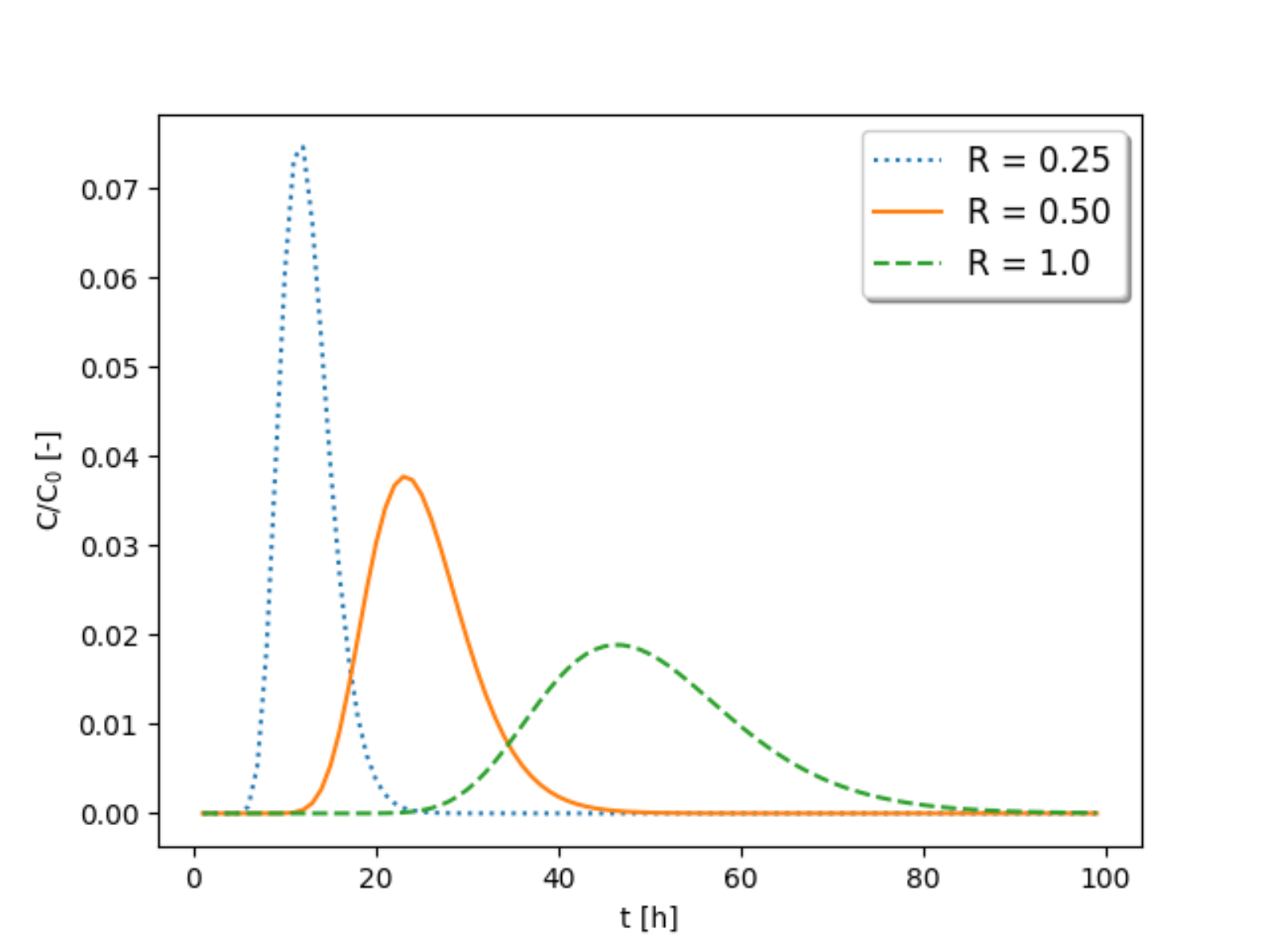}
\caption{Behavior of a pulse of a chemical for a $100.0~cm$ long saturated horizontal column with flow velocity $v = 2.0~cm~h^{-1}$ and different retardation factors.}
\label{fig:ch10_fig4}
\end{figure}

The retardation coefficient, as the name indicates, controls the rate of retardation or a delay in the transport of the chemical throughout the column and will tend to increase the width of the pulse. The effect of the retardation coefficient in Equation \ref{eq:ch10_eq27} is illustrated for $R_d = 0.25$, $0.50$ and $1.0$ (Figure \ref{fig:ch10_fig4}).

Note that for unsaturated porous media the retardation factor is a function of volumetric water content. There are numerous other cases and solutions that could be covered including unsaturated cases, presence of mobile and immobile fluid phases, partition among gas, solid and liquid phases, radioactive decay and others. For the time being the interested reader is encouraged to consult \cite{vangenuchtenalves82, bearcheng, fetteretal18, skaggsleij02}.
 
\section{Transport of gases}

The framework for discussing gas transport in porous media is the same as that of general chemical transport discussed in the previous section. Technically, gas transport could have been included with chemical transport, especially for gases that partition between liquid and gas phases and can be adsorbed onto solids. In this section we will provide a brief discussion of the specifics of gas transport. As with other types of mass transport, gases can be transported by advection and diffusion. Gas diffusion is controlled by Fick's law, one of the general transport laws in physics discussed in previous sections. For a binary mixture of gases, the mole flux of the component of interest is given by \cite{webb06}
\begin{equation}
	\mathbf{J}_g^{\text{mole}}  =  -C_g  \mathbf{D}_{g(ij)} \nabla \chi_g  
\end{equation}
in which $C_g$ is the concentration of the gas, $\mathbf{D}_g$ is the diffusion coefficient tensor and $\chi_g$ is the mole fraction of the component gas of interest. This relationship can be written in terms of a mass flux as 
\begin{equation}
	\mathbf{J}_g^{\text{mass}}  =  -\rho_g  \mathbf{D}_{g(ij)} \nabla \omega_g  
\end{equation}
in which $\omega_g$ is the mass fraction of the component of interest. Now $\mathbf{J}_g$ and $\mathbf{J}_g^{\text{mole}}$ indicate the flux relative to the average mole or mass velocity of the mixture and not to stationary coordinates \cite{birdetal02, webb06}. 
Supposing that the diffusion coefficient is constant throughout the medium, it does not need to be treated as a tensor and
\begin{equation}
	\mathbf{J}_g^{\text{mass}}  =  -\rho_g  D_g \nabla \omega_g  
\end{equation}
As we have discussed for water and chemicals, the general law for a volume devoid of solid phase is not valid when the diffusion or advection is taking place within a porous medium. If the porous medium was completely devoid of liquid, i.e. completely dry, the solid particles would still occupy part of the volume and generate a complex and tortuous pore space through which the gas would need to navigate. For a completely saturated porous medium, the gas phase would need to diffuse through the liquid, the rates being often orders of magnitude lower than that in air. If the pore space is partially saturated with water, only a fraction of the pore space inversely proportional to water content is available for diffusion and advection. Thus, Fick's law needs to be adapted to porous media. Under such conditions the diffusion coefficient is replaced by an effective diffusion coefficient   
\begin{equation}
	D_g^* = \beta D_g
\end{equation}
in which $\beta$ is a porous media factor
\begin{equation}
	\beta = \phi S_g \tau 
\end{equation}
where $\phi$ is the porosity, $S_g$ is the effective gas saturation, from zero to one, and $\tau$ is the tortuosity factor. The tortuosity factor can be written as 
\begin{equation}
	\tau = \phi^{1/3} S_g^{7/3} 
\end{equation}
which for a media completely saturated with gas simplifies to 
\begin{equation}
	\tau = \phi^{1/3}  
\end{equation}

In soil physics it is more common that the gradient is expressed in terms of concentration, which is effectively the density multiplied by the mass fraction, thus
\begin{equation}
	\mathbf{J}_g  =  - D_g^* \nabla C_g  
\end{equation}
For gas diffusion in soil, the soil gas diffusion coefficient is $ D_g^*=\tau D_g$ and the tortuosity if often expressed as $ \tau = 0.66 \theta_a $ in which $\theta_a$ is the soil volumetric air content \cite{kirkhampowers, juryhorton04}.

For advective gas transport, Darcy's law is valid \cite{juryhorton04, webb06}\footnote{The validity of Darcy's law for a compressible fluid should not be taken lightly as it has important mathematical implications.}

\begin{equation}
	\mathbf{q}_g  =  - \frac{\kappa_g}{\mu_g} (\nabla P_g - \rho_g \mathbf{g}) 
\end{equation}
here written in terms of the permeability of the media to the gas ($\kappa_g$), the viscosity of the gas ($\mu_g$), the density of the gas ($\rho_g$) and the pressure gradient ($\nabla P_g$) and in which, as defined before, $q_g$ is the flux density, but for the gas in this case. The formula can be also written  in terms of a gas conductivity $K_g = k_g/\mu_g$ parameter  
\begin{equation}
	\mathbf{q}_g  =  - K_g (\nabla P_g - \rho_g \mathbf{g}) 
\end{equation}
In general, the gravity term can be neglected because the density of most gases is low, thus 
\begin{equation}
	\mathbf{q}_g  =  - K_g \nabla P_g  
\end{equation}
for pressure in length units. 

Because advective flow of gases is usually not an important transport mechanism in soil physics and groundwater hydrology, except in very specific conditions, transport of gases in porous media is modeled as a diffusion process\footnote{Beware that this might not be the case when modeling natural gas reservoirs.}. We can now rewrite the general transport equation for a non-adsorbing gas for a generalized direction $z$, neglecting advective transport, as 

\begin{equation}
	\frac{\partial (\theta_g C_g)}{\partial t} = \frac{\partial }{\partial z} \mathbf{J}_g  +  \theta_g \rho_{g} \Gamma_g
\end{equation}

\begin{equation}
	\frac{\partial (\theta_g C)}{\partial t} = \frac{\partial }{\partial z} (\mathbf{D}^*_g \frac{\partial C^g}{\partial x}) +  \theta_g \rho_{g} \Gamma_g  
\end{equation}
and if the volumetric air content and the soil gas diffusion coefficient are constant 
\begin{equation}
	\theta_g \frac{\partial C_g}{\partial t} = D^*_g \frac{\partial^2 C_g}{\partial z^2}  +  \theta_g \rho_{g} \Gamma  
\end{equation}
In many cases this is a very simple differential equation whose techniques of solution were generally shown in Chapter \ref{ch1}. Thus, for transport of a inert gas (no sources of sinks) through a column of length $L$ saturated with inert gas with concentration $C_0$ at the inlet and zero at the outlet, the solution is \cite{juryhorton04}
\begin{equation}
	C_g(z) = C_0 (1 - \frac{z}{L})  
\end{equation}
while for steady-state transport, in which the gas concentration does not vary with time (i.e. $ \partial C_g/\partial t = 0 $) with a consumption term (e.g. oxygen in soil environment) the solution is \cite{juryhorton04}
\begin{equation}
	C_g(z) = C_0  +  \frac{\theta_g \rho_g \Gamma}{2 D^*_g} (z^2 + 2Lz)  
\end{equation}
Because the calculations are simple the previous derivations are left as an exercise to reader.

\section{Transport of heat}
The same conservation principles for mass applied in the previous sections generally apply for heat transport. However, heat transport is a form of energy transport and some particularities need to be consider. Like before, heat can be transported by diffusion and advection, but in the case of advection, heat is transported with transport of a solid, gas or liquid.

The amount of heat stored in a material is given by \cite{kirkhampowers}
\begin{equation}
	Q  = \rho c V T 
\end{equation}
in which $Q$ is the amount of thermal energy stored in the material, in joules or calories, $\rho$ is the density of the material in [M L$^{-3}$], $c$ is the mass based specific heat capacity or the heat capacity $C_h$ divided by the amount of mass of the material, $V$ is the volume of material and $T$ is the temperature. Because the heat capacity is given in energy per temperature, the mass based specific heat capacity is in energy times temperature$^{-1}$  mass$^{-1}$, representing thus the amount of energy necessary to increase the temperature of one unit mass of the material in one unit of temperature, usually in (calories) cal $g^{-1}$  $^{o}C^{-1}$ or in $J$ $kg^{-1}$ $K^{-1}$. 

As with the other transport laws discussed throughout this material, thermal conduction law is given by  
\begin{equation}
	\mathbf{J}_T  = -\frac{\kappa_T}{\rho c} \nabla T
\end{equation}
in which $\kappa_T$ is the thermal conductivity. Defining now a thermal diffusivity term as 
\begin{equation}
	D_T  = -\frac{\kappa_T}{\rho c} 
\end{equation}
\begin{equation}
	\mathbf{J}_T  = - D_T \nabla T
\end{equation}
This is Fourier's law of heat conduction and it is analogous to Fick's law for gases, Darcy's law for fluids and Ohm's law for electricity. The driver gradient is now not a gradient in terms of concentration or potential but in terms of temperature. As before, we can define a general conservation equation for heat \cite{juryhorton04} 
\begin{equation}
	\frac{\partial T}{\partial t} = - 	\frac{\partial J_T}{\partial z} + \gamma_T 	
\end{equation}
in which $\gamma_T$ is a general heat source or sink term. Following the procedures for the other conservation laws, and considering the thermal diffusivity term as constant, the heat flux $J_T$ can be replaced into the conservation equation, and ignoring heat sources and sinks 
\begin{equation}
	\frac{\partial T}{\partial t} = D_T 	\frac{\partial^2 T}{\partial z^2} 	
\end{equation}

\begin{figure}[h!]
\centering
 \includegraphics[width=0.80\textwidth]{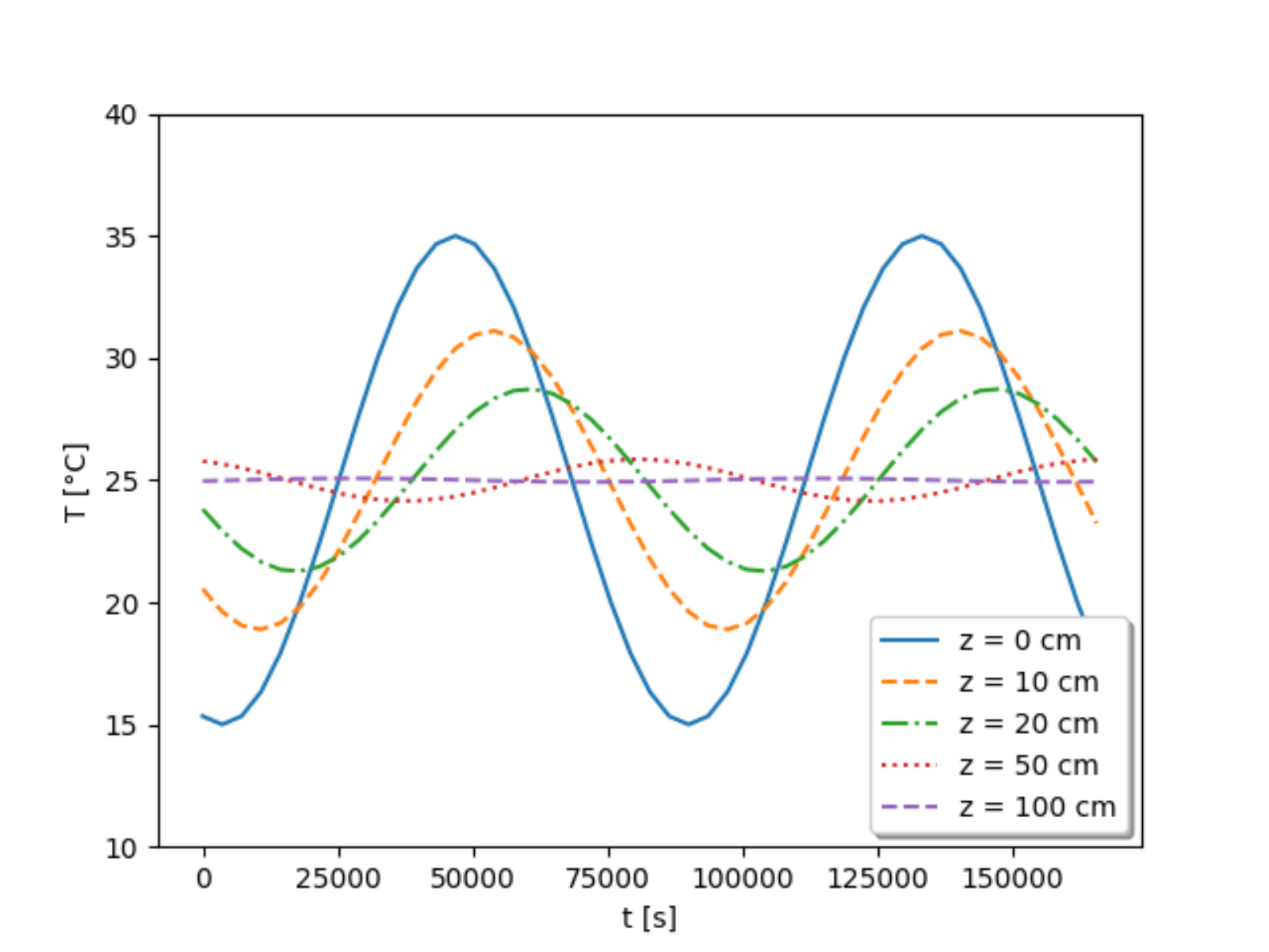}
	\caption{Variation of temperature with time over a period of two days up to a depth of $100~ cm$ for a soil with $D_T = 0.015~cm^2~s^{-1}$ .}
\label{fig:ch10_fig5}
\end{figure}

Because the temperature at surface ($z = 0$) varies sinusoidally due to daily and annual variations of temperature and solar heat influx,  a solution for temperature variation in depth is of the form \cite{kirkhampowers}

\begin{equation}
	T(z, t) = T_a + A \exp{(-\sqrt{\frac{\omega}{2D_T}}z)} \sin{[\omega t - \sqrt{\frac{\omega}{2D_T}}z + \xi]}  	
\end{equation}
in which $z$ is the depth within the material or porous media, $t$ is time, $T_a$ is the average temperature, $A$ is the amplitude of the sinusoidal wave components, $\omega = 2\pi/P$, where $P$ is the period, and $\xi $ is the phase constant. The temperature variation with depth up to $100.0~cm$, for a soil with thermal diffusivity $D_T =  0.015~cm^2~s^{-1}$, phase constant $\xi = -7\pi/12$,  $T_a = 25.0~^{o} C$ and $A = 10.0~^{o} C$ is illustrated in Figure \ref{fig:ch10_fig5}, where $\omega$ was calculated considering a period of $24~h$ or $86400~ s$. Note that the amplitude of temperature is damped with depth, being close to zero at $100~ cm$ and maximum at surface.      

%\section{Onsager reciprocal relations}

\section{List of symbols for this chapter}

\begin{longtable}{ll}
	$\mathbf{J}_{adv} $ & Advective flux density for a dissolved solid being transported\\
	$\theta $ & Volumetric water content\\
	$\mathbf{v} $ & Fluid velocity vector \\
	$C $ & Dissolved chemical concentration\\
	$D_d $ & Generalized diffusion coefficient \\
	$\mathbf{J}_{diff} $ & Diffusive flux density for a dissolved solid being transported\\
	$\mathbf{T}(\theta) $ & Tortuosity tensor\\
	$\mathbf{D}^*(\theta) $ & Coefficient of diffusion for a dissolved solid being transported in porous media\\
	$\mathbf{D} $ & Dispersion tensor in porous media\\
	$\mathbf{D}_{h} $ & Hydrodynamic dispersion tensor\\
	$\mathbf{J}_{tot} $ & Total flux density for a dissolved solid being transported in porous media\\
	$\mathbf{J}_{disp} $ & Dispersive flux density for a dissolved solid being transported in porous media\\
	$F $ & Mass of chemical absorbed per unit mass of solid phase\\
	$\rho_f $ & Fluid density \\
	$\rho $ & Bulk density of the porous media\\
	$\Gamma $ & Rate of creation or destruction of the chemical in the liquid phase\\
	$\Gamma_s $ & Rate of creation of destruction of the chemical adsorbed into the solid phase\\
	$v_x $ & Velocity in a generalized $x$ direction\\
	$D_h $ & Hydrodynamic dispersion coefficient\\
	$D $ & Dispersion coefficient corrected to volumetric water content $= D_h/\theta$\\
	$\phi $ & Total porosity \\
	$K_d $ & Coefficient of partition\\
	$R_d $ & Retardation term \\
	$t $ & Time \\
	$\mathbf{J}_{g}^{mole} $ & Mole flux for a gaseous component\\
	$\mathbf{D}_{g(ij)} $ & Gas diffusion coefficient tensor\\
	$ \chi_g$ & Mole fraction of the gaseous component of interest \\
	$ \omega $ & Mass fraction of the component of interest \\
	$\rho_g $ & Density of the gas \\
	$ D_g$ & Gas diffusion coefficient \\
	$ D^*_g $ & Effective gas diffusion coefficient in porous media \\
	$ \beta $ & Porous media factor \\
	$ S_g $ & Effective gas saturation \\
	$ \tau $ & Tortuosity factor \\
	$ \kappa_g $ & Permeability of the media to a gas \\
	$ \mu_g $ & Viscosity of the gas \\
	$ P_g $ & Gas pressure \\
	$ \mathbf{g} $ & Acceleration of gravity \\
	$ \mathbf{q}_g $ & Advective flux density of a gas in porous media \\
	$ K_g $ & Gas conductivity \\
	$\Gamma_g $ & Rate of creation of destruction of the gas\\
	$C_g $ & Concentration of the gas\\
	$ L $ & Length of a column filled with porous media \\
	$\theta_g$ & Gas volumetric content \\
	$ Q $ & Energy stored in the form of heat \\
	$ c $ & Specific heat capacity \\
	$ V$ & Volume of a material \\
	$ T$ & Temperature \\
	$\mathbf{J}_{T} $ & Heat flux \\
	$ \kappa_T $ & Thermal conductivity \\
	$ D_T $ & Thermal diffusivity \\
	$ \omega $ & $ = 2 \pi / P $, where $P$ is the period \\
	$ T_a$ & Average temperature \\
	$ A $ & Sinusoidal wave amplitude \\
	$ \xi $ & Phase constant 
\end{longtable}

% !TEX TS-program = pdflatex
% !TEX encoding = UTF-8 Unicode

% Example of the Memoir class, an alternative to the default LaTeX classes such as article and book, with many added features built into the class itself.

\chapter{Numerical Methods}
\label{ch11}

\section{Justification for numerical methods}
As we have seen throughout this book, many of the models that describe physical phenomena are in the form of partial differential equations which might not have have simple analytical solutions for modeling real world phenomena, and very often might not have any analytical solutions at all. When analytical solutions are not available, numerical approximations are an extremely useful tool in modeling real world phenomena. With the availability and processing power of computers in current times, numerical methods can be easily implemented using computational tools. Large computational codes, both open source and proprietary, are used for modeling subsurface flow and transport in soils, geological deposits and watersheds. Even larger models for matter and energy balance are available for Earth's scale modeling, but these are somewhat beyond the scope of this book. 

One example of numerical model that all of us encounter in daily life is weather prediction. Weather prediction is based on a complicated set of differential equations, based on a large set of parameters which cannot be solved analytically for larger domains such as a city, state, country or the whole world. For weather prediction the variables available are the past observations and current observations. These values will be then the \emph{initial conditions} of the model, as we have seen for the solution of many differential equations. The domains might also be subjected to a set of prescribed boundary conditions. Given past predictions, and boundary and initial conditions, a numerical model can predict the weather in the future. This type of prediction based on initial and boundary conditions using numerical solutions of differential equations is called \emph{forward modeling}. The stability, accuracy and range of the predictions is dependent on a series of factors including accuracy of past and current predictions, nature of the differential equations being solved, numerical method being used and many others.   

Another type of modeling frequently used in soil physics and groundwater hydrology is called \emph{inverse modeling}. In inverse modeling we have a set of observations and we wish to fit a particular empirical model to data to make predictions in the same system from which the data was obtained or in similar systems. Inverse modeling can be used to make predictions in time and/or space but it is fundamentally different from numerical forward modeling because the empirical data is necessary \emph{a priori}. One example is a dataset of pairs water content and matric potential values which is fit to a water retention curve equation. Fitting is usually achieved using statistical techniques such as least squares, maximum likelihood or Bayesian regression, among others. The parameterized model can be then used to monitor water content from matric potential data in soils, vadose zone or other media for agricultural applications and environmental modeling.

Both types of models can be improved, calibrated and validated using independent data. In the weather prediction model, once data has been collected, it can be compared to the numerical predictions and error indicators can be computed. The model used and computation technique can be adapted accordingly. In the case of the inverse model, once the water retention equation is fit to data, it can be compared to data collected in nearby sites using the same or similar laboratory techniques. Once a large enough data set is compiled, stochastic estimation of the empirical parameters that are representative of the area based on more or less arbitrary statistical and mathematical indicators can be determined. Over the next sections we will provide an extremely condensed account of curve fitting techniques and numerical methods used in soil physics and groundwater hydrology. For the time being only finite differences schemes will discussed for numerical methods for solving differential equations, because they are somewhat simpler from the mathematical point of view and can be easily implemented in most programming languages.

%%\section{Analytical versus numerical solutions}

%%\section{Inverse and  direct methods}

\section{Curve fitting}
The equations presented in this material are used to model real world data. Consider for example the water retention curve and soil mechanical resistance equations. In the first case data from the water retention curve can be obtained using the gravimetric method for water content, and pressure plates or tensiometers for soil matric potential. For the soil mechanical resistance to penetration equation the soil resistance can be obtained by cone penetrometers, bulk density by the core method and water content by the gravimetric method. After data collection, the equations can be fit to data using statistical or mathematical techniques. Most of the equations encountered in soil physics and groundwater hydrology are nonlinear\footnote{The use of the term nonlinear here is not related to the classification of partial differential equations.}. The meaning of \emph{nonlinear} is not directly related to the shape of the curve when data is plotted, nonlinear is related to the parameters of the equation. Consider a general linear equation, essentially a first degree polynomial
\begin{equation}
y = a x + b
\end{equation}
where $y$ is the dependent variable, $x$ is the independent variable and $a$ and $b$ are fitting parameters. For pairs of $y$ and $x$ the procedure to fit a linear regression model is exact. In other words, there are exact analytical formulas that can be used to obtain the values of $a$ and $b$ \footnote{A theoretical analysis is beyond the scope of this material and can be found in introductory statistics and generalized linear models textbooks.} %%In this case 
\begin{equation}
a = \frac{\sum (x_i - \bar{x})(y_i - \bar{y})}{\sum (x_i - \bar{x})^2}
\end{equation}
\begin{equation}
b = \frac{\sum y_i - a \sum x_i }{n}
\end{equation}
in which $x_i$ and $y_i$ are the $i^{th}$ observation and $\bar{x}$ and $\bar{y}$ the averages for the variables $x$ and $y$, and $n$ is the number of observations.

%%\begin{equation}
%%a = \frac{y-y_0}{x-x_0}
%%\end{equation}
%%\begin{equation}
%%b = y_0 - a x_0
%%\end{equation}             %%%%% This does not make sense, you need to present the statistical formulas for linear regression
For polynomials of degree greater than one, although the shape of the curve of a plot is not linear, the fitting procedure is still exact and the equation is not considered a nonlinear equation.    The definition of a nonlinear regression model is related to nonlinearity in relation to the parameters. In other words, a true nonlinear function is nonlinear with respect to its parameters. Consider a second degree polynomial. 
\begin{equation}
y = a x^2 + b x + c
\end{equation}
The partial derivatives with respect to the parameters are 
\begin{equation}
\frac{\partial y}{\partial a} = x^2
\end{equation}
\begin{equation}
\frac{\partial y}{\partial b} = x
\end{equation}
None of the derivatives are functions of the parameters. Consider now a water retention function of the form 
\begin{equation}
\Theta = \frac{A}{A + |\psi_m|^B}
\end{equation}
in which $\Theta = S_e$. The partial derivatives with respect to the parameters are   
\begin{equation}
\frac{\partial \Theta}{\partial A} = \frac{|\psi_m|^B}{(A + |\psi_m|^B)^2}
\end{equation}
\begin{equation}
\frac{\partial \Theta}{\partial B} = -\frac{A|\psi_m|^B \log{|\psi_m|}}{(A + |\psi_m|^B)^2}
\end{equation}
It is easy to see that the partial derivatives with respect to the each parameter are functions of the parameters. These types of functions cannot be fit to data by straightforward analytical and computational methods as in the case of linear and polynomial regression. Nonlinear regression methods are required to fit nonlinear models. Nonlinear regression methods rely on numerical methods based on successive approximations using \emph{iterative} methods such as the Newton-Raphson, Gauss-Newton and Levenberg-Marquardt
methods. In each case an initial guess of the values of the parameters is provided by the user and the procedure tries to converge to an optimal parameter vector where the sum of the squares of the differences between the observed and predicted $y$ values is minimized.  This is a general description of one of the most common methods for fitting nonlinear regression models, \emph{nonlinear least squares} \cite{bateswatts88}. Although there are other methods for fitting nonlinear models such as  maximum likelihood, Bayesian estimation and other robust methods \cite{bateswatts88}, nonlinear least squares is still the most common methods in commercial and noncommercial software packages. A generalized nonlinear model can be written as a function of not only the independent vector  $\mathbf{x} $ but also of a parameter vector $\mathbf{\Omega} $  \cite{bateswatts88} 
\begin{equation}
y_i = f(\mathbf{x}; \mathbf{\Omega}) + \epsilon_i  
\end{equation}
in which $\epsilon_i$ is random error term with mathematical expectation zero\footnote{In other words, the average is zero.}. The procedure used must find a least squares estimate of $\mathbf{\Omega} $, denoted by $\hat{\mathbf{\Omega}} $ which minimizes the sum of squares  \cite{bateswatts88, seberwild03}
\begin{equation}
SSQ(\mathbf{\Omega}) =   ||  \mathbf{y} -  \mathbf{x} \mathbf{\Omega} ||^2 = \sum_{i=1}^ {n} [y_i - f(x_i; \mathbf{\Omega})]^2 
\end{equation}
In simple terms, $y_i$ are the measured values and $f(x_i; \mathbf{\Omega})$ are the values predicted by the function. Thus, $y_i - f(x_i; \mathbf{\Omega})$ are the differences between measured and predicted values, often called residuals or errors which are sought to be minimized.

Because of their numerical nature, nonlinear regression models can be unstable and \emph{convergence} to an optimal solution might not always happen. As the complexity of the model and the number of parameters increases, the likelihood that the search procedure will diverge and not find the minimum on the parameters surface increases. Because of that, a topic of research in statistics, numerical methods and computation is to find algorithms with adequate properties for nonlinear regression. The search procedures discussed here try to find an optimal parameter vector by an iterative procedure of the type \cite{bateswatts88, seberwild03}
\begin{equation}
\mathbf{\Omega} ^{(\phi+1)} = \mathbf{\Omega} ^{(\phi)} + \boldsymbol{ \delta }^{(\phi)}  
\end{equation}
In which $\phi$ can be understood as the iteration number, thus the estimate of the $\mathbf{\Omega}$ vector at $\phi+1$ is conditionally based on the estimate at $\phi$ plus a step term $\boldsymbol{\delta}$ at the iteration $\phi$.
For the Gauss-Newton method \cite{bateswatts88}
\begin{equation}
\boldsymbol{\delta}^{(\phi)} = - (\mathbf{J}^{(\phi)T} \mathbf{J}^{(\phi)})^{-1} \mathbf{J}^{(\phi)T} \mathbf{r}^{(\phi)T}   
\end{equation}
Where $\mathbf{J}$ is the Jacobian matrix estimates, the Jacobian is the matrix of the derivatives with respect to the parameters, $T$ indicates a transposed vector or matrix, $-1$ indicates the inverse of the matrix and $\mathbf{r}$ is a vector of residuals, i.e. observed minus predicted values at a given iteration level. Obviously there is much more to this, although the actual application to a simple function might not be very complicated from the programming standpoint. 

A simple implementation of the Gauss-Newton method will be provided next. Consider the full implementation of the iterative procedure for the parameter vector using the Gauss-Newton step
\begin{equation}
\mathbf{\Omega} ^{(\phi+1)} = \mathbf{\Omega} ^{(\phi)} - (\mathbf{J}^{(\phi)T} \mathbf{J}^{(\phi)})^{-1} \mathbf{J}^{(\phi)T} \mathbf{r}^{(\phi)T}
\end{equation}
Suppose now we want to fit a simple water retention function of the form 
\begin{equation}
\theta = \frac{a}{|\psi_m|^b}
\end{equation}
in which $\theta $ is the volumetric water content and $\psi_m$ is the matric potential, in generalized units not relevant for this example, and $a$ and $b$ are fitting parameters.   We wish to fit the model to the fictitious data set

\begin{equation*}
\theta = \{ 0.4,	0.38,	0.35,	0.3,	0.28,	0.23,	0.22,	0.21  \}
\end{equation*}
\begin{equation*}
\psi_m = \{ -1,	-5,	-10,	-50,	-100,	-1000,	-5000,	-10000     \}
\end{equation*}
$\boldsymbol{\theta}$ and  $\boldsymbol{\psi}_m$ are vectors containing the observed water content and matric potential estimates. We start the Gauss-Newton method with an initial guess for the parameter vector, say $a = 0.5$ and $b = 0.2$. Ideally the initial parameter vector should be defined taking into consideration the physical meaning of the parameters and/or the user should plot predicted and estimated values in the same graph and try to estimate the approximate best fit visually, so the parameter vector is close to the optimum values. The parameter vector at the zeroth iteration contains the initial guess, thus 
\begin{equation}
\boldsymbol{\Omega}^{(0)} = 
\begin{bmatrix}
 0.50\\
0.20
\end{bmatrix}
\end{equation}
The Jacobian matrix is the matrix containg the partial derivatives of the residuals vector  
\begin{equation}
\mathbf{r} = \boldsymbol{\theta}- \hat{\boldsymbol{\theta}}
\end{equation}
with respect to the parameters, where $\hat{\boldsymbol{\theta}}$ is the vector containing the model estimates based on the parameter vector. In simpler term, for each observation $i$, the residuals function is
\begin{equation}
r_i = \theta_i- \frac{a}{|\psi_{m,i}|^b}
\end{equation}
Thus the partial derivatives with respect to the parameters are 
\begin{equation}
\frac {\partial r_i}{\partial a} = -|\psi_{m,i}|^{-b}
\end{equation}
\begin{equation}
\frac {\partial r_i}{\partial b} = a|\psi_{m,i}|^{-b}\log{|\psi_{m,i}|}
\end{equation}
The Jacobian matrix is then populated as 
\begin{equation}
J_{i,2} = \{\frac {\partial r_i}{\partial a}, \frac {\partial r_i}{\partial b} \}
\end{equation}
Thus\footnote{We are approximating everything to four decimals whenever possible to save space, however, it is almost never a good idea to work with low precision approximations in numerical methods.}
\begin{equation}
\mathbf{J}^{(0)} = 
\begin{bmatrix}
-1.00000 & 	0.00000 \\
-0.72478 & 0.58324  \\
-0.63096 & 	0.72642 \\
-0.45731 & 	0.89449 \\
-0.39811 & 0.91668 \\
-0.25119 & 0.86757 \\
-0.18206 & 	0.77530 \\
-0.15849 & 0.72987
\end{bmatrix}
\end{equation}
with transpose
\begin{equation}
\begin{split}
 & \mathbf{J}^{(0)T} =  \\ 
&
\begin{bmatrix}
-1.0000&	-0.7248&	-0.6310&	-0.4573&	-0.3981&	-0.2512&	-0.1821&	-0.1585 \\
0.0000&	0.5832&	0.7264&	0.8945&	0.9167&	0.8676&	0.7753&	0.7299
\end{bmatrix}
\end{split}
\end{equation}
The product $\mathbf{J}^{(0)T}\mathbf{J}^{(0)}$ is then a $ 2 \times 2$ matrix\footnote{At this point it might convenient for the reader to review an introductory \emph{linear algebra} book.}
\begin{equation}
\mathbf{J}^{(0)T}\mathbf{J}^{(0)} = 
\begin{bmatrix}
2.4124 &	-2.1298 \\
-2.1298 & 4.3948
\end{bmatrix}
\end{equation}
With inverse likewise calculated using linear algebra procedures
\begin{equation}
(\mathbf{J}^{(0)T}\mathbf{J}^{(0)})^{-1} = 
\begin{bmatrix}
0.7245 &	0.3511 \\
0.3511 &	0.3977
\end{bmatrix}
\end{equation}
The residuals vector is 
\begin{equation}
\mathbf{r}^{(0)} = 
\begin{bmatrix}
-0.1000 \\
0.0176 \\
0.0345 \\
0.0713 \\
0.0809 \\
0.1044 \\
0.1290 \\
0.1308 \\
\end{bmatrix}
\end{equation}
resulting in a Gauss-Newton step vector
\begin{equation*}
\boldsymbol{\delta}^{0} =  (\mathbf{J}^{(0)T} \mathbf{J}^{(0)})^{-1} \mathbf{J}^{(0)T} \mathbf{r}^{(0)T}
\end{equation*}
of
\begin{equation}
\boldsymbol{\delta}^{(0)} = 
\begin{bmatrix}
0.1107 \\
0.1582
\end{bmatrix}
\end{equation}
Noting that 
\begin{equation}
\mathbf{\Omega} ^{(1)} = \boldsymbol{\Omega} ^{(0)} - \boldsymbol{\delta}^{(0)}
\end{equation}
results in 
\begin{equation}
\mathbf{\Omega} ^{(1)} = 
\begin{bmatrix}
0.3893 \\
0.0418
\end{bmatrix}
\end{equation}
with $SSQ = 0.067$. Repeating the procedure for five iterations results in $a = 0.4099$ and $b = 0.0761$ and $SSQ = 0.00075715$. For comparison, a numerical procedure from a popular free software spreadsheet resulted in the same values after 40 iterations.

The Levenberg-Marquardt method improves upon the Gauss-Newton method allowing for search in two direction and for singular or ill conditioned matrices \cite{bateswatts88, seberwild03}
\begin{equation}
\boldsymbol{\delta}^{(\phi)} = - (\mathbf{J}^{(\phi)T} \mathbf{J}^{(\phi)}   + \eta^{(\phi)} \mathbf{D}^{(\phi)} )^{-1} \mathbf{J}^{(\phi)T} \mathbf{r}^{(\phi)T}   
\end{equation}
in which $\mathbf{D}^{(\phi)}$ is a diagonal matrix with positive elements and $\eta$  is a damping factor.
 The Levenberg-Marquardt procedure is usually efficient for Brutsaert-van Genuchten type equations, although the Gauss-Newton procedure is usually adequate if the number of parameters is reduced and if the initial guess is reasonably close to optimal values. A program for fitting the Brutsaert-van Genuchten to soil water retention data is presented in Appendix \ref{app:appB} (Figure \ref{ch11_fig1}). The Levenberg-Marquardt is invoked from a program module and thus was not written from scratch. 
\begin{figure}[ht]
\centering
 \includegraphics[width=0.80\textwidth]{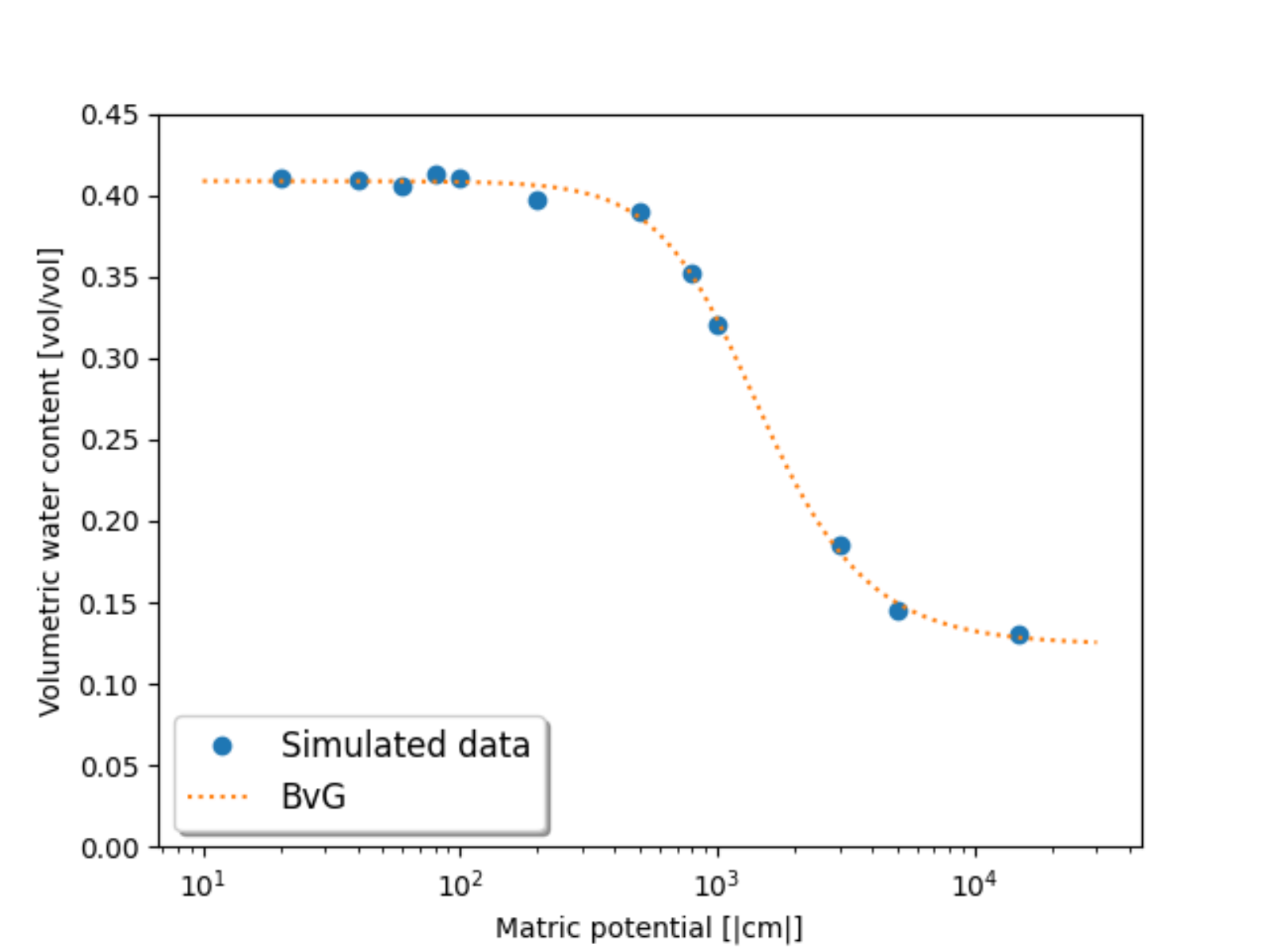}
	\caption{Brutsaert-van Genuchten equation (BvG) fit to simulated water retention data from  the program in Appendix \ref{app:appB}.}
\label{ch11_fig1}
\end{figure}

The procedures described above can be also used in soil mechanical resistance equations of the form 
\begin{equation}
SR = a \rho^b \theta^c   
\end{equation}
Note that the partial derivatives indicate that this function is nonlinear
\begin{equation}
\frac{\partial SR}{\partial a} = \rho^b \theta^c   
\end{equation}
\begin{equation}
\frac{\partial SR}{\partial b} = ba\theta^c\rho^{(b-1)}    
\end{equation}
\begin{equation}
\frac{\partial SR}{\partial c} = ca\rho^b \theta^{(c-1)}   
\end{equation}
However, this equation is \emph{intrinsically linear} since it can be transformed into a linear function. In this case the transformation is achieved by applying logarithms at both sides
\begin{equation}
\log{SR} = \log{a} + b \log{\rho} + c \log{ \theta}  
\end{equation}
which can be fitted by multiple linear regression with fitting parameters $a$, $b$ and $c$. Note, however, that linearization of intrinsically linear functions can result in undesirable statistical properties \cite{bateswatts88, leao2005}
 
\section{Finite differences methods for solving differential equations}

\subsection{Taylor series}

In order to approximate differential equations using finite difference methods, it is necessary to use a mathematical tool called \emph{Taylor series}. Taylor series allow us to expand a infinitely differentiable function around a point using its derivatives. It is a fundamental tool in physics and mathematics, having applications in several different areas, especially in the approximation of functions which are too difficult or impossible to solve analytically. Consider a real of imaginary valued function $f(x)$ which we want to expand around the point $a$, in which $a$ is a real or complex number. The Taylor series expansion is 
\begin{equation}
	f(x) = f(a) + \frac{(x-a)}{1!} \frac{d }{dx} f(a) + \frac{(x-a)^2}{2!} \frac{d^2 }{dx^2} f(a) + \frac{(x-a)^3}{3!}\frac{d^3 }{dx ^3} f(a)+ ... 
\end{equation}	
which can be written in summation form as 
\begin{equation}
	f(x) = \sum_{0}^{\infty} \frac{(x-a)^n}{n!} \frac{d^n }{dx ^n}  f(a) 
\end{equation}	
An important case of the Taylor series is when the expansion is done around zero, being called \emph{Maclaurin series}
\begin{equation}
	f(x) = f(0) + \frac{x}{1!} \frac{d  }{dx} f(0)+ \frac{x^2}{2!} \frac{d^2 }{dx^2}f(0)  + \frac{x^3}{3!}\frac{d^3 }{dx ^3} f(0)+ ... 
\end{equation}	
For most applications in physics, the higher order terms (higher powers) are small and can be neglected, thus resulting in a simple and often precise approximation of mathematical functions. This was particularly important before the advent of electronic calculators and digital computers, when the estimations of functions which today are viewed as trivial such as the value of $\pi$ where difficult tasks. Examples of Taylor series approximations for a few simple functions are given below \cite{symon}
\begin{equation}
	e^x = 1 + x + \frac{x^2}{2} +  \frac{x^3}{6} + \frac{x^4}{24} + ... 
\end{equation}	
\begin{equation}
	\log(1 + x) = x - \frac{x^2}{2} + \frac{x^3}{3} - \frac{x^4}{4} + ... 
\end{equation}	
\begin{equation}
	(1 + x)^n = 1 + nx + \frac{n(n-1)}{2} x^2 +  \frac{n(n-1)(n-2)}{6} x^3 + ...
\end{equation}

\subsection{Jacobi method for 2D Laplace equation}

\begin{figure}[ht]
\centering
 \includegraphics[width=0.80\textwidth]{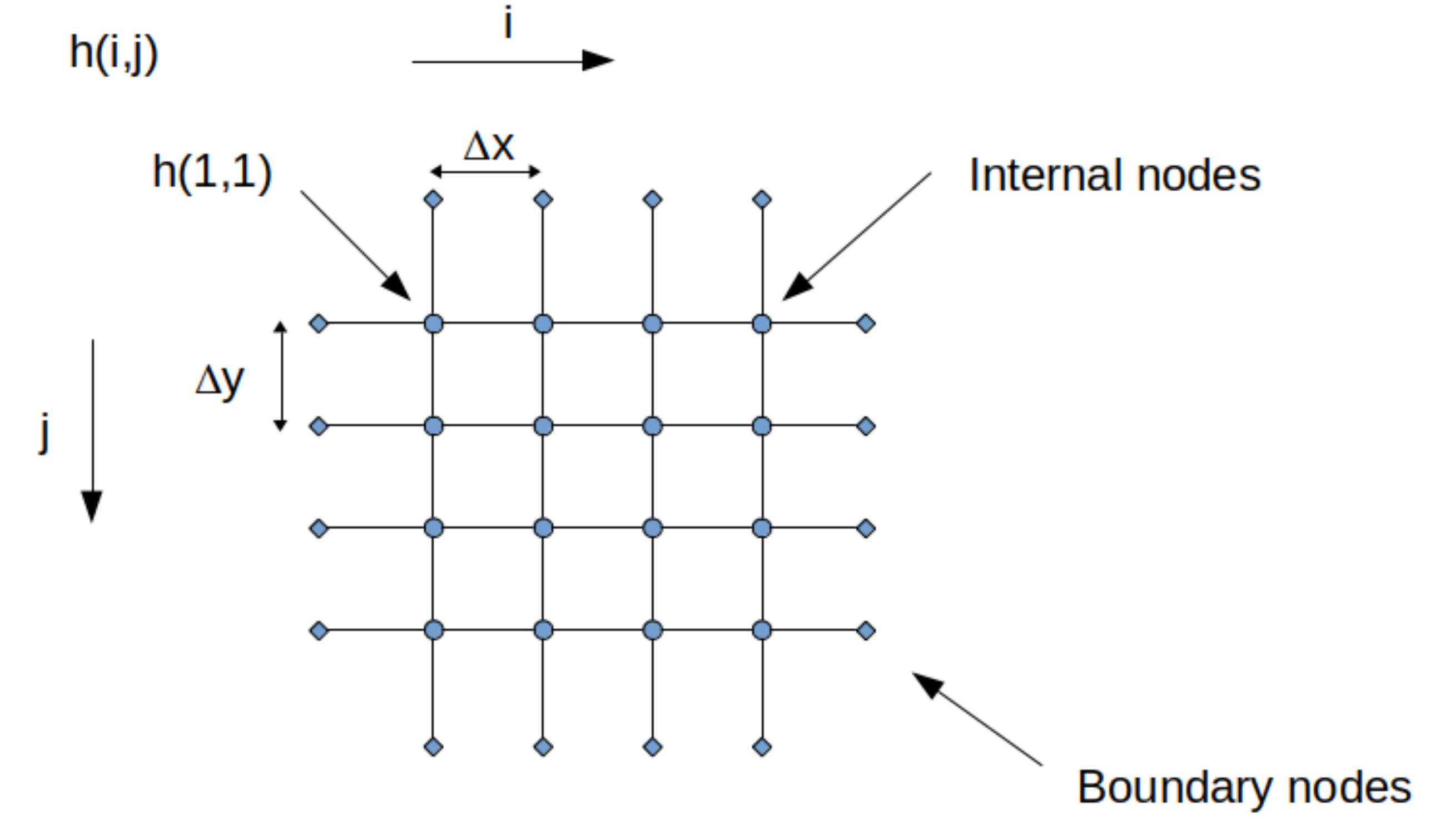}
	\caption{Discretization for the finite differences solution of the 2D Laplace equation.}
\label{ch11_fig2}
\end{figure}

We rewrite Laplace's equation for total water potential for a two dimensional domain 
\begin{equation*}
 \frac{\partial^2 h}{\partial x^2} + \frac{\partial^2 h}{\partial y^2} = 0
\end{equation*}
As we have discussed, this equation does have an analytical solution in terms of series. However, the numerical solution of this equation provides a convenient introduction to finite difference methods. Our goal here is to approximate the second order partial derivatives within a domain. Let's define the domain as a region in space, here in the horizontal plane, with dimensions $L_x$ and $L_y$. The domain can be discretized in the form of a regular grid with nodes spaced $\Delta x$ in the $x$ direction and $\Delta y$ in the $y$ direction (Figure \ref{ch11_fig2}). 

\begin{figure}[ht]
\centering
 \includegraphics[width=0.80\textwidth]{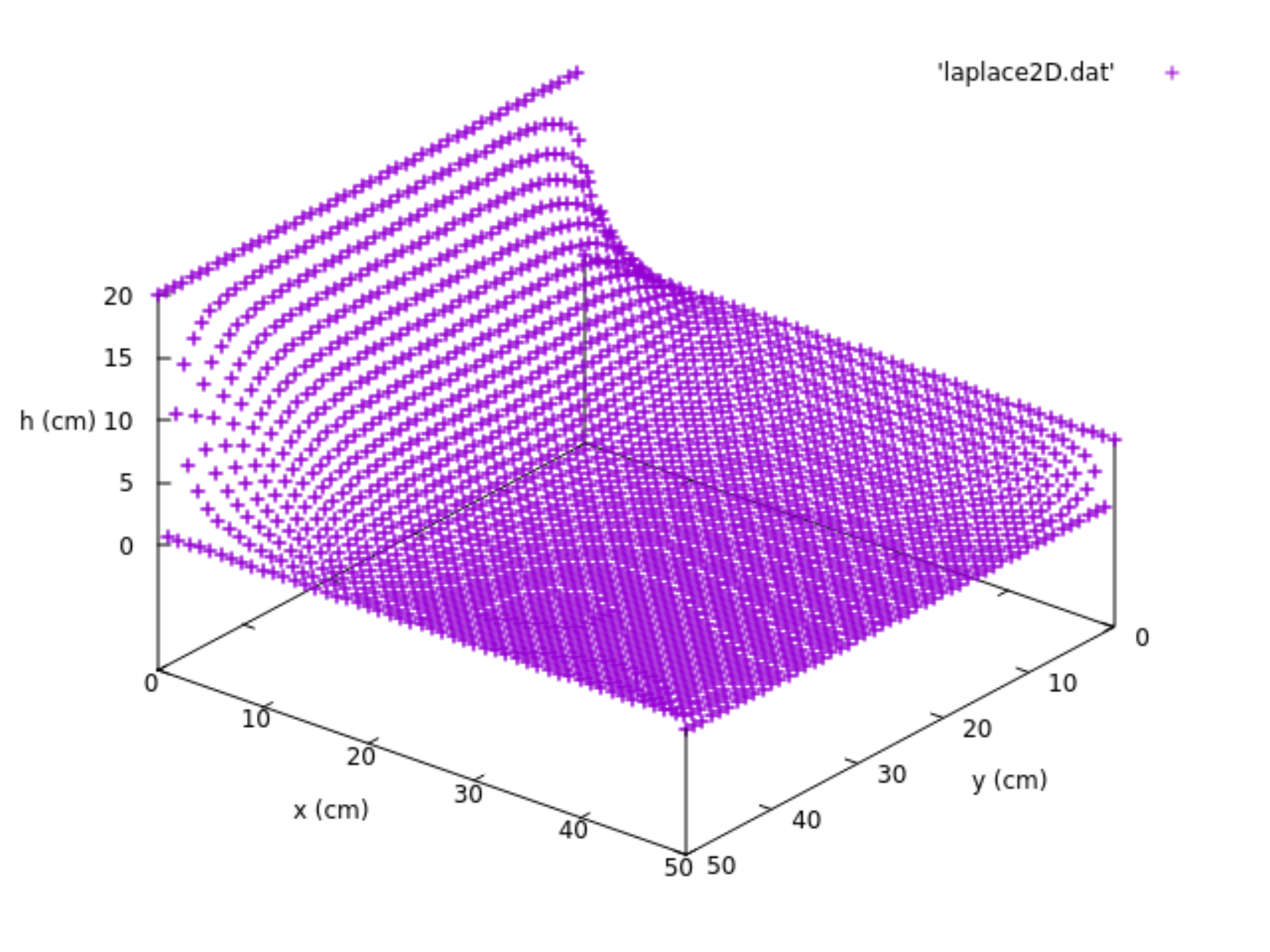}
	\caption{Numerical solution of the  2D Laplace equation from the program in Appendix \ref{app:appC}.}
\label{ch11_fig3}
\end{figure}

The second order partial derivatives in the $x$ direction can be approximated between two adjacent points,  where the next point is located in the positive $x$ direction of $h^j_i$, by \emph{Taylor series expansion} \cite{hoffman01}
\begin{equation}
\begin{split}
	h^j_{i+1} = & h^j_{i} + (x^j_{i+1} - x^j_{i}) \frac{\partial h}{\partial x} + \frac{(x^j_{i+1} - x^j_{i})^2}{2!} \frac{\partial^2 h}{\partial x^2} + \\ & \frac{(x^j_{i+1} - x^j_{i})^3}{3!} \frac{\partial^3 h}{\partial x^3} + ...
\end{split}
\label{eq:ch11_eq20}
\end{equation}
and for points is located in the negative $x$ direction of $h(i,j)$ by
\begin{equation}
\begin{split}
h^j_{i-1} = & h^j_{i} - (x^j_{i-1} - x^j_{i}) \frac{\partial h}{\partial x} + \frac{(x^j_{i-1} - x^j_{i})^2}{2!} \frac{\partial^2 h}{\partial x^2} - \\  & \frac{(x^j_{i-1} - x^j_{i})^3}{3!} \frac{\partial^3 h}{\partial x^3} + ...
\end{split}
\label{eq:ch11_eq21}
\end{equation}
Adding the two equations and noting that for our regular grid 
\begin{equation}
(x^j_{i+1} - x^j_{i}) =  \Delta x   
\end{equation}
\begin{equation}
(x^j_{i-1} - x^j_{i}) = -\Delta x   
\end{equation}

results in
\begin{equation}
h^j_{i+1} + h^j_{i-1} = 2h^j_{i}  + (\Delta x)^2 \frac{\partial^2 h}{\partial x^2} +  \mathcal{O}((\Delta x)^3)
\end{equation}
In which $\mathcal{O}((\Delta x)^n)$ are higher order terms represented using the big ``o'' notation $\mathcal{O}$. If we ignore error terms with order greater than two, which tends to be increasingly smaller as the order of the term increases, the approximation to the second order derivative in $x$ is    
\begin{equation}
 \frac{\partial^2 h}{\partial x^2} = \frac{h^j_{i+1} + h^j_{i-1} - 2h^j_{i}}{(\Delta x)^2} 
\end{equation}
The same procedure results in the approximation in $y$
\begin{equation}
 \frac{\partial^2 h}{\partial y^2} = \frac{h^{j+1}_{i} + h^{j-1}_{i} - 2h^{j}_{i}}{(\Delta y)^2} 
\end{equation}
Replacing the approximations into the Laplace equation and solving for $h^j_i $ allows us to approximate the values of the total potential inside the domain - given the boundary conditions of fixed total potentials at the upper, lower, right and left boundaries - using the formula
\begin{equation}
 h^j_{i} = \frac{1}{4} [h^j_{i+1} + h^j_{i-1} + h^{j+1}_{i} + h^{j-1}_{i}]
\end{equation}
The solution is obtained by numerical methods. A program for solving the 2D Laplace equation using the Jacobi iterative method is presented in Appendix \ref{app:appC}. A surface plot \footnote{The figure was produced with Gnuplot: \url{http://www.gnuplot.info/}.} of the solution for $h_{left} =  20~cm$, $h_{right} =  1~cm$,  $h_{top} =  5~cm$ and $h_{bottom} =  0~cm$ is presented in Figure \ref{ch11_fig3}.

\subsection{Explicit method for 1D advective-dispersion}

\begin{figure}[ht]
\centering
 \includegraphics[width=0.80\textwidth]{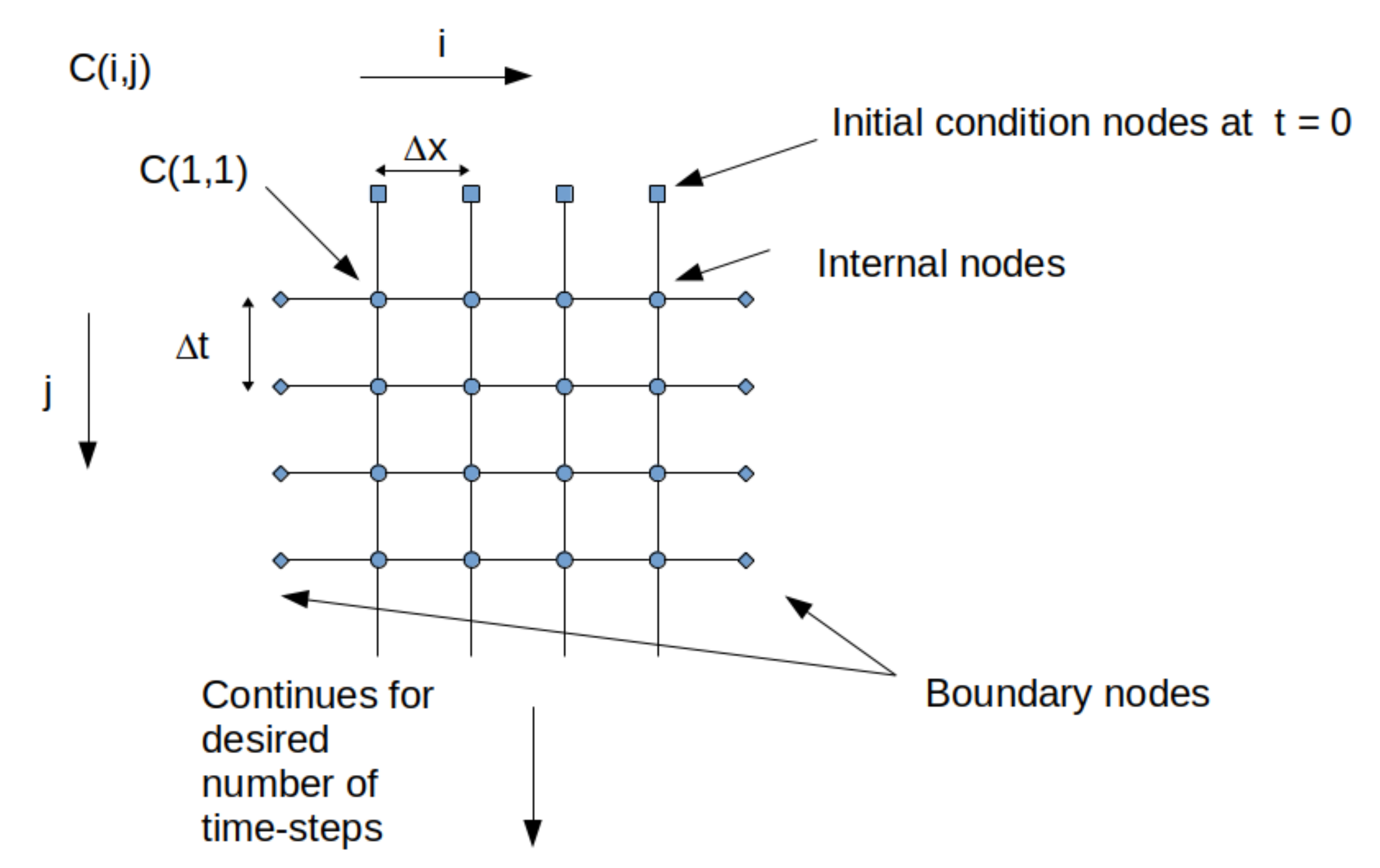}
	\caption{Discretization for the finite differences solution of the 1D advective-dispersive equation.}
\label{ch11_fig4}
\end{figure}

The discretization of an 1D partial differential equation in time and space follows a similar scheme as 2D discretization in space. In diffusion type equations (e.g. advective-dispersive, Richards-Richardson, heat equation) besides the second order partial derivative in space, there is a first order derivative in time. Recall the 1D advective-dispersive equation for constant flux velocity and diffusion coefficient
\begin{equation}
\frac {\partial C}{\partial t} = D \frac {\partial C}{\partial x} - v \frac {\partial^2 C}{\partial x^2}
\end{equation}
We already know from the previous section that the second order partial derivatives in space can be approximated by
\begin{equation}
 \frac{\partial^2 C}{\partial x^2} = \frac{C^j_{i+1} + C^j_{i-1} - 2C^j_{i}}{(\Delta x)^2} 
\end{equation}
Obviously for concentration $C$ in this case. To find the approximation of first order partial derivatives, the Taylor series expansion equations as in Equations \ref{eq:ch11_eq20} \ref{eq:ch11_eq21} are subtracted.  Ignoring higher order terms and solving for the partial derivatives of interest, the discretization of the first order partial derivatives in space and time for the advective-dispersive equation are
\begin{equation}
 \frac{\partial C}{\partial x} = \frac{C^j_{i+1} + C^j_{i-1}}{(2\Delta x)} 
\end{equation}
\begin{equation}
 \frac{\partial C}{\partial t} = \frac{C^j_{i+1} + C^j_{i-1}}{(2\Delta t)} 
\end{equation}
The discretization is now in space and time, where $i$ is the increment in
space and $j$ the increment in time. Care must be taken in writing programs
regarding the order of $i$ and $j$ in loops.   For the discretization in time,
the condition at $t=0$ is now the initial contition and the conditions in the
inlet and outlet of the 1D domain are the boundary conditions (Figure \ref{ch11_fig4}).

\begin{figure}[ht]
\centering
 \includegraphics[width=0.80\textwidth]{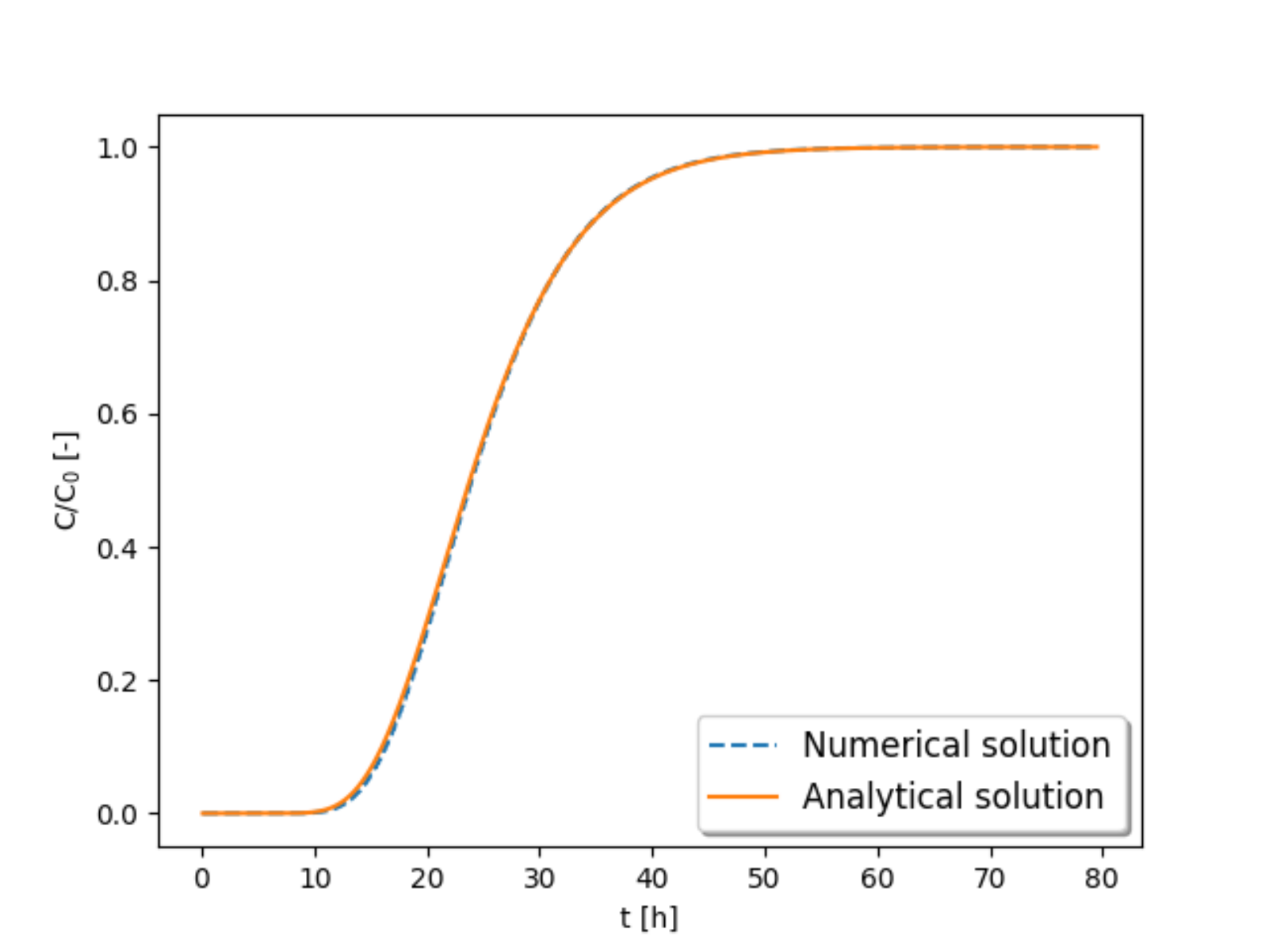}
	\caption{Comparison of numerical and analytical solutions of the 1D advective-dispersive equation from the program in Appendix \ref{app:appD}.}
\label{ch11_fig5}
\end{figure}

By replacing the partial derivatives in time and space, the finite differences equation to be solved for the 1D advective-dispersive equation is
\begin{equation}
C^j_{i} = \Delta t C^{j-1}_{i}  +  D\frac{\Delta t}{\Delta x^2}  (C^{j-1}_{i-1}   - 2C^{j-1}_{i} + C^{j-1}_{i+1} ) -   v\frac{\Delta t }{2 \Delta x} (C^{j-1}_{i+1} - C^{j-1}_{i-1}  ) 
\end{equation}
One of the issues of the numerical solution of the advective-dispersive equation is that it is valid for infinitely long columns. To circumvent this issue a program was written to solve the equation for a long column, but the outlet concentration measurements were made a fixed point along the length $x$. The comparison of the numerical and analytical solutions of the advective-dispersive equation for concentration measured at $50~cm$ from the inlet is presented in Figure \ref{ch11_fig5}. The column length for the numerical simulations was $L = 100~cm$, the flow velocity was $v = 2.0~cm~h^{-1} $, the dispersion coefficient $D = D_h/\theta = 5.0~cm^2~h^{-1} $ and the simulation time was $100~h$. The program is listed in the Appendix  \ref{app:appD}.

\subsection{Explicit method for 1D Richardson-Richards equation}
\begin{figure}[ht]
\centering
 \includegraphics[width=0.80\textwidth]{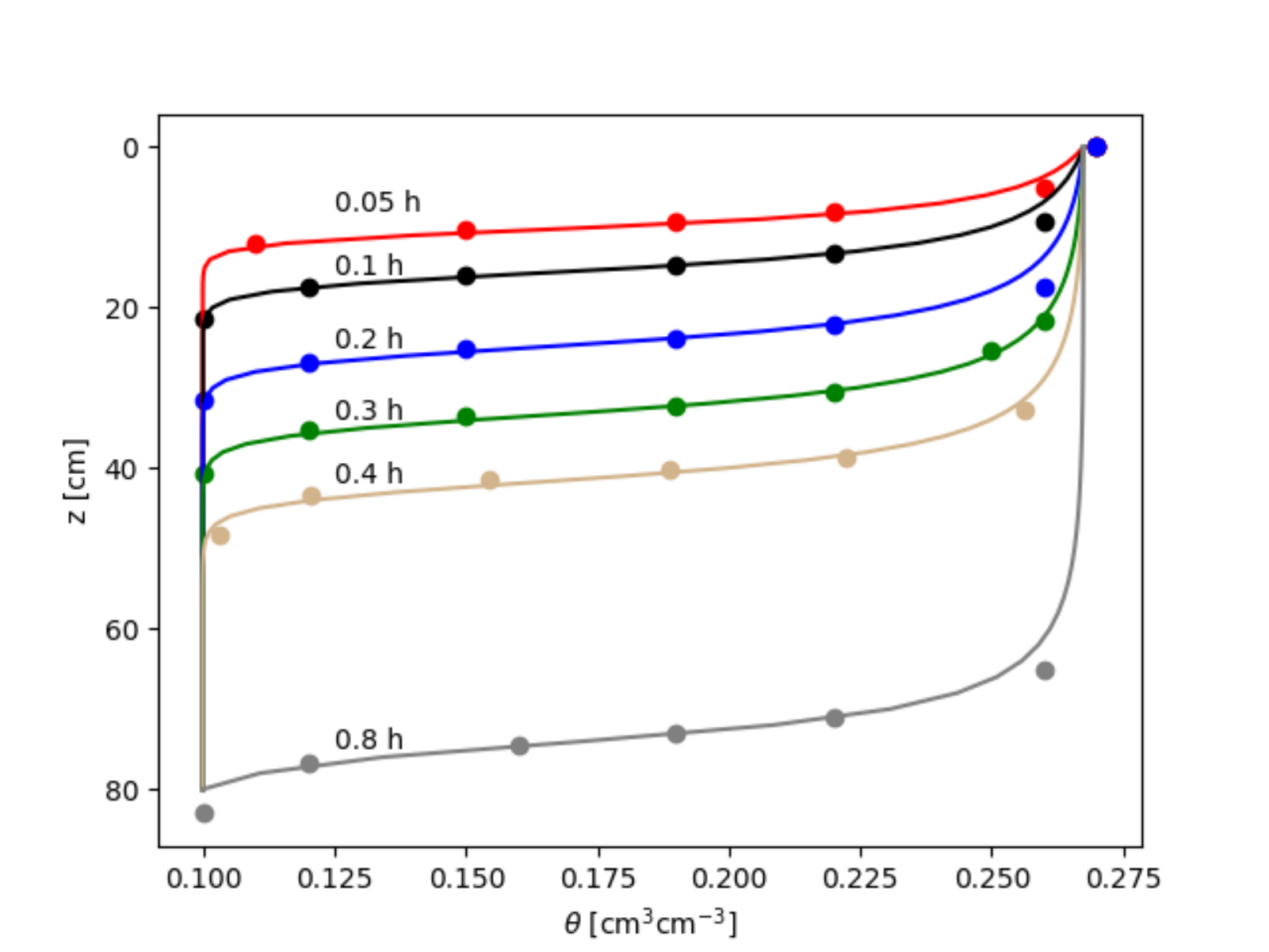}
	\caption{Comparison of numerical and quasi-analytical solutions of the  1D Richardson-Richards equation from the program in Appendix \ref{app:appE}. The quasi-analytical solution follows Philip's method and is presented in \cite{haverkamp77}.}
\label{ch11_fig6}
\end{figure}

By the same principles illustrated for the Laplace and advective-dispersive equations, the Richardson-Richards equation can be discretized to be solved by explicit methods. Recall the 1D matric potential (here represented as $\psi$ for simplicity) form of the Richardson-Richards equation in the general coordinate $z$ \cite{haverkamp77}

\begin{align*}
	C^*(\psi)\frac{\partial \psi}{\partial t} =  \frac{\partial}{\partial z} [K(\psi) (\frac{\partial \psi}{\partial z} + 1 )]  
\end{align*}
The main issue here is that the unsaturated hydraulic conductivity is not constant and cannot be placed outside of the derivative terms. The hydraulic capacity function also needs to be defined for this equation to be solved numerically. Following the discussion in Chapter \ref{ch9} we will define the following water retention and unsaturated hydraulic conductivity functions \cite{haverkamp77}

\begin{equation}
	\theta(\psi) =   \theta_r + \frac{\alpha (\theta_s - \theta_r)}{\alpha + |\psi|^{\beta}}
\end{equation}
\begin{equation}
	K(\psi) =   \frac{A}{A + |\psi|^B}
\end{equation}
Recalling that the hydraulic capacity function is 
\begin{equation}
	C^*(\psi) =   [\alpha h_m  (\theta_s - \theta_r) \psi^{\beta -1}] \frac{\beta (\alpha + \psi ^ {\beta})}{\alpha + \psi^{\beta}} 
\end{equation}
The discretization for the explicit solution is \cite{haverkamp77}
\begin{equation}
\begin{split}
	{\psi}^{j+1}_{i} = & {\psi}^{j}_{i} + \frac{\Delta t}{{C^*}^{j}_{i} \Delta z}  [ K^{j}_{i+1/2}  (\frac{{\psi}^{j}_{i+1} - {\psi}^{j}_{i}}{\Delta z} -1 ) \\& - K^{j}_{i-1/2}  (\frac{{\psi}^{j}_{i} - {\psi}^{j}_{i-1}}{\Delta z} -1 ) ]
\end{split}
\end{equation}
where the unsaturated hydraulic conductivity at intermediary point in nodes is 
\begin{equation}
	 K^{j}_{i+1/2}  = \frac{K^{j}_{i+1}+ K^{j}_{i}}{2} 
\end{equation} 
\begin{equation}
	 K^{j}_{i-1/2}  = \frac{K^{j}_{i}+ K^{j}_{i-1}}{2} 
\end{equation} 
Thus, the numerical solution can be achieved by applying the discretization schemes above while simultaneously evaluating the functions $K(\psi)$ and $^*C(\psi)$ at the $\psi^j_i$ points. The grid is analogous to the one described for the advective-dispersive equation. A program for solving the 1D Richardson-Richards equation using an explicit scheme is presented in Appendix \ref{app:appE}. The results for the parameters $\alpha = 1.611 \times 10^{6}~cm^{-1}$, $\theta_s = 0.287~cm^3~cm^{-3} $, $\theta_r = 0.075~cm^3~cm^{-3}= $ and $\beta = 3.96$ are compared to the Philip's quasi-analytical solution presented by \cite{haverkamp77} for an $80~cm$ 1D soil column initially at $\psi = -61.5~cm$ ($t=0$) and with $\psi = -20.73~cm$ at soil surface and $\psi = -61.5~cm$ at the bottom of the column (or profile) for $t > 0$ (Figure \ref{ch11_fig6}). The unsaturated hydraulic conductivity parameters were $ A = 0.00944$ and $B = 4.74$.   

\subsection{Implicit methods}
Although explicit methods are mathematically simpler and easier to implement from the computational point of view, they are notoriously unstable for most real world problems. The discretization scheme is then limited by the stability of the method. For most application \emph{implicit methods} are preferred, these methods result in a set of equations which needs to be solved using matrix manipulation algorithms. In the implicit discretization of the 1D equation, there is no simple equation which can be inserted into a code to solve the problem, a simultaneous solution for the different nodes is sought. Two popular choices of discretization are the Backward Time Centered Space (BTCS) 
\begin{equation}
	(-\frac{\gamma}{\Delta x^2}) u^{j+1}_{i-1} + (\frac{1}{\Delta
	t} + \frac{2\gamma}{\Delta x^2}) u^{j+1}_{i} + (-\frac{\gamma}{\Delta
	x^2}) u^{j+1}_{i+1} = \frac{1}{\Delta t} u^{j}_{i}  
\end{equation} 
and the Crank-Nicolson scheme \cite{crank75} \footnote{A very good introduction to these methods can be found in \url{https://web.cecs.pdx.edu/~gerry/class/ME448/} from which these discretization schemes were transcribed.}.
\begin{equation}
\begin{split}
	(-\frac{\gamma}{2\Delta x^2}) u^{j+1}_{i-1} + (\frac{1}{\Delta t} + & \frac{\gamma}{\Delta x^2}) u^{j+1}_{i} + (-\frac{\gamma}{2\Delta x^2}) u^{j+1}_{i+1} =  \\  
	& \frac{\gamma}{2\Delta x^2}) u^{j}_{i-1} + (\frac{1}{\Delta t} - \frac{\gamma}{\Delta x^2}) u^{j}_{i} + \frac{\gamma}{2\Delta x^2} u^{j}_{i+1}
\end{split}
\end{equation} 
In which $u^j_i$ is the temperature at each grid point, $\gamma = \kappa_T/ (\rho c_p)$ is the thermal diffusivity and the other parameters are the same as previously defined. 

These numerical schemes can be applied to the Richardson-Richards and other partial differential equations. One common strategy of solution is the diagonal matrix solver, often called the \emph{Thomas algorithm} \cite{carnahanetal90}. Technically, the explicit scheme can be thought as an special case of the discretization of the heat equation called the Backward Time Centered Space (BTCS). The Crank-Nicolson and other more complex schemes are recommended over BTCS and (Forward Time Ceneterd Space) FTCS for real world applications. The main advantages of simpler methods, specially the FTCS is that it can be used as a teaching tool for a first introduction to finite differences methods, as it is done in most numerical methods courses and books. For many problems other even more complicated schemes, both mathematically and computationally are employed, namely \emph{Finite Elements}, \emph{Finite Volume} and \emph{Lattice Boltzmann} methods \footnote{A simple introduction to these should be included in further editions.}. Most modern computational fluid mechanics, soil physics and groundwater hydrology numerical codes are based on one or more of these methods.

%\subsection{Implicit Crank-Nicolson method for 1D heat equation}

%\section{Finite elements}

%\section{Finite volume}

%\section{Lattice Boltzmann}

\section{Symbols in this chapter}

\begin{longtable}{ll}
	$ y $ & Generic dependent variable\\
	$ x $ & Generic independent variable\\
	$ a, b, c, d, e  $ & Generic fitting parameters \\
	$ y_0  $ & Generic $y$ coordinate \\
	$ x_0  $ & Generic $x$ coordinate \\
	$ \Theta $ & Effective saturation $ = (\theta - \theta_s)/(\theta_s - \theta_r)$\\
	$ \theta $ & Volumetric water content\\
	$ \theta_s $ & Saturation volumetric water content\\
	$ \theta_r $ & Residual  volumetric water content\\
	$ A, B $ & Unsaturated hydraulic conductivity function empirical parameters \\
	$ \psi_m $ & Matric potential in unspecified units\\
	$ SSQ $ & Sum of squares of the residuals \\
	$ \boldsymbol{\Omega} $ & Parameter vector \\
	$ \mathbf{ y} $ & Observed dependent variable values vector \\
	$ \mathbf{ x} $ & Observed independent variable values vector \\
	$ \boldsymbol{ \delta} $ & Step in nonlinear regression procedures \\ %, i.e. Gauss-Newton, Levenberg-Marquardt, ... \\
	$ \mathbf{ J} $ & Jacobian matrix \\
	$ \mathbf{ r} $ & Predicted values vector  \\ % check this
	$ \mathbf{ D} $ & Diagonal matrix  \\
	$  \phi $ & Iteration level  \\
	$ \eta $ & Damping factor  \\
	$ SR $ & Soil mechanical resistance to penetration   \\
	$ \rho $ & Bulk density \\
	$ h $ & Total potential in length units \\
	$ i, j$ & Integer variables for step in space and/or time \\ 
	$ \Delta x,  \Delta y, \Delta t $ & Discretization interval in time and space  \\
	$ C $ & Concentration \\
	$ v $ & Flow velocity \\
	$ D $ & Dispersion coefficient corrected to volumetric water content $= D_h/\theta$\\
	$ t $ & Time \\
	$ h_m $ & Matric potential in length units  \\
	$ \psi $ & Matric potential in length units  \\
	$ C^*(h_m) $ & Hydraulic capacity function $ = \partial \theta/\partial h_m$  \\
	$ \theta(h_m) $ & Water retention function  \\
	$ K(h_m) $ & Unsaturated hydraulic conductivity function \\
	$ \alpha,  \beta $ & Water retention function fitting parameters \\
	$ u$ & Temperature at specified grid points \\
	$ y^j_i$ & Specifies the value of the property at $i$ and $j$ for any function $y$ \\
	$ \gamma $ &  Thermal diffusivity \\
	$ \kappa_T $ & Thermal conductivity \\
	$ c_p$ & Specific heat capacity
\end{longtable}

% !TEX TS-program = pdflatex
% !TEX encoding = UTF-8 Unicode

% Example of the Memoir class, an alternative to the default LaTeX classes such as article and book, with many added features built into the class itself.

\appendix

\chapter*{Appendix A - Greek alphabet}
\label{app:appA}

$\alpha$, $A $: alpha  \\
$\beta$, $B $: beta \\
$\gamma$, $\Gamma $: gamma  \\
$\delta$, $\Delta $: delta \\
$\epsilon$,$\varepsilon$, $E $: epsilon   \\
$\zeta$, $ Z $: zeta \\
$\eta$, $H $: eta \\
$\theta$, $\Theta $: theta  \\
$\iota$, $I $: iota \\
$\kappa$, $ K $: kappa  \\
$\lambda$, $\Lambda $: lambda \\
$\mu$, $ M $: mu  \\
$\nu$, $ N $: nu  \\
$\xi$, $ \Xi $: xi  \\
$o$, $ O $: omicron  \\
$\pi$, $ \Pi $: pi \\
$\rho$, $ R $: rho  \\
$\sigma$, $ \varsigma$, $ \Sigma $: sigma \\
$\tau$, $ T $: tau \\
$\upsilon$, $ Y $: upsilon  \\
$\phi$, $\varphi$, $ \Phi $: phi  \\
$\chi$, $X$: chi  \\
$\psi$,  $ \Psi $: psi \\
$\omega$, $\Omega$, $ \Phi $: omega  \\

\chapter*{Appendix B - Program for fitting the Brutsaert-van Genuchten equation using the Levenberg-Marquardt method}
\label{app:appB}

\begin{spverbatim}
## T.P. Leão in Oct 5 2021, light revision in Aug  29 2022 
## Fits the Brutsaert-van Genuchten equation to data using the Leverberg-Marquardt algorithm
## Language: Python 3.8.10
## This code does not check for regression diagnostics

### Import the necessary packages
import numpy as np
from scipy.optimize import curve_fit, least_squares
import matplotlib.pyplot as plt

### Data
psi = np.array([20, 40,	60, 80, 100, 200, 500, 800, 1000, 3000, 5000, 15000])
theta = np.array([0.41,	0.409, 0.406, 0.41268, 0.41045, 0.39708, 0.3899, 0.3525, 0.3203, 0.18478, 0.14497, 0.13])

### Define initial parameters (Initial guess for the nonlinear regression procedure)
thetas = 0.4
thetar = 0.1
alpha = 0.001
n = 2.0
m = 0.5

### Defines the nonlinear regression model to be fit
def vg(x, qs, qr, alpha, n, m):
     return qr + (qs - qr)/(1+ (alpha*psi)**n)**m

### Puts the initial parametes into the "InitialParams" vector
InitPar = [thetas, thetar, alpha, n, m]

### Curve fitting procedure - "lm" is the Leverberg-Marquardt algorithm
popt, pcov = curve_fit(vg, psi, theta, p0=InitPar, method='lm')

### Creates values of psi for plotting the predicted data    
xd2 = np.linspace(10, 30000, 10000)

### Uses the parameters from the fit to create the data for the predicted plot
thetap2 = popt[1] + (popt[0] - popt[1])/(1+ (popt[2]*xd2)**popt[3])**popt[4] 

### computes the standard deviation of the parameters
perr = np.sqrt(np.diag(pcov))

print("Best fit paramaters, thetas, thetar, alpha, n and m \n", popt)     
print("Covariance of the parameters \n", popt)     
print("Standard deviation of the parameters \n", perr)

### Plot of the data and best fit line
plt.plot(psi, theta, "o", label = "Simulated data")
plt.plot(xd2, thetap2, ":", label = "van Genuchten (1980)")
plt.ylim(0, 0.45)
plt.ylabel('Volumetric water content [vol/vol]')
plt.xlabel('Matric potential [|cm|]')

legend = plt.legend(loc='lower left', shadow=True, fontsize='large')
plt.xscale("log")
plt.show()
\end{spverbatim}

\chapter*{Appendix C - Program for solving the 2D Laplace equation}
\label{app:appC}

\begin{spverbatim}
// Solution to the laplace equation 2D using an explicit finite differences scheme
// T.P. Leao on 5 out 2021 light revision on Aug 29 2022
// Language: C++

#include<iostream>
#include<cmath>
#include<iomanip>
#include<stdlib.h>

int main(){

// define the simulation parameters

double  dx = 5.0;  
double  dy = 5.0; 
double  X = 100.00;  // x length  
double  Y = 100.00;  // y length

int  nx = X/dx+1;
int  ny = Y/dy+1;

int i, j;

// define the grid 

double h[nx][ny];  

// initialize the array with zeros

for(i=0; i < nx; i++){
for(j=0; j < ny; j++){
	 h[i][j] = 0.0;
}}


std::cout << "-------------------------------------" <<  std::endl;

// define the boundary conditions

for(i=0; i < nx; i++){
		h[i][0] = 5.0; 
		h[i][ny-1] = 1.0; 
	}

	for(j=1; j < ny; j++){
		h[0][j] = 20.0; 
		h[nx-1][j] = 0; 
	}

// Solves the system 

double d[1000], diff;
double tol = 1e-5;



std::cout << "Iterations and diff " << std::endl;	
for (int iter = 0; iter <100; iter++) { // Maximum number of iterations 
	for(i=1; i < nx-1; i++){
		for(j=1; j < ny-1; j++){
			 h[i][j] = 0.25 * (h[i+1][j] + h[i-1][j] + h[i][j+1] + h[i][j-1]);
			 d[iter] = h[1][1];
			 diff = d[iter] - d[iter-1]; 	
			}  
		 } if (diff < tol) {break;} 
		   std::cout << iter << "\t" <<   diff << std::endl; // Print iteration and difference between sucessive iterations at h[1][1]	  
	}
	

std::cout << "Solution in gnuplot format " << std::endl;

for(i=0; i < nx; i++){ 
	for(j=0; j < ny; j++){
			 std::cout << std::setprecision(6) << h[i][j] << "\n"; // Prints solution in terminal
		}std::cout << std::endl;} 

 
return 0;

}
\end{spverbatim}

\chapter*{Appendix D - Program for solving the 1D advective-dispersive equation}
\label{app:appD}

\begin{spverbatim}
// Solution to the advective dispersive 1D equation using an explicit method
// T.P. Leao 01 out 2021 light revision 29 aug 2022
// Language: C++

#include<iostream>
#include<cmath>
#include<iomanip>
#include<stdlib.h>

int main(){

// define the simulation parameters

double dx = 1.0;  
double dtau = 0.1; 
double TMAX = 500.0; 
double L = 100.0;  
double D = 5;
double V = 2.0;

int  nx = L/dx+1;
int  ntau = TMAX/dtau+1;

int i, j;

// define the grid 

double C[ntau][nx];  

// initialize the array with zeros

for(i=0; i < nx; i++){
for(j=0; j < ntau; j++){
	 C[j][i] = 0.0;
}}


std::cout << "-------------------------------------" <<  std::endl;

// define the initial conditions

for(i=0; i < nx; i++){
	C[0][i] = 0.0; 
}

// define the boundary conditions 
//
for(j=1; j < ntau; j++){
	C[j][0] = 1.0;
	C[j][nx-1] = 0.0;
}

// this is the explicit solution 
for(j=0; j < ntau-1; j++){
for(i=1; i < nx-1; i++){
		 C[j+1][i] = C[j][i] + dtau/pow(dx,2)*D*(C[j][i-1] -2*C[j][i] +C[j][i+1]) - V *dtau/(2*dx)*(C[j][i+1] - C[j][i-1]); 
} } 
	

std::cout << " Explicit solution to the 1D advective-dispersion equation in time   ";
std::cout << "\n";
std::cout << " _____________________________________________   ";
std::cout << "\n";


 for(j=0; j < ntau; j++){ std::cout << "|  "; 
for(i=0; i < nx; i++){
		 std::cout << C[j][i] << " " ;
		 }
} 

 
return 0;

}
\end{spverbatim}

\chapter*{Appendix E - Program for solving the 1D Richardson-Richards equation using explicit method}
\label{app:appE}

\begin{spverbatim}
// Solves the 1D richards equatio using and explicit finite differences scheme
// This code has been validaded using Haverkamp 1977 analytical solutions using Philip's model 
// The discretization also follows Haverkamp 77
// In 2 out 2021 - TPL, light revision 29 aug 2022
// Language: C++

#include<iostream>
#include<cmath>
#include <fstream>
#include <cstdlib> 

// static parameters for Warwick hydraulic functions, see Zarba 1988 MIT Masters 
	#define	thetas  0.287
	#define	thetar  0.075
	#define	alpha  1.611e06
	#define	beta1  3.96

	#define	Ks 9.44e-3 
	#define	A  1.175e06
	#define	B  4.74

using namespace std;

// define the functions for water retention and hydraulic conductivity
double theta(double h){
	double theta;
	theta = alpha * (thetas-thetar)/(alpha+ pow(abs(h), beta1)) + thetar;
	return theta;
}

double K(double h){
	double K;
	K = Ks * A/(A + (pow(abs(h), B)));
	return K;
}

double C(double h){
	double C;
	C = alpha*(thetas-thetar) * pow(abs(h), beta1-1) * beta1/((alpha+ pow(abs(h), beta1))* (alpha+ pow(abs(h), beta1)));
	return C;
}

int main(){


double dz = 1.0;
double dt = 0.4;

int j, i;

double L = 80.0;
double tmax = 360;

int nx = L/dz + 1;
int ntau = tmax/dt + 1;

double h[ntau][nx];//, K[ntau][nx], C[ntau][nx], q[ntau][nx];

for (j = 1; j < ntau; j++){
	h[j][0] = -20.73;
	h[j][nx-1] = -61.5;
	}


for (i = 0; i < nx; i++){
	h[0][i] = -61.5;
}

for (j=0; j < ntau-1; j++){
	for (i=1; i < nx-1; i++){

h[j+1][i] = h[j][i] + dt*1.0/(C(h[j][i])*dz)   * (  (K(h[j][i+1])+K(h[j][i]) )/2.0 * ((h[j][i+1]-h[j][i])/dz  -1.0 )
-  (K(h[j][i])+K(h[j][i-1]) )/2.0 * ((h[j][i]-h[j][i-1])/dz  - 1.0 ));

	}
}


for (j = 0; j < ntau; j++){
for (i = 0; i < nx; i++){
cout  <<  i << "\t" << h[j][i]<<"\n";
} cout << endl;
}

// saves output to *.dat file 

	 std::ofstream outdata; 
	 std::ofstream output;
	 outdata.open("rich1d.dat");

for (int j=0; j < ntau; j++){outdata <<  std::endl;	//std::cout <<  j << std::endl;
for (int i=0; i < nx; i++){

outdata <<  h[j][i] <<"\n";
	}	

} 

outdata.close();


cout << endl;
cout << "---------------------------" << endl;
cout << "Results saved to rich1d.dat" << endl;


return 0;
}\end{spverbatim}


\addcontentsline{toc}{chapter}{Bibliography}

\begin{thebibliography}{}


\bibitem{acheson}
Acheson, D.J. 1990. \textit{Elementary fluid dynamics}. Oxford University Press.  

\bibitem{alazard}
Alazard, T.; Lazar, O. 2020. Paralinearization of the muskat equation and application to the Cauchy problem. \textit{Archive for 
Rational Mechanics and Analysis}. 237:545–583. doi.org/10.1007/s00205-020-01514-6  

\bibitem{andraskiscanlon}
Andraski, B.J.; Scanlon, B.R. 2002. Thermocouple psychormetry. \emph{In.}  Dane, J.H.; Topp, G.C. (eds.) \textit{Methods of Soils Analysis: Part 4 - Physical Methods}.  pp. 609-642, Soil Science Society of America, Madison Wisconsin, USA.   

\bibitem{aris}
Aris, R.J. 1962. \textit{Vectors, tensors and the basic equations of fluid mechanics}. Dover Publications, Inc. 286p.  

\bibitem{asmar}
Asmar, N.H. 2000. \textit{Partial Differential Equations with Fourier Series and Boundary Value Problems}. 2nd ed., Pearson Education Inc.

\bibitem{babcock}
Babcock, K.L. 1963. Theory of the chemical properties of soil colloidal systems at equilibrium. \textit{Hilgardia}. 34(11):417-542. doi.org/10.3733/hilg.v34n11p417   

\bibitem{babcockoverstreet}
Babcock, K.L.; Overstreet, R. 1955. Thermodynamics of soil moisture. \textit{Soil Science}. 80(4):257-364. doi.org/10.1097/00010694-195510000-00002   

\bibitem{barenblattetal}
Barenblatt, G.I.; Entov, V.M.; Ryzhik, V.M. 1990. \textit{Theory of fluid flows through natural rocks}. Theory and Applications of Transport in Porous Media, Volume 3. Kluwer  

\bibitem{batchelor}
Batchelor, G.K. 1967. \textit{An introduction to fluid dynamics}. Cambridge University Press, New York.  

\bibitem{bateswatts88}
Bates, D.M.; Watts, D.G. 1988. \textit{Nonlinear regression analysis and its applications}. John Wiley \& Sons.  365p.

\bibitem{bear}
Bear, J. 1972. \textit{Dynamics of fluids in porous media}. Dover Publications, Inc. New York. 764p.  

\bibitem{bearbachmat}
Bear, J.; Bachmat, Y. 1990. \textit{Introduction to modeling transport phenomena in porous media}. Theory and Application of Transport in Porous Media. Volume 4.  Kluwer Academic Publishers. 553p.  

\bibitem{bearcheng}
Bear, J.; Cheng, A.H.-D. 2010. \textit{Modeling groundwater flow and contaminant transport}. Theory and Application of Transport in Porous Media. Volume 23.  Springer. 834p.  

\bibitem{birdetal02}
Bird, R.B.; Stewart, W.E.; Lightfoot, E.N. 2002. \textit{Transport phenomena}. 2nd ed. John Wiley \& Sons Inc. 895p.  

\bibitem{boas}
Boas, M.L. 1980. \textit{Mathematical methods for the physical sciences}. 3rd ed. Wiley India Pvt. Ltd., New Delhi.  

\bibitem{boltfrissel}
Bolt, G.H.; Frissel, M.J. 1960. Thermodynamics of soil moisture. \textit{NJAS - Wagening Journal of Life Sciences}. 8(1):57-78. doi: 10.18174/njas.v8i1.17663   

\bibitem{boltzmann1894}
Boltzmann, L. 1894. Zur  Integration  der  diffusionsgleichung bei variablen diffusionskoeffizienten. \textit{Annalen der Physik}.  289(13):959 - 964. doi: 10.1002/andp.18942891315   

\bibitem{borchardt}
Borchardt, G. 1989. Smectites. \emph{In.}  Dixon, J.B.; Weed, S.B. (eds.) \textit{Minerals in Soil Environments}. 2nd ed., pp. 675-727, Soil Science Society of America, Madison Wisconsin, USA.   

\bibitem{bragg}
Bragg, W.H.; Bragg, W.L. 1913. The reflection of X-rays by crystals. \textit{Proceedings of the Royal Society A}. 88(605):428-438. doi.org/10.1098/rspa.1913.0040 

\bibitem{brookscorey}
Brooks, R.H.; Corey, A.T. 1964. Hydraulic properties of porous media. Hydrology Papers. Colorado State University. Fort Collins, Colorado. 

\bibitem{brown}
Brown, G.O. 2002. Henry Darcy and the making of a law. \textit{Water Resources Research}. 38(7):1106, doi: 10.1029/2001WR000727  

\bibitem{bruceklute56}
Bruce, R.R.; Klute, A. 1956. The measurement of soil moisture diffusivity. \textit{Soil Science Society of America Proceedings}. 20(4):458-462. doi: 10.2136/ssaj1956.03615995002000040004x 

\bibitem{brutsaert}
Brutsaert, W. 1966. Probability laws for pore-size distributions. \textit{Soil Science}. 101(2):85-92. doi: 10.1097/00010694-196602000-00002 

\bibitem{buckingham}
Buckingham, E. 1907. Studies on the movement of soil moisture. United States Department of Agriculture. Washington. Government Printing Office. 61p. 

\bibitem{burdine}
Burdine, N.T. 1953. Relative permeability calculations from pore size distribution date. \textit{Journal of Petroleum Technology}. 5(03):71-78. doi: 10.2118/225-G
 
\bibitem{callen}
Callen, H.B. 2005. \textit{Thermodynamics and an introduction to thermostatistics}. 2nd ed. Wiley India Pvt. Ltd., New Delhi.  

\bibitem{campbell}
Campbell, G.S. 1974. A simple method for determining unsaturated conductivity from moisture retention data. \textit{Soil Science}. 117(6):311-314. doi: 10.1097/00010694-197406000-00001   

\bibitem{carnahanetal90}
Carnahan, B.; Luther, H.A.; Wilkes, J.O. 1990. \textit{Applied numerical methods}. Krieger Publishing Company.  604p. 

\bibitem{childscollisgeorge48}
Childs, E.C.; Collis-George, N. 1948. The geometry of the soil-water equilibria. \textit{Discussions of the Faraday Society}. 3:78-85. doi: 10.1039/df9480300078 

\bibitem{childscollisgeorge50}
Childs, E.C.; Collis-George, N. 1950. The permeability of porous materials. \textit{The Royal Society Publishing}. 201(1066):392-405. doi: 10.1097/00010694-197406000-00001   

\bibitem{craig}
Craig, R.F. 2004. \textit{Craig's soil mechanics}. 7th ed. Taylor \& Francis. 447p.   

\bibitem{crank75}
Crank, J. 1975. \textit{The mathematics of diffusion}. 2nd ed. Clarendon Press. 414p.   

\bibitem{danetopp}
Dane, J.H.; Topp, G.C. (eds.) \textit{Methods of Soils Analysis: Part 4 - Physical Methods}.  Soil Science Society of America, Madison Wisconsin, USA.   

\bibitem{darcy}
Darcy, H. 1856. \textit{Les fontaines publiques de la ville de dijon}. Librarie des Corps Impériaux des Ponts et Chaussées et des Mines.  

\bibitem{das}
Das, B.M. 2008. \textit{Advanced soil mechanics}. 3rd ed. Taylor \& Francis. 567p.  

\bibitem{defay}
Defay, R.; Prigogine, I.; Bellemans, A. 1966. \textit{Surface tension and adsorption}. Longmans Green and Co. Ltd.  

\bibitem{dixon}
Dixon, J.B. 1989. Kaolin and serpentine group minerals. \emph{In.}  Dixon, J.B.; Weed, S.B. (eds.) \textit{Minerals in Soil Environments}. 2nd ed., pp. 467-525, Soil Science Society of America, Madison Wisconsin, USA.   

\bibitem{essington03}
Essington, M.E. 2003. \textit{Soil and water chemistry: An integrative approach}. CRC Press. 534p. 

\bibitem{fetter}
Fetter, C.W. 2000. \textit{Applied hydrogeology}. 4th ed. CRC Preentice Hall. 598p. 

\bibitem{fetteretal18}
Fetter, C.W.; Boving, T.; Kreamer, D. 2018. \textit{Contaminant hydrogeology}. 3rd ed. Waveland Press, Inc. 647p. 

\bibitem{gast}
Gast, R.G. 1977. Surface and colloid chemistry. \emph{In.}  Dixon, J.B.; Weed, S.B. (eds.) \textit{Minerals in Soil Environments}. pp. 27-74, Soil Science Society of America, Madison Wisconsin, USA.   

\bibitem{geeor}
Gee, G.W.; Or, D. 2002. Particle size analysis. \emph{In.}  Dane, J.H.; Topp, G.C. (eds.) \textit{Methods of Soils Analysis: Part 4 - Physical Methods}.  pp. 255-294, Soil Science Society of America, Madison Wisconsin, USA.   

\bibitem{groeneveltparlange}
Groenevelt, P.H.; Parlange, J.-H. 1974. Thermodynamic stability of swelling soils. \textit{Soil Science}. 118(1):1-5. doi.org/10.1097/00010694-197407000-00001 

\bibitem{guggenheim}
Guggenheim, E.A. 1967. \textit{Thermodynamics: An advanced treatment for chemists and physicists}. Elsevier Science Publishers B.V. 

\bibitem{grim}
Grim, R.E. 1953. \textit{Clay mineralogy}. McGraw-Hill. 384p. 

\bibitem{hammond}
Hammond, C. 2015. \textit{The basics of crystallography and diffraction}. International Union of Crystallography Book Series. 519p.  

\bibitem{hignettevett}
Hignett, C.; Evett, S.R. 2002. Neutron thermalization. \emph{In.}  Dane, J.H.; Topp, G.C. (eds.) \textit{Methods of Soils Analysis: Part 4 - Physical Methods}.  pp. 501-521, Soil Science Society of America, Madison Wisconsin, USA.   

\bibitem{hoffman01}
Hoffman, J.D. 2001. \textit{Numerical methods for engineers and scientists}. 2nd edition revised and expanded. Marcel Dekker, Inc. 823p.  

\bibitem{juryhorton04} 
Jury, W.A.; Horton, R. 2004. \textit{Soil physics}, 6th ed., John Wiley and Sons, Inc. 370p.

\bibitem{jurysposito85}
Jury, W.A.; Sposito, G. 1985.  Field calibration and validation of solute transport models for the unsaturated zone. \emph{Soil Science Society of America Journal}. 49:1331-1341.   doi: 10.2136/sssaj1985.03615995004900060002x

\bibitem{thomas} 
Hass, J.R.; Heil, C.; Weir, M.D. 2017. \textit{Thomas' Calculus Early Transcendentals}, 14th ed., Addison-Wesley Longman Publishing Co., Inc., USA.

\bibitem{haverkamp77} 
Haverkamp, R.; Vauclin, M.; Touma, T.; Wierenga, P.J.; Vachaud, G. 1977. A comparison of numerical simulation models for one-dimensional infiltration. \textit{Soil Science Society of America Journal}. 41:285-294. doi: 10.2136/sssaj1977.03615995004100020024x

\bibitem{huber} 
Huber, M.L.;  Perkins, R.A.; Laesecke, A.;  Friend, D.G.;  Sengers, J.V.;  Assael, M.J.;   Metaxa, I.N.; Vogel, E.;  Mares, R.;  Miyagawa, K. 2009.
New International Formulation for the Viscosity of H$_2$O. \textit{Journal of Physical and Chemical Reference Data}. 38(2):101-125. doi:10.1063/1.3088050

\bibitem{israelachvili} 
Israelachvili, J.N. 2010. \textit{Intermolecular and surface forces}, 3rd ed., Elsevier. 674p.

\bibitem{kirkhampowers}
Kirkham, D.; Powers, W.L. 1972. \textit{Advanced soil physics}. Wiley-Interscience. 534p.  

\bibitem{klute52}
Klute, A. 1952. A numerical method for solving the flow equation for water in unsaturated materials. \textit{Soil Science}. 73(2):105-116. doi: 10.1097/00010694-195202000-00003

\bibitem{knightraats}
Knight, J.; Raats, P. 2016. The contributions of Lewis Fry Richardson to drainage theory, soil physics, and the soil-plant-atmosphere continuum. \textit{Geophysical Research Abstracts}. 18:EGU2016-10980-1  

\bibitem{kosugui} 
Kosugui, K. 1994. Three-parameter lognormal distribution model for soil water retention. \textit{Water Resources Research}. 30(4):891-901. doi: 10.1029/93WR02931

\bibitem{landau}
Landau, L.D.; Lifshitz, E.M. 1987. \textit{Fluid mechanics}. Course on theoretical physics. Volume 6. 2nd ed., Elsevier Buttersworth-Heinemann, Oxford. 539p.  

\bibitem{landaulifshitz1960}
Landau, L.D.; Lifshitz, E.M. 1960. \textit{Electrodynamics of continuous media}. Course on theoretical physics. Volume 8. Pergamon Press. 417p.  

\bibitem{leao2005}
Leão, T.P.; Silva, A.P.; Perfect, E.; Tormena, C.A. 2005. An algorithm for calculating the least limiting water range of soils. \textit{Agronomy Journal}. 97(4):1210-1215. doi:10.2134/agronj2004.0229

\bibitem{leao2013}
Leão, T.P.; Guimarães, T.L.B.; Figueiredo, C.C.; Busato, J.G.; Breyer, H.S. 2013. On critical coagulation concentration theory and grain size analysis of oxisols. \textit{Soil Science Society of America Journal}. 77:1955-1964. doi:10.2136/sssaj2013.06.0211    

\bibitem{leaotuller14}
Leão, T.P.; Tuller, M. 2014. Relating soil specific surface area, water film thickness, and water vapor adsorption. \textit{Water Resources Research}. 50:7873-7885. doi:10.1002/2013WR014941   

\bibitem{leao2019}
Leão, T.P. 2019. Water retention and penetration resistance equations for the least limiting water range. \textit{Scientia Agricola}. 76(2):172-178. doi:10.1590/1678-992x-2017-0280   

\bibitem{leao2020}
Leão, T.P. 2020. Modeling magnetic minerals effect on water content estimation in porous media. \textit{Progress in Electromagnetics Research C}. 106:215-228. doi: 10.2528/PIERC20081405   

\bibitem{leaoetal2020}
Leão, T.P.; da Costa, B.D.F.; Bufon, V.B.; Aragón, F.F.H. 2020. Using time domain reflectometry to estimate water content of three soil orders under savanna in Brazil. \textit{Geoderma Regional}. 21:e00280. doi: 10.1016/j.geodrs.2020.e00280   

\bibitem{munson}
Munson, B.R.; Young, D.F.; Okiishi, T.H., Huebsch, W.W. 2009. \textit{Fundamentals of Fluid Mechanics}. 6th ed., John Wiley \& Sons, Inc., Hoboken, NJ.

\bibitem{lamb}
Lamb, H. 1932. \textit{Hydrodynamics}. 6th ed., Dover Publications, New York.

\bibitem{mualem} 
Mualem, Y. 1976. A new model for predicting the hydraulic conductivity of unsaturated porous media. \textit{Water Resources Research}. 12:513-522. doi: 10.1029/WR012i003p00513

\bibitem{muskat29} 
 Muskat, M. 1929. The Continuous Spectra of Hydrogen Like Atoms. Dissertation (Ph.D.), California Institute of Technology.  

\bibitem{muskat46} 
 Muskat, M.; Wyckoff, R.D. 1946. \textit{The Flow of Homogeneous Fluids Through Porous Media}. New York: McGraw-Hill. ISBN 978-0934634168. 

\bibitem{myslinska} 
Myslinska, E. 2003. Classification of organic soils for engineering geology. \textit{Geological Quarterly}. 47(1):39-42. doi: 10.1007/s10040-009-0565-5

\bibitem{nielsenetal72} 
Nielsen, D.R.; Jackson, R.D.; Cary, J.W.; Evans, D.D. 1972. \textit{Soil water}. American Society of Agronomy and Soil Science Society of America. 175p. 

\bibitem{nitaobear} 
Nitao, J.J.; Bear, J. 1996. Potentials and their role in transport in porous media. \textit{Water Resources Research}. 32(2):225-250. doi: 10.1029/95WR02715

\bibitem{nutting} 
Nutting, P.G. 1930. Physical analysis of oil sands. \textit{AAPG Bulletin}. 14:1337-1349. doi: 10.1306/3D932938-16B1-11D7-8645000102C1865D

\bibitem{ogatabanks61} 
Ogata, A.; Banks, R.B. 1961. Differential equation of longitudinal dispersion in porous media. Fluid Movement in Earth Materials. Geological Survey Professional Paper 411-A. United States Government Printing Office, Washington. 7p.

\bibitem{pellicer} 
Pellicer, J.; Manzanares, J.A.; Mafé, S.  1995. The physical description of elementary surface phenomena: thermodynamics versus mechanics. \textit{American Journal of Physics}. 63(3):542-547. doi.org/10.1119/1.17866

\bibitem{piazza} 
Piazza, R. 2014. Settled and unsettled issues in particle settling. \textit{Reports on Progress in Physics}. 77(5):056602. doi.org/10.1088/0034-4885/77/5/056602

\bibitem{perrier} 
Perrier, E.; Rieu, M.; Sposito, G.; de Marsily, G. 1996. Models of the water retention curve for soils with a fractal pore size distribution. \textit{Water Resources Research}. 32(10):3025-3031. doi.org/10.1029/96WR01779 

\bibitem{richards31} 
Richards, L.A. 1931. Capillary conduction of liquids through porous mediums. \textit{Journal of Applied Physics}. 1(5):318-333. doi: 10.1063/1.1745010 

\bibitem{richardson} 
Richardson, L.F. 1922.  \textit{Weather prediction by numerical process}. Cambridge at University Press. 236p.:w

\bibitem{robinsonfriedman}
Robinson, D.A.; Friedman, S.P. 2003.  A method for measuring the solid particle permittivity or electrical conductivity of rocks, sediments, and granular materials. \emph{Journal of Geophysical Research}. 108:2076.   doi:10.1029/2001JB000691
 
\bibitem{sauty80}
Sauty, J.P. 1980. An analysis of hydrodispersive transfer in aquifers. \emph{Water Resources Research}. 16(1):145-158. doi:  10.1029/WR016i001p00145

\bibitem{scanlonetal}
Scanlon, B.R.; Andraski, B.J.; Bilskie, J. 2002.  Miscellaneous methods for measuring matric or water potential. \emph{In.}  Dane, J.H.; Topp, G.C. (eds.) \textit{Methods of Soils Analysis: Part 4 - Physical Methods}.  pp. 643-670, Soil Science Society of America, Madison Wisconsin, USA.   

\bibitem{seberwild03}
Seber, G.A.F.; Wild, C.J. 2003.  \textit{Nonlinear regression}. Wiley-Interscience. 768p.  

\bibitem{simmons} 
Simmons, C.T. 2008. Henry Darcy (1803–1858): Immortalised by his scientiﬁc legacy. \textit{Hydrogeology Journal}. 16:1023-1038. doi.org/10.1007/s10040-008-0304-3 

\bibitem{skaggsleij02}
Skaggs, T.H.; Leij, F.J. 2002. Solute transport: Theoretical background. \emph{In.}  Dane, J.H.; Topp, G.C. (eds.) \textit{Methods of Soils Analysis: Part 4 - Physical Methods}.  pp. 1352-1380, Soil Science Society of America, Madison Wisconsin, USA.   

\bibitem{sobieskitrykozko} 
Sobieski, W.; Trykozko, A. 2014. Darcy's and forchheimer's laws in practice. Part 1. The experiment. \textit{Technical Sciences}. 14:321-335. doi.org/10.1007/s10040-009-0565-5

\bibitem{soilsurvey} 
Soil Science Division Staff. 2017. \textit{Soil survey manual}. USDA Handbook 18. Government Printing Office, Washington D.C.

\bibitem{sophocleous} 
Sophocleous, M. 2017. Understanding and explaining surface tension and capillarity: an introduction to fundamental physics for water professionals. \textit{Hydrogeology Journal}. 18:811-821. doi.org/10.1007/s10040-009-0565-5

\bibitem{sparks}
Sparks, D.L. 2003. \textit{Environmental Soil Chemistry}. 2nd ed., Elsevier Science. 352p. 

\bibitem{sposito81} 
Sposito, G. 1981. \textit{Thermodynamics of soil solutions}. Oxford Clarendon Press. 223p.

\bibitem{sposito84} 
Sposito, G. 1981. \textit{The surface chemistry of soils}. Oxford Univeristy Press. 234p.

\bibitem{stokes}
Stokes, G.G. 1901. \textit{Mathematical and Physical Papers}. Volume III, Cambridge University Press.  

\bibitem{symon}
Symon, K.R. 1960. \textit{Mechanics}. 2nd ed., Addison-Wesley Publishing Company, Inc. 557p. 

\bibitem{tanaka} 
Tanaka, M.; Girard, G.;  Davis, R.;  Peuto,  A.;  Bignell, N. 2001. Recommended table for the density of water between 0 $^{o}$C and 40 $^{o}$C based on recent experimental reports.
\textit{Metrologia}. 38:301-309.

\bibitem{taylorashcroft}
Taylor, S.A.; Ashcroft, G.L. 1972. \textit{Physical edaphology: The physics of irrigated and nonirrigated soils}. W.H. Freeman and Company, San Francisco.  

\bibitem{tulleretal99} 
Tuller, M.; Or, D.; Dudley, L.M. 1999. Adsorption and capillary condensation in porous media: Liquid retention and interfacial configurations in angular pores.
\textit{Water Resources Research}. 35(7):1949-1964. doi: 10.1029/1999WR900098

\bibitem{vangenuchten} 
van Genuchten, M.Th. 1980. A closed-form equation for predicting the hydraulic conductivity of unsaturated soils. \textit{Soil Science Society of America Journal}. 44:892-898. doi: 10.2136/sssaj1980.03615995004400050002x

\bibitem{vangenuchtenalves82} 
van Genuchten, M.Th.; Alves, W.J. 1982. Analytical solutions of the one-dimensional convective-dispersive solute transport equation. U.S. Department of Agriculture, Technical Bulletin No. 1661, 151 p. 

\bibitem{vonhippel}
von Hippel, A. 1954. \emph{Dielectrics and waves}. Wiley, Chapman \& Hall. 284p.  

\bibitem{webb06}
	Webb, S.W. 2006.  Gas transport mechanisms. \emph{In.}  Ho, C.K.; Webb, S.W. (eds.) \textit{Gas transport in porous media}.  pp. 5-26, Theory and applications of transport in porous media. Volume 20. Springer.   

\bibitem{wyckoffetal} 
Wyckoff, R.D.; Botset, H.G.; Muskat, M.M.; Reed, D.W. 1933. The measurement of the permeability of porous media for homogeneous fluids. \textit{Review of Scientific Instruments}. 4(7):394-405. doi: 10.1063/1.1749155

\bibitem{youngsisson}
Young,M.H.; Sisson, J.B. 2002. Tensiometry. \emph{In.}  Dane, J.H.; Topp, G.C. (eds.) \textit{Methods of Soils Analysis: Part 4 - Physical Methods}.  pp. 575-608, Soil Science Society of America, Madison Wisconsin, USA.   

\end{thebibliography}
\end{document}